\documentclass[a4paper,11pt]{article}
\pdfoutput=1
\usepackage{jheppub}

\usepackage{graphicx}
\usepackage[figuresright]{rotating}
\usepackage{bm,amsmath,amssymb}
\usepackage[mathscr]{eucal}
\usepackage{multirow}
\usepackage{xcolor}
\usepackage{mathrsfs}
\usepackage{mathtools}
\usepackage{slashed}
\usepackage{graphics}
\usepackage{graphicx}
\usepackage{subfigure}
\usepackage{dsfont}
\usepackage{longtable}
\usepackage{bbm} 
\usepackage{float}
\usepackage{tcolorbox}
\usepackage{lipsum}
\usepackage{cancel}
\usepackage{enumerate}
\definecolor{lcolor}{rgb}{0.,0.0,0.}
\definecolor{citcolor}{rgb}{0,0.,0.5}

\newcommand{\Mcal}{\mathcal{M}}
\newcommand{\Ncal}{\mathcal{N}}
\newcommand{\Tcal}{\mathcal{T}}
\newcommand{\Scal}{\mathcal{S}}
\newcommand{\Ccal}{\mathcal{C}}
\newcommand{\Pcal}{\mathcal{P}}
\newcommand{\Rcal}{\mathcal{R}}
\newcommand{\Ical}{\mathcal{I}}
\newcommand{\Jcal}{\mathcal{J}}
\newcommand{\vect}[1]{\boldsymbol{#1}_{\perp}}
\newcommand{\kt}{\vect{k}}
\newcommand{\kgt}{\boldsymbol{k_{g\perp}}}
\newcommand{\pt}{\vect{p}}
\newcommand{\Pt}{\vect{P}}

\newcommand{\qt}{\vect{q}}
\newcommand{\lt}{\vect{l}}
\newcommand{\vt}{\vect{v}}
\newcommand{\at}{\vect{a}}
\newcommand{\bt}{\vect{b}}
\newcommand{\Kt}{\vect{K}}
\newcommand{\ltone}{\boldsymbol{l_{1\perp}}}
\newcommand{\lttwo}{\boldsymbol{l_{2\perp}}}
\newcommand{\ltthre}{\boldsymbol{l_{3\perp}}}

\newcommand{\Lttwox}{\boldsymbol{L_{2x\perp}}}
\newcommand{\Lttwoy}{\boldsymbol{L_{2y\perp}}}
\newcommand{\Ltthrex}{\boldsymbol{L_{3x\perp}}}

\newcommand{\wt}{\vect{w}}
\newcommand{\wtbar}{\vect{\bar w}}
\newcommand{\xt}{\vect{x}}
\newcommand{\yt}{\vect{y}}
\newcommand{\zt}{\vect{z}}
\newcommand{\ut}{\vect{u}}

\newcommand{\rt}{\vect{r}}
\newcommand{\Rt}{\vect{R}}
\newcommand{\Lt}{\vect{L}}
\newcommand{\rtone}{\boldsymbol{r_{1\perp}}}
\newcommand{\rttwo}{\boldsymbol{r_{2\perp}}}
\newcommand{\rxyt}{\boldsymbol{r}_{xy}}

\newcommand{\rwyt}{\boldsymbol{r}_{wy}}

\newcommand{\rzxt}{\boldsymbol{r}_{zx}}
\newcommand{\rzyt}{\boldsymbol{r}_{zy}}
\newcommand{\rzytp}{\boldsymbol{r}_{z'y'}}

\newcommand{\rxytp}{\boldsymbol{r}_{x'y'}}
\newcommand{\rxxtp}{\boldsymbol{r}_{xx'}}
\newcommand{\ryytp}{\boldsymbol{r}_{yy'}}
\newcommand{\rxypt}{\boldsymbol{r}_{xy'}}
\newcommand{\rzypt}{\boldsymbol{r}_{zy'}}
\newcommand{\rzxpt}{\boldsymbol{r}_{zx'}}
\newcommand{\rzxtp}{\boldsymbol{r}_{z'x'}}
\newcommand{\rxpyt}{\boldsymbol{r}_{x'y}}
\newcommand{\rwytp}{\boldsymbol{r}_{w'y'}}
\newcommand{\et}{\vect{\epsilon}}
\newcommand{\Kcal}{\mathcal{K}}
\newcommand{\ptj}{\boldsymbol{p}_J}
\newcommand{\ptk}{\boldsymbol{p}_{K}}
\newcommand{\RtS}{\boldsymbol{R}_{\rm SE}}
\newcommand{\RtV}{\boldsymbol{R}_{\rm V}}
\newcommand{\RtR}{\boldsymbol{R}_{\rm R}}

\newcommand{\der}{\mathrm{d}}
\newcommand{\dijet}{2\textrm{jet}}
\newcommand{\Tr}{\mathrm{Tr}}

\title{Dijet impact factor in DIS at next-to-leading order in the Color Glass Condensate}
\author[a]{Paul Caucal,}
\emailAdd{pcaucal@bnl.gov}
\author[a,b,c]{Farid Salazar,}
\emailAdd{farid.salazarwong@stonybrook.edu}
\author[a]{Raju Venugopalan}
\emailAdd{raju@bnl.gov}

\affiliation[a]{Physics Department, Brookhaven National Laboratory, Upton, NY 11973, USA}
\affiliation[b]{Physics Department, Stony Brook University, Stony Brook, NY 11794, USA}
\affiliation[c]{Center for Frontiers in Nuclear Science (CFNS), Stony Brook University,
Stony Brook, NY 11794, USA}

\abstract{We compute the next-to-leading order impact factor for inclusive dijet production in deeply inelastic electron-nucleus scattering at small $x_{\rm Bj}$. Our computation, performed in the framework of the Color Glass Condensate effective field theory, includes all real and virtual contributions in the gluon shock wave background of all-twist lightlike Wilson line correlators. We demonstrate explicitly that the rapidity evolution of these correlators, to leading logarithmic accuracy, is described by the JIMWLK Hamiltonian. 
When combined with the  next-to-leading order JIMWLK Hamiltonian, our results for the impact factor improve the accuracy of the inclusive dijet cross-section to  $\mathcal{O}(\alpha_s^2\ln(x_f/x_{\rm Bj}))$, where $x_f$ is a rapidity factorization scale. These results are an essential ingredient in assessing the
discovery potential of inclusive dijets to uncover the physics of gluon saturation at the Electron-Ion Collider.
}
 
\begin{document}
\maketitle
\newpage 
\section{Introduction}

High energy deeply inelastic scattering experiments at HERA, in the kinematics of fixed large squared  momentum transfer $Q^2$, and small Bjorken $x_{\rm Bj}$, revealed the rapid proliferation of gluons that carry small momentum fractions $x$ inside the proton \cite{Abramowicz:1998ii}. At sufficiently small $x$ (or high energies $x \geq x_{\rm Bj}\sim Q^2/s$, for fixed $Q^2$ and large squared center-of-mass energies $s$), the nonlinear dynamics of quantum chromodynamics (QCD) leads to the screening and recombination of  gluons. These emergent many-body effects can tame the growth
in the corresponding gluon distribution function~\cite{Gribov:1984tu,Mueller:1985wy}, a phenomenon known as gluon saturation. Its discovery  and characterization is one of the primary goals of the future Electron-Ion Collider (EIC) \cite{Accardi:2012qut,Aschenauer:2017jsk,AbdulKhalek:2021gbh}. In this saturation regime, gluons  attain large occupation numbers $\sim 1/ \alpha_s$, for which the  appropriate description is in terms of strong classical fields~\cite{McLerran:1993ni,McLerran:1993ka,McLerran:1994vd}. The Color Glass Condensate (CGC) is an effective field theory (EFT) describing the properties of these overoccupied small-$x$ gluons   and it has been employed to study numerous observables in electron-nucleus, proton-nucleus and nucleus-nucleus collisions~\cite{Iancu:2003xm,Gelis:2010nm,Kovchegov:2012mbw,Albacete:2014fwa,Blaizot:2016qgz,Morreale:2021pnn}.

In the CGC EFT, the high energy scattering of color charged particles off the small-$x$ gluon fields is encoded in effective vertices that resum 
the multiple scatterings of these colored charges off this background field and are expressed in terms of lightlike Wilson lines. Analogously to  the operator product expansion, physical observables can be written as convolutions of perturbatively calculable (process dependent) impact factors with that of correlators of these lightlike Wilson lines. Their $n$-point correlators obey a set of coupled nonlinear renormalization group equations, the B-JIMWLK equations. The leading order kernel of these evolution equations  resums $\alpha_s^n \ln^n(1/x)$  contributions, from each order in perturbation theory, to cross-sections to leading logarithmic (LL) accuracy \cite{Balitsky:1995ub,JalilianMarian:1996xn,JalilianMarian:1997dw,Kovner:2000pt,Iancu:2000hn,Iancu:2001ad,Ferreiro:2001qy}. Likewise, the NLO kernel resums $\alpha_s^{n+1}\ln^n(1/x)$ contributions at  next-to-leading logarithmic (NLL) accuracy \cite{Balitsky:2008zza,Kovchegov:2006vj,Balitsky:2013fea,Kovner:2013ona,Kovner:2014lca,Lublinsky:2016meo}. The lowest two-point ``dipole" correlator in this hierarchy, for large $N_c$ and atomic mass number $A\gg 1$, satisfies the  Balitsky-Kovchegov (BK) equation describing the evolution of the fully inclusive DIS cross-section in the high energy limit~\cite{Balitsky:1995ub,Kovchegov:1999yj}. The evolution kernel of this equation has recently been computed to NNLO in the planar limit of ${\cal N}=4$ super Yang-Mills theory~\cite{Caron-Huot:2016tzz}, with results that could be applied towards eventual full QCD computations at this order.

For precision computations of physical processes, one also needs high order computations of the corresponding impact factors, commensurate with the increasing accuracy of the evolution equations. Significant progress has been made in this direction for a variety of processes at NLO in the CGC \cite{Balitsky:2012bs,Beuf:2011xd,Beuf:2016wdz,Beuf:2017bpd,Hanninen:2017ddy,Ducloue:2017ftk,Beuf:2021qqa,Boussarie:2016bkq,Boussarie:2016ogo,Boussarie:2019ero,Mantysaari:2021ryb,Chirilli:2011km,Chirilli:2012jd,Altinoluk:2014eka,Stasto:2013cha,Roy:2019cux,Roy:2019hwr,Iancu:2020mos}. As an example of work in this direction, the inclusive DIS cross-section has been computed recently by combining the corresponding full NLO impact factor with 
the dominant subset~\cite{Iancu:2015joa} of
NLO contributions to the BK kernel and compared to HERA data \cite{Beuf:2020dxl}.

A notable absence amongst the existing computations is the NLO impact factor for inclusive dijet/dihadron production in deep-inelastic scattering (DIS) at small $x_{\rm Bj}$. This DIS process off the proton, and in large nuclei, is of great phenomenological interest at the future EIC.  The  computation of the impact factor to NLO accuracy can provide  novel information on the effects of gluon saturation in  back-to-back 
dihadron/dijets \cite{Zheng:2014vka}, the study of the Weizsäcker-Williams gluon distribution \cite{Dumitru:2015gaa,Dumitru:2018kuw} and in the extraction of a fundamental building block of high energy QCD - the quadrupole correlator of Wilson lines at small $x$ \cite{Mantysaari:2019hkq,Boussarie:2021lkb}.

In this paper, we compute the inclusive production of dijets in electron-nucleus collisions  at NLO within the CGC EFT. We will follow the strategy for the NLO computation of inclusive photons+dijets in \cite{Roy:2019cux,Roy:2019hwr} by performing our computation using covariant perturbation theory following momentum space Feynman rules with CGC effective vertices that represent the many-body dynamics inherent in the gluon shock wave\footnote{For earlier applications of this particular momentum space approach in the context of proton-nucleus collisions, we refer the reader to \cite{Blaizot:2004wv,Benic:2016uku}.}.
Our results for the inclusive DIS cross-section are of $\mathcal{O}(\alpha_s^2 \ln(x_f/x_{\rm Bj}))$ accuracy if the NLO impact factor result is combined with NLL BK/JIMWLK evolution equations. Here $x_f$ represents a scale that separates contributions to the impact factor from that of the rapidity evolution of the target nucleus at high energies and plays a role analogous to the factorization scale in collinear factorization  computations. 

Since the process we consider is simpler than the inclusive photon+dijet computation, we will obtain results that are significantly more tractable analytically. In particular, we are able to work out the Dirac algebra and the internal momentum integration of each contribution. We will show that all divergences (soft, collinear and ultraviolet) cancel at one loop 
order and shall demonstrate JIMWLK factorization of our result when the real or virtual gluon is ``slow", namely, when it carries a small longitudinal momentum fraction relative to the virtual photon. This allows us to isolate and obtain explicit expressions for the NLO impact factor, which will be computed numerically in the future to make concrete predictions for experiments at the EIC.

Another motivation for our work, besides its strong phenomenological relevance, is to explore the power and efficiency of CGC computations using the techniques that we have developed that employ covariant perturbation theory in contrast to lightcone perturbation theory (LCPT) employed by many of the NLO computations in the literature. While at NLO order the computations are of comparable complexity, the situation will likely be different at NNLO~\cite{Bassetto:1991ue}. 

The paper is organized as follows. For the discussion to be self-contained, we review in  Sec.\,\ref{sec:LO_review} the basic elements of the CGC EFT and employ them in the computation of inclusive dijet production in DIS at leading order. In Sec.\,\ref{sec:gen-strat}, we write down the Feynman diagrams for real and virtual (self energy and vertex) contributions at NLO  and outline the general strategy for the computation of these diagrams. In addition, we briefly point out the connection to lightcone perturbation theory.  We  proceed to real gluon emission contributions in Sec.\,\ref{sec:real}, where we compute the triple parton production $q\bar{q}$ + $g$ amplitude. Virtual gluon contributions are discussed in Sec.\,\ref{sec:virtual}, where  we first consider the self energy contributions followed by vertex corrections. We show how ultraviolet divergences associated with these contributions cancel. An infrared divergence survives the sum of all virtual contributions, which will cancel with those in real emissions after we introduce appropriate jet 
functions, leading to an infrared and collinear safe inclusive dijet cross-section. We extract the slow gluon divergence of all the real and virtual contributions in Sec.\,\ref{sec:slow_gluon}, and show that the net  result satisfies JIMWLK factorization. In Sec.\,\ref{sec:small_cone}, we implement the small-cone algorithm to show explicitly that our final result, as noted, is IR and collinear safe. We present a compact final expression for the inclusive dijet NLO impact factor  of longitudinally polarized virtual photons in Sec.\,\ref{sub:final}. (The more elaborate expressions for transversely polarized photons are presented in appendix~\ref{app:final-res-trans}.) 
In Sec.\,\ref{sec:NLO_conclusion}, we conclude with a summary and outlook. 

The paper is supplemented by appendices which are useful to the reader interested in the details of the computation.  Appendix~\ref{app:convention} summarizes our conventions and the Feynman rules in the CGC effective field theory. As noted, appendix~\ref{app:final-res-trans} contains the NLO impact factor for transversely polarized photons. In appendix~\ref{app:dirac}, we provide useful formulas for the calculation of the Dirac algebra in the Feynman amplitudes. Appendix~\ref{app:contour} presents  examples of the computation of relevant contour integrals and appendix~\ref{app:transverse-int} provides all the necessary transverse momentum integrals in our calculation. In appendix~\ref{app:jdotjcross}, we study analytically the two transverse momentum integrals that appear in the free vertex correction after the shock wave. Finally, appendices~\ref{app:R1}, \ref{app:SE1}, \ref{app:V1} and \ref{appp:V3} provide details respectively of the calculation of the diagrams labeled $\rm R2$, $\rm SE2$, $\rm V2$ and $\rm V3$.

\section{Review of the leading order dijet cross-section}

\label{sec:LO_review}

In this section, we will review the general formalism of the CGC effective field theory and outline the computation of the leading order dijet cross-section in this framework, already computed in \cite{Dominguez:2011wm}. 

\subsection{The CGC effective field theory}

The Color Glass Condensate is an effective field theory that describes the Regge limit of QCD. It is formulated in terms of stochastic color sources $\rho^a_A$ which represent the large $x$ degrees of freedom inside the target $A$ (a proton or a large nucleus) and a classical gauge field $A_{\rm cl}^\mu$ created by these sources, which represents the small-$x$ gluons that carry high occupancy number. Sources and fields are related by the Yang-Mills equations $[D_\mu,F^{\mu\nu}]=J^\nu$ where $J^\mu$ is the 4-current associated with the large $x$ sources. For a fast moving target along the $+$ lightcone direction, this current is independent of $x^+$:
\begin{equation}
\label{eq:CGC-current}
    J^\mu(x^-,\xt)=\delta^{\mu+}\rho_A(x^-,\xt) \,.
\end{equation}
The solution of the Yang-Mills equations in Lorenz gauge $\partial_\mu A^\mu_{\rm cl}=0$ is
\begin{align}
    A^+_{\rm cl}(x)=\alpha(x^-,\xt),\quad A_{\rm cl}^-=0,\quad A_{\rm cl}^i=0\label{eq:dijet-LO-Acl-lorentzgauge} \,,
\end{align}
with $\alpha(x^-,\xt)$ a solution of the Poisson equation $\nabla_\perp^2\alpha=-\rho_A$.
For the present calculation, it is  convenient to work in the ``wrong" lightcone gauge $A^-_{\rm cl}=0$ \cite{Gelis:2005pt}. As shown in Eq.\,\eqref{eq:dijet-LO-Acl-lorentzgauge}, the classical solution in this gauge is identical to the solution in Lorenz gauge.  Even though the wrong lightcone gauge, in contrast to the lightcone gauge $A_{\rm cl}^+=0$, does not provide a simple partonic interpretation of the target wavefunction, it simplifies tremendously the form of the propagators inside the background field for a fast moving projectile in the ``minus" lightcone direction. These propagators were computed previously in \cite{McLerran:1994vd,Balitsky:1995ub,Ayala:1995hx,Balitsky:2001mr,McLerran:1998nk,Ayala:1995kg}; the corresponding 
effective quark and gluon propagators are identical~\cite{Hentschinski:2018rrf,Bondarenko:2021rbp} to the quark-quark-reggeon and gluon-gluon-reggeon propagators in Lipatov's reggeon effective field theory~\cite{Bondarenko:2017ern,Hentschinski:2020rfx}. 
\begin{center}
\begin{figure}[H]
    \centering
    \includegraphics[width=0.8\textwidth]{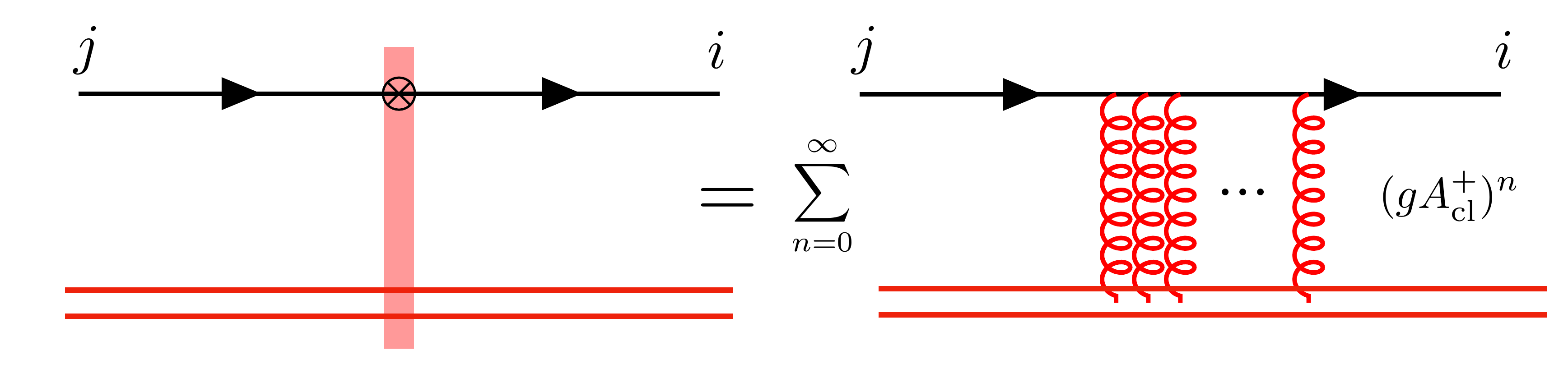}
    \caption{The effective vertex of the dressed quark propagator on the l.h.s of the figure represents the multiple scattering off the classical gauge fields 
    $A_{\rm cl}^+ = O(1/g)$ of the nucleus shown on the r.h.s. The $n=0$ term represents the free quark propagator.}
    \label{fig:fig1}
\end{figure}
\end{center}
The dressed eikonal propagator of a quark in the classical background field of the nuclear target (depicted in Fig.\,\ref{fig:fig1}) is given by\,\cite{McLerran:1994vd,Ayala:1995hx,Ayala:1995kg,Balitsky:1995ub,Balitsky:2001mr}
\begin{equation}
S_{ij}(l,l') =  S^0(l)_{ik}\,\mathcal{T}^{q}_{kl}(l,l')\,S^0_{lj} (l')
\,,
\label{eq:dressed-quark-mom-prop2}
\end{equation}
where $i,j$ are the color indices for the outgoing and incoming quark which respectively carry momenta $l$ and $l'$, and $S^0_{ij}(l)$ is the quark free propagator. The effective quark-gluon vertex represented by a cross is given by 
\begin{align}
    \mathcal{T}^q_{ij}(l,l') =  (2\pi) \delta(l^--l'^-) \gamma^- \mathrm{sgn}(l^-) \int \der^2\xt e^{-i(\lt-\lt')\cdot \xt} V_{ij}^{\mathrm{sgn}(l^-)}(\xt)\label{eq:dijet-NLO-cgc-quark-propagator} \,,
\end{align}
where the Wilson line in the fundamental representation is given by the following path ordered exponential along the lightcone time of the projectile $x^-$:
\begin{align}
    V_{ij}(\xt) = \mathcal{P} \ \exp \left ( ig \int_{-\infty}^\infty \der z^- A^{+,a}_{\mathrm{cl}}  (z^-,\xt) t^a_{ij}  \right ) \,,
\end{align}
where $t^a$ are the generators of SU(3) in the fundamental representation, and the
superscript $\mathrm{sgn}(l^-)$ in Eq.\,\eqref{eq:dijet-NLO-cgc-quark-propagator} denotes whether the color matrix or its inverse (Hermitian conjugate). As in the free propagator, the dressed propagator for the antiquark is obtained by following the fermion line.

The Wilson line resums to all orders multiple interactions between the projectile and the small-$x$ gluons in the target, and ensures the unitarization of the cross-section in the high-energy limit. This dressed propagator, and the corresponding dressed gluon propagator, are represented in the standard momentum space Feynman rules with the effective CGC vertex, symbolized respectively by a cross or a dot in diagrams. These rules are summarized in Appendix~\ref{app:convention}. 

In the CGC effective field theory, a path integral for any observable ${\mathcal O}$ at small $x$ is first computed for the charge configuration $\rho_A$ of larger $x$ sources (drawn from a stochastic distribution $W_Y[\rho_A]$ of such sources) that is static on the dynamical time scales of the small $x$ gauge fields: 
\begin{equation}
    \langle\mathcal{O}[\rho_A]\rangle_{Y}=\int D\rho_A\, W_Y[\rho_A]\,\mathcal{O}[\rho_A]\,.
    \label{eq:CGC-expectation value}
\end{equation}
The expression on the r.h.s for $ {\mathcal O}[\rho_A]$ implicitly contains the QCD path integral in the presence of these sources. In the gluon saturation regime, the path integral is dominated by the classical ``shock wave" configurations $A_{\rm cl}(\rho_A)\sim 1/\sqrt{\alpha_s}$ with lightcone momenta $k^+\ll P^+$ (where $P^+\rightarrow \infty$ is the lightcone momentum of the nucleus) or small $x= \Lambda^+/P^+\ll 1$. As we noted in Eq.~(\ref{eq:dijet-LO-Acl-lorentzgauge}), they are  determined by solving the Yang-Mills equations for the eikonal sources $J^\mu = \delta^{\mu+}\rho_A(x_\perp)\delta(x^-)$ in Eq.~(\ref{eq:CGC-current}) with  $k^+ > \Lambda^+$ that are localized at rapidities above $Y=\ln(\Lambda^+/P^+)$. 

Quantum corrections to the CGC shock wave classical fields, specifically the small fluctuations propagator, are computed in the shock wave background and are seen to diverge in rapidity; they are however small as long as the window in rapidity is small, given by  $\alpha_s \ln({\Lambda^\prime}^+/\Lambda^+)\ll 1$. These can be absorbed in the charge configuration $\rho_A\rightarrow \rho_A^\prime$ at the new scale ${\Lambda^\prime}^+$ and the process iterated through a self-similar Wilsonian renormalization group (RG) procedure, as the scale ${\Lambda^\prime}^+$ (or equivalently, the corresponding rapidity) is varied. In particular, such quantum corrections to the operator ${\mathcal O}$ can, by an integration by parts in Eq.~(\ref{eq:CGC-expectation value}), be expressed as the RG evolution of $W_Y[\rho_A]$ with the change in the rapidity scale that separates sources from fields. This RG equation, to LL accuracy is precisely the JIMWLK equation which generates the Balitsky-JIMWLK hierarchy for the $n$-point Wilson line correlators; sub-leading quantum corrections $\alpha_s^2 \ln({\Lambda^\prime}^+/\Lambda^+)$ to $\langle {\mathcal O}\rangle$ can likewise be reexpressed in terms of the NLO JIMWLK equation\footnote{We refer the reader to \cite{Iancu:2016vyg,Ducloue:2017ftk,Ducloue:2019ezk,Liu:2020mpy} for discussions of the choice of the evolution rapidity variable $Y$ in NLO calculations.}.

\subsection{Outline of the LO computation}
\label{sub:LO}
\begin{center}
    \begin{figure}[H]
    \centering
    \includegraphics[width=0.55\textwidth]{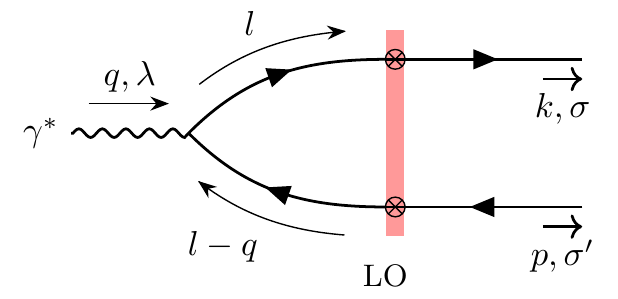}
    \caption{Leading order contribution to the amplitude for dijet production. The cross symbol on the quark and antiquark legs refers to the CGC quark effective vertex, as represented in Fig.~\ref{fig:fig1} and defined in  Eq.~\eqref{eq:dijet-NLO-cgc-quark-propagator}.}
    \label{fig:feyLO}
    \end{figure}
\end{center}
We will work in the dipole frame, where the virtual photon exchanged between the electron and the target is left moving with a large ``minus" lightcone component $q^->0$ of its four momentum and a vanishing transverse momentum $\qt = \vect{0}$. We denote the longitudinal polarization as $\lambda =0$, and the two transverse polarizations as $\lambda = \pm 1$. The helicities of the quark and antiquark are denoted $\sigma$ and $\sigma'$ respectively. We shall also define $z_q=k^-/q^-$ and $z_{\bar q}=p^-/q^-$, the ``minus" lightcone momentum fraction of the quark and the antiquark relative to the virtual photon respectively. We neglect the masses of the quark, the antiquark and the electron throughout the calculation. The notation for the kinematic variables used in this paper are summarized table~\ref{tab:kinematics_dijets}.
\begin{table}[h]
\centering
\caption{ Kinematic variables} 
\label{tab:kinematics_dijets}
\begin{tabular}{ll}
\hline \hline
$P_A $ & nucleus four-momentum \\
$P_n $ & nucleon four-momentum \\
$k_e \ (k_e')$ & incoming (outgoing) electron four-momentum \\ 
$q=k_e-k_e'$ & virtual photon four-momentum \\
$k,p$ & quark (antiquark) four-momentum\\ 
$z_{q}, z_{\bar{q}}$  & quark (antiquark) longitudinal momentum fraction relative to $q^-$ \\
$\eta_{q}, \eta_{\bar{q}}$ & quark (antiquark) rapidity \\
$\kt, \pt$ &  quark (antiquark) transverse momentum \\
$s=(P_n+k_e)^2$ & nucleon-electron system center of momentum energy squared  \\
$W^2=(P_n+q)^2$ & nucleon-virtual photon system center of momentum energy squared\\
$m^2_n = P_n^2$ & nucleon invariant mass squared\\
$Q^2=-q^2$ & virtuality squared of the exchanged photon \\
\hline \hline
\end{tabular}
\end{table}

Using standard momentum space Feynman rules, together with the effective vertices in the presence of the classical background field detailed in the previous subsection, one can easily write the scattering amplitude\footnote{We have not included explicitly the internal color indices of propagators and outgoing particles, and the momenta of the external particle $(q,k,p)$ on the l.h.s. of Eq.\,\eqref{eq:dijetNLO-Scal-LO}.} of the LO diagram shown Fig.~\ref{fig:feyLO}:
\begin{align}
    \Scal^{\lambda\sigma\sigma'}_{\mathrm{LO}} &= \!\! \int\!\! \frac{\der^4 l}{(2\pi)^4}\bar{u}(k,\sigma) \Tcal^q(k,l) S^0(l) \left(-i e e_f \slashed{\epsilon}(q,\lambda) \right) S^0(l-q)\Tcal^q(l-q,-p) v(p,\sigma')\label{eq:dijetNLO-Scal-LO} \,.
\end{align}
The effective CGC vertices include all possible scatterings of the quark or antiquark off the target, including the possibility of no-scattering which has to be subtracted to obtain the physical amplitude. As shown in \cite{Roy:2018jxq}, this can be done systematically by subtracting from  Eq.~\eqref{eq:dijetNLO-Scal-LO} a term in which all the Wilson lines inside the effective CGC vertices are set to unity. Factorizing  further  the overall delta function from ``minus" lightcone momentum conservation\footnote{In mathematical terms, $(2\pi) \delta(k^- + p^- - q^-) \Mcal_{\mathrm{LO}} = \mathcal{S}_{\mathrm{LO}} - \mathcal{S}_{\mathrm{LO}}[A_{\mathrm{cl}}=0]$. An overall delta function in the ``minus" lightcone momentum is always present as a consequence of the eikonal interactions of the effective CGC propagators.}, the reduced amplitude $\Mcal^{\lambda\sigma\sigma'}_{\rm LO}$ is given by 
\begin{align}
    \Mcal^{\lambda\sigma\sigma'}_{\mathrm{LO}} = \frac{ee_f q^-}{\pi} \int \der^2 \xt \der^2 \yt e^{-i \kt \cdot \xt } e^{-i \pt \cdot \yt } \Ccal_{\mathrm{LO}}(\xt,\yt)  \mathcal{N}_{\mathrm{LO}}^{\lambda\sigma\sigma'}(\rxyt) \,,
\end{align}
where we introduce the leading order color structure,
\begin{align}
    \Ccal_{\mathrm{LO}}(\xt,\yt) = V(\xt) V^\dagger(\yt) - \mathbbm{1}  \,,
\end{align}
and the leading order perturbative factor
\begin{align}
    \mathcal{N}_{\mathrm{LO}}^{\lambda\sigma\sigma'} (\rxyt) &= - i (2q^-) \int \frac{\der^4 l}{(2\pi)^2} e^{i \lt \cdot \rxyt} \frac{ N_{\mathrm{LO}}^{\lambda\sigma\sigma'}(l)   \delta(k^- - l^-) }{(l^2 + i \epsilon)((l-q)^2+i \epsilon)}\label{eq:dijets-LO-pert} \,.
\end{align}
Throughout this paper, we will represent the difference between two transverse coordinates with the notation $\boldsymbol{r}_{xy}=\vect{x}-\vect{y}$ and its magnitude as $r_{xy}=|\vect{x}-\vect{y}|$.

When computing the NLO diagrams, we will also decompose the amplitude in terms of its color structure and a similar perturbative factor. In the latter, the numerator $N_{\rm LO}$  contracts spinors and Dirac matrices, 
\begin{align}
    N_{\mathrm{LO}}^{\lambda\sigma\sigma'}(l)=[\bar{u}(k,\sigma)\mathcal{D}_{\rm LO}^{\lambda}(l)v(p,\sigma')]\label{eq:dijet-LO-DiracLO}
    =\frac{1}{(2q^-)^2} \left[ \bar{u}(k,\sigma) \gamma^- \slashed{l} \slashed{\epsilon}(q,\lambda) (\slashed{l}-\slashed{q})\gamma^- v(p,\sigma') \right].
\end{align}
For a longitudinally polarized virtual photon ($\slashed{\epsilon}(q,\lambda=0) = \frac{Q}{q^-}\gamma^-$) we find,
\begin{align}
    \mathcal{D}_{\rm LO}^{\lambda = 0 }(l) &= -Q\  \frac{l^-}{q^-}\left( 1- \frac{l^-}{q^-}\right) \frac{\gamma^-}{q^-}\,.
\end{align}
Likewise, in the transversely polarized case ($\slashed{\epsilon}(q,\lambda=\pm 1) = -\gamma^i \et^{\lambda,i}$) we have
\begin{align}
    \mathcal{D}_{\rm LO}^{\lambda = \pm 1 }(l) &= \frac{\et^{\lambda} \cdot \lt }{2} \left[\left( 1- \frac{2l^-}{q^-} \right) - \lambda \Omega  \right] \frac{\gamma^-}{q^-} \,,
\end{align}
where $\Omega=\frac{i}{2}[\gamma^1,\gamma^2]$.

The subsequent computation of $\Ncal^{\lambda\sigma\sigma'}_{\rm LO}$ is straightforward. We use Cauchy's theorem to perform the $l^+$ contour integration while the remaining $\lt$ integral is expressed in terms of modified Bessel functions of the second kind $K_i(z)$. For a longitudinally polarized virtual photon, one obtains 
\begin{align}
     \mathcal{N}_{\mathrm{LO}}^{\lambda = 0,\sigma \sigma'} (\rt) 
     &=-z_qz_{\bar q}QK_0(\bar Q r_{xy})\frac{[\bar{u}(k,\sigma)\gamma^- v(p,\sigma')]}{q^-} \\
     &
     = -2(z_q z_{\bar{q}})^{3/2}  Q K_0(\bar{Q} r_{xy})\delta^{\sigma,-\sigma'}\,,
\end{align}
with $\bar{Q}^2=z_qz_{\bar q}Q^2$. For the transverse polarization case, the leading order perturbative factor reads,
\begin{align}
    \Ncal^{\lambda=\pm1,\sigma \sigma'}_{\rm LO}(\rxyt)&=\frac{i\bar Q\et^{\lambda} \cdot \rxyt}{2r_{xy}}K_1(\bar Q r_{xy})\left\{\bar{u}(k,\sigma)\left[ (z_{\bar q}-z_q) -\lambda \Omega \right] \frac{\gamma^-}{q^-} v(p,\sigma')\right\}\\
    &=2 (z_qz_{\bar q})^{1/2} \frac{i\bar{Q}\et^{\lambda}\cdot\rxyt}{r_{xy}}K_1(\bar{Q}r_{xy})\Gamma_{\gamma^*_\mathrm{T} \to q\bar{q}}^{\sigma,\lambda}(z_q,z_{\bar q})\delta^{\sigma,-\sigma'}\,,
\end{align}
where $\Gamma_{\gamma^*_{\rm T} \to q\bar q}$ is the spin-helicity dependent splitting vertex defined as
\begin{equation}
    \Gamma_{\gamma^*_\mathrm{T} \to q\bar{q}}^{\sigma,\lambda}(z_1,z_2)=z_2\delta^{\sigma,\lambda}-z_1\delta^{\sigma,-\lambda}\,.
    \label{eq:dijet-NLO-spinhel-splitting}
\end{equation}
A more detailed discussion is provided in appendix~\ref{app:dirac}.

The next step is to compute the differential cross-section for the production of a  $q\bar{q}$ pair in the collision of a virtual photon $\gamma^*_{\lambda}$ with a nucleus $A$, given by
\begin{align}
    \left.\frac{\der \sigma^{\gamma_{\lambda}^*+A\to q\bar{q}+X}}{ \der^2 \kt \der^2 \pt \der \eta_q \der \eta_{\bar{q}}}\right|_{\rm LO}  =  \frac{1}{4 (2\pi)^6} \frac{1}{2q^-} (2\pi) \delta(k^- + p^- - q^-) \!\!\!\!\! \sum_{\rm \sigma \sigma', colors} \!\!\!\!\! \left \langle \mathcal{M}^{\lambda \sigma \sigma'\dagger}_{\mathrm{LO}}[\rho_A] \mathcal{M}^{\lambda\sigma \sigma'}_{\mathrm{LO}} [\rho_A] \right \rangle_{Y}\,,
    \label{eq:dijet-cross-section-LO}
\end{align}
where the sum over colors amounts to taking the trace over the product of Wilson lines (at cross-section level). The rapidities\footnote{Note that in the massless limit, the rapidity and pseudorapidity variables are identical.} of the quark and antiquark jets are given by
\begin{align}
    \eta_q = \ln\left(\sqrt{2} z_q q^-/k_\perp \right) \,, \quad \eta_{\bar{q}} = \ln\left(\sqrt{2} z_{\bar{q}} q^-/p_\perp \right)\,,
\end{align}
where they are defined such that positive rapidities correspond to 
particles propagating in the virtual photon-going direction.

When taking the squared modulus of the amplitude, one has to take care of the square of the $\delta$-function for ``minus" lightcone momentum conservation by constructing a properly normalized wave packet for the incoming virtual photon \cite{Gelis:2002ki}. The $\langle ...\rangle_{Y}$ notation in the Eq.\,\eqref{eq:dijet-cross-section-LO} stands for the CGC averaging over all possible charge configurations inside the target at rapidity scale $Y$.
Introducing the following measure with Fourier phases,
\begin{align}
    \der \Pi_{\mathrm{LO}} = \der^2 \xt \der^2 \yt \der^2 \xt' \der^2 \yt' e^{-i \kt \cdot (\xt -\xt')} e^{-i \pt \cdot (\yt - \yt') }\,,
    \label{eq:dijet-NLO-LOdiffmeasure}
\end{align}
the final result for the differential cross-section can be expressed 
as\footnote{In this manuscript, we focus on the ``diagonal" terms where the polarization $\lambda$ is the same in the amplitude and in the conjugate amplitude. The off-diagonal terms are important when considering correlations with the electron plane~\cite{Mantysaari:2020lhf}.}
\begin{align}
    \left.\frac{\der \sigma^{\gamma_{\lambda}^*+A\to q\bar{q}+X}}{ \der^2 \kt \der^2 \pt \der \eta_q \der \eta_{\bar{q}}}\right|_{\rm LO}  \!\!\!\!\!\!\! =  \frac{\alpha_{\mathrm{em}} e_f^2 N_c}{(2\pi)^6}  \delta(1-z_q- z_{\bar{q}})  \int \der \Pi_{\mathrm{LO}}  \Xi_{\mathrm{LO}}(\xt,\yt;\xt',\yt')  \Rcal_{\mathrm{LO}}^{\lambda}(\rxyt,\rxytp)\label{eq:dijet-LO-cross-section} \,.
\end{align}
The final factor in this expression is the sum over the quark and antiquark helicities of the square of the perturbative factor, defined as 
\begin{align}
    \Rcal_{\mathrm{LO}}^{\lambda}(\rxyt,\rxytp)=\sum_{\sigma\sigma'} \Ncal_{\rm LO}^{\lambda\sigma \sigma' \dagger}(\rxyt)\Ncal_{\rm LO}^{\lambda \sigma \sigma'}(\rxytp) \,.
\end{align}
We find for longitudinally and transversely polarized photons:
\begin{align}
    \Rcal_{\mathrm{LO}}^{\mathrm{L}}(\rxyt,\rxytp) &=  8 z_q^3 z_{\bar{q}}^3  Q^2 K_0(\bar{Q} r_{xy}) K_0(\bar{Q} r_{x'y'}) \,, \\
    \Rcal_{\mathrm{LO}}^{\mathrm{T}}(\rxyt,\rxytp) &=  2 z_q z_{\bar{q}} \left[z_q^2 + z_{\bar{q}}^2 \right]  \frac{\rxyt \cdot \rxytp}{r_{xy} r_{x'y'}}  \bar{Q}^2K_1(\bar{Q} r_{xy}) K_1(\bar{Q}r_{x'y'})\label{eq:dijet-NLO-TLO} \,,
\end{align}
where for the transversely polarized photon, we average over both polarizations $\lambda=\pm 1$.

The dynamics of strongly correlated gluons inside the target is encoded in the nonperturbative expression $\Xi_{\rm LO}$, defined in terms of Wilson line correlators as 
\begin{align}
    \Xi_{\mathrm{LO}}(\xt,\yt;\xt', \yt')  =\left \langle Q(\xt,\yt;\yt', \xt') - D(\xt, \yt) -  D(\yt', \xt') + 1 \right \rangle_{Y} \,,
\end{align}
where the dipole $D$ and quadrupole $Q$ operators are defined as
\begin{align}
    &D_{xy}=D(\xt,\yt)=\frac{1}{N_c}\Tr\left( V(\xt)V^\dagger(\yt)\right) \,,\\
    &Q_{xy,y'x'}=Q(\xt,\yt;\yt',\xt')=\frac{1}{N_c}\Tr \left(V(\xt)V^\dagger(\yt)V(\yt')V^\dagger(\xt')\right) \,.
\end{align}
The LO expression in Eq.~\eqref{eq:dijet-LO-cross-section} for inclusive dijet production was first derived in \cite{Dominguez:2011wm}. 

Thus far, we have focused on the hadronic part of the dijet cross-section in DIS. For completeness, we will now explain  how the DIS cross-section can be obtained from the subprocess $\gamma^{*}_{\lambda}+A\to q\bar q +X$.  The leptonic part is encoded in the longitudinal and transverse photon fluxes defined as
\begin{align}
    f_{\lambda=\mathrm{L}}(x_{\rm Bj}, Q^2)&=\frac{\alpha_{\mathrm{em}}}{\pi Q^2 x_{\rm Bj}}(1-y)\,,\\
    f_{\lambda=\mathrm{T}}(x_{\rm Bj}, Q^2)&=\frac{\alpha_{\mathrm{em}}}{2\pi Q^2 x_{\rm Bj}}[1+(1-y)^2]\,,
\end{align}
where $y=Q^2/(s\, x_{\rm Bj})$ denotes the inelasticity, and $s$ is the center of mass energy squared of the collision. For fixed $s$, the final expression for the LO  $e+A\to e'+q\bar q+X$ cross-section is given by
\begin{equation}
    \frac{\der\sigma^{e+A\to e'+q\bar q+X}}{\der x_{\rm Bj} \der Q^2\der^2\kt\der^2\pt\der\eta_q\der\eta_{\bar q}}=\sum_{\lambda=\mathrm{L,T}}f_{\lambda}(x_{\rm Bj}, Q^2) \ \frac{\der \sigma^{\gamma_{\lambda}^*+A\to q\bar{q}+X}}{ \der^2 \kt \der^2 \pt \der \eta_q \der \eta_{\bar{q}}} \,.
\end{equation}
At NLO in the strong coupling constant, an identical convolution occurs between the lepton and hadron tensors, with the former remaining unchanged. We will therefore focus on the latter in the following sections.

\section{General strategy for the NLO computation}
\label{sec:gen-strat}
In this section, we present the general strategy that we will follow for the calculation of the Feynman diagrams that contribute to the dijet NLO impact factor. Our discussion is meant to guide the reader through  the computations detailed in the next sections and to highlight some of their  general features.

\begin{center}
\begin{figure}[H]
    \centering
    \includegraphics[width=0.8\textwidth]{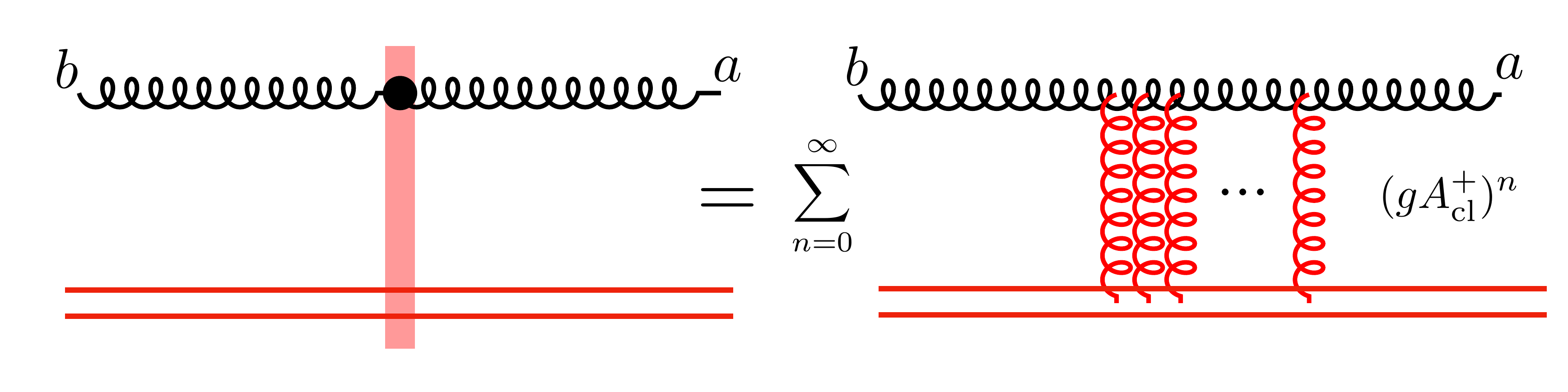}
    \caption{The effective vertex of the dressed gluon propagator on the l.h.s of the figure represents multiple scattering off the classical gauge fields 
    $A_{\rm cl}^+ = \mathcal{O}(1/g)$ of the nucleus shown on the r.h.s. The $n=0$ term represents the free gluon propagator.}
    \label{fig:dressed-gluon}
\end{figure}
\end{center}

At NLO in the inclusive dijet computation, one has both real and virtual gluon emission. 
The key ingredient is the dressed gluon propagator (shown in Fig.~\ref{fig:dressed-gluon}), which in the wrong lightcone gauge has a structure very similar to the dressed quark propagator. It is given by  ~\cite{McLerran:1994vd,Ayala:1995hx,Ayala:1995kg,Balitsky:1995ub,McLerran:1998nk,Balitsky:2001mr},
\begin{equation}
G_{\mu \nu;ab}(l,l') =  G^{0}_{\mu \rho;ac}(l)\,\mathcal{T}^{g;\rho \sigma,cd}(l,l')\,G^{0}_{\sigma \nu;db} (l') \, ,
\label{eq:dressed-gluon-mom-prop}
\end{equation}
where $\mu , \nu $ and $a,b$ are the Lorentz and adjoint color indices for the outgoing and incoming gluon which respectively carry momenta $l$ and $l'$, and $G^{0}_{\mu\nu;ab}(l)$ is the gluon free propagator.
The effective vertex, represented by a filled circle, is given by 
\begin{align}
    \mathcal{T}^g_{\mu\nu,ab}(l,l') &=  -(2\pi) \delta(l^--l'^-) (2l^-)\,  g_{\mu\nu}\, \mathrm{sgn}(l^-) \int \der^2\vect{z} e^{-i(\lt-\lt')\cdot \vect{z}} U_{ab}^{\mathrm{sgn}(l^-)}(\vect{z})\,,
    \label{eq:dijet-NLO-cgc-gluon-propagator}
\end{align}
where the Wilson line $U(\zt)$ lives in the \textit{adjoint} representation of $\textrm{SU}(3)$.
\begin{align}
    U_{ab}(\xt) = \mathcal{P} \ \exp \left ( ig \int_{-\infty}^\infty \der z^- A^{+,c}_{\mathrm{cl}}  (z^-,\xt) T^c_{ab}  \right ) \,,
\end{align}
where $T^a$ are generators of $\textrm{SU}(3)$ in the adjoint representation. This effective vertex also encodes multiple scattering effects to all orders.

\begin{figure}[tbh]
\centering
\includegraphics[width=1\textwidth]{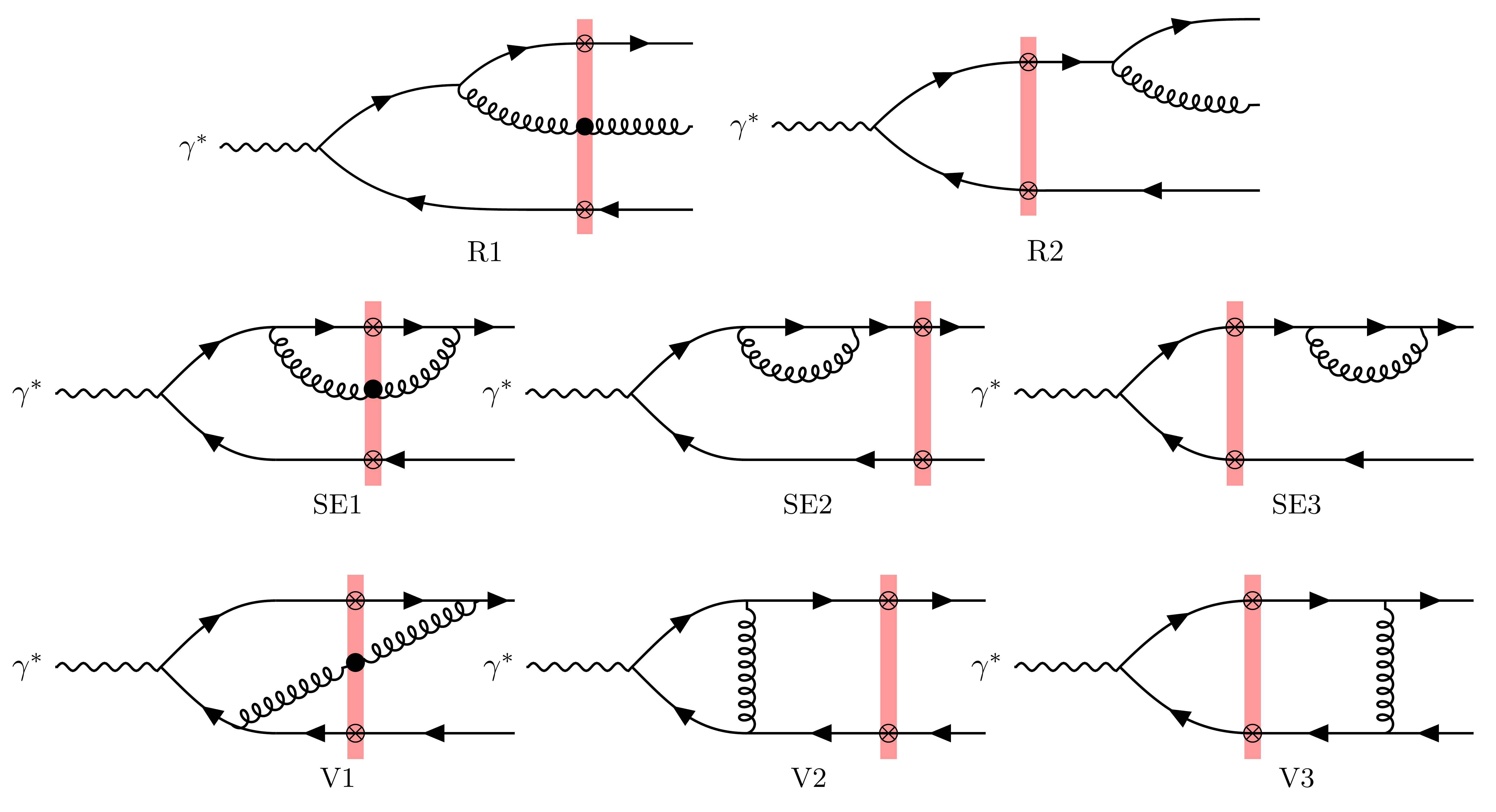}
\caption{Feynman diagrams that appear in the production of dijets at NLO. Top: real gluon emission diagrams. Middle: self energy  diagrams. Bottom: vertex correction diagrams. As in the LO case, the crossed dot denotes the effective quark CGC vertex. The bullet (full circle) denotes the effective CGC gluon vertex, which for dijet production only appears at NLO. Diagrams obtained from $q \leftrightarrow \bar{q}$ interchange are not shown.}
\label{fig:NLO-dijet-all-diagrams}
\end{figure}

The Feynman graphs necessary for inclusive dijet production at NLO  are gathered in Fig.~\ref{fig:NLO-dijet-all-diagrams}. The real diagrams are represented in the first line of  Fig.~\ref{fig:NLO-dijet-all-diagrams} while the virtual contributions are shown in the bottom two lines. (For brevity's sake, we omitted drawing the diagrams obtained by quark-antiquark interchange. The corresponding diagrams will henceforth be labeled with a ``prime" label, such as $\mathrm{R}1\rightarrow\mathrm{R}1'$, where $\mathrm{R}1'$ correspond to the diagram in which the gluon is emitted from the antiquark before it scatters the shock wave.) In principle, we have then 14 diagrams to compute, since diagram $\mathrm{V}2$ and $\mathrm{V}3$ are invariant under this transformation. As symmetry arguments enable one to infer the diagrams linked by  $(q\leftrightarrow\bar q)$ interchange, only the diagrams represented in Fig.~\ref{fig:NLO-dijet-all-diagrams} have to be computed explicitly. For the cross-section at order $\alpha_s$, one separately takes the modulus square of the real amplitudes and the product between the virtual amplitudes and the complex conjugation of the LO amplitude. Schematically, one has
\begin{equation}
    \left.\der\sigma^{\gamma^*A\to q\bar q+X}\right|_{\rm NLO}\propto \left(\Mcal_{\rm virtual}\Mcal^{*}_{\rm LO}+c.c.\right)+\int\der\Omega_g \  \Mcal_{\rm real}\Mcal_{\rm real}^{*} \,,
\end{equation}
where $\der\Omega_g$ is the differential gluon phase space. The real part of the cross-section is thus a sum of 16 terms, but only 6 of those need to be computed explicitly as the other 10 are related either by $(q\leftrightarrow\bar q)$ symmetry or by complex conjugation. The contribution of virtual diagrams to the cross-section contain in total 20 terms, when one accounts for complex conjugation. 

The physical amplitude for a given diagram will be given by the convolution of its color structure $\Ccal$ and its perturbative factor $\Ncal$: \begin{align}
    \Mcal=\frac{ee_fq^-}{\pi} \int \der^2 \xt \der^2 \yt \der^2 \zt \Pcal(\xt,\yt,\zt) \Ccal(\xt,\yt,\zt)\Ncal(\xt,\yt,\zt)\,,
    \label{eq:dijet-NLO-general-structure}
\end{align}
where $\Pcal(\xt,\yt,\zt)$ denote the Fourier phases
\begin{align}
    \Pcal(\xt,\yt,\zt) = \begin{cases} 
      e^{-i(\kt \cdot \xt + \pt \cdot \yt)} & \mathrm{real \ contributions}\,, \\
      e^{-i(\kt \cdot \xt + \pt \cdot \yt + \kgt \cdot \zt)} & \mathrm{virtual \ contributions}\,.
   \end{cases}
\end{align}
The transverse coordinate $\zt$ corresponds to the location of the emitted real or virtual gluon while crossing the shock wave. 

The bulk of the computation is to find explicit expressions for the perturbative factors $\Ncal$, which are obtained after internal (loop) momentum integrations;  they have the generic form:
\begin{align}
    \Ncal(\xt,\yt,\zt) = \int_{l_1,l_2,...} \frac{N_{\rm Dirac}}{D_1 D_2 ... D_k} \delta_{\rm eikonal} e^{i \ltone \cdot (...) + i \lttwo \cdot (...) + ...}\,,
\end{align}
where $N_{\rm Dirac}$ contains the Dirac structure which depends on the internal momenta\footnote{For convenience, we will drop the subscript {\rm Dirac} in the following sections.}. The factors $D_j$ are the propagator denominators, which are quadratic function of the momenta with a appropriate $+i\epsilon$ prescription. The eikonal delta functions (one or more) are denoted by $\delta_{\rm eikonal}$. The dependence on $\xt$, $\yt$ and $\zt$ is fully contained in the phases (not explicitly shown).

In order to proceed with the loop integration, we first perform the ``minus" lightcone momentum integrals using the delta functions arising from the eikonal vertices. For the real gluon emission contributions, this is sufficient to fix all the ``minus" lightcone momenta in terms of the external variables $k^-$, $p^-$ and $k_g^-$.
\begin{align}
    \Ncal_{\rm R} = \int_{\ltone,\lttwo,...} \!\!\!\!\!\!\!\!\!\!\!\!\!\!\!\!\! e^{i \ltone \cdot (...) + i \lttwo \cdot (...) + ...} \int_{l_1^+,l_2^+,...} \frac{N_{\rm Dirac}}{D_1 D_2 ... D_k} \,. \label{eq:generic_pert_factor_R}
\end{align}
For virtual contributions, there will be a remaining integration (without loss of generality, we will call it $l^-$), corresponding to the longitudinal momentum of the virtual gluon. This integral  develops a divergence at $l^- \to 0$, which will be regulated with a cut-off $\Lambda^-_0$.
\begin{align}
    \Ncal_{\rm V} = \int_{\Lambda^-_0} \der l^- \int_{\ltone,\lttwo,...} \!\!\!\!\!\!\!\!\!\!\!\!\!\!\!\!\! e^{i \ltone \cdot (...) + i \lttwo \cdot (...) + ...} \int_{l_1^+,l_2^+,...} \frac{N_{\rm Dirac}}{D_1 D_2 ... D_k}\,. \label{eq:generic_pert_factor_V}
\end{align}
As in the leading order case, the ``plus" lightcone momentum integration is performed  using Cauchy's theorem of residues. However unlike the LO case, the Dirac structure $N_{\rm Dirac}$ in the numerator of the integrals will generally depend on $l_1^+$, $l_2^+$, ... and thus the integration must be done with care. Fortunately, it is always possible to decompose the Dirac structure as follows:
\begin{align}
    N_{\rm Dirac} &= N_{\mathrm{reg}} + D_1 N_{\mathrm{inst}1} +  D_2 N_{\mathrm{inst}2} + ... + D_k N_{\mathrm{inst}k} 
    \label{eq:generic_dirac_expansion} \,,
\end{align}
where $N_{\mathrm{reg}}$ and $N_{\mathrm{inst}j}$'s are independent of any ``plus" lightcone momentum. All the ``plus" lightcone momentum dependence is contained in the prefactors $D_j$, which are precisely the factors\footnote{We should point out that not all terms $N_{\mathrm{inst}}$ are nonzero, and only a subset of them will contribute to the final result. Another caveat is that when a diagram has a ``double propagator", such as  the free self energy before the shock wave (in which the propagator squared $(l_1^2+i\epsilon)^2$ appears in the denominator), the regular term  also has a prefactor which cancels one of the propagators in the denominator. This is discussed below Eq.\,\eqref{eq:dijet-NLO-SE1-dirac-dec}).} 
that enter the denominator in Eqs.\,\eqref{eq:generic_pert_factor_R} and \eqref{eq:generic_pert_factor_V}.

Inserting the expansion in Eq.\,\eqref{eq:generic_dirac_expansion} into Eq.\,\eqref{eq:generic_pert_factor_V} we find for the virtual contributions (and analogous results for real contributions)
\begin{align}
    \Ncal_{\rm V} = \Ncal_{\mathrm{V},\mathrm{reg}} + \Ncal_{\mathrm{V},\mathrm{inst}1} + \Ncal_{\mathrm{V},\mathrm{inst}2} + ... + \Ncal_{\mathrm{V},\mathrm{inst}k}\,, \label{eq:generic_pert_factor_expansion}
\end{align}
where
\begin{align}
    \Ncal_{\mathrm{V},\mathrm{reg}} &= \int_{\Lambda^-_0} \der l^- \int_{\ltone,\lttwo,...} \!\!\!\!\!\!\!\!\!\!\!\!\!\!\!\!\! e^{i \ltone \cdot (...) + i \lttwo \cdot (...) + ...} \Ical_{\rm reg}(\ltone,\lttwo,...; l^-)  N_{\rm reg} (\ltone,\lttwo,...; l^-)\,, \nonumber \\
    \Ncal_{\mathrm{V},\mathrm{inst}j} &= \int_{\Lambda^-_0} \der l^- \int_{\ltone,\lttwo,...} \!\!\!\!\!\!\!\!\!\!\!\!\!\!\!\!\! e^{i \ltone \cdot (...) + i \lttwo \cdot (...) + ...} \Ical_{\mathrm{inst}j}(\ltone,\lttwo,...; l^-)  N_{\mathrm{inst}j} (\ltone,\lttwo,...; l^-)\,, \label{eq:generic_pert_factor_V2}
\end{align}
and
\begin{align}
    \Ical_{\mathrm{reg}}(\ltone,\lttwo,...; l^-) &= \int_{l_1^+,l_2^+,...}\frac{1}{D_1 D_2 ... D_k} \,,\label{eq:genetic_pole_structure_regular} \\
    \Ical_{\mathrm{inst}j}(\ltone,\lttwo,...; l^-) &= \int_{l_1^+,l_2^+,...}\frac{\cancel{D_j}}{D_1 D_2 ...\cancel{D_j} ... D_k}\label{eq:genetic_pole_structure_inst} \,.
\end{align}
Note that Eqs.\,\eqref{eq:genetic_pole_structure_regular} and \eqref{eq:genetic_pole_structure_inst} have different pole structures. General identities for such contour integrals are provided in Appendix~\ref{app:contour}.

The terms in the r.h.s. of Eq.\,\eqref{eq:generic_pert_factor_expansion} correspond to the contribution from different diagrams in lightcone perturbation theory (LCPT). More precisely, the computation of the NLO impact factor within LCPT involves regular propagators and instantaneous quark, antiquark or gluon propagators. These diagrams are in one-to-one correspondence with the instantaneous perturbative factors in Eq.\,\eqref{eq:generic_pert_factor_expansion}.

Finally, the remaining transverse momentum integrals in Eqs.\,\eqref{eq:generic_pert_factor_V2} are performed analytically. When a given diagram develops a UV divergence, the  transverse momentum integrations  are  carried out in  dimensional  regularization by going to $2-\varepsilon$ dimension, both in the internal momentum integration and in the transverse coordinate integration in Eq.~\eqref{eq:dijet-NLO-general-structure}. This is because of the fact that simple power counting in internal (loop) momenta might not reveal the presence of a UV divergence, as we will see in the discussion of the dressed self energy in Sec.~\ref{subsub:SE2}. Therefore one also needs to perform the coordinate integrations in $2-\varepsilon$ dimensions, which could result in a UV divergence. 

As explained at the beginning of this section, we do not compute explicitly the Feynman amplitudes obtained from the graphs in Fig.~\ref{fig:NLO-dijet-all-diagrams} by quark-antiquark interchange. Once the amplitude for the ``quark diagram" is known, the ``antiquark" amplitude  can be obtained straightforwardly from the following transformations:
\begin{enumerate}[i]
    \item interchange the quark and antiquark four-momenta $k^\mu\leftrightarrow p^\mu$ \,,
    \item interchange their transverse 
    coordinates $\xt\leftrightarrow\yt$ \,,
    \item flip the sign of the helicities $\sigma\to-\sigma,\sigma'\to-\sigma'$ \,,
    \item take the Hermitian conjugate of the color structure $\Ccal(\xt,\yt,\zt)\!\!\to\Ccal^\dagger(\yt,\xt,\zt)$ \,,
    \item and for the real amplitudes, we observe that an additional overall minus sign is required.
\end{enumerate}

\section{Dijet at NLO: real corrections}
\label{sec:real}

We begin our NLO computation by computing the real corrections, specifically,  the triple parton production amplitude $\gamma^*A \to q \bar{q} +g +X $. These were previously computed in \cite{Ayala:2016lhd,Ayala:2017rmh} using spinor helicities techniques. We have checked that our results agree with those obtained by the authors of \cite{Ayala:2016lhd,Ayala:2017rmh}. Our interest of rederiving these results in our approach is that it enables us to use the same notation and techniques as the virtual corrections which are novel\footnote{More specifically, we will use identities with Dirac matrices that allow us to separate regular and instantaneous contributions, which will be later generalized to obtain expressions in $4-\varepsilon$ dimensions.}. In addition, at the end of this section we will highlight a connection between gluon emission before and after the shock wave, which was also pointed out in the context of dijet production in p-A collisions in \cite{Iancu:2020mos}. Not least, the slow gluon limit of our results are straightforward to extract; as discussed in section~\ref{sec:slow_gluon}, when combined with the corresponding virtual corrections, they give rise to the JIMWLK rapidity evolution equations.

Our final results for the diagrams $\mathrm{R}1$ and $\mathrm{R}2$ are given by Eqs.~\eqref{eq:dijet-NLO-R2-amplitude}-\eqref{eq:dijet-NLO-R2-color}-\eqref{eq:dijet-NLO-R2-Npert-long-final}-\eqref{eq:dijet-NLO-R2-Npert-trans-final} and Eqs.~\eqref{eq:dijet-NLO-R1alternate}-\eqref{eq:dijet-NLO-R1-color}-\eqref{eq:dijet-NLO-R1-Npert-long-final}-\eqref{eq:dijet-NLO-R1-Npert-trans-final} respectively, while diagrams $\mathrm{R}1'$ and $\mathrm{R}2'$ can be obtained from $\mathrm{R}1$ and $\mathrm{R}2$ from quark-antiquark interchange, as explained in the previous section.

\subsection{Real gluon emission before the shock wave}

\begin{figure}[tbh]
\centering
\includegraphics[width=0.5\textwidth]{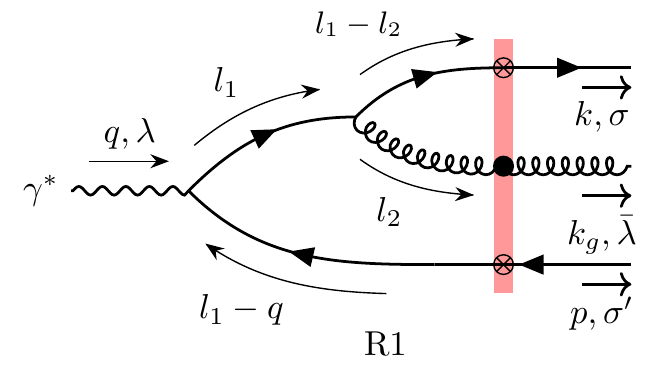}
\caption{Real gluon emission from quark before scattering from shock wave.}
\label{fig:NLO-dijet-R1}
\end{figure}

We denote by $k_g^\mu$ and $\bar{\lambda}$ the four momentum and the polarization of the outgoing gluon. In particular, we call $z_g=k_g^-/q^-$ the longitudinal momentum fraction of the gluon with respect to the longitudinal momentum of the virtual photon.
The scattering amplitude for $q\bar{q}$ + gluon emission from quark before scattering from shock wave is given by
\begin{align}
    \Scal^{\lambda\bar{\lambda}\sigma\sigma'}_{\mathrm{R}1} = \int \frac{\der^4 l_1}{(2\pi)^4} \frac{\der^4 l_2}{(2\pi)^4}  & \left[\bar{u}(k,\sigma) \Tcal^q(k,l_1-l_2) S^0(l_1-l_2) \left(ig t^a \gamma^\alpha \right) S^0(l_1) \left(-ie e_f \slashed{\epsilon}(q,\lambda) \right)  \right.\nonumber  \\ 
    &\!\!\!\!\!\!\!\!\!\!\!\!\!\! \times\left. S^0(l_1 - q) \Tcal^q(l_1-q,-p) v(p,\sigma')\right]  \epsilon^{*}_{\rho}(k_g,\bar{\lambda}) \Tcal^{g,\rho\beta}_{ba}(k_g,l_2) G^{0}_{\beta\alpha }(l_2)  \,.
\end{align}
Subtracting the noninteracting piece and factoring the overall $2\pi \delta(q^- - k^- - p^- - k_g^-)$, we obtain the physical amplitude
\begin{align}
    \Mcal^{\lambda\bar{\lambda}\sigma\sigma'}_{\mathrm{R}1} = \frac{ee_f q^-}{\pi} &\int \der^2 \xt \der^2 \yt \der^2 \zt e^{-i \kt \cdot \xt } e^{-i \pt \cdot \yt } e^{-i \kgt \cdot \zt } \nonumber \\
    &\times \Ccal_{\mathrm{R}1}(\xt,\yt,\zt) \mathcal{N}_{\mathrm{R}1}^{\lambda\bar{\lambda}\sigma\sigma'}(\rxyt,\rzxt)\label{eq:dijet-NLO-R2-amplitude} \,,
\end{align}
with the color structure (employing the Fierz identity) is denoted by 
\begin{align}
    \Ccal_{\mathrm{R}1}(\xt,\yt,\zt) = \left[  V(\xt)V^\dagger(\zt ) t^a V(\zt) V^\dagger(\yt) -t^a \right]\label{eq:dijet-NLO-R2-color} \,,
\end{align}
and the perturbative factor
\begin{align}
    \mathcal{N}_{\mathrm{R}1}^{\lambda\bar{\lambda}\sigma\sigma'}(\rxyt,\rzxt) &= -g(2q^-) \int \frac{\der^4 l_1}{(2\pi)^3} \frac{\der^4 l_2}{(2\pi)^2}e^{i\ltone \cdot \rxyt+i\lttwo\cdot\rzxt}\nonumber\\
    &\times\frac{(2l_2^-) N_{\mathrm{R}1}^{\lambda\bar{\lambda}\sigma\sigma'}(l_1,l_2) \delta(l_1^- - q^- + p^-) \delta(k_g^- - l_2^-) }{((l_1-l_2)^2 + i \epsilon )(l_1^2 + i \epsilon)((l_1 -q)^2 + i \epsilon)(l_2^2 + i \epsilon)}   \label{eq:dijet-NLO-real2-pert} \,.
\end{align}
The Dirac structure of this diagram which appears in the numerator of this integral reads as,
\begin{align}
    N_{\mathrm{R}1}^{\lambda\bar{\lambda}\sigma\sigma'} = \frac{1}{(2q^-)^2} \left[ \bar{u}(k,\sigma) \gamma^- (\slashed{l}_1 - \slashed{l}_2) \slashed{\epsilon}^*(l_2,\bar{\lambda}) \slashed{l}_1 \slashed{\epsilon}(q,\lambda) (\slashed{l}_1 -\slashed{q}) \gamma^- v(p,\sigma') \right] \,.
\end{align}
The outgoing gluon polarization vector has been turned into an internal polarization vector thanks to the identity $\epsilon_{\rho}^{*}(k_g,\bar\lambda) g^{\rho\beta} \Pi_{\beta\alpha}(l_2)=-\epsilon^{*}_{\alpha}(l_2,\bar \lambda)$, see Appendix \ref{app:dirac}.

\subsubsection*{Calculation of the Dirac structure $N_{\mathrm{R}1}$}

The integrations over $l_1^-$ and $l_2^-$ are trivial due to the presence of the delta functions $\delta(l_1^- - q^- + p^-)$ and $\delta(l_2^- - k_g^-)$ which enforce
\begin{equation}
    l_1^-=q^-(1-z_{\bar q}),\qquad l_2^-=q^-z_g \,.
\end{equation}
Using equation \eqref{eq:gluon_emi_quark_beforeSW} given in Appendix \ref{app:dirac} to simplify the component of the Dirac structure coming from the gluon emission from the quark before the shock wave, one can express $N^{\lambda\bar{\lambda}\sigma\sigma'}_{\mathrm{R}1}$ as
\begin{equation}
    N_{\mathrm{R}1}(l_1,l_2)=N_{\mathrm{R}1,\rm reg}(l_1,l_2)+l_1^2N_{\mathrm{R}1,q \rm inst}(l_1,l_2) \,.
\end{equation}
This decomposition follows the general strategy of the computation of the Dirac structure outlined in section~\ref{sec:gen-strat}. Indeed, the second term corresponds to the instantaneous quark contribution in LCPT, since the $l_1^2$ factor in the numerator cancels the identical term from the quark propagator. After some elementary algebra, the regular and instantaneous Dirac structure can be expressed as
\begin{align}
    N_{\mathrm{R}1,\mathrm{reg}}^{\lambda\bar{\lambda}\sigma\sigma'}&= \frac{2\Lttwox \cdot  \et^{\bar{\lambda}*}}{x}  \left\{ \bar{u}(k,\sigma) \left[\left(1 - \frac{x}{2}\right) + \bar{\lambda} \frac{x}{2} \Omega \right] \mathcal{D}_{\rm LO}^\lambda(l_1) v(p,\sigma') \right\} \label{eq:dijet-NLO-real2-dirac-reg} \,, \\
    N_{\mathrm{R}1,q\mathrm{inst}}^{\lambda\bar{\lambda}\sigma\sigma'} &= -\frac{(1-x) \et^{\bar{\lambda}*,i}}{4(q^-)^2} \left[ \bar{u}(k,\sigma) \gamma^i\gamma^- \slashed{\epsilon}(q,\lambda)(\slashed l_1-\slashed q)\gamma^- v(p,\sigma') \right] \label{eq:dijet-NLO-real2-dirac-ins} \,,
\end{align}
where $x = z_g/(1-z_{\bar{q}})$, $\Lttwox = \lttwo- x \ltone$, and $\mathcal{D}_{\rm LO}^\lambda(l)$ was defined in Eq.\,\eqref{eq:dijet-LO-DiracLO}. Recall too that $\Omega=\frac{i}{2}[\gamma^1,\gamma^2]$.

Let us consider the longitudinally polarized case with $\slashed{\epsilon}(q,\lambda=0) = \frac{Q}{q^-} \gamma^-$. Observe that the instantaneous piece vanishes since $(\gamma^-)^2=0$:
\begin{align}
    N_{\mathrm{R}1,q\mathrm{inst}}^{\lambda=0,\bar{\lambda}\sigma\sigma'} = 0\,.
\end{align}
For the regular piece in Eq.\,\eqref{eq:dijet-NLO-real2-dirac-reg}, we obtain after a little bit of algebra,
\begin{align}
    N_{\mathrm{R}1,\mathrm{reg}}^{\lambda=0,\bar{\lambda}\sigma\sigma'}
    &= -\frac{ z_{\bar{q}}(1-z_{\bar{q}})Q}{z_g}  \left\{ \bar{u}(k,\sigma) \left[ \left( 2z_q + z_g\right)  +  z_g \bar{\lambda} \Omega \right] \frac{\gamma^-}{q^-} v(p,\sigma')  \right\} \Lttwox\cdot  \et^{\bar{\lambda}*}\,.
\end{align}
For the transversely polarized virtual photon, with $\slashed{\epsilon}(q,\lambda = \pm 1) = -\gamma^j \et^{\lambda,j}$, the instantaneous terms result in
\begin{align}
    N_{\mathrm{R}1,q\mathrm{inst}}^{\lambda=\pm1,\bar{\lambda}\sigma\sigma'}&= 
    -\frac{z_q z_{\bar{q}}}{2(1-z_{\bar{q}})} \left\{\bar{u}(k,\sigma) \left[1 -\bar{\lambda} \Omega \right]  \frac{\gamma^-}{q^-} v(p,\sigma') \right\}\delta^{\lambda,\bar{\lambda}}\,,
\end{align}
and the regular term can be written as
\begin{align}
    N_{\mathrm{R}1,\mathrm{reg}}^{\lambda=\pm1,\bar{\lambda}\sigma\sigma'}&= \frac{(\Lttwox \cdot  \et^{\bar{\lambda}*})(\ltone \cdot \et^{\lambda})}{2 z_g} \nonumber \\
    &\times \left\{ \bar{u}(k,\sigma) \left[\left(2z_q + z_g\right) + z_g \bar{\lambda} \Omega \right] \left[(2z_{\bar{q}}-1) -\lambda \Omega \right] \frac{\gamma^-}{q^-} v(p,\sigma') \right\}\,.
\end{align}
\subsubsection*{Pole structure for the regular and instantaneous pieces}
Note that the Dirac structures in Eqs.\,\eqref{eq:dijet-NLO-real2-dirac-reg} and \eqref{eq:dijet-NLO-real2-dirac-ins} are independent of $l_1^+$ or $l_2^+$.
 The decomposition of the Dirac structure translates into a similar decomposition of the perturbative factor (Eq.\,\eqref{eq:dijet-NLO-real2-pert}) as 
\begin{align}
   \mathcal{N}_{\mathrm{R}1}(\rxyt,\rzxt) = \mathcal{N}_{\mathrm{R}1,\mathrm{reg}}(\rxyt,\rzxt) + \mathcal{N}_{\mathrm{R}1,q\mathrm{inst}}(\rxyt,\rzxt) \,,
\end{align}
where
\begin{align}
    \mathcal{N}_{\mathrm{R}1,\mathrm{reg}}(\rxyt,\rzxt)&=\frac{g}{2\pi}\int\frac{\der^2\ltone\der^2\lttwo}{(2\pi)^2}e^{i\ltone \cdot \rxyt+i\lttwo\cdot\rzxt}N_{\mathrm{R}1,\rm reg}\Ical_{\mathrm{R}1,\rm reg} \,,\\
    \mathcal{N}_{\mathrm{R}1,q\mathrm{inst}}(\rxyt,\rzxt)&=\frac{g}{2\pi}\int\frac{\der^2\ltone\der^2\lttwo}{(2\pi)^2}e^{i\ltone \cdot \rxyt+i\lttwo\cdot\rzxt}N_{\mathrm{R}1,q\rm inst}\Ical_{\mathrm{R}1,q\rm inst} \,.  
\end{align}
The pole structure is included in the two $l^+$ integrals $\Ical_{\mathrm{R}1\rm reg}$ and $\Ical_{\mathrm{R}1,q\rm inst}$ which are respectively defined to be 
\begin{align}
    \Ical_{\mathrm{R}1,\rm reg}&=\int\frac{\der l_1^+}{(2\pi)}\frac{\der l_2^+}{(2\pi)}\frac{-(2q^-)(2l_2^-)}{((l_1-l_2)^2+i\epsilon)(l_1^2+i\epsilon)((l_1-q)^2+i\epsilon)(l_2^2+i\epsilon)}\label{eq:dijet-NLO-real2-pert-reg} \,,\\
   \Ical_{\mathrm{R}1,q\rm inst}&=\int\frac{\der l_1^+}{(2\pi)}\frac{\der l_2^+}{(2\pi)}\frac{-(2q^-)(2l_2^-)}{((l_1-l_2)^2+i\epsilon)((l_1-q)^2+i\epsilon)(l_2^2+i\epsilon)}\label{eq:dijet-NLO-real2-pert-ins} \,.
\end{align}
Note that the pole structures in Eqs.\,\eqref{eq:dijet-NLO-real2-pert-reg} and \eqref{eq:dijet-NLO-real2-pert-ins} differ; performing the integrals using Cauchy's theorem, and closing the contour in the upper half plane, they read respectively:
\begin{align}
    \Ical_{\mathrm{R}1,\rm reg} &= \frac{1}{z_q} \frac{\Theta(z_g)\Theta(1-z_{\bar{q}}-z_g)}{\left( z_{\bar{q}}(1-z_{\bar{q}}) Q^2 + \ltone^2 \right) \left(Q^2 + \frac{\ltone^2}{z_{\bar{q}}} + \frac{(\ltone - \lttwo)^2}{z_q} + \frac{\lttwo^2}{z_g} \right)}\label{eq:dijet-NLO-R2-polestucture-reg} \,,\\
 \Ical_{\mathrm{R}1,q\rm inst}&= -\frac{1}{z_q z_{\bar{q}}}\frac{\Theta(z_g)\Theta(1-z_{\bar{q}}-z_g)}{ \left(Q^2 + \frac{\ltone^2}{z_{\bar{q}}} + \frac{(\ltone - \lttwo)^2}{z_q} + \frac{\lttwo^2}{z_g} \right)}.\label{eq:dijet-NLO-R2-polestucture-ins}
\end{align}
\subsubsection*{Transverse momentum integration: longitudinal photon}

For the longitudinally polarized photon, there is no instantaneous contribution and the  regular term can be expressed as, 
\begin{align}
    \mathcal{N}_{\mathrm{R}1,\mathrm{reg}}^{\lambda=0,\bar{\lambda}\sigma\sigma'}&(\rxyt,\rzxt) = - \frac{g}{2\pi} \frac{ z_{\bar{q}}(1-z_{\bar{q}})Q}{z_g z_q}  \Theta(z_g)\Theta(1-z_{\bar{q}}-z_g)  \nonumber\\
    & \times \left\{ \bar{u}(k,\sigma) \left[ \left( 2z_q + z_g\right)  +  z_g \bar{\lambda} \Omega \right] \frac{\gamma^-}{q^-} v(p,\sigma')  \right\}   \nonumber  \nonumber \\
    &  \times \int\frac{\der^2\ltone\der^2\lttwo}{(2\pi)^2}  \frac{(\Lttwox \cdot  \et^{\bar{\lambda}*})\  e^{i\ltone \cdot \rxyt+i\lttwo\cdot\rzxt}  }{\left( z_{\bar{q}}(1-z_{\bar{q}}) Q^2 + \ltone^2 \right) \left(Q^2 + \frac{\ltone^2}{z_{\bar{q}}} + \frac{(\ltone - \lttwo)^2}{z_q} + \frac{\lttwo^2}{z_g} \right)}\,.
\end{align}
Remarkably, the $\ltone$ and $\lttwo$ integration can be performed analytically~\cite{Beuf:2011xd}, giving the very compact result:
\begin{align}
    &\int \frac{\der^2 \ltone \der^2 \lttwo}{(2\pi)^2} \frac{ \left( \lttwo^i - \frac{z_g}{(1-z_{\bar{q}})} \ltone^i \right) e^{i\ltone \cdot \rxyt+i\lttwo\cdot\rzxt} }{\left( z_{\bar{q}}(1-z_{\bar{q}}) Q^2 + \ltone^2 \right) \left(Q^2 + \frac{\ltone^2}{z_{\bar{q}}} + \frac{(\ltone - \lttwo)^2}{z_q} + \frac{\lttwo^2}{z_g} \right)} \nonumber \\
    &= \frac{i z_g z_q}{1-z_{\bar{q}}} \frac{\rzxt^i}{\rzxt^2} K_0(Q X_{\rm R}) \,,\label{eq:dijet-NLO-R2-transverse-integral}
\end{align}
with $X_{\rm R}$ defined by
\begin{equation}
    X^2_{\rm R}=z_qz_{\bar q}\rxyt^2+z_qz_g\rzxt^2+z_{\bar q}z_g\rzyt^2 \,.\label{eq:dijet-NLO-R2-XR-def}
\end{equation}
This formula is derived in Appendix \ref{app:transverse-int}. The parameter $X_{\rm R}$ can be  interpreted as the effective 
transverse size of the $q\bar qg$ dipole when it crosses the shock wave. It plays a role analogous to the quantity $z_qz_{\bar q}r_{xy}$ in the LO photon wavefunction. 

Gathering all these results, one ends up with the following expression for the perturbative factor:
\begin{align}
    \mathcal{N}_{\mathrm{R}1,\mathrm{reg}}^{\lambda=0,\bar{\lambda}\sigma\sigma'}(\rxyt,\rzxt)&=  \frac{ig}{\pi}  \frac{\rzxt \cdot \et^{\bar{\lambda}*}}{\rzxt^2} (-z_q z_{\bar{q}}) Q K_0(Q X_{\rm R}) \Theta(z_g)\Theta(1-z_{\bar{q}}-z_g)  \nonumber\\
    &\times\frac{1}{2z_q}\left\{ \bar{u}(k,\sigma) \left[ \left( 2z_q + z_g\right)  +  z_g \bar{\lambda} \Omega \right] \frac{\gamma^-}{q^-} v(p,\sigma')  \right\} \,.
\end{align}
The remaining contraction with the quark and antiquark spinors is performed in the last paragraph of this section.

\subsubsection*{Transverse momentum integration: transversely polarized photon}
We begin with the instantaneous contribution given by 
\begin{align}
    \mathcal{N}_{\mathrm{R}1,q\mathrm{inst}}^{\lambda=\pm 1,\bar{\lambda}\sigma\sigma'}(\rxyt,\rzxt) &= \frac{g}{2\pi}
    \frac{1}{2(1-z_{\bar{q}})} \Theta(z_g)\Theta(1-z_{\bar{q}}-z_g)   \nonumber \\
    & \times \left\{\bar{u}(k,\sigma) \left[1 -\bar{\lambda} \Omega \right]  \frac{\gamma^-}{q^-} v(p,\sigma') \right\}\delta^{\lambda,\bar{\lambda}} \nonumber \\
    &\times \int \frac{\der^2 \ltone \der^2 \lttwo}{(2\pi)^2} \frac{e^{i\ltone \cdot \rxyt+i\lttwo\cdot\rzxt} }{\left(Q^2 + \frac{\ltone^2}{z_{\bar{q}}} + \frac{(\ltone - \lttwo)^2}{z_q} + \frac{\lttwo^2}{z_g} \right)}\,.
\end{align}
The transverse momentum integration of the instantaneous term can be performed as well using
\begin{align}
    \int \frac{\der^2 \ltone \der^2 \lttwo}{(2\pi)^2} \frac{e^{i\ltone \cdot \rxyt+i\lttwo\cdot\rzxt} }{\left(Q^2 + \frac{\ltone^2}{z_{\bar{q}}} + \frac{(\ltone - \lttwo)^2}{z_q} + \frac{\lttwo^2}{z_g} \right)} &=z_gz_qz_{\bar q}\frac{QK_1(QX_{\rm R})}{X_{\rm R}}\label{eq:dijet-NLO-R2transverse-inst} \,.
\end{align}
Gathering these results, the instantaneous perturbative factor can be written as
\begin{align}
    \mathcal{N}_{\mathrm{R}1,q\mathrm{inst}}^{\lambda  =\pm 1,\bar{\lambda}\sigma\sigma'}(\rxyt,\rzxt)&=\frac{g}{\pi}\frac{z_qz_{\bar q}z_g}{4(1-z_{\bar q})}\frac{QK_1(QX_{\rm R})}{X_{\rm R}} \Theta(z_g)\Theta(1-z_{\bar{q}}-z_g) \nonumber \\
    & \times \left\{\bar{u}(k,\sigma) \left[1 -\bar{\lambda} \Omega \right]  \frac{\gamma^-}{q^-} v(p,\sigma') \right\}\delta^{\lambda,\bar{\lambda}} \,.
\end{align}
The regular term reads
\begin{align}
    \mathcal{N}_{\mathrm{R}1,\mathrm{reg}}^{\lambda=\pm 1,\bar{\lambda}\sigma\sigma'}  &(\rxyt,\rzxt)  = \frac{g}{2\pi} \frac{1}{2 z_g z_q}  \Theta(z_g)\Theta(1-z_{\bar{q}}-z_g)  \nonumber \\
    & \times \left\{ \bar{u}(k,\sigma) \left[\left(2z_q + z_g\right) + z_g \bar{\lambda} \Omega \right] \left[(2z_{\bar{q}}-1) -\lambda \Omega \right] \frac{\gamma^-}{q^-} v(p,\sigma') \right\} \nonumber \\
    &  \times \int\frac{\der^2\ltone\der^2\lttwo}{(2\pi)^2}  \frac{(\Lttwox \cdot  \et^{\bar{\lambda}*})(\ltone \cdot \et^{\lambda})\  e^{i\ltone \cdot \rxyt+i\lttwo\cdot\rzxt}  }{\left( z_{\bar{q}}(1-z_{\bar{q}}) Q^2 + \ltone^2 \right) \left(Q^2 + \frac{\ltone^2}{z_{\bar{q}}} + \frac{(\ltone - \lttwo)^2}{z_q} + \frac{\lttwo^2}{z_g} \right)} \,.
\end{align}
Using the same trick as previously, we obtain
\begin{align}
    \int \frac{\der^2 \ltone \der^2 \lttwo}{(2\pi)^2} &\frac{ \left( \lttwo^i - \frac{z_g}{(1-z_{\bar{q}})} \ltone^i \right) \ltone^l e^{i\ltone \cdot \rxyt+i\lttwo\cdot\rzxt}  }{\left( z_{\bar{q}}(1-z_{\bar{q}}) Q^2 + \ltone^2 \right) \left(Q^2 + \frac{\ltone^2}{z_{\bar{q}}} + \frac{(\ltone - \lttwo)^2}{z_q} + \frac{\lttwo^2}{z_g} \right)}\nonumber\\
    &= -z_g z_q z_{\bar{q}} \frac{\rzxt^i}{\rzxt^2} \left[  \rxyt^l +  \frac{z_g}{(1-z_{\bar{q}})} \rzxt^l \right] \frac{Q K_1(QX_{\rm R})}{X_{\rm R}}\,.
\end{align}
Hence we find that the perturbative factor for the regular term is 
\begin{align}
    \Ncal_{\mathrm{R}1,\mathrm{reg}}^{\lambda=\pm 1,\bar{\lambda}\sigma\sigma'} (\rxyt,\rzxt) &= \frac{i g}{\pi}  \frac{\rzxt \cdot \et^{\bar{\lambda}*}}{\rzxt^2}  \frac{i z_q z_{\bar{q}}Q \,\RtR \cdot \et^{\lambda} }{2 X_{\rm R}} K_1(QX_{\rm R}) \Theta(z_g)\Theta(1-z_{\bar{q}}-z_g) \nonumber \\
    & \times \frac{1}{2 z_q} \left\{ \bar{u}(k,\sigma) \left[\left(2z_q + z_g\right) + z_g \bar{\lambda} \Omega \right] \left[(2z_{\bar{q}}-1) -\lambda \Omega \right] \frac{\gamma^-}{q^-} v(p,\sigma') \right\}\,,
\end{align}
where $\RtR$ is the size of the $q \bar q$ dipole before the emission of the gluon, defined to be 
\begin{equation}
    \RtR = \rxyt+\frac{z_g}{z_g+z_q}\rzxt\,.
\end{equation}

\subsubsection*{Spinor contractions}
It is possible to simplify further our expressions for the perturbative factor by performing the contraction with the spinors $\bar{u}(k,\sigma)$ and $v(p,\sigma')$.  We will provide here expressions that sum both regular and instantaneous pieces. Using the formulas given in Appendix \ref{app:dirac}, one gets 
\begin{align}
    \mathcal{N}_{\mathrm{R}1}^{\lambda=0,\bar{\lambda}\sigma\sigma'}(\rxyt,\rzxt)&=  \frac{ig}{\pi}  \frac{\rzxt \cdot \et^{\bar{\lambda}*}}{\rzxt^2} \frac{-2(z_qz_{\bar q})^{3/2}}{z_q} QK_0(QX_{\rm R})\delta^{\sigma,-\sigma'} \Gamma_{q\to qg}^{\sigma,\bar \lambda}(z_q,1-z_{\bar q})\label{eq:dijet-NLO-R2-Npert-long-final} \,,
\end{align}
for a longitudinally polarized photon. 

For a transversely polarized photon, one can combine the regular and instantaneous terms using $\rzxt^i\rzxt^i/\rzxt^2=1$:
\begin{align}
    &\Ncal^{\lambda=\pm1,\bar{\lambda}\sigma\sigma'}_{\mathrm{R}1}(\rxyt,\rzxt)=\frac{ig}{\pi}\frac{\rzxt^i}{\rzxt^2}\frac{2(z_qz_{\bar q})^{3/2}}{z_q}\frac{iQK_1(QX_{\rm R})}{X_{\rm R}} \delta^{\sigma,-\sigma'}\nonumber\\
    &\times\left\{\Gamma_{q\to qg}^{\sigma,\bar \lambda}(z_q,1-z_{\bar q})\Gamma_{\gamma^{*}_{\rm T} \to q\bar q} ^{\sigma,\lambda}(1-z_{\bar q},z_{\bar q})(\RtR\cdot\et^\lambda)\et^{\bar\lambda*,i}-\frac{z_qz_g\rzxt^i}{2(1-z_{\bar q})}\delta^{\sigma,\lambda}\delta^{\lambda,\bar{\lambda}}\right\}. \label{eq:dijet-NLO-R2-Npert-trans-final}
\end{align}
The spin-helicity dependent splitting vertex has been defined in Eq.~\eqref{eq:dijet-NLO-spinhel-splitting}. One notices the appearance of another such vertex coming from the splitting of the quark into a quark-gluon pair. Such a splitting is naturally related to the $\gamma^*_{\rm T} \to q\bar q$ splitting by crossing symmetry; the transformation $z_2\to-z_2$ (where $z_2$ is the longitudinal momentum fraction of the quark before it emits the gluon) and $\bar\lambda\to-\bar\lambda$:
\begin{align}
    \Gamma_{q\to qg}^{\sigma,\bar \lambda}(z_1,z_2)=-\Gamma_{\gamma^*_{\rm T} \to q\bar q}^{\sigma,-\bar\lambda}(z_1,-z_2)=z_1\delta_{\sigma,\bar\lambda}+z_2\delta_{\sigma,-\bar\lambda} \,. \label{eq:Gammaqtoqg}
\end{align}
We should also point out that when $z_g\to0$, $\Gamma_{q\to qg}^{\sigma,\bar \lambda}(z_q,1- z_{\bar q})\to z_q$, which ensures that the $1/z_q$ factor cancels in this limit.
The expressions in Eq. \eqref{eq:dijet-NLO-R2-Npert-long-final} and Eq.~\eqref{eq:dijet-NLO-R2-Npert-trans-final} are our final results for the perturbative factor in diagram $\mathrm{R}1$. 

\subsection{Real gluon emission after the shock wave}
\label{sub:R1}

\begin{figure}[tbh]
\centering
\includegraphics[width=0.5\textwidth]{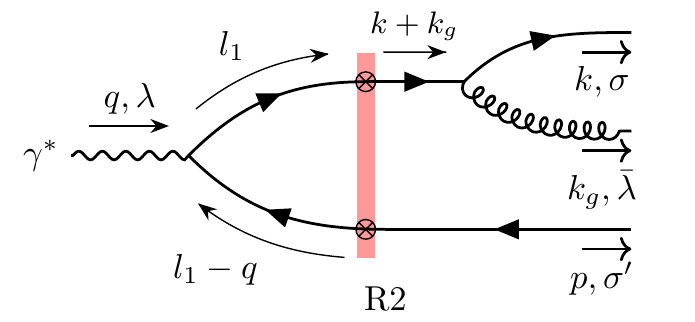}
\caption{Real gluon emission from the quark after scattering off the shock wave.}
\label{fig:NLO-dijet-R2}
\end{figure}

We turn now to diagram $\mathrm{R}2$ in which the real gluon is emitted after the shock wave. The details of the calculation are provided in Appendix~\ref{app:R1}. As explained in section~\ref{sec:gen-strat}, the amplitude can be organized as
\begin{align}
    \Mcal^{\lambda\bar{\lambda}\sigma\sigma'}_{\mathrm{R}2} = \frac{ee_fq^-}{\pi} \int \der^2 \wt \der^2 \yt e^{-i (\kt + \kgt )\cdot \wt} e^{-i \pt \cdot \yt } \Ccal_{\mathrm{R}2}(\wt,\yt)  \mathcal{N}_{\mathrm{R}2}^{\lambda\bar{\lambda}\sigma\sigma'}(\rwyt) \,,
\end{align}
with the color structure
\begin{align}
    \Ccal_{\mathrm{R}2}(\wt,\yt) = \left[ t^a V(\wt)V^\dagger(\yt) -t^a \right]\label{eq:dijet-NLO-R1-color} \,.
\end{align}
Since the radiated gluon does not scatter off the shock wave, the coordinate space integral involves only the transverse coordinates $\wt$ (to be distinguished from $\xt$ as discussed below) and $\yt$ of the quark and the antiquark when they cross the shock wave. 

We find for the perturbative factors,
\begin{align}
    \Ncal^{\lambda=0,\bar{\lambda}\sigma\sigma'}_{\mathrm{R}2}(\rwyt)&=2g\frac{\left( \kgt - \frac{z_g}{z_q} \kt \right)\cdot\et^{\bar\lambda*}}{\left( \kgt - \frac{z_g}{z_q} \kt\right)^2} 
    z_{\bar q}(1-z_{\bar q})QK_0(\bar{Q}_{\mathrm{R}2}\, r_{wy}) \nonumber \\ 
    & \times \frac{1}{2z_q}\left \{\bar{u}(k,\sigma) \left[(2z_q+z_g)+z_g\bar\lambda\Omega\right]\frac{\gamma^-}{q^-} v(p,\sigma') \right \}\label{eq:dijet-NLO-R1-perturbative} \,,\\
    \Ncal^{\lambda=\pm1,\bar{\lambda}\sigma\sigma'}_{\mathrm{R}2}(\rwyt)&=-2g\frac{\left( \kgt - \frac{z_g}{z_q} \kt \right)\cdot\et^{\bar\lambda*}}{\left( \kgt - \frac{z_g}{z_q} \kt\right)^2} 
    \frac{i\bar{Q}_{\rm {R}2}\et^{\lambda}\cdot \rwyt }{2r_{wy}}K_1(\bar{Q}_{\mathrm{R}2}\, r_{wy}) \nonumber \\ 
    & \times\frac{1}{2z_q} \left \{\bar{u}(k,\sigma) \left[(2z_q+z_g)+z_g\bar\lambda\Omega\right] \left[(2z_{\bar q}-1)-\lambda\Omega\right]\frac{\gamma^-}{q^-}v(p,\sigma') \right \}\label{eq:dijet-NLO-R1-perturbative-trans} \,,
\end{align}
where $\bar{Q}_{\mathrm{R}2}^2 = z_{\bar{q}} (1-z_{\bar{q}}) Q^2$. After contraction with the quark and antiquark spinors, one gets
\begin{align}
    \Ncal^{\lambda=0,\bar{\lambda}\sigma\sigma'}_{\mathrm{R}2}(\rwyt)&=4g\frac{\left(\kgt - \frac{z_g}{z_q} \kt \right)\cdot\et^{\bar\lambda*}}{\left(\kgt - \frac{z_g}{z_q} \kt\right)^2}\frac{(z_qz_{\bar q})^{1/2}z_{\bar q}(1-z_{\bar q})}{z_q}\,Q K_0(\bar{Q}_{\mathrm{R}2} r_{wy}) \nonumber\\
    &\times\Gamma_{q\to qg}^{\sigma,\bar\lambda}(z_q,1-z_{\bar q})\delta^{\sigma,-\sigma'} \,,\\
    \Ncal^{\lambda=\pm1,\bar{\lambda}\sigma\sigma'}_{\mathrm{R}2}(\rwyt)&=-4g\frac{\left(\kgt - \frac{z_g}{z_q} \kt \right)\cdot \et^{\bar{\lambda}*}}{\left(\kgt - \frac{z_g}{z_q} \kt \right)^2}  \frac{(z_q z_{\bar q})^{1/2}}{z_q} \, \frac{i\bar{Q}_{\mathrm{R}2} \rwyt\cdot \et^{\lambda}}{ r_{wy}} K_1(\bar{Q}_{\mathrm{R}2} r_{wy})\nonumber\\     &\times\Gamma_{q\to qg}^{\sigma,\bar\lambda}(z_q,1-z_{\bar q})
\Gamma_{\gamma^*_\mathrm{T} \to q\bar{q}}^{\sigma,\lambda}(1-z_{\bar{q}},z_{\bar{q}}) \delta^{\sigma ,-\sigma'} \,,
\end{align}
respectively for longitudinal and transverse virtual photons. 

\subsubsection*{An alternative expression for the perturbative factors} 
The above expressions are sufficient; however, it is useful to provide an alternative expression that more closely resembles the result for real gluon emission before shock wave. Towards this aim, we introduce an additional transverse coordinate integration using the identity,
\begin{align}
    \frac{\left(\kgt^i - \frac{z_g}{z_q} \kt^i \right)}{\left(\kgt - \frac{z_g}{z_q} \kt\right)^2}=\frac{i}{2\pi}\frac{z_q}{z_q+z_g} \int \der^2 \Rt  \frac{\Rt^i}{\Rt^2} e^{-i \left(\frac{z_q\kgt-z_g\kt}{z_q+z_g}\right)\cdot \Rt} \,,
\end{align}
to transform the relative quark-gluon transverse momentum appearing in Eq.~\eqref{eq:dijet-NLO-R1-perturbative} into an integral over the extra transverse coordinate $\Rt$. After a change of variables
\begin{align}
    \Rt    =\zt-\xt \ , \qquad \wt= \frac{z_q \xt + z_g \zt}{z_q + z_g} \,,
\end{align}
one gets
\begin{align}
    \Mcal^{\lambda\bar{\lambda}\sigma\sigma'}_{\mathrm{R}2} \!\!= \! \frac{ee_fq^-}{\pi}\!\! \int \! \der^2 \xt \der^2 \yt \der^2 \zt  e^{-i \kt \cdot \xt } e^{-i \pt \cdot \yt } e^{- i \kgt \cdot \zt} \Ccal_{\mathrm{R}2}(\wt,\yt)  \mathcal{N}_{\mathrm{R}2}^{\lambda\bar{\lambda}\sigma\sigma'}(\rwyt,\rzxt)\label{eq:dijet-NLO-R1alternate} \,.
\end{align}
The alternative form for the perturbative factor depends now on the two transverse coordinates $\rwyt$ and $\rzxt$, and can be expressed as 
\begin{align}
    \mathcal{N}_{\mathrm{R}2}^{\lambda=0,\bar{\lambda}\sigma\sigma'}(\rwyt,\rzxt) &= \frac{(-ig)}{\pi} \frac{\rzxt \cdot\et^{\bar{\lambda}*}}{\rzxt^2} \frac{- 2(z_qz_{\bar q})^{3/2}}{z_q} Q K_0(\bar{Q}_{\mathrm{R}2} r_{wy}) \Gamma_{q\to qg}^{\sigma,\bar\lambda}(z_q,1-z_{\bar q}) \delta^{\sigma ,-\sigma'},  \label{eq:dijet-NLO-R1-Npert-long-final} \\
    \mathcal{N}_{\mathrm{R}2}^{\lambda=\pm1,\bar{\lambda}\sigma\sigma'}(\rwyt,\rzxt) &=\frac{(-ig)}{\pi}\frac{\rzxt\cdot \et^{\bar{\lambda}*}}{\rzxt^2}  \frac{2(z_q z_{\bar q})^{1/2}}{1-z_{\bar q}}  \frac{i\bar{Q}_{\mathrm{R}2} \rwyt\cdot \et^{\lambda}}{ r_{wy}} K_1(\bar{Q}_{\mathrm{R}2} r_{wy}) \nonumber\\
    &\times \Gamma_{q\to qg}^{\sigma,\bar\lambda}(z_q,1-z_{\bar q}) \Gamma_{\gamma^*_\mathrm{T} \to q\bar{q}}^{\sigma,\lambda}(1-z_{\bar{q}},z_{\bar{q}}) \delta^{\sigma ,-\sigma'}.\label{eq:dijet-NLO-R1-Npert-trans-final}
\end{align}
This has a nice pictorial representation in the language of LCPT. While $\rzxt$ is the size of the quark-gluon pair, the transverse coordinate $\rwyt$ is the size of the $q\bar q$ dipole right after the photon splits. On the other hand, the transverse coordinate $\rxyt$ corresponds, for this diagram, to the size of the $q\bar q$ after  the emission of the gluon, and thus differs from $\rwyt$ due to the subsequent recoil of the pair.
 
\subsubsection*{Relation between diagrams $\mathrm{R}1$ and $\mathrm{R}2$} Writing the amplitude for $\mathrm{R}2$ as in Eq.~\eqref{eq:dijet-NLO-R1alternate} presents the additional advantage of highlighting the connection between gluon emission before and after the shock wave. The perturbative factor for the amplitudes in diagrams 1 and 2 can be written as 
\begin{align}
    \Ncal_{\textrm{R}n}^{\lambda\bar{\lambda}\sigma\sigma'}(\xt,\yt\,\zt) = \frac{\rzxt^i}{\rzxt^2} \widetilde{\Ncal}_{\textrm{R}n}^{\lambda\bar{\lambda}\sigma\sigma',i}(\xt,\yt\,\zt) \,.
\end{align}
With this, it is not difficult to show that the perturbative factor for  gluon emission after the shock wave can be recovered from that for gluon emission before the shock wave from the relation,
\begin{align}
     \lim_{\substack{\xt \to \wt \\ \zt \to \wt}}\widetilde{\Ncal}_{\textrm{R}1}^{\lambda\bar{\lambda}\sigma\sigma',i}(\xt,\yt\,\zt) =- \widetilde{\Ncal}_{\textrm{R}2}^{\lambda\bar{\lambda}\sigma\sigma',i}(\xt,\yt\,\zt)\,.
\end{align}
This also holds true for the color structures,
\begin{align}
    \lim_{\substack{\xt \to \wt \\ \zt \to \wt}}\Ccal_{\textrm{R}1}(\xt,\yt\,\zt) = \Ccal_{\textrm{R}2}(\xt,\yt\,\zt)  \,.
\end{align}
These relations imply therefore  that we can obtain the amplitude $\Mcal_{\mathrm{R}2}^{\lambda}$ from $\Mcal_{\mathrm{R}1}^{\lambda}$ by taking the limits $\xt, \zt \to  \wt$, except in the phases and in the gluon emission kernel $\rzxt^i / \rzxt^2$. A similar observation was made in \cite{Iancu:2020mos} for the computation of dijet production in p-A collisions. The same holds true when gluon emission is off the antiquark. Notice that this observation also implies that the sum of real emission diagrams is free of short distance (UV) divergences in the limit $\zt \to \xt,\yt$.

\section{Dijet at NLO: virtual corrections}
\label{sec:virtual}

We now turn to the virtual corrections to the dijet cross-section and the calculation of the perturbative factor $\Ncal^\lambda$ defined in Eq.~\eqref{eq:dijet-NLO-general-structure} for each diagram.
As some diagrams develop a UV divergence, we will use conventional dimensional regularization of the transverse integration to extract the UV pole. It means that all the transverse coordinates and gamma matrix representations are analytically continued to $2-\varepsilon$ dimension.

We shall not detail the calculation of each diagram. Only the dressed self energies and dressed vertex corrections are computed in details in the following subsections. For the other diagrams, we provide additional details in the appendices \ref{app:SE1}, \ref{app:V1}, \ref{appp:V3} or refer to extant results in the literature.

\subsection{Self energy diagrams}

Let us consider first the self energy diagrams which are shown in the second line of Fig.\,\ref{fig:NLO-dijet-all-diagrams}. They can be divided into two classes according to whether the virtual gluon scatters off the shock wave or not. For the diagram with dressed gluon propagator, the final result for the amplitude is given by Eqs.\,\eqref{eq:dijet-NLO-MSE1}-\eqref{eq:dijet-NLO-CSE1}-\eqref{eq:dijet-NLO-SE1-perturbative-final}-\eqref{eq:dijet-NLO-SE1-transverse-inst}-\eqref{eq:dijet-NLO-SE1-transverse-reg} (see also our discussion on UV and finite pieces). The free self energy before the shock wave is given by Eqs,\,\eqref{eq:dijet-NLO-SE2-final}-\eqref{eq:dijet-NLO-SE2-final-transverse}. While the self energy after the shock wave vanishes (for massless quarks) we show the nature of this cancellation in Eq.\,\eqref{eq:dijet-NLO-SE3-final}.

\subsubsection{Dressed gluon propagator}
\label{subsub:SE2}

\begin{figure}[tbh]
\centering
\includegraphics[width=0.5\textwidth]{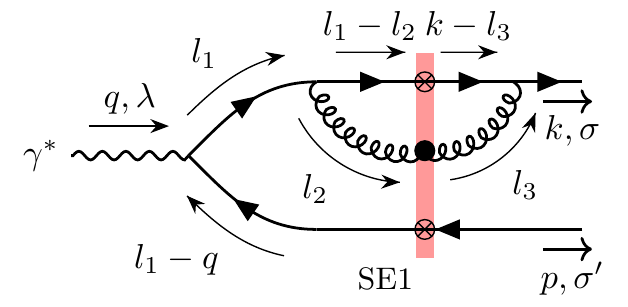}
\caption{Gluon self energy with gluon crossing the shock wave}
\label{fig:NLO-dijet-SE1}
\end{figure}

For the dressed gluon self energy shown in Fig.~\ref{fig:NLO-dijet-SE1}, the extraction of its UV divergence and finite part constitutes one of the principal results of this paper. Given that its computation is new and quite subtle, we will provide details of some of the  intermediate steps in the computation.

The amplitude of diagram $\mathrm{SE}1$ reads
\begin{align}
    \Scal_{\mathrm{SE}1}^{\lambda\sigma\sigma'} &=  \mu^{3\varepsilon}\int \frac{\der^{4-\varepsilon} l_1}{(2\pi)^{4-\varepsilon}} \frac{\der^{4-\varepsilon} l_2}{(2\pi)^{4-\varepsilon}} \frac{\der^{4-\varepsilon} l_3}{(2\pi)^{4-\varepsilon}} \left[ \bar{u}(k,\sigma) (ig \gamma^\mu t^a) S^0(k-l_3) \Tcal^q(k-l_3,l_1-l_2) \right. \nonumber  \\ &\left. \times S^0(l_1-l_2)  (ig \gamma^\nu t^b) S^0(l_1) 
    (-i e e_f \slashed{\epsilon}(q,\lambda)) S^0(l_1-q) \Tcal^q(l_1-q,-p) v(p,\sigma') \right] \nonumber \\
    & \times G^{0,ac}_{\mu\rho}(l_3)\Tcal^{g,\rho\sigma}_{cd}(l_3,l_2)  G^{0,db}_{\sigma\nu}(l_2) \,.
\end{align}
After subtraction of the noninteracting contribution and extracting the overall delta function $2\pi \delta(q^- - k^- - p^-)$, we find the physical amplitude
\begin{align}
    \Mcal^{\lambda\sigma\sigma'}_{\mathrm{SE}1} \!\!= \! \frac{e e_f q^-}{\pi} \!\mu^{-3\varepsilon} &\int \der^{2-\varepsilon} \xt \der^{2-\varepsilon} \yt \der^{2-\varepsilon} \zt e^{-i \kt \cdot \xt } e^{-i \pt \cdot \yt } \nonumber\\
    &\times\Ccal_{\mathrm{SE}1}(\xt,\yt,\zt) \Ncal_{\mathrm{SE}1}^{\lambda\sigma\sigma'}(\rxyt,\rzxt) \,.
    \label{eq:dijet-NLO-MSE1}
\end{align}
The color structure of this diagram depends on the transverse coordinate $\zt$ of the gluon through the shock wave,
\begin{align}
    \Ccal_{\mathrm{SE}1}(\xt,\yt,\zt)
    &= \left[ t^a V(\xt) V^\dagger(\zt) t_a V(\zt) V^\dagger(\yt) -t^a t_a \right]\,,
\label{eq:dijet-NLO-CSE1}
\end{align}
and the perturbative factor reads
\begin{align}
    \Ncal_{\mathrm{SE}1}^{\lambda\sigma\sigma'}(\rxyt,\rzxt) &= \frac{ g^2 }{(2q^-)}   \mu^{3\varepsilon}\int \frac{\der^{4-\varepsilon} l_1}{(2\pi)^{3-\varepsilon}} \frac{\der^{4-\varepsilon} l_2}{(2\pi)^{3-\varepsilon}} \frac{\der^{4-\varepsilon} l_3}{(2\pi)^{3-\varepsilon}} e^{i \ltone \cdot \rxyt} e^{i(\lttwo -\ltthre)\cdot \rzxt}   \nonumber\\
    &\hspace{-1cm}\times\frac{-i(2 l_3^-)(2q^-)^2N_{\mathrm{SE}1}^{\lambda\sigma\sigma'}(l_1,l_2,l_3)\delta(k^--l_1^-)\delta(l_3^- - l_2^-) }{\left[ (l_3-k)^2 + i \epsilon \right] \left[ (l_2-l_1)^2 + i \epsilon \right] \left[ l_1^2 + i \epsilon \right] \left[ (l_1-q)^2 + i \epsilon \right] \left[ l_3^2 + i \epsilon \right] \left[ l_2^2 + i \epsilon \right] }
    \label{eq:dijet-NLO-SE2-pert} \,,
\end{align}
where the Dirac numerator is given by
\begin{align}
    N_{\mathrm{SE}1}^{\lambda\sigma\sigma'} \!\! =\! \frac{1}{(2q^-)^2} \! \left[ \bar{u}(k,\sigma) \gamma^\mu (\slashed{k}-\slashed{l}_3)\gamma^- (\slashed{l_1}-\slashed{l}_2) \gamma_\nu \slashed{l}_1 \slashed{\epsilon}(q,\lambda) (\slashed{l_1}-\slashed{q}) \gamma^- v(p,\sigma')  \right] \Pi_{\mu\rho}(l_3) \Pi^{\rho\nu} (l_2)
    \label{eq:dijet-NLO-SE2-dirac-1} \,.
\end{align}

\subsubsection*{Calculation of the Dirac numerator $N_{\mathrm{SE}1}$}

Before we proceed with the Dirac numerator computation, note that the integration over $l_1^-$ and $l_2^-$ can be easily done with the delta functions resulting in $l_2^- = l_3^- $ and $l_1^- = k^-$.
Using the identity Eqs.\,\eqref{eq:gluon_tensor_squared_decomp} to express the product of the two gluon polarizations tensors in terms of gluon polarization vectors, and then Eqs.\,\eqref{eq:gluon_abs_quark_afterSW-ddim},\,\eqref{eq:gluon_emi_quark_beforeSW-ddim} to simplify the gluon absorption and emission parts of the Dirac structure in Eq.\,\eqref{eq:dijet-NLO-SE2-dirac-1}, we can write the latter as
\begin{equation}
        N_{\mathrm{SE}1} = N_{\mathrm{SE}1,\mathrm{reg}} +  l_1^2 N_{\mathrm{SE}1,q\mathrm{inst}}
    \label{eq:dijet-NLO-SE2-dirac-2},
\end{equation}
with 
\begin{align}
    N_{\mathrm{SE}1,\mathrm{reg}}^{\lambda\sigma\sigma'}  &=-\frac{4\Ltthrex^i \Lttwox^k}{x^2} \left\{ \bar{u}(k,\sigma)\left[\left(1-x+\left(1-\frac{\varepsilon}{2}\right)\frac{x^2}{2}\right)\delta^{ik}\right.\right.\nonumber\\
    &\hspace{4.5cm}\left.\left.+\left(x-\left(1-\frac{\varepsilon}{2}\right)\frac{x^2}{2}\right)\omega^{ik}\right] \mathcal{D}_{\rm LO}^\lambda(l_1) v(p,\sigma')  \right\}  \label{eq:dijet-NLO-SE2-dirac-reg} \,, \\
    N_{\mathrm{SE}1,q\mathrm{inst}}^{\lambda\sigma\sigma'}  &=  \frac{(1-x) \Ltthrex^i}{2(q^-)^2 x} \left\{ \bar{u}(k,\sigma) \left[\left(1-\frac{x}{2}\right)\delta^{ij}+\frac{x}{2}\omega^{ij}\right]  \gamma^j \gamma^-  \slashed{\epsilon}(q,\lambda) (\slashed{l_1}-\slashed q) \gamma^- v(p,\sigma')  \right\} \,, \label{eq:dijet-NLO-SE2-dirac-qins}
\end{align}
where $x=z_g/z_q$, $\Ltthrex = \ltthre - x \kt$ and $\Lttwox = \lttwo - x \ltone$, and $\omega^{ij}=\frac{1}{2}[\gamma^i,\gamma^j]$. One notices that the leading order Dirac structure $\mathcal{D}_{\rm LO}^\lambda(l_1) = \gamma^-\slashed l_1\slashed \epsilon(q,\lambda)(\slashed l_1-\slashed q)\gamma^-$ factorizes, as a consequence of the topology of the self energy diagram. Since we are working in  dimensional regularization, we have included the $\mathcal{O}(\varepsilon)$ dependence of the Dirac numerator that results from the identity Eq.\,\eqref{eq:omega-product}.

We first consider the longitudinally polarized case with $\slashed{\epsilon}(q,\lambda) = \frac{Q}{q^-}\gamma^-$. 
Observe that the instantaneous piece in Eq.\,\eqref{eq:dijet-NLO-SE2-dirac-qins} vanishes  since $(\gamma^-)^2=0$:
\begin{align}
    N_{\mathrm{SE}1,q\mathrm{inst}}^{\lambda=0,\sigma\sigma'} = 0\,.
\end{align}
For the regular piece in Eq.\,\eqref{eq:dijet-NLO-SE2-dirac-reg}, we can write it as 
\begin{align}
    N_{\mathrm{SE}1,\mathrm{reg}}^{\lambda=0,\sigma\sigma'}   =  \frac{4 z_q^3 z_{\bar{q}} Q}{z_g^2} &\left \{\left[1-\frac{z_g}{z_q}+ \left(1-\frac{\varepsilon}{2}\right)\frac{z_g^2}{2z_q^2} \right] \frac{\left[\bar{u}(k,\sigma) \gamma^- v(p,\sigma') \right]}{q^-} \delta^{ik}\right.\nonumber\\
    &\left.+ \left[\frac{z_g}{z_q}-\left(1-\frac{\varepsilon}{2}\right) \frac{z_g^2}{2z_q^2} \right] \frac{\left[\bar{u}(k,\sigma)\omega^{ik}\gamma^- v(p,\sigma') \right]}{q^-} \right \} \Ltthrex^i \Lttwox^k \,. \label{eq:dijet-NLO-SE2-dirac-reg-final}
\end{align}

For a transversely polarized virtual photon, the instantaneous term contributes, and one obtains,
\begin{equation}
    N^{\lambda=\pm1,\sigma\sigma'}_{\mathrm{SE}1,q \rm inst}=\frac{z_{\bar q}(z_g-z_q)^2}{z_gz_q}\Ltthrex\cdot\et^{\lambda}\left\{\bar{u}(k,\sigma)(1-\lambda\Omega)\frac{\gamma^-}{q^-}v(p,\sigma')\right\} \,. \label{eq:dijet-NLO-SE2-diracinst}
\end{equation}
This instantaneous contribution is free of any ultraviolet divergence so that we do not have to keep track of the  $\varepsilon$ dependencies in the Dirac algebra and we can use $\omega^{ij}=-i\epsilon^{ij}\Omega$. Finally, the regular term for a transversely polarized virtual photon can be written as
\begin{align}
    N^{\lambda=\pm1,\sigma\sigma'}_{\mathrm{SE}1,\rm reg} &=-\frac{2 z_q^2}{z_g^2}\Ltthrex\cdot\Lttwox \left[1-\frac{z_g}{z_q}+ \left(1-\frac{\varepsilon}{2}\right)\frac{z_g^2}{2z_q^2} \right]\nonumber\\
    &\times\left\{\bar{u}(k,\sigma)[(z_{\bar q}-z_q)\delta^{lm}+\omega^{lm}]\frac{\gamma^-}{q^-} v(p,\sigma')\right\}\et^{\lambda,l}\ltone^m+\Ltthrex^i\Lttwox^j[...]\omega^{ij}[...]\,.
    \label{eq:dijet-NLO-SE2-dirac-reg-final-transverse}
\end{align}
To keep the r.h.s compact, the argument of the last term proportional to $\Ltthrex^i\Lttwox^j\omega^{ij}$ is not shown since it vanishes anyway after transverse  momentum integration.

\subsubsection*{Pole structure of the regular and instantaneous piece}

The next step of the calculation consists in performing the ``plus" lightcone momentum integration associated with $l_1$, $l_2$ and $l_3$. Since neither $N_{\mathrm{SE}1,\mathrm{reg}}$ nor $N_{\mathrm{SE}1,q\mathrm{inst}}$ depend explicitly on any ``plus" component, the integration can easily be done using standard techniques of contour integration and the Cauchy theorem. Using the decomposition in Eq.\,\eqref{eq:dijet-NLO-SE2-dirac-2} we can express Eq.\,\eqref{eq:dijet-NLO-SE2-pert} as 
\begin{equation}
        \Ncal_{\mathrm{SE}1}(\rxyt,\rzxt) = \Ncal_{\mathrm{SE}1,\mathrm{reg}}(\rxyt,\rzxt) + \Ncal_{\mathrm{SE}1,\bar{q} \rm inst}(\rxyt,\rzxt)\,,
\end{equation}
with 
\begin{align}
    \Ncal_{\mathrm{SE}1,\mathrm{reg}} (\rxyt,\rzxt)&=   \frac{g^2}{2} \int \der z_g \mu^{3\varepsilon}\int \frac{\der^{2-\varepsilon}\ltone}{(2\pi)^{2-\varepsilon}}\frac{\der^{2-\varepsilon}\lttwo}{(2\pi)^{2-\varepsilon}}\frac{\der^{2-\varepsilon}\ltthre}{(2\pi)^{2-\varepsilon}} e^{i \ltone \cdot \rxyt+i(\lttwo -\ltthre) \cdot \rzxt}  \nonumber\\
    &\times\Ical_{\mathrm{SE}1,\mathrm{reg}} N_{\mathrm{SE}1,\mathrm{reg}}\,,
    \label{eq:dijet-NLO-SE2-pert-reg}\\
    \Ncal_{\mathrm{SE}1,q\mathrm{inst}} (\rxyt,\rzxt)&=   \frac{g^2}{2}  \int \der z_g\mu^{3\varepsilon}\int \frac{\der^{2-\varepsilon}\ltone}{(2\pi)^{2-\varepsilon}}\frac{\der^{2-\varepsilon}\lttwo}{(2\pi)^{2-\varepsilon}}\frac{\der^{2-\varepsilon}\ltthre}{(2\pi)^{2-\varepsilon}} e^{i \ltone \cdot \rxyt +i(\lttwo -\ltthre) \cdot \rzxt}\nonumber\\
    &\times\Ical_{\mathrm{SE}1,q\mathrm{inst}} N_{\mathrm{SE}1,q\mathrm{inst}} \,,
    \label{eq:dijet-NLO-SE2-pert-qins}
\end{align}
where we encounter the  pole structures
\begin{align}
    \Ical_{\mathrm{SE}1,\mathrm{reg}}&=  \int \frac{\der l_1^+}{(2\pi)}\int\frac{\der l_2^+}{(2\pi)}  \int\frac{\der l_3^+}{(2\pi)} \frac{  -i(2q^-)^2 (2l_3^-)}{ (l_3-k)^2 (l_2-l_1)^2 l_1^2  (l_1-q)^2  l_3^2  l_2^2 } \,,
    \label{eq:dijet-NLO-SE2-l+reg} \\
    \Ical_{\mathrm{SE}1,q\mathrm{inst}}&=  \int \frac{\der l_1^+}{(2\pi)} \int \frac{\der l_2^+}{(2\pi)}\int \frac{\der l_3^+}{(2\pi)} \frac{ -i (2q^-)^2 (2l_3^-)}{ (l_3-k)^2  (l_2-l_1)^2 (l_1-q)^2  l_3^2  l_2^2 } \,.
    \label{eq:dijet-NLO-SE2-l+qins}
\end{align}
To keep these integrals compact, we have omitted the $+i\epsilon$ prescription which fixes the location of the poles in the complex plane.
Employing equations Eqs.\,\eqref{eq:contour_ll'_generic}, \eqref{eq:contour_ll'_generic3}
we can write the integrals in Eq.\,\eqref{eq:dijet-NLO-SE2-l+reg} and \eqref{eq:dijet-NLO-SE2-l+qins} as (see more details in appendix~\ref{subsub:contour-SE1})
\begin{align}
    \Ical_{\mathrm{SE}1,\mathrm{reg}}&= -\frac{z_g}{z_q^2}\frac{\Theta(z_g)\Theta(z_q-z_g)  }{\left( \ltone^2 +  \bar{Q}^2 \right)\left[\omega_{\mathrm{SE}1}\left( \ltone^2 +\bar{Q}^2 \right)+\Lttwox^2\right]\Ltthrex^2} 
    \label{eq:dijet-NLO-SE2-l+reg-2} \,,\\
   \Ical_{\mathrm{SE}1,\bar{q}\mathrm{inst}} &=  \frac{z_g}{z_q^2 z_{\bar{q}}}  \frac{ \Theta(z_g) \Theta(z_q - z_g)}{\left[ \omega_{\mathrm{SE}1} \left(\ltone^2 + \bar{Q}^2  \right)+\Lttwox^2 \right] \Ltthrex^2} \,,
    \label{eq:dijet-NLO-SE2-l+qbarins-2}
\end{align}
where we introduced the kinematic variables
\begin{align}
    \bar{Q}^2 =  z_q z_{\bar{q}} Q^2 \,\,\,\,\,{\rm and}\,\,\,\,\,
    \omega_{\mathrm{SE}1} =\frac{z_g (z_q- z_g)}{z_q^2 z_{\bar{q}}}\,.
\end{align}

\subsubsection*{Transverse momentum integration}
We are now ready to perform the integration over the transverse components of the internal momenta.
Given that the integrals depend only on $\ltone$, $\Lttwox$ and $\Ltthrex$, we perform a change of variable in terms of these transverse vectors. The phase in \eqref{eq:dijet-NLO-SE2-pert-reg} and \eqref{eq:dijet-NLO-SE2-pert-qins} reads then
\begin{align}
    e^{i \ltone \cdot \rxyt}e^{i(\lttwo-\ltthre) \cdot \rzxt}=e^{-i\frac{z_g}{z_q}\kt \cdot \rzxt}e^{i\ltone\cdot \left(\rxyt+\frac{z_g}{z_q}\rzxt \right)}e^{i(\Lttwox-\Ltthrex)\cdot\rzxt} \,.
\end{align}

\noindent {\it Longitudinal polarization case}\\

For a longitudinally polarized photon, there is no instantaneous contribution since the Dirac structure vanishes $N_{\mathrm{SE}1,q\mathrm{inst}}^{\lambda=0} = 0$. In the regular piece, the Dirac structure in the numerator contains two terms with two different tensor structures, one proportional to $\delta^{ik}\Ltthrex^i\Lttwox^k$ and the other proportional to $\omega^{ik}\Ltthrex^i\Lttwox^k$. The latter vanishes since the integration over $\Ltthrex$ and $\Lttwox$ is proportional to $\rzxt^i\rzxt^k\omega^{ik}=0$. Therefore using Eq.~\eqref{eq:dijet-NLO-SE2-l+reg-2} and Eq.~\eqref{eq:dijet-NLO-SE2-dirac-reg-final}, we find the regular contribution to be
\begin{align}
    \Ncal^{\lambda=0,\sigma\sigma'}_{\mathrm{SE}1,\mathrm{reg}}\!\! &=  \! \frac{-g^2}{2} \! \int_0^{z_q} \! \frac{ \der z_g}{z_g}4z_qz_{\bar q}Q\left[1-\frac{z_g}{z_q}+\left(1-\frac{\varepsilon}{2}\right)\frac{z_g^2}{2z_q^2}\right]\frac{[\bar{u}(k,\sigma)\gamma^-v(p,\sigma')]}{q^-}\delta^{ik}e^{-i\frac{z_g}{z_q}\kt \cdot \rzxt}\nonumber\\
    &\times\left\{ \mu^{\varepsilon}\int \frac{\der^{2-\varepsilon} \ltone}{(2\pi)^{2-\varepsilon}}\frac{e^{i\ltone \cdot \left(\rxyt+\frac{z_g}{z_q} \rzxt \right)}}{ \ltone^2 +  \bar{Q}^2}\times\mu^\varepsilon\int
    \frac{\der^{2-\varepsilon} \Lttwox}{(2\pi)^{2-\varepsilon}}\frac{\Lttwox^i e^{i\Lttwox\cdot\rzxt}}{\omega_{\mathrm{SE}1}\left( \ltone^2 +\bar{Q}^2 \right)+\Lttwox^2}\right.\nonumber\\
    &\left.\times \mu^\varepsilon\int\frac{\der^{2-\varepsilon} \Ltthrex}{(2\pi)^{2-\varepsilon}}\frac{\Ltthrex^ke^{-i\Ltthrex \cdot \rzxt}}{\Ltthrex^2}\right\}\,.
    \label{eq:dijet-NLO-SE2-transverse-int1}
\end{align}
Using the formulas Eq.~\eqref{eq:Transverse_eps_2} and Eq.~\eqref{eq:Transverse_eps_3} from Appendix~\ref{app:transverse-int}, one can perform the transverse integrals in $2-\varepsilon$ dimensions with the result 
\begin{align}
    \Ncal^{\lambda=0,\sigma\sigma'}_{\mathrm{SE}1,\mathrm{reg}} &=   -\frac{\alpha_s}{\pi^2} \int_0^{z_q}\frac{ \der z_g}{z_g}z_qz_{\bar q}Q\left[1-\frac{z_g}{z_q}+\left(1-\frac{\varepsilon}{2}\right)\frac{z_g^2}{2z_q^2}\right]\frac{[\bar{u}(k,\sigma)\gamma^-v(p,\sigma')]}{q^-}\nonumber\\
    &\times\mu^{\varepsilon}\frac{-\varepsilon\Gamma\left(-\frac{\varepsilon}{2}\right)}{2^{2+\varepsilon/2}(2\pi)^{-3\varepsilon/2}}\frac{(\mu^2\rzxt^2)^\varepsilon}{\rzxt^2}e^{-i\frac{z_g}{z_q}\kt \cdot \rzxt}\nonumber\\
    &\times\int_0^{\infty}\frac{\der s}{s^{1-\varepsilon/2}}e^{-s\bar{Q}^2}\exp\left[-\frac{\left(\rxyt+\frac{z_g}{z_q}\rzxt \right)^2}{4s}\right]\Gamma\left(1-\frac{\varepsilon}{2},\frac{\omega_{\mathrm{SE}1}\rzxt^2}{4s}\right)\,,
    \label{eq:dijet-NLO-SE1-perturbative-final}
    \end{align}
with $\Gamma(a,x)$ the incomplete gamma function defined by
\begin{equation}
    \Gamma(a,x)=\int_x^\infty\der t \ t^{a-1}e^{-t}\,,
\end{equation}
which is related to the gamma function by
\begin{align}
    \Gamma(a)= \Gamma(a,0) \,.
\end{align}
The expression in Eq.~\eqref{eq:dijet-NLO-SE1-perturbative-final} is our final result for the regular perturbative factor in the longitudinally polarized case, specifying the exact dependence on $\varepsilon$.\\ 

\noindent {\it Transverse polarization case}\\

We turn now to the transverse momentum integration for a transversely polarized virtual photon. We need to consider the instantaneous contribution in this case. As it is free of UV divergences, we can directly set $\varepsilon=0$ and perform the transverse integration in two dimensions. Inserting Eq.~\eqref{eq:dijet-NLO-SE2-diracinst} and \eqref{eq:dijet-NLO-SE2-l+qbarins-2} inside \eqref{eq:dijet-NLO-SE2-pert-qins}, one finds
\begin{align}
    &\Ncal^{\lambda=\pm1,\sigma\sigma'}_{\mathrm{SE}1,q \rm inst} =   \frac{g^2}{2} \int_0^{z_q}\der z_g\frac{(z_g-z_q)^2}{z_q^3}\et^{\lambda,i}\left\{\bar{u}(k,\sigma)(1-\lambda\Omega)\frac{\gamma^-}{q^-}v(p,\sigma')\right\}e^{-i\frac{z_g}{z_q}\kt \cdot \rzxt}\nonumber\\
    &\times\left\{ \int \frac{\der^{2} \ltone}{(2\pi)^{2}}\int
    \frac{\der^{2} \Lttwox}{(2\pi)^{2}}\frac{e^{i\ltone \cdot \left(\rxyt+\frac{z_g}{z_q}\rzxt \right)} e^{i\Lttwox\cdot\rzxt}}{\omega_{\mathrm{SE}1}\left( \ltone^2 +\bar{Q}^2 \right)+\Lttwox^2}\int\frac{\der^{2} \Ltthrex}{(2\pi)^{2}}\frac{\Ltthrex^ie^{-i\Ltthrex \cdot \rzxt}}{\Ltthrex^2}\right\}\,.
    \label{eq:dijet-NLO-SE1-transverse-inst}
\end{align}
Using the formula Eq.~\eqref{eq:dijet-NLO-transint-inst} in Appendix~\ref{app:transverse-int}, this integral can be expressed in terms of modified Bessel functions as
\begin{align}
    \Ncal^{\lambda=\pm1,\sigma\sigma'}_{\mathrm{SE}1,q \rm inst}\!\! &=-\frac{\alpha_s}{\pi^2}\int_0^{z_q}\der z_g \ \frac{i(z_g-z_q)^2}{4z_q^3}e^{-i\frac{z_g}{z_q}\kt \cdot \rzxt}\frac{\bar{Q}}{\sqrt{\RtS^2+\omega_{\mathrm{SE}1}\rzxt^2}}K_1\left(\bar{Q}\sqrt{\RtS^2+\omega_{\mathrm{SE}1}\rzxt^2}\right)\nonumber\\
    &\times \frac{\rzxt\cdot\et^\lambda}{\rzxt^2}\left\{\bar{u}(k,\sigma)(1-\lambda\Omega)\frac{\gamma^-}{q^-} v(p,\sigma')\right\}\,,
\end{align}
where $\RtS=\rxyt+\frac{z_g}{z_q}\rzxt$, is the size of the $q\bar q$ dipole before the emission of the virtual gluon.
Performing the contraction over spinors, this expression can be further simplified to read,
\begin{align}
    \Ncal^{\lambda=\pm1,\sigma\sigma'}_{\mathrm{SE}1,q\rm inst}&=-\frac{\alpha_s}{\pi^2}\int_0^{z_q}\der z_g \ e^{-i\frac{z_g}{z_q}\kt \cdot \rzxt}\frac{(z_g-z_q)^2z_{\bar q}}{z_q^2}\delta^{\sigma,-\sigma'}\delta^{\sigma,\lambda}\frac{\et^\lambda\cdot\rzxt}{\rzxt^2}\frac{i\bar QK_1\left( QX_{\rm V}\right)}{X_{\rm V}}\,,
\end{align}
with
\begin{equation}
    X_{\rm V}^2=z_{\bar q}(z_q-z_g)\rxyt^2+z_g(z_q-z_g)\rzxt^2+z_{\bar q}z_g\rzyt^2\,.
\end{equation}
The parameter $X_{\rm V}$ has a geometric interpretation similar to $X_{\rm R}$ (see for example  Eq.~\eqref{eq:dijet-NLO-R2-XR-def}) in diagram $\mathrm{R}1$. It is the effective transverse size of the virtual $q\bar q g$ dipole when it crosses the shock wave.

We end this subsection with the computation of the regular perturbative factor for the  transversely polarized photon. As in the longitudinal case, the term proportional to $\Ltthrex^i\Lttwox^k\omega^{ik}$ does not contribute. Proceeding then similarly, and employing the integral Eq.~\eqref{eq:Transverse_eps_4} in Appendix~\ref{app:transverse-int}, one obtains 
\begin{align}
    \Ncal^{\lambda=\pm1,\sigma\sigma'}_{\mathrm{SE}1,\rm reg}&=\frac{\alpha_s}{\pi^2}\int_0^{z_q}\frac{\der z_g}{z_g}\left[1-\frac{z_g}{z_q}+\left(1-\frac{\varepsilon}{2}\right)\frac{z_g^2}{2z_q^2}\right]\mu^\varepsilon\frac{-i\varepsilon\Gamma\left(-\frac{\varepsilon}{2}\right)}{2^{4+\varepsilon/2}(2\pi)^{-3\varepsilon/2}}\nonumber\\
    &\times e^{-i\frac{z_g}{z_q}\kt \cdot \rzxt}\frac{(\mu^2\rzxt^2)^\varepsilon}{\rzxt^2}\int_0^\infty\frac{\der s}{s^{2-\varepsilon/2}}e^{-s\bar{Q}^2}e^{-\frac{\RtS^2}{4s}}\Gamma\left(1-\frac{\varepsilon}{2},\frac{\omega_{\mathrm{SE}1}\rzxt^2}{4s}\right)\nonumber\\
    &\times\et^{\lambda,l}\RtS^m\left\{\bar{u}(k,\sigma)[(z_{\bar q}-z_q)\delta^{lm}+\omega^{lm}]\frac{\gamma^-}{q^-}v(p,\sigma')\right\}\,.
    \label{eq:dijet-NLO-SE1-transverse-reg}
\end{align}
This concludes the computation of the regular perturbative factors in $4-\varepsilon$ dimensions.

\subsubsection*{UV divergent and finite pieces}

The perturbative factor $\Ncal_{\mathrm{SE}1}$ is convergent in 4 dimensions ($\varepsilon=0$). Yet the amplitude $\Mcal_{\mathrm{SE}1}$ is UV divergent. This divergence appears within the $\zt$ integral because of the $1/\rzxt^2$ factor. In this section, we will first extract the UV pole and then choose a suitable subtraction term which enables one to express the finite piece in a compact way. To illustrate the method, we choose to focus on the longitudinally polarized case; the extension to a transversely polarized incoming photon is  straightforward.

In order to isolate the finite term, we use a UV divergent subtraction term which captures the leading singularity as $\rzxt\to0$, or equivalently, as $\zt\to\xt$. In dimensional regularization, this singularity becomes a $1/\varepsilon$ pole. The UV singular part of the self energy crossing the shock wave is unique up to finite terms. We thus have the freedom to choose the UV divergent piece of the diagram in several ways. In mathematical terms, this can be  expressed as 
\begin{equation}
    \Mcal_{\mathrm{SE}1}=\underbrace{\mathcal{M}_{\mathrm{SE}1}-\mathcal{M}_{\mathrm{SE}1,\rm UV}}_{\rm finite}+\mathcal{M}_{\mathrm{SE}1,\rm UV} \,,
\end{equation}
with $\mathcal{M}_{\mathrm{SE}1,\rm UV}$ 
chosen in such a way that the first two terms give a convergent $\zt$ integral and such that the integration over $\zt$ in the UV divergent term can be computed analytically in $2-\varepsilon$ dimensions. In order to simplify the discussion of the slow gluon limit of the dressed self energy, an additional requirement is that the UV subtraction term should subtract the UV divergence for all values of $z_g$ without bringing an additional infrared singularity.

Following \cite{Hanninen:2017ddy}, we present one possible choice of the UV subtraction term which satisfies these conditions. When $\zt$ and $\xt$ are close to each other, the color structure of the self energy crossing the shock wave reduces to the color structure of the free self energies thanks to the unitarity of Wilson lines. We thus replace $\Ccal_{\mathrm{SE}1}$ by $C_F\,\Ccal_{\rm LO}$ inside the amplitude. Considering also Eq.~\eqref{eq:dijet-NLO-SE1-perturbative-final} in the limit $\rzxt\to0$, we can approximate
\begin{equation}
    e^{-i\frac{z_g}{z_q}\kt \cdot \rzxt} e^{-\frac{\RtS^2}{4s}}\Gamma\left(1-\frac{\varepsilon}{2},\frac{\omega_{\mathrm{SE}1}\rzxt^2}{4s}\right)\simeq e^{-\frac{\rxyt^2}{4s}}\Gamma\left(1-\frac{\varepsilon}{2}\right)e^{-\frac{\rzxt^2}{2\xi}}\,.
\end{equation}
The last exponential factor is harmless in the $\rzxt\to0$ limit, but ensures that no infrared singularity is introduced as $\rzxt^2\to\infty$. It depends on a parameter $\xi$ which will be fixed later. 

The freedom of choosing $\xi$ simply reflects the fact that the UV subtraction term is not unique; the change in $\mathcal{M}_{\mathrm{SE}1}-\mathcal{M}_{\mathrm{SE}1,\rm UV}$ induced by a variation of $\xi$ is compensated for by the change of the finite component of $\mathcal{M}_{\mathrm{SE}1,\rm UV}$. Note that using a $z_g$ independent $\xi$ is a sufficient condition for this Gaussian factor to cut off the infrared region of the $\zt$ integral for all $z_g$ values. To sum up, our choice for $\Mcal_{\mathrm{SE}1,\mathrm{UV}}$ is
\begin{align}
    &\mathcal{M}^{\lambda=0,\sigma\sigma'}_{\mathrm{SE}1,\rm UV}= \frac{ee_fq^-}{\pi}\mu^{-2\varepsilon}\int\der^{2-\varepsilon}\xt\int\der^{2-\varepsilon}\yt e^{-i\kt \cdot \xt-i\pt \cdot \yt}\Ccal_{\rm LO}(\xt,\yt)\nonumber\\
    &\times\frac{-\alpha_sC_F}{\pi^2}\int_0^{z_q}\frac{ \der z_g}{z_g}z_qz_{\bar q}Q\left[1-\frac{z_g}{z_q}+\left(1-\frac{\varepsilon}{2}\right)\frac{z_g^2}{2z_q^2}\right]\frac{[\bar{u}(k,\sigma)\gamma^-v(p,\sigma')]}{q^-}\frac{-\varepsilon\Gamma\left(-\frac{\varepsilon}{2}\right)}{2^{2+\varepsilon/2}(2\pi)^{-3\varepsilon/2}}\nonumber\\
    &\times\int\der^{2-\varepsilon}\rzxt\int_0^{\infty}\frac{\der s}{s^{1-\varepsilon/2}}e^{-s\bar{Q}^2}e^{-\frac{\rxyt^2}{4s}}\Gamma\left(1-\frac{\varepsilon}{2}\right)\exp\left(-\frac{\rzxt^2}{2\xi}\right)\frac{(\mu^2\rzxt^2)^\varepsilon}{\rzxt^2}\,.
\end{align}
It is possible to perform the $\rzxt$ integral in $2-\varepsilon$ dimensions, as well as the integral over the Schwinger parameter $s$, leading to
\begin{align}
    \mathcal{M}^{\lambda=0\sigma\sigma'}_{\mathrm{SE}1,\rm UV}    &=\frac{ee_fq^-}{\pi}\mu^{-2\varepsilon}\int\der^{2-\varepsilon}\xt\der^{2-\varepsilon}\yt e^{-i\kt \cdot \xt-i\pt \cdot \yt}\Ccal_{\rm LO}(\xt,\yt)\Ncal^{\lambda=0,\sigma\sigma'}_{\rm LO,\varepsilon}(\rxyt)\nonumber\\
    & \times\frac{\alpha_sC_F}{2\pi}\left\{\left(2\ln\left(\frac{z_q}{z_0}\right)-\frac{3}{2}\right)\left(\frac{2}{\varepsilon}+\ln(2\pi\mu^2\xi)\right)-\frac{1}{2}+\mathcal{O}(\varepsilon)\right\}\,,
\label{eq:dijet-NLO-SE2-UVfinal-long}
\end{align}
with $\varepsilon>0$.
In this expression, the leading order perturbative factor computed in Sec.~\ref{sub:LO} is generalized to $4-\varepsilon$ dimension using the integral Eq.\,\eqref{eq:dijet-LO-wf-dimreg-long},
\begin{align}
    \Ncal_{\rm LO,\varepsilon}^{\lambda=0,\sigma\sigma'}(\rxyt)=-z_qz_{\bar q}Q\left(\frac{\bar Q}{2\pi r_{xy} \mu^2}\right)^{-\varepsilon/2}K_{-\varepsilon/2}\left(\bar{Q}r_{xy} \right)\frac{[\bar{u}(k,\sigma)\gamma^- v(p,\sigma')]}{q^-}\,,
    \label{eq:dijet-NLO-LOddim}
\end{align}
which factors it out from the terms inside the curly bracket. Writing the result in this way will enable us to demonstrate that the finite (rational) terms coming from the product between the $\mathcal{O}(\varepsilon)$ term in the expansion of Eq.\,\eqref{eq:dijet-NLO-LOddim} and the $\varepsilon$ pole cancel at cross-section level. 

Finally, since the difference $\mathcal{M}_{\mathrm{SE}1}-\mathcal{M}_{\mathrm{SE}1,\rm UV}$ is UV finite, one can freely take the limit $\varepsilon\to0$. Remarkably, one can find an analytic expression for the $s$ integral in Eq.\,\eqref{eq:dijet-NLO-SE1-perturbative-final} when $\varepsilon=0$ (see the discussion of the formula Eq.~\eqref{eq:Transverse_int_4} in Appendix \ref{app:transverse-int}). Hence the finite piece of the dressed self energy within our UV subtraction scheme reads,
\begin{align}
    \left.\mathcal{M}^{\lambda=0,\sigma\sigma'}_{\mathrm{SE}1}\right|_{\rm UV-fin.} & =\frac{ee_fq^-}{\pi}\frac{\alpha_s}{\pi^2}\int \der^2\xt\der^2\yt e^{-i\kt \cdot \xt-i\pt \cdot \yt}\int_0^{z_q}\frac{\der z_g}{z_g}(-2z_qz_{\bar q})\bar Q\delta^{\sigma,-\sigma'}\nonumber\\
    & \times\left[1-\frac{z_g}{z_q}+\frac{z_g^2}{2z_q^2}\right]
    \int \frac{\der^2\zt }{\rzxt^2}\left\{e^{-i\frac{z_g}{z_q}\kt \cdot \rzxt}K_0\left(QX_{\rm V}\right)\Ccal_{\mathrm{SE}1}(\xt,\yt,\zt)\right.\nonumber\\
    &\left.-e^{-\frac{\rzxt^2}{2\xi}}K_0\left(\bar{Q} r_{xy}\right)C_F\Ccal_{\rm LO}(\xt,\yt)\right\}\,.
  \label{eq:dijet-NLO-SE2-finite_term_2}
    \end{align}
For a transversely polarized virtual photon (see Eq.\,\ref{eq:dijet-NLO-SE1-transverse-reg}), one finds similarly,
\begin{align}
    \mathcal{M}^{\lambda=\pm1,\sigma\sigma'}_{\mathrm{SE}1,\rm UV}    &=\frac{ee_fq^-}{\pi}\mu^{-2\varepsilon}\int\der^{2-\varepsilon}\xt\der^{2-\varepsilon}\yt e^{-i\kt \cdot \xt-i\pt \cdot \yt}\Ccal_{\rm LO}(\xt,\yt)\Ncal^{\lambda=\pm1,\sigma\sigma'}_{\rm LO,\varepsilon}(\rxyt)\nonumber\\
    & \times\frac{\alpha_s C_F}{2\pi}\left\{\left(2\ln\left(\frac{z_q}{z_0}\right)-\frac{3}{2}\right)\left(\frac{2}{\varepsilon}+\ln(2\pi\mu^2\xi)\right)-\frac{1}{2}+\mathcal{O}(\varepsilon)\right\}\,,
   \label{eq:dijet-NLO-SE2-UVfinal-trans}
\end{align}
with the LO perturbative factor in $4-\varepsilon$ dimensions given by
\begin{align}
    \Ncal^{\lambda=\pm1,\sigma\sigma'}_{\rm LO,\varepsilon}(\rxyt)&=\frac{i}{2}\frac{\bar{Q}\et^{\lambda,i}\rxyt^j}{r_{xy}}\left(\frac{\bar Q}{2\pi r_{xy}\mu^2}\right)^{-\varepsilon/2}\!\!\!\!\! K_{1-\varepsilon/2}(\bar{Q}r_{xy})\nonumber\\
    &\times\frac{[\bar{u}(k,\sigma)((z_{\bar q}-z_q)\delta^{ij}+\omega^{ij})\gamma^-v(p,\sigma')]}{q^-}\,.
\end{align}
Given that there is an ambiguity in the analytic continuation of the Levi-Civita tensor to $4-\varepsilon$ dimensions, we restricted ourselves to using the identity in  Eq.~\eqref{eq:app-B-gigj} to simplify the Dirac structure. In the end, the limit $\varepsilon\to0$ exists so this perturbative factor can be evaluated in 4 dimensions.
The UV finite piece then is given by 
\begin{align}
    &\left.\mathcal{M}^{\lambda=\pm1,\sigma\sigma'}_{\mathrm{SE}1}\right|_{\rm UV-fin.}=\nonumber\\
    &\frac{ee_fq^-}{\pi}\frac{\alpha_s}{\pi^2}\int \der^2\xt\der^2\yt e^{-i\kt \cdot \xt-i\pt \cdot \yt}\left[2z_qz_{\bar q}\Gamma_{\gamma^*_\mathrm{T} \to q\bar{q}}^{\sigma,\lambda}(z_q,z_{\bar q})\delta^{\sigma,-\sigma'}\right]\int_0^{z_q}\frac{\der z_g}{z_g}\left[1-\frac{z_g}{z_q}+\frac{z_g^2}{2z_q^2}\right]\nonumber\\
    &\times\int\frac{\der^2\zt}{\rzxt^2}\left\{e^{-i\frac{z_g}{z_q}\kt \cdot \rzxt}\frac{i\bar Q\et^\lambda\cdot\RtS}{X_{\rm V}}K_1\left(QX_{\rm V}\right)\Ccal_{\mathrm{SE}1}-e^{-\frac{\rzxt^2}{2\xi}}\frac{iQ\et^\lambda\cdot\rxyt}{r_{xy}}K_1(\bar Qr_{xy})C_F\Ccal_{\rm LO}\right\}\nonumber\\
    &-\frac{ee_fq^-}{\pi}\frac{\alpha_s}{\pi^2}\int \der^2\xt\der^2\yt e^{-i\kt \cdot \xt-i\pt \cdot \yt}\left[\delta^{\sigma,-\sigma'}\delta^{\sigma,\lambda}\right]\int_0^{z_q}\der z_g \ \frac{(z_g-z_q)^2z_{\bar q}}{z_q^2}\nonumber\\
    &\times\int\der^2\zt e^{-i\frac{z_g}{z_q}\kt \cdot \rzxt}\frac{\et^\lambda\cdot\rzxt}{\rzxt^2}\frac{i\bar QK_1\left( QX_{\rm V}\right)}{X_{\rm V}}\Ccal_{\rm SE1}\,,
    \label{eq:dijet-NLO-SE2-finite_term_2-trans}
\end{align}
where we have omitted the transverse coordinate dependencies of the color structures $\Ccal_{\mathrm{SE}1}$ and $\Ccal_{\rm LO}$ for compactness.

\subsubsection{Free gluon propagator}
We now consider the self energy diagrams with the free gluon propagator, either before or after the shock wave. As we shall see, the self energies after the shock wave vanish in dimensional regularization. 

\begin{figure}[tbh]
    \centering
    \includegraphics[width=0.48\textwidth]{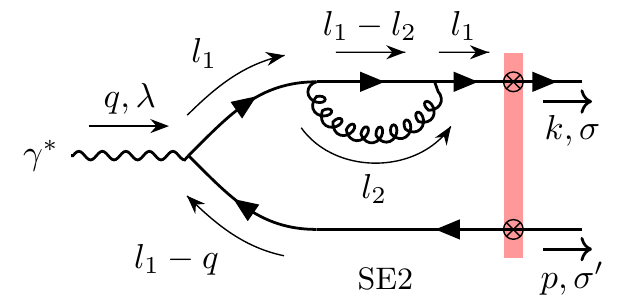}
    \hfill
    \includegraphics[width=0.48\textwidth]{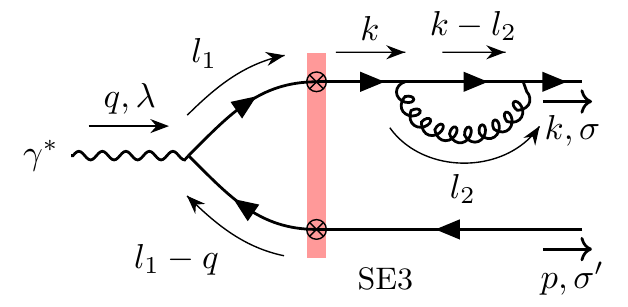}
    \caption{Free self energy before the shock wave (left) and after (right).}
    \label{fig:NLO-dijet-SE23}
\end{figure}

\subsubsection*{Self energy before the shock wave} We start with a brief discussion of the self energy before the shock wave, whose Feynman diagram is pictured in  Fig.~\ref{fig:NLO-dijet-SE23}-left. It is equal to the sum of all the self energy diagrams contributing to the light-front wavefunction of the $q\bar q$ Fock component inside an incoming virtual photon, \textit{including therefore the instantaneous quark and antiquark diagrams}. This calculation has been done previously in \cite{Beuf:2016wdz,Hanninen:2017ddy} within the LCPT framework. We rederived these results in standard covariant perturbation theory and demonstrated the equivalence between the two approaches. Intermediate steps are provided in Appendix~\ref{app:SE1}. 

After integration over the internal momenta $l_1$ and $l_2$ in $d=4-\varepsilon$ dimensions, the subtracted amplitude reads
\begin{align}
    \Mcal^{\lambda=0,\sigma\sigma'}_{\mathrm{SE}2}&=\frac{ee_fq^-}{\pi}\mu^{-2\varepsilon}\int\der^{2-\varepsilon}\xt\der^{2-\varepsilon}\yt e^{-i\kt \cdot \xt-i\pt \cdot \yt}\Ccal_{\rm LO}(\xt,\yt)\Ncal^{\lambda=0,\sigma\sigma'}_{\rm LO,\varepsilon}(\rxyt)\nonumber\\
    &\times\frac{\alpha_sC_F}{2\pi}\left\{\left(-2\ln\left(\frac{z_q}{z_0}\right)+\frac{3}{2}\right)\left(\frac{2}{\varepsilon}+\frac{1}{2}\ln\left(\frac{\bar{Q}^2\rxyt^2}{4}\right)+\gamma_E-\ln\left(\frac{z_qQ^2}{\tilde\mu^2}\right)\right)\right.\nonumber\\
    &\left.+\left(\frac{1}{2}+3-\frac{\pi^2}{3}-\ln^2\left(\frac{z_q}{z_0}\right)\right)+\mathcal{O}(\varepsilon)\right\}\,,
    \label{eq:dijet-NLO-SE2-final}
\end{align}
for a longitudinally polarized virtual photon, and
\begin{align}
    &\Mcal^{\lambda=\pm1,\sigma\sigma'}_{\mathrm{SE}2}=\frac{ee_fq^-}{\pi}\mu^{-2\varepsilon}\int\der^{2-\varepsilon}\xt\der^{2-\varepsilon}\yt e^{-i\kt \cdot \xt-i\pt \cdot \yt}\Ccal_{\rm LO}(\xt,\yt)\Ncal^{\lambda=\pm1,\sigma\sigma'}_{\rm LO,\varepsilon}(\rxyt)\nonumber\\
    &\times\frac{\alpha_sC_F}{2\pi}\left\{\left(-2\ln\left(\frac{z_q}{z_0}\right)+\frac{3}{2}\right)\left(\frac{2}{\varepsilon}+\frac{1}{2}\ln\left(\frac{\bar{Q}^2\rxyt^2}{4}\right)+\gamma_E-\ln\left(\frac{z_qQ^2}{\tilde\mu^2}\right)-\frac{1}{r_{xy}\bar{Q}}\frac{K_0(\bar{Q}r_{xy})}{K_1(\bar{Q}r_{xy})}\right)\right.\nonumber\\
    &\left.+\left(\frac{1}{2}+3-\frac{\pi^2}{3}-\ln^2\left(\frac{z_q}{z_0}\right)\right)+\mathcal{O}(\varepsilon)\right\}\,,
    \label{eq:dijet-NLO-SE2-final-transverse}
\end{align}
for a transversely polarized virtual photon. In these formulas, we use the notation $\tilde\mu^2=4\pi e^{-\gamma_E}\mu^2$.

These expressions exhibit two types of divergences. The pole in $1/\varepsilon$ comes from the UV divergence of the $\lttwo$ integral and factorizes from the LO amplitude. There is also a slow gluon logarithmic divergence when the lower cut-off $z_0=\Lambda^-_0/q^-$ for the $z_g$ integration of the gluon goes to $0$. In the last line, we keep separate the finite $1/2$ term coming from the product between the $\mathcal{O}(\varepsilon)$ term in $N^{\lambda\sigma\sigma'}_{\mathrm{SE}2}$ and the $1/\varepsilon$ pole. Such a term arises in  dimensional regularization. We will check explicitly that all such finite terms cancel at the level of the cross-section.

When compared to the longitudinal case, there is an additional term in Eq.~\eqref{eq:dijet-NLO-SE2-final-transverse} which depends on $\rxyt$ via modified Bessel functions. At first sight, this term looks a little bit odd. Indeed, it contributes to the slow gluon logarithmic divergence and we expect this divergence to depend on the polarization of the virtual photon via the leading order perturbative factor only. We shall see that it cancels against a similar term in the vertex correction before the shock wave. One should thus interpret with caution the results for each individual diagram, since many nontrivial cancellations occur only once they are combined together.

\subsubsection{Self energy after the shock wave} 
\label{sec:SE-afterSW}

In the limit of massless quarks, the quark or antiquark self energy after the shock wave vanishes in dimensional regularization. It is nevertheless enlightening to understand this statement more deeply. When we sum  the virtual diagrams (self energy and vertex contributions) with the exception of the self energy after the shock wave, we observe that a UV divergence survives. This is at first glance surprising since typically one expects UV divergences to cancel\footnote{The physical reason for the cancellation of UV divergences is that (i) quarks are  treated as being massless,  and (ii) the quark electric charge is not affected by QCD corrections. As a result, UV renormalization is not required at this order in perturbation theory~\cite{Schwartz:2014sze}.} and only an IR divergence to remain. This IR divergence in turn is expected to cancel with the collinear divergence in the real emission cross-section after proper definition of an IR safe jet  observable.  

In this subsection, we will show that the self energy after the shock wave vanishes in dimensional regularization because it contains {\it both} a UV pole and an IR pole with the same prefactor but with opposite signs. Thus  one can  use this UV pole to cancel the surviving UV pole from the sum of the other virtual diagrams and one then ends up keeping the IR divergence, 
thereby resolving the apparent conundrum stated above and in line with expectations from perturbative QCD.

More concretely, the subtracted amplitude for $\mathrm{SE}3$ in $d=4-\varepsilon$ dimensions, and in the massless limit, is 
\begin{align}
    \Mcal^{\lambda\sigma\sigma'}_{\mathrm{SE}3}=\frac{ee_fq^-}{\pi}\mu^{-2\varepsilon}\int \der^{2-\varepsilon}\xt\der^{2-\varepsilon}\yt e^{-i\kt \cdot \xt-i\pt \cdot \yt}\Ccal_{\mathrm{SE}3}(\xt,\yt)\Ncal^{\lambda\sigma\sigma'}_{\mathrm{SE}3}(\rxyt)\,,
    \label{eq:SE-afterSW-amplitude}
\end{align}
with the color structure
\begin{equation}
    \Ccal_{\mathrm{SE}3}(\xt,\yt)=C_F\left[V(\xt)V^\dagger(\yt)-\mathbbm{1}\right]\,,
    \label{eq:SE-afterSW-color}
\end{equation}
and the perturbative factor

\begin{align}
    \Ncal^{\lambda\sigma\sigma'}_{\mathrm{SE}3}(\rxyt)&=-4\alpha_s\Ncal^{\lambda\sigma\sigma'}_{\rm LO,\varepsilon}(\rxyt)\int_0^{z_q}\frac{\der z_g}{z_g}\left[1-\frac{z_g}{z_q}+\left(1-\frac{\varepsilon}{2}\right)\frac{z_g^2}{2z_q^2}\right] \mu^\varepsilon\int\frac{\der^{2-\varepsilon}\Lttwox}{(2\pi)^{2-\varepsilon}}\frac{1}{\Lttwox^2}\,,
\end{align}
where $z_g=l_2^-/q^-$ is the longitudinal momentum fraction of the gluon inside the loop.
The remaining transverse momentum integral is both UV and IR divergent in $2$ dimensions. In dimensional regularization, one takes care of such integral by introducing an arbitrary scale $\Lambda$ to divide the UV and IR regions:
\begin{align}
    \mu^\varepsilon\int\frac{\der^{2-\varepsilon}\Lttwox}{(2\pi)^{2-\varepsilon}}\frac{1}{\Lttwox^2}&=\frac{(4\pi)^{\varepsilon/2}\mu^\varepsilon}{(2\pi)\Gamma\left(1-\frac{\varepsilon}{2}\right)}\left\{\int_0^\Lambda\frac{\der L_{2x\perp}}{L_{2x\perp}^{1-\varepsilon}}+\int_\Lambda^\infty\frac{\der L_{2x\perp}}{L_{2x\perp}^{1-\varepsilon}}\right\} \nonumber \\
    &=\frac{1}{4\pi}\left(\frac{2}{\varepsilon_{\rm UV}}-\frac{2}{\varepsilon_{\rm IR}}\right)+\mathcal{O}(\varepsilon) \,,
\end{align}
where $\varepsilon=\varepsilon_{\rm UV}>0$ in the UV divergent term and $\varepsilon=\varepsilon_{\rm IR}<0$ in the IR divergent one. Setting formally $\varepsilon_{\rm UV}=\varepsilon_{\rm IR}=\varepsilon$, one sees that this transverse momentum integral vanishes, meaning that the full perturbative factor for $\textrm{SE}_{3}$,
\begin{align}
    \Ncal^{\lambda\sigma\sigma'}_{\mathrm{SE}3}(\rxyt)&=-\frac{\alpha_s}{2\pi}\Ncal^{\lambda\sigma\sigma'}_{\rm LO,\varepsilon}(\rxyt)\left(\frac{2}{\varepsilon_{\rm UV}}-\frac{2}{\varepsilon_{\rm IR}}\right)\left\{2\ln\left(\frac{z_q}{z_0}\right)-\frac{3}{2}\right\}\,,
    \label{eq:dijet-NLO-SE3-final}
\end{align}
is identically zero. The price to pay is that the nature of the divergence, either infrared or ultraviolet, is lost when one takes the limit $\varepsilon_{\rm IR}=\varepsilon_{\rm UV}$. It also explains how the apparent UV pole that we will obtain at the end of our computation of all the virtual amplitudes can be ``turned into" an IR pole when combined with the self energy after the shock wave. 

The final result for this amplitude is given by
\begin{align}
    \Mcal^{\lambda\sigma\sigma'}_{\mathrm{SE}3}&=\frac{ee_fq^-}{\pi}\mu^{-2\varepsilon}\int \der^{2-\varepsilon}\xt\der^{2-\varepsilon}\yt e^{-i\kt \cdot \xt-i\pt \cdot \yt}\Ccal_{\mathrm{LO}}(\xt,\yt) \nonumber \\
    &\times -\frac{\alpha_s C_{\rm F}}{2\pi}\Ncal^{\lambda\sigma\sigma'}_{\rm LO,\varepsilon}(\rxyt)\left(\frac{2}{\varepsilon_{\rm UV}}-\frac{2}{\varepsilon_{\rm IR}}\right)\left\{2\ln\left(\frac{z_q}{z_0}\right)-\frac{3}{2}\right\} \,.
    \label{eq:SE-afterSW-amplitude-final}
\end{align}

\subsection{Vertex diagrams}
\label{sec:vertex_contributions}

We continue our computation of the virtual diagrams with the vertex corrections which are listed in the third line of Fig.~\ref{fig:NLO-dijet-all-diagrams}. As in the case of the self energy diagrams, the virtual gluon can interact with the shock wave or propagate freely. For the diagram with dressed gluon propagator, the final result for the amplitude is given in Eqs.~ \eqref{eq:dijet-NLO-MV3}-\eqref{eq:dijet-NLO-Ccal-V3}-\eqref{eq:dijet-NLO-virtual-V3-final}-\eqref{eq:dijet-NLO-V3-inst-final}-\eqref{eq:dijet-NLO-virtual-V3-final-trans}. The free vertex correction before and after shock wave are respectively given by Eqs.\,\eqref{eq:dijet-NLO-V1-final-long}-\eqref{eq:dijet-NLO-V1-final-trans} and Eqs.\,\eqref{eq:dijet-NLO-virtual-V2-pert-final-spin}-\eqref{eq:dijet-NLO-virtual-V2-pert-final-trans-spin}.

\subsubsection{Dressed gluon propagator}
    \label{sub:V3}
    \begin{figure}[tbh]
    \centering
    \includegraphics[width=0.5\textwidth]{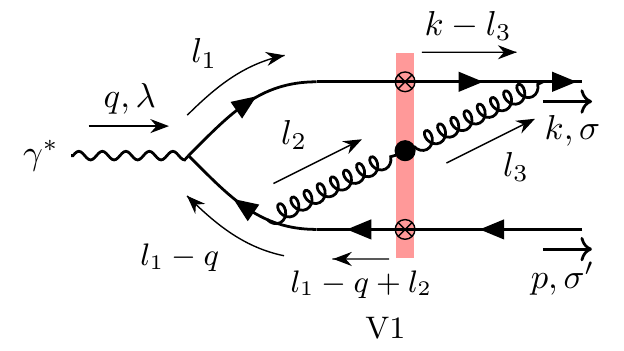}
    \caption{Vertex correction from the gluon crossing the shock wave.}
\label{fig:NLO-dijet-V1}
\end{figure}

We detail now the calculation of the vertex correction with the gluon crossing the shock wave and shall derive a compact expression for this diagram.  The scattering amplitude for this process is given by
\begin{align}
    S_{\mathrm{V}1}^{\lambda\sigma\sigma'} &=  \int \frac{\der^{4} l_1}{(2\pi)^{4}} \frac{\der^{4} l_2}{(2\pi)^{4}} \frac{\der^{4} l_3}{(2\pi)^{4}}  \left[\bar{u}(k,\sigma) (ig\gamma^\mu t^a) S^0(k-l_3) \Tcal^q(k-l_3,l_1) S^0(l_1) (-i e e_f \slashed{\epsilon}(q,\lambda)) \right. \nonumber \\
    &\left. \times S^0(l_1-q) (ig \gamma^\nu t^b) S^0(l_1-q+l_2) \Tcal^q(l_1-q+l_2,-p) v(p,\sigma')\right] \nonumber \\
    & \times G^{0,ac}_{\mu \rho}(l_3)\Tcal^{g,\rho \sigma}_{cd}(l_3,l_2)G^{0,db}_{\sigma \nu}(l_2)\,.
\end{align}
Note that we have written the amplitude now in $d=4$ dimensions; this is because, as we shall see, this diagram does not have ultraviolet divergences.
After subtraction of the noninteracting piece and factoring an overall delta function $2\pi \delta(q^- - p^- - k^-)$, we find the physical amplitude
\begin{align}
    \Mcal_{\mathrm{V}1}^{\lambda\sigma\sigma'} = \frac{e e_f q^-}{\pi}\int \der^{2} \xt \der^{2} \yt \der^{2} \zt e^{- i \kt \cdot \xt- i \pt \cdot \yt} \Ccal_{\mathrm{V}1}(\xt,\yt,\zt) \Ncal^{\lambda\sigma\sigma'}_{\mathrm{V}1}(\rxyt,\rzyt)\,,
    \label{eq:dijet-NLO-MV3}
\end{align}
with the color structure
\begin{align}
    \Ccal_{\mathrm{V}1}(\xt,\yt,\zt) & = \left[t^a V(\xt) t^b V^\dagger(\yt) U_{ab}(\zt) - t^at_a\right] \nonumber \\
    &= \left[ t^a V(\xt) V^\dagger(\zt) t_a V(\zt) V^\dagger(\yt)  -t^a t_a
    \right]\,,
    \label{eq:dijet-NLO-Ccal-V3}
\end{align}
and the perturbative factor
\begin{align}
    \Ncal^{\lambda\sigma\sigma'}_{\mathrm{V}1}&=  \frac{g^2}{(2q^-)}  \int \frac{\der^{4} l_1}{(2\pi)^{3}}
    \frac{\der^{4} l_2}{(2\pi)^{3}}\frac{\der^{4} l_3}{(2\pi)^{3}} e^{i \ltone \cdot \rxyt}e^{i\lttwo \cdot \rzyt}e^{-i\ltthre \cdot \rzxt}\nonumber\\
    &\times\frac{ -i  (2q^-)^2 (2l_3^-) \delta(l_3^- - l_2^-)\delta(k^- - l_3^- - l_1^-) N^{\lambda\sigma\sigma'}_{\mathrm{V}1}(l_1,l_2,l_3)}{\left[(l_3-k)^2 + i \epsilon \right] \left[l_1^2 + i \epsilon \right] \left[(l_1-q)^2 + i \epsilon \right]\left[(l_2-q+l_1)^2 + i \epsilon \right] \left[l_2^2 + i \epsilon \right]\left[l_3^2 + i \epsilon \right]}\,,
    \label{eq:dijet-NLO-V3-pert}
\end{align}
where the Dirac structure is given by
\begin{align}
    N_{\mathrm{V}1}^{\lambda\sigma\sigma'} &= \frac{1}{(2q^-)^2}\left[\bar{u}(k,\sigma) \gamma^\mu (\slashed{k}-\slashed{l}_3) \gamma^- \slashed{l}_1  \slashed{\epsilon}(q,\lambda) (\slashed{l}_1 - \slashed{q} ) \gamma_\nu (\slashed{l}_1 - \slashed{q} + \slashed{l}_2) \gamma^- v(p,\sigma') \right]\nonumber\\
    &\times\Pi_{\mu\rho}(l_3) \Pi^{\rho\nu} (l_2)\,.
    \label{eq:dijet-NLO-V3-dirac-1}
\end{align}

\subsubsection*{Dirac structure}
As usual, the integration over $l_1^-$ and $l_2^-$ can be easily done with the delta functions that enforce $l_2^- = l_3^-$ and $l_1^- = k^- - l_3^-$. Using the identities in Eqs.\,\eqref{eq:gluon_tensor_squared_decomp}\, \eqref{eq:gluon_abs_quark_afterSW} and \eqref{eq:gluon_emi_antiquark_beforeSW} from Appendix~\ref{app:dirac} we can express the Dirac structure in Eq.\,\eqref{eq:dijet-NLO-V3-dirac-1} as
\begin{equation}
    N_{\mathrm{V}1} = N_{\mathrm{V}1,\mathrm{reg}} +  (l_1 -q)^2 N_{\mathrm{V}1,\bar{q}\mathrm{inst}}\,,
    \label{eq:dijet-NLO-V3-dirac-2}\\
\end{equation}
with
\begin{align}
    N_{\mathrm{V}1,\mathrm{reg}}^{\lambda\sigma\sigma'} &= \frac{4\Ltthrex^i \Lttwoy^k}{xy}\left\{\bar{u}(k,\sigma) \left[\left(1-\frac{x}{2}\right)\delta^{ij}-i\frac{x}{2}\epsilon^{ij} \Omega \right] \mathcal{D}_{\rm LO}^\lambda(l_1) \right.\nonumber\\
    &\times \left.\left[\left(1-\frac{y}{2}\right) \delta^{kj}-i\frac{y}{2}\epsilon^{kj}\Omega \right] v(p,\sigma') \right\}\,, \label{eq:dijet-NLO-V3-dirac-reg}\\
    N_{\mathrm{V}1,\bar{q}\mathrm{inst}}^{\lambda\sigma\sigma'} &= \frac{(1-y)}{2x(q^-)^2} \left\{\bar{u}(k,\sigma) \left[\left(1-\frac{x}{2}\right)\delta^{ik}-i\frac{x}{2}\epsilon^{ik}\Omega\right] \gamma^- \slashed{l}_1 \slashed{\epsilon}(q,\lambda) \gamma^- \gamma^k v(p,\sigma')  \right\} \Ltthrex^i\,,
    \label{eq:dijet-NLO-V3-dirac-qbarins}
\end{align}
where $x= z_g/z_q$, $\Ltthrex^i = \ltthre^i - x \kt^i$, $y = z_g/(z_{\bar{q}}+z_g)$ and $\Lttwoy^i = \lttwo^i + y \ltone^i $.
In the longitudinally polarized case, the instantaneous piece in Eq.\,\eqref{eq:dijet-NLO-V3-dirac-qbarins} vanishes, using again$(\gamma^-)^2=0$,
\begin{align}
    N_{\mathrm{V}1,\bar{q}\mathrm{inst}}^{\lambda=0,\sigma\sigma'} = 0\,.
    \label{eq:dijet-NLO-V3-dirac-qbarins-long}
\end{align}
For the regular piece in Eq.\,\eqref{eq:dijet-NLO-V3-dirac-reg} one obtains, after some Dirac algebra,
\begin{align}
    N_{\mathrm{V}1,\mathrm{reg}}^{\lambda=0,\sigma\sigma'}&=-  \frac{4 z_q (z_{\bar{q}}+ z_g)^2 (z_q - z_g) Q}{z_g^2} \left\{ \left[1  - \frac{z_g}{2z_q} - \frac{z_g}{2(z_{\bar{q}}+z_g)} \right] \frac{\left[\bar{u}(k,\sigma) \gamma^- v(p,\sigma') \right]}{q^-} \delta^{ik}\right.\nonumber\\
    &\left.-i\left[\frac{z_g}{2z_q} - \frac{z_g}{2(z_{\bar{q}}+z_g)} \right] \frac{\left[\bar{u}(k,\sigma) \gamma^- \Omega v(p,\sigma') \right]}{q^-} \epsilon^{ik}\right\}\Ltthrex^i \Lttwoy^k\,.
    \label{eq:dijet-NLO-V3-dirac-reg-long}
\end{align}
The transverse polarization case is worked out similarly, giving
\begin{align}
       N^{\lambda=\pm1,\sigma\sigma'}_{\mathrm{V}1,\bar q \rm inst}  & = -\frac{z_qz_{\bar q}(z_q-z_g)}{z_g(z_g+z_{\bar q})}\left\{\bar{u}(k,\sigma)(1+\lambda\Omega)\frac{\gamma^-}{q^-}v(p,\sigma') \right\} (\Ltthrex \cdot \et^{\lambda})\,,
       \label{eq:dijet-NLO-V3-Diracins}
\end{align}
for the instantaneous antiquark piece, and 
\begin{align}
    N^{\lambda=\pm1,\sigma\sigma'}_{\mathrm{V}1,\mathrm{reg}}
   & =\frac{2 z_q(z_{\bar{q}}+z_g)}{z_g^2} \Ltthrex^i \Lttwoy^k  (\ltone \cdot \et^\lambda)\nonumber\\
   &\times\left \{ \left[1-\frac{z_g}{2z_q} -\frac{z_g}{2(z_{\bar{q}}+z_g)} \right] \left[\bar{u}(k,\sigma)((1-2z_q+2z_g)-\lambda\Omega)\frac{\gamma^-}{q^-}v(p,\sigma')\right] \delta^{ik} \right. \nonumber \\
    &\left. -i \left[ \frac{z_g}{2 z_q} -\frac{z_g}{2(z_{\bar{q}} + z_g)} \right] \left[ \bar{u}(k,\sigma)((1-2z_q+2z_g)\Omega-\lambda)\frac{\gamma^-}{q^-}v(p,\sigma')
    \right] \epsilon^{ik} \right \} \,,
\end{align}
for the regular Dirac piece.

\subsubsection*{Pole structure of the instantaneous and regular terms}

The Dirac numerator does not depend on $l_1^+$, $l_2^+$ and $l_3^+$. One can then perform the ``plus" ligthcone momentum integration using Cauchy's theorem.
Using the decomposition in Eq.\,\eqref{eq:dijet-NLO-V3-dirac-2} we can express Eq.\,\eqref{eq:dijet-NLO-V3-pert} as 
\begin{equation}
    \Ncal_{\mathrm{V}1}= \Ncal_{\mathrm{V}1,\mathrm{reg}} + \Ncal_{\mathrm{V}1,\bar{q} \rm inst}\,,
\end{equation}
with 
\begin{align}
    \Ncal_{\mathrm{V}1,\mathrm{reg}} &=   \frac{g^2}{2} \int \der z_g\int \frac{\der^{2} \ltone}{(2\pi)^{2}}
    \frac{\der^{2} \lttwo}{(2\pi)^{2}}\frac{\der^{2} \ltthre}{(2\pi)^{2}}  e^{i \ltone \cdot \rxyt+i\lttwo \cdot \rzyt-i\ltthre \cdot \rzxt} \Ical_{\mathrm{V}1,\mathrm{reg}} N_{\mathrm{V}1,\mathrm{reg}}\,,
    \label{eq:dijet-NLO-V3-pert-reg}\\
    \Ncal_{\mathrm{V}1,\bar{q}\mathrm{inst}} &=   \frac{g^2}{2}  \int \der z_g\int \frac{\der^{2} \ltone}{(2\pi)^{2}}
    \frac{\der^{2} \lttwo}{(2\pi)^{2}}\frac{\der^{2} \ltthre}{(2\pi)^{2}}  e^{i \ltone \cdot \rxyt+i\lttwo \cdot \rzyt-i\ltthre \cdot \rzxt} \Ical_{\mathrm{V}1,\bar{q}\mathrm{inst}} N_{\mathrm{V}1,\bar{q} \mathrm{inst}} \,,
    \label{eq:dijet-NLO-V3-pert-qbarins}
\end{align}
where we encounter the following pole structures (omitting again the $+i\epsilon$ prescription for the propagators):
\begin{align}
    \Ical_{\mathrm{V}1,\mathrm{reg}}&=\int \frac{\der l_1^+}{(2\pi)}\frac{\der l_2^+}{(2\pi)} \frac{\der l_3^+}{(2\pi)} \frac{-i (2q^-)^2 (2l_3^-)}{(l_3-k)^2 l_1^2(l_1-q)^2 (l_2-q+l_1)^2 l_2^2 l_3^2 }  \nonumber \\
    &= - \frac{z_g }{(z_{\bar{q}} + z_g)z_q} \frac{\Theta(z_g)\Theta(z_q - z_g)}{\left(\ltone^2 + \Delta_{\mathrm{V}1}^2 \right) \left[ \omega_{\mathrm{V}1} \left(\ltone^2 + \Delta_{\mathrm{V}1}^2 \right) + \Lttwoy^2 \right] \Ltthrex^2} \,, 
    \label{eq:dijet-NLO-V3-l+reg-2}\\
    \Ical_{\mathrm{V}1,\bar{q}\mathrm{inst}}&= \int \frac{\der l_1^+}{(2\pi)} \frac{\der l_2^+}{(2\pi)} \frac{\der l_3^+}{(2\pi)} \frac{  -i (2q^-)^2 (2l_3^-)}{(l_3-k)^2 l_1^2 (l_2-q+l_1)^2l_2^2 l_3^2} \nonumber \\
    &=   \frac{z_g}{(z_{\bar{q}} + z_g)(z_q-z_g)z_q} \frac{ \Theta(z_g)\Theta(z_q - z_g)}{\left[  \Lttwoy^2 + \omega_{\mathrm{V}1} \left( \ltone^2 + \Delta_{\mathrm{V}1}^2 \right) \right] \Ltthrex^2} \,.
    \label{eq:dijet-NLO-V3-l+qbarins-2}
\end{align}
The computation of these integrals is done using Cauchy's theorem as outlined in Appendix~\ref{app:contour}. The definitions of the kinematic parameters which appear in the energy denominators (in the language of LCPT) are
\begin{align}
    \Delta_{\mathrm{V}1}^2 &= (z_q - z_g)(z_{\bar{q}}+ z_g) Q^2 \,, \\
    \omega_{\mathrm{V}1} &= \frac{z_g z_{\bar{q}}}{(z_q - z_g)(z_{\bar{q}}+z_g)^2}\,.
\end{align}

\subsubsection*{Transverse momentum integration}
Before we proceed with the transverse momentum integration, we will, as previously,  write the expressions for the relevant phases in terms of the momenta $\Ltthrex$ and $\Lttwoy$: 
\begin{align}
    e^{i \ltone \cdot \rxyt}e^{i\lttwo \cdot \rzyt}e^{-i\ltthre \cdot \rzxt}
    &= e^{-i \frac{z_g}{z_q} \kt \cdot \rzxt }e^{i \ltone \cdot \left(\rxyt - \frac{z_g}{(z_{\bar{q}}+z_g)} \rzyt\right)}e^{i\Lttwoy\cdot \rzyt}e^{-i\Ltthrex\cdot \rzxt} \,. \label{eq:dijet-NLO-V3-phases}
\end{align}
We will separately discuss the transverse momentum integrals for  longitudinally polarized and transversely polarized virtual photons.

\noindent{\it Longitudinal polarization} \\
Clearly, the instantaneous contribution (Eq.\,\eqref{eq:dijet-NLO-V3-pert-qbarins}) again vanishes since the corresponding Dirac structure is identically zero (Eq.\,\eqref{eq:dijet-NLO-V3-dirac-qbarins-long}):
\begin{align}
    \Ncal^{\lambda=0,\sigma\sigma'}_{\mathrm{V}1,\bar{q}\mathrm{inst}} = 0\,.
\end{align}
The regular piece is found by inserting Eqs.\,\eqref{eq:dijet-NLO-V3-dirac-reg-long},\eqref{eq:dijet-NLO-V3-l+reg-2} and \eqref{eq:dijet-NLO-V3-phases} in Eq.\,\eqref{eq:dijet-NLO-V3-pert-reg}; we find
\begin{align}
    \Ncal^{\lambda=0,\sigma\sigma'}_{\mathrm{V}1,\mathrm{reg}}&= \frac{g^2}{2}   \int_0^{z_q} \frac{\der z_g}{z_g} e^{-i \frac{z_g}{z_q} \kt \cdot \rzxt } 4 (z_{\bar{q}}+ z_g) (z_q - z_g) Q\nonumber\\
    &\times\left\{ \left[1  - \frac{z_g}{2z_q} - \frac{z_g}{2(z_{\bar{q}}+z_g)} \right] \frac{\left[\bar{u}(k,\sigma) \gamma^- v(p,\sigma') \right]}{q^-} \delta^{ik} \right. \nonumber \\
    & \left. -i\left[\frac{z_g}{2z_q} - \frac{z_g}{2(z_{\bar{q}}+z_g)} \right] \frac{\left[\bar{u}(k,\sigma) \gamma^- \Omega v(p,\sigma') \right]}{q^-} \epsilon^{ik}\right\} \nonumber \\
    &\times \int \frac{\der^{2} \ltone}{(2\pi)^{2}} \frac{e^{i \ltone \cdot \left(\rxyt - \frac{z_g}{(z_{\bar{q}}+z_g)} \rzyt\right)}}{\left(\ltone^2 + \Delta_{\mathrm{V}1}^2 \right)  }  \int \frac{\der^{2} \Lttwoy }{(2\pi)^{2}} \frac{\Lttwoy^i e^{i \Lttwoy \cdot \rzyt} }{\left[ \Lttwoy^2 + \omega_{\mathrm{V}1} \left(\ltone^2 + \Delta_{\mathrm{V}1}^2 \right)  \right]}\nonumber\\
    &\times  \int \frac{\der^{2} \Ltthrex }{(2\pi)^{2}}\frac{\Ltthrex^k e^{-i\Ltthrex\cdot \rzxt} }{\Ltthrex^2} \,.
\end{align}
The transverse momentum integrals are performed using the formulas Eq.\,\eqref{eq:Transverse_int_3} and Eq.\,\eqref{eq:Transverse_int_4} in Appendix~\ref{app:transverse-int}. After contracting the remaining gamma matrices with the  spinors, one obtains the  compact expression,
\begin{align}
    \Ncal^{\lambda=0,\sigma\sigma'}_{\mathrm{V}1,\mathrm{reg}}&=\frac{\alpha_s}{\pi^2}\int_0^{z_q}\frac{\der z_g}{z_g}e^{-i\frac{z_g}{z_q}\kt \cdot \rzxt}(2z_qz_{\bar q})\bar Q\delta^{\sigma,-\sigma'}K_0\left(QX_V\right)\left(1-\frac{z_g}{z_q}\right)\left(1+\frac{z_g}{z_{\bar q}}\right)\nonumber\\
    &\times \left\{ \left[1  - \frac{z_g}{2z_q} - \frac{z_g}{2(z_{\bar{q}}+z_g)} \right]  \frac{\rzxt \cdot \rzyt}{\rzxt^2 \rzyt^2}+i\sigma\left[ \frac{z_g}{2z_q} - \frac{z_g}{2(z_{\bar{q}}+z_g)} \right]  \frac{\rzxt \times \rzyt}{\rzxt^2 \rzyt^2}\right\}\label{eq:dijet-NLO-virtual-V3-final},
\end{align}
with the $q\bar qg$ dipole effective size $X_{\rm V}$ defined by
\begin{equation}
    X_{\rm V}^2=z_{\bar q}(z_q-z_g)\rxyt^2+z_g(z_q-z_g)\rzxt^2+z_g z_{\bar q}\rzyt^2 \,,
\end{equation}
as for the quark dressed  self energy.

\noindent {\it Transverse polarization}\\ For a transversely polarized virtual photon, one needs to compute the instantaneous antiquark term. Inserting the expressions for the instantaneous Dirac structure Eq.\,\eqref{eq:dijet-NLO-V3-Diracins}, and that for the contour integral in  Eq.\,\eqref{eq:dijet-NLO-V3-l+qbarins-2}, inside the equation \eqref{eq:dijet-NLO-V3-pert-qbarins}, one gets 
\begin{align}
    &\Ncal^{\lambda=\pm1,\sigma\sigma'}_{\mathrm{V}1,\bar{q}\mathrm{inst}}= -\frac{g^2}{2}   \int_0^{z_q} \der z_g \ e^{-i \frac{z_g}{z_q} \kt \cdot \rzxt } \frac{z_{\bar{q}}}{(z_g + z_{\bar{q}})^2} \left\{\bar{u}(k,\sigma)(1+\lambda\Omega)\frac{\gamma^-}{q^-}v(p,\sigma')\right\}\et^{\lambda,i} \nonumber\\
    &\times \left\{\int \frac{\der^{2} \ltone}{(2\pi)^{2}}\int \frac{\der^{2} \Lttwoy }{(2\pi)^{2}}\frac{ e^{i \ltone \cdot \left(\rxyt - \frac{z_g}{(z_{\bar{q}}+z_g)} \rzyt\right)} e^{i \Lttwoy \cdot \rzyt} }{\left[ \Lttwoy^2 + \omega_{\mathrm{V}1} \left(\ltone^2 + \Delta_{\mathrm{V}1}^2 \right)  \right]} \int \frac{\der^{2} \Ltthrex }{(2\pi)^{2}}\frac{\Ltthrex^i e^{-i\Ltthrex\cdot \rzxt} }{\Ltthrex^2} \right \}\,.
\end{align}
The three transverse momentum integrals can be performed analytically using Eq.\,\eqref{eq:Transverse_int_3} and Eq.\,\eqref{eq:dijet-NLO-transint-inst} in Appendix~\ref{app:transverse-int}. The antiquark instantaneous term finally reads
\begin{align}
    \Ncal^{\lambda=\pm1,\sigma\sigma'}_{\mathrm{V}1,\bar{q}\mathrm{inst}}= \frac{\alpha_s}{\pi^2}   \int_0^{z_q} \der z_g \ e^{-i \frac{z_g}{z_q} \kt \cdot \rzxt }\frac{z_{\bar q}(z_q-z_g)}{z_g+z_{\bar q}}\delta^{\sigma,-\sigma'}\delta^{\sigma,-\lambda}\frac{\rzxt\cdot\et^\lambda}{\rzxt^2} \frac{i \bar QK_1\left(QX_V\right)}{X_V}\,.
    \label{eq:dijet-NLO-V3-inst-final}
\end{align}
For the regular term, the steps are the same as in the longitudinal case. The only difference comes from the additional $\ltone$ dependence of the Dirac numerator. The relevant transverse momentum integral with this additional factor is given in Eq.~\eqref{eq:Transverse_int_5} of Appendix~\ref{app:transverse-int}. One ends up with an expression which looks very similar in structure to Eq.\,\eqref{eq:dijet-NLO-virtual-V3-final}, 
\begin{align}
    \Ncal^{\lambda=\pm1,\sigma\sigma'}_{\mathrm{V}1,\mathrm{reg}} \!\!& = \frac{\alpha_s}{\pi^2}\int_0^{z_q}\frac{\der z_g}{z_g}e^{-i\frac{z_g}{z_q}\kt \cdot \rzxt}(-2z_qz_{\bar q})\delta^{\sigma,-\sigma'}\Gamma_{\gamma^*_\mathrm{T} \to q\bar{q}}^{\sigma,\lambda}(z_q-z_g,z_{\bar q}+z_g)\nonumber\\
    &\times (\RtV\cdot\et^\lambda)\frac{i\bar QK_1\left(QX_V\right)}{X_V} \left(1-\frac{z_g}{z_q}\right)\left(1+\frac{z_g}{z_{\bar q}}\right) \nonumber \\
    & \times \left\{ \left[1  - \frac{z_g}{2z_q} - \frac{z_g}{2(z_{\bar{q}}+z_g)} \right]  \frac{\rzxt \cdot \rzyt}{\rzxt^2 \rzyt^2}+i\sigma\left[ \frac{z_g}{2z_q} - \frac{z_g}{2(z_{\bar{q}}+z_g)} \right]  \frac{\rzxt \times \rzyt}{\rzxt^2 \rzyt^2}\right\}\,,
    \label{eq:dijet-NLO-virtual-V3-final-trans}
\end{align}
with the initial size of the $q\bar q $ dipole (before gluon emission) defined by
\begin{equation}
    \RtV=\rxyt-\frac{z_g}{z_{\bar q}+z_g}\rzyt \,.
\end{equation}
This concludes our computation of the dressed vertex correction.

\subsubsection{Free gluon propagator before shock wave}

\begin{figure}[tbh]
    \centering
    \includegraphics[width=0.48\textwidth]{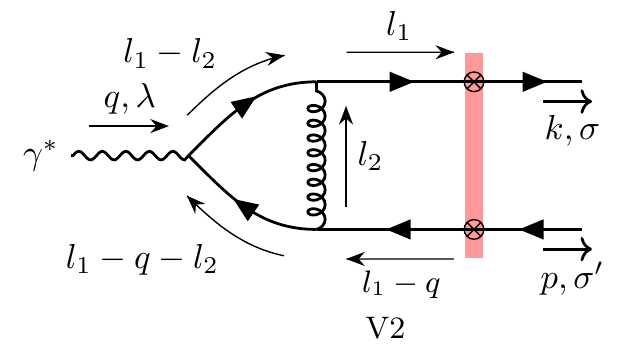}
    \hfill
    \includegraphics[width=0.48\textwidth]{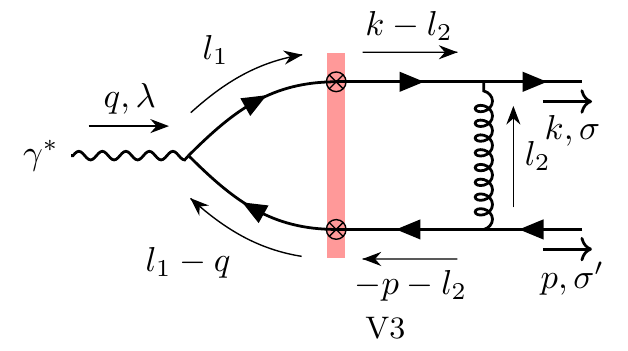}
    \caption{Free vertex correction before the shock wave (left) and after (right).}
    \label{fig:NLO-dijet-V23}
\end{figure}

The free vertex correction before the shock wave was computed  in \cite{Beuf:2016wdz,Hanninen:2017ddy} using LCPT. In our framework, the left diagram in Fig.~\ref{fig:NLO-dijet-V23} corresponds to the sum of all the vertex corrections to the lightcone wavefunction of the incoming virtual photon. We have checked these results and the main steps of the computation are outlined in Appendix~\ref{app:V1} . We quote here only the final result:
\begin{align}
    \Mcal^{\lambda\sigma\sigma'}_{\mathrm{V}2} = \frac{e e_f q^-}{\pi}  \mu^{-2\varepsilon}\int \der^{2-\varepsilon} \xt \der^{2-\varepsilon} \yt e^{-i \kt \cdot \xt } e^{-i \pt \cdot \yt } \Ccal_{\mathrm{V}2}(\xt,\yt)\, \Ncal_{\mathrm{V}2}^{\lambda\sigma\sigma'}(\rxyt) \,,
\end{align}
with the color structure identical to the LO color structure, up to a $C_F$ factor (just as is the case for the free self energies),
\begin{align}
    \Ccal_{\mathrm{V}2}(\xt,\yt) = C_F \left[V(\xt) V^\dagger(\yt) -\mathbbm{1} \right]\,.
\end{align}
The perturbative factor for a longitudinally polarized photon reads,
\begin{align}
    \mathcal{N}^{\lambda=0,\sigma\sigma'}_{\mathrm{V}2}(\rxyt)&=\frac{\alpha_s}{2\pi}\mathcal{N}^{\lambda=0,\sigma\sigma'}_{\rm LO,\varepsilon}(\rxyt)\left\{\left(\frac{2}{\varepsilon}+\ln\left(\frac{\tilde\mu^2}{\bar{Q}^2}\right)\right)\left[\ln\left(\frac{z_q}{z_0}\right)+\ln\left(\frac{z_{\bar q}}{z_0}\right)-\frac{3}{2}\right]\right.\nonumber\\
    &+\ln^2\left(\frac{z_q}{z_0}\right)+\ln^2\left(\frac{z_{\bar q}}{z_0}\right)+\frac{1}{2}\ln^2\left(\frac{z_q}{z_{\bar q}}\right)+\frac{\pi^2}{2}\nonumber\\
    &\left.+\left(2\ln\left(\frac{z_{\bar q}}{z_0}\right)-\frac{3}{2}\right)\ln(z_{ q})+\left(2\ln\left(\frac{z_q}{z_0}\right)-\frac{3}{2}\right)\ln(z_{\bar q})-\frac{7}{2}-\frac{1}{2}+\mathcal{O}(\varepsilon)\right\}\label{eq:dijet-NLO-V1-final-long}\,,
    \end{align}
and for a transversely polarized photon, 
\begin{align}
    \mathcal{N}^{\lambda=\pm1,\sigma\sigma'}_{\mathrm{V}2}(\rxyt)&=\frac{\alpha_s}{2\pi}\mathcal{N}^{\lambda=\pm1,\sigma\sigma'}_{\rm LO,\varepsilon}(\rxyt) \nonumber \\
    &\times \left\{\left(\frac{2}{\varepsilon}+\ln\left(\frac{\tilde\mu^2}{\bar{Q}^2}\right)-\frac{2}{r_{xy}\bar Q}\frac{K_0(\bar Qr_{xy})}{K_1(\bar Qr_{xy})}\right)\left[\ln\left(\frac{z_q}{z_0}\right)+\ln\left(\frac{z_{\bar q}}{z_0}\right)-\frac{3}{2}\right]\right.\nonumber\\
    & +\ln^2\left(\frac{z_q}{z_0}\right)+\ln^2\left(\frac{z_{\bar q}}{z_0}\right)
   +\frac{1}{2}\ln^2\left(\frac{z_q}{z_{\bar q}}\right)+\frac{\pi^2}{2}\nonumber\\
   &\left.+\left(2\ln\left(\frac{z_{\bar q}}{z_0}\right)-\frac{3}{2}\right)\ln(z_{ q})+\left(2\ln\left(\frac{z_q}{z_0}\right)-\frac{3}{2}\right)\ln(z_{\bar q})-\frac{7}{2}-\frac{1}{2}+\mathcal{O}(\varepsilon)\right\}\,.
   \label{eq:dijet-NLO-V1-final-trans}
    \end{align}
The last finite term in these expressions corresponds to the regularization scheme dependent term from the $\varepsilon$ dependence of the Dirac algebra. As discussed for the case of the gluon self energy before the shock wave, such scheme dependent terms cancel at cross-section level. This particular diagrams has a UV divergence in $4$ dimension, manifest as a pole in $1/\varepsilon$, and a slow divergence which is cured by the cut-off $z_0=\Lambda^-_0/q^-$.

\subsubsection{Free gluon propagator after shock wave} 
\label{sub:V2}

This diagram is free of UV divergences; it is therefore enough to compute it directly in $4$ dimensions. The calculation is detailed in Appendix~\ref{appp:V3}. We provide here the final expressions and additional comments. The physical amplitude reads
\begin{align}
    \Mcal^{\lambda\sigma\sigma'}_{\mathrm{V}3}=\frac{ee_fq^-}{\pi}\int\der^{2}\xt\der^{2}\yt \,e^{-i\kt \cdot \xt -i\pt \cdot \yt}\Ccal_{\mathrm{V}3}(\xt,\yt)\Ncal^{\lambda\sigma\sigma'}_{\mathrm{V}3}(\rxyt)\,,
\end{align}
has the color structure
\begin{equation}
    \Ccal_{\mathrm{V}3}(\xt,\yt)=t^a V(\xt)V^\dagger(\yt)t_a-C_F\mathbbm{1}\,.
\end{equation}
Observe that it is different from the LO expression, despite being independent of $\zt$.
The perturbative factor is written as an integral over the longitudinal momentum fraction $z_g$ of the virtual gluon. In contrast to the other diagrams, we have not found closed analytic expressions for the transverse momentum integrals. The perturbative factor involves the functions,
\begin{align}
    \Jcal_{\odot}(\rt,\Kt,\Delta)&=\int\frac{\der^2 \lt}{(2\pi)}\frac{2\lt \cdot \Kt \ e^{i\lt \cdot \rt}}{\lt^2\left[(\lt-\Kt)^2-\Delta^2 - i \epsilon\right]}\,,
    \label{eq:Jdot-def}\\
    \Jcal_{\otimes}(\rt,\Kt,\Delta)&=\int\frac{\der^2 \lt}{(2\pi)}\frac{(-i)\lt \times \Kt \ e^{i\lt \cdot \rt}}{\lt^2\left[(\lt-\Kt)^2-\Delta^2- i \epsilon\right]}\,,
    \label{eq:Jtimes-def}
\end{align}
for all $\rt, \Kt$ and $\Delta^2 > \Kt^2$.
These transverse momentum integrals are computed and their structure analyzed in Appendix \ref{app:jdotjcross}. The final result contains scalar integrals over a Feynman parameter, as can be seen in Eqs.\,\eqref{eq:dijet-NLO-app-Jcrossfinal} and \eqref{eq:dijet-NLO-app-Jdotfinal2}. The function $\Jcal_{\otimes}$ does not have  singularities, unlike $\Jcal_{\odot}$ which diverges near $\Delta^2=\Kt^2$, which occurs in the slow gluon limit $z_g\to 0$. We will return to this issue in the next section. 

In terms of these two functions, the perturbative factor for a longitudinally polarized photon reads
\begin{align}
    \Ncal^{\lambda=0,\sigma\sigma'}_{\mathrm{V}3}(\rxyt)&=\frac{\alpha_s}{\pi}\int_0^{z_q}\frac{\der z_g}{z_g}(-2)(z_qz_{\bar q})^{3/2}\delta^{\sigma,-\sigma'}QK_0(\bar{Q}_{\mathrm{V3}} r_{xy})\left(1-\frac{z_g}{z_q}\right)\left(1+\frac{z_g}{z_{\bar q}}\right)\nonumber\\
    &\times \left\{\left[(1+z_g)\left(1-\frac{z_g}{z_q}\right)\right]e^{i(\Pt+z_g(\kt+\pt)) \cdot \rxyt}K_0(-i\Delta_{\mathrm{V}3}r_{xy})\right.\nonumber\\
    &-\left[1-\frac{z_g}{2z_q}+\frac{z_g}{2z_{\bar q}}-\frac{z_g^2}{2z_qz_{\bar q}}\right]e^{i\frac{z_g}{z_q}\kt \cdot \rxyt}\Jcal_{\odot}\left(\rxyt,\left(1-\frac{z_g}{z_q}\right)\Pt,\Delta_{\mathrm{V}3}\right)\nonumber\\
    &\left.+\sigma\left[\frac{z_g}{z_q}-\frac{z_g}{z_{\bar q}}+\frac{z_g^2}{z_qz_{\bar q}}\right]e^{i\frac{z_g}{z_q}\kt \cdot \rxyt}\Jcal_{\otimes}\left(\rxyt,\left(1-\frac{z_g}{z_q}\right)\Pt,\Delta_{\mathrm{V}3}\right)\right\}\nonumber\\
    &+(q\leftrightarrow\bar q)\,,
    \label{eq:dijet-NLO-virtual-V2-pert-final-spin}
\end{align}
with
 \begin{align}
    \bar{Q}_{\mathrm{V}3}^2&=(z_q - z_g)(z_{\bar{q}} + z_g) Q^2 \nonumber \,.\\
    \Pt&=z_{\bar q}\kt-z_q\pt \,,\\
    \Delta_{\mathrm{V}3}^2&=\left(1-\frac{z_g}{z_q}\right)\left(1+\frac{z_g}{z_{\bar q}}\right)\Pt^2\,.
\end{align}
In the argument of the $K_0$ Bessel function of this expression, $\Delta_{\mathrm{V}3}$ is positive, due to the location of the poles in the integral over the virtual gluon momentum (see details in App.~\ref{appp:V3}).
One recognizes $\Pt$ as the relative transverse momentum between the two jets.
Note that $\Delta^2_{\mathrm{V}3}\to\Pt^2$ when $z_g \to 0$.
In Eq.~\eqref{eq:dijet-NLO-virtual-V2-pert-final-spin}, the quark-antiquark interchange amounts to the following transformations: $z_q\leftrightarrow z_{\bar q}$, $\Pt\to -\Pt$, $\kt \leftrightarrow \pt$ and $\rxyt\to-\rxyt$.

Finally, for completeness, we will state here the result (worked out in App.~\ref{appp:V3}) for the transversely polarized virtual photon:
\begin{align}
    \Ncal^{\lambda=\pm1,\sigma\sigma'}_{\mathrm{V}3}(\rxyt)&=\frac{\alpha_s}{\pi}\int_0^{z_q}\frac{\der z_g}{z_g}2(z_qz_{\bar q})^{1/2} \delta^{\sigma,-\sigma'}\Gamma_{\gamma^*_\mathrm{T} \to q\bar{q}}^{\sigma,\lambda}(z_q-z_g,z_{\bar q}+z_g)\frac{i\bar{Q}_{\mathrm{V3}}\rxyt\cdot\et^\lambda}{r_{xy}}K_1(\bar{Q}_{\mathrm{V3}} r_{xy})\nonumber\\
    &\times \left\{\left[(1+z_g)\left(1-\frac{z_g}{z_q}\right)\right]e^{i(\Pt+z_g(\kt+\pt)) \cdot \rxyt}K_0(-i\Delta_{\mathrm{V}3}r_{xy})\right.\nonumber\\
    &-\left[1-\frac{z_g}{2z_q}+\frac{z_g}{2z_{\bar q}}-\frac{z_g^2}{2z_qz_{\bar q}}\right]e^{i\frac{z_g}{z_q}\kt \cdot \rxyt}\Jcal_{\odot}\left(\rxyt,\left(1-\frac{z_g}{z_q}\right)\Pt,\Delta_{\mathrm{V}3}\right)\nonumber\\
    &\left.+\sigma\left[\frac{z_g}{z_q}-\frac{z_g}{z_{\bar q}}+\frac{z_g^2}{z_q z_{\bar q}}\right]e^{i\frac{z_g}{z_q}\kt \cdot \rxyt}\Jcal_{\otimes}\left(\rxyt,\left(1-\frac{z_g}{z_q}\right)\Pt,\Delta_{\mathrm{V}3}\right)\right\}\nonumber\\
    &+(q\leftrightarrow\bar q)\,.
    \label{eq:dijet-NLO-virtual-V2-pert-final-trans-spin}
\end{align}
In spite of the different spin-helicity structure and the leading order photon wave function, one observes a strong similarity of this expression with the longitudinal polarization result; indeed, the factor within the curly brackets is identical.

\subsection{Combining the UV divergent virtual diagrams}
\label{sub:virtual-combined}

We will now summarize here our results for the virtual amplitudes. We found that the dressed vertex corrections ($\mathrm{V}1,\mathrm{V}1'$), and the vertex correction with gluon after the shock wave $(\mathrm{V}3)$ are free of UV divergences. On the other hand, the dressed self energies ($\mathrm{SE}1,\mathrm{SE}1'$), the free self energies before shock wave ($\mathrm{SE}2,\mathrm{SE}2'$), and the vertex correction before shock wave $(\mathrm{V}2)$ have a $1/\varepsilon$ UV pole ($\varepsilon>0$). Given that they share the same color structure as the LO amplitude (up to a multiplicative $C_F$ factor), it is advantageous to combine all these UV singular contributions, and define:
\begin{align}
    \Mcal_{\rm UV}&=\Mcal_{\mathrm{V}2}+\left(\Mcal_{\mathrm{SE}1, \rm{UV}}+\Mcal_{\mathrm{SE}2}+q\leftrightarrow\bar q\right)\label{eq:dijet-NLO-MUV-final-def}\\
    &=\frac{ee_fq^-}{\pi}\mu^{-2\varepsilon}\int\der^{2-\varepsilon}\xt\der^{2-\varepsilon}\yt e^{-i\kt \cdot \xt-i\pt \cdot \yt}\Ccal_{\rm LO}(\xt,\yt)\Ncal_{\rm LO,\varepsilon}(\rxyt)\nonumber\\
    &\times \frac{\alpha_sC_F}{2\pi}\left\{\left(\ln\left(\frac{z_q}{z_0}\right)+\ln\left(\frac{z_{\bar q}}{z_0}\right)-\frac{3}{2}\right)\left(\frac{2}{\varepsilon}-2\gamma_E-\ln\left(\frac{\rxyt^2\tilde{\mu}^2}{4}\right)+2\ln(2\pi\mu^2\xi)\right)\right.\nonumber\\
    &\hspace{1.5cm}\left.+\frac{1}{2}\ln^2\left(\frac{z_{\bar q}}{z_q}\right)-\frac{\pi^2}{6}+\frac{5}{2}-\frac{1}{2}\right\}\,,
    \label{eq:dijet-NLO-MUV-final-xi}
\end{align}
where $\varepsilon>0$, and the second equality comes the combination of Eqs.~\eqref{eq:dijet-NLO-SE2-UVfinal-long}-\eqref{eq:dijet-NLO-SE2-final}-\eqref{eq:dijet-NLO-V1-final-long} for a longitudinal photon and Eqs.~\eqref{eq:dijet-NLO-SE2-UVfinal-trans}-\eqref{eq:dijet-NLO-SE2-final-transverse}-\eqref{eq:dijet-NLO-V1-final-trans} for a transverse photon.  

If in addition we include the contribution from self energies with gluon after the shock wave $(\mathrm{SE}3,\mathrm{SE}3')$ in Eq.\,\eqref{eq:SE-afterSW-amplitude-final}, which formally vanished in dimensional regularization, we see that the surviving pole is infrared, $\Mcal_{\rm UV}\rightarrow\Mcal_{\rm IR}$,
\begin{align}
    \Mcal_{\rm IR}&=\Mcal_{\mathrm{V}2}+\left(\Mcal_{\mathrm{SE}1, \rm UV}+\Mcal_{\mathrm{SE}2}+\Mcal_{\mathrm{SE}3}+q\leftrightarrow\bar q\right)\label{eq:dijet-NLO-MIR-final-def}\\
    &=\frac{ee_fq^-}{\pi}\mu^{-2\varepsilon}\int\der^{2-\varepsilon}\xt\der^{2-\varepsilon}\yt e^{-i\kt \cdot \xt-i\pt \cdot \yt}\Ccal_{\rm LO}(\xt,\yt)\Ncal_{\rm LO,\varepsilon}(\rxyt)\nonumber\\
    &\times \frac{\alpha_sC_F}{2\pi}\left\{\left(\ln\left(\frac{z_q}{z_0}\right)+\ln\left(\frac{z_{\bar q}}{z_0}\right)-\frac{3}{2}\right)\left(\frac{2}{\varepsilon}-2\gamma_E-\ln\left(\frac{\rxyt^2\tilde{\mu}^2}{4}\right)+2\ln(2\pi\mu^2\xi)\right)\right.\nonumber\\
    &\hspace{1.5cm}\left.+\frac{1}{2}\ln^2\left(\frac{z_{\bar q}}{z_q}\right)-\frac{\pi^2}{6}+\frac{5}{2}-\frac{1}{2}\right\}\,,
    \label{eq:dijet-NLO-MIR-final-xi}
\end{align}
where $\varepsilon<0$ indicates the infrared nature of the divergence.

The results in Eqs.\,\eqref{eq:dijet-NLO-MUV-final-xi}-\eqref{eq:dijet-NLO-MIR-final-xi} are valid for both longitudinally and transversely polarized virtual photons. In other words, the dependence on the polarization of the photon enters only through the leading order perturbative factor $\Ncal_{\rm LO,\varepsilon}$. This is nontrivial given that diagrams $\mathrm{SE}2$ and $\mathrm{V}2$ do not independently satisfy  this property because of the term proportional to $K_0(\bar Qr_{xy})/K_1(\bar Qr_{xy})$ in Eq.\,\eqref{eq:dijet-NLO-SE2-final-transverse} and \eqref{eq:dijet-NLO-V1-final-trans}.
Another important point relates to the cancellation of the double logarithmic divergence  $\ln^2(z_0)$. Even though each individual diagram exhibits such an unphysical divergence, the sum of the diagrams is free of it. Since such double logarithmic terms would violate the small-$x$ factorization into the JIMWLK evolution equation, it is a crucial result of our calculation.
Finally, the $-1/2$ term in Eq.\,\eqref{eq:dijet-NLO-MIR-final-xi} is the scheme dependent rational term in dimensional regularization \cite{Beuf:2016wdz} (see also \cite{Gelis:2019yfm} for an overview on the rational terms in the context of QCD loop calculations in $d$ dimensions and \cite{Hanninen:2017ddy} for a detailed discussion of the dimensional regularization scheme dependence of these rational terms in LCPT).

We remind the reader that the free parameter $\xi$ in Eq.\,\eqref{eq:dijet-NLO-MIR-final-xi} is arbitrary.
Anticipating in advance the discussion in the next section on the slow gluon limit, we choose
\begin{equation}
    \xi = \frac{\rxyt^2e^{\gamma_E}}{2}\,,
    \label{eq:dijet-NLO-xidef}
\end{equation}
in agreement with \cite{Hanninen:2017ddy}, leading to
\begin{align}
 \Mcal_{\rm IR}&=\frac{ee_fq^-}{\pi}\mu^{-2\varepsilon}\int\der^{2-\varepsilon}\xt\der^{2-\varepsilon}\yt e^{-i\kt \cdot \xt-i\pt \cdot \yt}\Ccal_{\rm LO}(\xt,\yt)\Ncal_{\rm LO,\varepsilon}(\rxyt)\nonumber\\
    &\times \frac{\alpha_sC_F}{2\pi}\left\{\left(\ln\left(\frac{z_q}{z_0}\right)+\ln\left(\frac{z_{\bar q}}{z_0}\right)-\frac{3}{2}\right)\left(\frac{2}{\varepsilon}+\ln(e^{\gamma_E}\pi\mu^2\rxyt^2)\right)+\frac{1}{2}\ln^2\left(\frac{z_{\bar q}}{z_q}\right)\right.\nonumber\\
    &\hspace{1.5cm}\left.-\frac{\pi^2}{6}+\frac{5}{2}-\frac{1}{2}\right\}\,.
    \label{eq:dijet-NLO-SE1+V1+SE2IR}
\end{align}
With the $\xi$ above, the UV regularized amplitude Eq.\,\eqref{eq:dijet-NLO-SE2-finite_term_2} for $\mathrm{SE}1$ can be expressed as
\begin{align}
    \left.\mathcal{M}^{\lambda=0,\sigma\sigma'}_{\mathrm{SE}1}\right|_{\rm UV-fin.}&= \mathcal{M}^{\lambda=0,\sigma\sigma'}_{\mathrm{SE}1} - \mathcal{M}^{\lambda=0,\sigma\sigma'}_{\mathrm{SE}1,\mathrm{UV}}\nonumber \\
    &=\frac{ee_fq^-}{\pi}\,\frac{\alpha_s}{\pi^2}\int \der^2\xt\der^2\yt e^{-i\kt \cdot \xt-i\pt \cdot \yt}\int_0^{z_q}\frac{\der z_g}{z_g}(-2z_qz_{\bar q})\bar Q\delta^{\sigma,-\sigma'}\nonumber\\
    &  \times \left[1-\frac{z_g}{z_q}+\frac{z_g^2}{2z_q^2}\right] \int \frac{\der^2\zt }{\rzxt^2}\left\{e^{-i\frac{z_g}{z_q}\kt \cdot \rzxt}K_0\left(QX_{\rm V}\right)\Ccal_{\mathrm{SE}1}(\xt,\yt,\zt) \right. \nonumber \\
    & \left. -e^{-\frac{\rzxt^2}{\rxyt^2e^{\gamma_E}}}K_0\left(\bar{Q} r_{xy}\right)C_F\Ccal_{\rm LO}(\xt,\yt)\right\}\,,
  \label{eq:dijet-NLO-SE2-finite_term_3}
\end{align}
and similarly for a transversely polarized virtual photon. The diagrams $\mathrm{V}1$ and $\mathrm{V}3$ are unchanged since they do not depend on $\xi$. 

Thus Eq.\,\eqref{eq:dijet-NLO-SE1+V1+SE2IR} combined with the expressions for $\mathrm{V}1,\mathrm{V}1'$, $\mathrm{V}3$, and the UV finite pieces of $\mathrm{SE}1,\mathrm{SE}1'$ in Eq.\,\eqref{eq:dijet-NLO-SE2-finite_term_3} contain all the virtual contributions to our NLO computation. 

\section{Slow gluon limit: JIMWLK factorization}

\label{sec:slow_gluon}

In this section, we  will examine the slow gluon limit of our results, corresponding to the logarithmic divergence of the cross-section as the longitudinal momentum fraction $z_g$ of the (real or virtual) gluon goes to $0$. This divergence is cured by introducing an arbitrary cut-off $z_0=\Lambda^-_0/q^-$, with $\Lambda^-_0$ the longitudinal momentum separating the fast gluon modes from the slow ones; the latter being described by the CGC classical field/shock wave. We will then demonstrate that the dependence on this cut-off ($z_0$, or equivalently $\Lambda^-_0$) can be absorbed into the JIMWLK evolution of the leading order dijet cross-section.

\subsection{Extracting the logarithmic slow  divergence}
At NLO, the real and virtual contributions to the differential cross-section can be generically written in the form
\begin{align}
    \der \sigma_{\mathrm{NLO}} = \int_{z_0}^{z} \frac{\der z_g}{z_g} f(z_g)
    \,,
    \label{z_g-expansion}
\end{align}
where the upper bound $z$ can be either $z_q$ or $z_{\bar{q}}$.

In this section, we will show that the integrand $f(z_g)$ has the  expansion:
\begin{align}
    f(z_g) = a_0 + \sum_{n=1}^{\infty} a_n z_g^n \,. 
    \label{eq:expansion_integrand}
\end{align}
It is worth pointing that the expansion above holds only for the sum of all contributions, and that individual contributions  will develop terms proportional to $\ln(z_g)$, which when inserted in Eq.\,\eqref{z_g-expansion} will generate squared logarithmic divergences. This property (Eq.\,\eqref{eq:expansion_integrand}) is essential to recover the JIMWLK factorization.

The $a_0$ will generate the leading slow gluon singularity.
In the same spirit as standard DGLAP factorization, we introduce a rapidity factorization scale $z_f=\Lambda_f^-/q^-$, and use this factorization scale to isolate the logarithmic divergence as
\begin{equation}
    \int_{z_0}^{z}\frac{\der z_g}{z_g}f(z_g)=a_0 \ln\left(\frac{z_f}{z_0}\right)+\int_0^{z}\frac{\der z_g}{z_g}\left[f(z_g)-a_0\Theta(z_f-z_g)\right] + \mathcal{O}(z_0)\,.
    \label{eq:dijet-NLO-slow-vs-finite}
\end{equation}
We will explicitly show that the first term on the right hand-side can be absorbed into the leading logarithmic JIMWLK evolution, while the second term on the right hand side will constitute the NLO impact factor. Note in order for the impact factor to be independent of $z_0$, we choose the lower limit in the second integral in the right-hand side of Eq.\,\eqref{eq:dijet-NLO-slow-vs-finite} to be $0$, instead of $z_0$. This approximation is valid up terms that are power suppressed in the high energy limit, of $\mathcal{O}(z_0)$.

\subsubsection{Virtual contributions}

We start our discussion of the slow gluon limit of the cross-section with the virtual corrections summarized at the end of section~\ref{sub:virtual-combined}. It is convenient to organize the calculation as follows. We first take the slow gluon limit at the amplitude level for the singular term with the $1/\varepsilon$ pole defined in Eq.\,\eqref{eq:dijet-NLO-MIR-final-def}. We then consider separately the UV finite part of $\mathrm{SE}1$, and the vertex corrections $\mathrm{V}1$ and $\mathrm{V}3$, which are also free of UV divergences.

\subsubsection*{The pole term} Using Eq.~\eqref{eq:dijet-NLO-SE1+V1+SE2IR}, it is straightforward to isolate the divergence, integrated between $z_0$ and $z_f$ (instead of integrating up to $z_q$ or $z_{\bar{q}}$). One finds
\begin{align}
    \left.\Mcal_{\rm IR}\right|_{\rm slow}&=\frac{ee_fq^-}{\pi}\mu^{-2\varepsilon}\int\der^{2-\varepsilon}\xt\der^{2-\varepsilon}\yt e^{-i\kt \cdot \xt-i\pt \cdot \yt}\Ccal_{\rm LO}(\xt,\yt)\Ncal_{\rm LO,\varepsilon}(\rxyt)\nonumber\\
    &\times\ln\left(\frac{z_f}{z_0}\right)\frac{\alpha_sC_F}{\pi}\left[\frac{2}{\varepsilon}+\ln(e^{\gamma_E}\pi\mu^2\rxyt^2)+\mathcal{O}(\varepsilon)\right]\,.
    \label{eq:dijet-NLO-slow-UVamplitude}
\end{align}
Since the JIMWLK evolution equation is known not to have IR divergences, we expect this $1/\varepsilon$ pole to cancel when combined with real emissions. Indeed, this would be the case, as the collinear and slow contribution to real emissions will generate a similar pole (see Eq.\,\eqref{eq:dijet-NLO-slow-xsection5} in a next section).

It is therefore useful to write the virtual slow gluon divergence at the cross-section level. Defining
\begin{align}
    \left.\frac{\der\sigma^{\gamma_{\lambda}^*+A\to q\bar{q}+X}}{\der^2\kt\der\eta_q\der^2\pt\der\eta_{\bar q}}\right|_{\rm IR\times \rm LO}&=\frac{1}{4(2\pi)^6}\frac{1}{2q^-}(2\pi)\delta(q^--k^--p^-)\nonumber\\
    &\times\sum_{\sigma\sigma',\mathrm{color}} \left \langle \mathcal{M}^{\lambda\sigma\sigma'\dagger}_{\rm IR}[\rho_A] \mathcal{M}^{\lambda\sigma\sigma'}_{\rm LO} [\rho_A] \right \rangle_{Y} \,,
\end{align}
and using Eq.\,\eqref{eq:dijet-NLO-slow-UVamplitude}, one gets
\begin{align}
    &\left.\frac{\der\sigma^{\gamma_{\lambda}^*+A\to q\bar{q}+X}}{\der^2\kt\der\eta_q\der^2\pt\der\eta_{\bar q}}\right|_{\rm IR\times \rm LO, slow}=\frac{\alpha_{\rm em}e_f^2N_c}{(2\pi)^6}\delta(1-z_q-z_{\bar q})\int\der\Pi_{\rm LO,\varepsilon}\Rcal_{\mathrm{LO},\varepsilon}^{\lambda}(\rxyt,\rxytp)\nonumber\\
    &\times\frac{\alpha_sC_F}{\pi}\ln\left(\frac{z_f}{z_0}\right)\Xi_{\rm LO}(\xt,\yt;\xt',\yt')\left[\frac{2}{\varepsilon}+\ln(e^{\gamma_E}\pi\mu^2\rxyt^2)+\mathcal{O}(\varepsilon)\right] \,,
    \label{eq:dijet-NLO-slow-xsection1}
\end{align}
where the $\varepsilon$ subscript for the differential measure and $\Rcal_{\mathrm{LO}}^{\lambda}$ accounts for the straightforward generalization of the definitions Eq.\,\eqref{eq:dijet-NLO-LOdiffmeasure} and \eqref{eq:dijet-NLO-TLO} to $d=4-\varepsilon$ dimensions. The complex conjugate to this term ($c.c.$) is obtained by changing $\xt\to\xt'$ and $\yt\to\yt'$ and performing complex conjugation,  a transformation that leaves $\Xi_{\rm LO}$ invariant.

\subsubsection*{UV finite piece of the dressed self energies} We turn now to the finite term of the dressed quark or antiquark self energies, whose very explicit expression is given in Eq.\,\eqref{eq:dijet-NLO-SE2-finite_term_3} for  longitudinally polarized photons and in Eq.\,\eqref{eq:dijet-NLO-SE2-finite_term_2-trans} for transversely polarized photons. In the $z_g\to0$ limit, one observes several simplifications: the effective $qqg$ dipole size $X_{\rm V}$ goes to $\bar Qr_{xy}$ and the phase in the exponential vanishes. The same simplifications occur for a transversely polarized virtual photon, with the additional limit $\RtS\to\rxyt$ when $z_g\to0$ (note that the instantaneous contribution vanishes as well). One obtains as a result the following expression for the logarithmic divergence:
\begin{align}
    \left.\mathcal{M}_{\mathrm{SE}1}\right|_{\rm UV- fin.,slow}&=\frac{ee_fq^-}{\pi}\,\frac{\alpha_s}{\pi^2}\ln\left(\frac{z_f}{z_0}\right)\int \der^2\xt\der^2\yt e^{-i\kt \cdot \xt-i\pt \cdot \yt} \Ncal_{\rm LO}(\rxyt)\nonumber\\
    &  \times\int \frac{\der^2\zt }{\rzxt^2}\left[\Ccal_{\mathrm{SE}1}(\xt,\yt,\zt)-e^{-\frac{\rzxt^2}{\rxyt^2e^{\gamma_E}}}C_F\,\Ccal_{\rm LO}(\xt,\yt)\right]\,.
    \label{eq:dijet-NLO-SE2-finite_term-slow}
\end{align}
From the perspective of proving JIMWLK factorization of the slow gluon divergence, it is convenient to further simplify this expression using the  identity \cite{Hanninen:2017ddy},
\begin{equation}
    \int\der^2\zt \ \left[\frac{\rxyt^2}{\rzxt^2\rzyt^2}-\frac{1}{\rzxt^2}e^{-\frac{\rzxt^2}{\rxyt^2e^{\gamma_E}}}-\frac{1}{\rzyt^2}e^{-\frac{\rzyt^2}{\rxyt^2e^{\gamma_E}}}\right]=0\,.
    \label{eq:dijet-NLO-slow-jimwlk-identity}
\end{equation}
This equation is derived  in Appendix~\ref{sub:gluonkernel} and is crucial to simplify the slow gluon limit of our results. This is the reason why the choice of $\xi$ in  Eq.\,\eqref{eq:dijet-NLO-xidef} is particularly convenient. One finds then
\begin{align}
    &\left.\mathcal{M}_{\mathrm{SE}1}\right|_{\rm UV- fin.\rm,slow} + \left.\mathcal{M}_{\mathrm{SE}1'}\right|_{\rm UV- fin.\rm,slow} \nonumber \\
    & =\frac{ee_fq^-}{\pi}\frac{\alpha_s}{\pi^2}\ln\left(\frac{z_f}{z_0}\right)\int \der^2\xt\der^2\yt e^{-i\kt \cdot \xt-i\pt \cdot \yt}
    \Ncal_{\rm LO}(\rxyt)\nonumber\\
    &\times\int\der^{2}\zt\left[ \frac{1}{\rzxt^2}\Ccal_{\mathrm{SE}1}(\xt,\yt,\zt) +\frac{1}{\rzyt^2}\Ccal_{\mathrm{SE}1'}(\xt,\yt,\zt)-\frac{\rxyt^2}{\rzxt^2\rzyt^2}C_F\Ccal_{\rm LO}(\xt,\yt)\right]\,.
\label{eq:dijet-NLO-SE2-finite_term-slow2}
\end{align}
The color factor $\Ccal_{\mathrm{SE}1'}$ of the antiquark dressed self energy
\begin{equation}
    \Ccal_{\mathrm{SE}1'}(\xt,\yt,\zt)=V(\xt)V^\dagger(\zt)t^aV(\zt)V^\dagger(\yt)t_a-C_F\mathbbm{1}\,,
\end{equation}
is related to $\Ccal_{\mathrm{SE}1}$ defined in Eq.\,\eqref{eq:dijet-NLO-CSE1} by  $\xt\leftrightarrow\yt$ interchange and taking the Hermitian conjugate of this expression. 

Finally, we obtain for the slow gluon limit the result,
\begin{align}
    &  \left.\frac{\der\sigma^{\gamma_{\lambda}^*+A\to q\bar{q}+X}}{\der^2\kt\der\eta_q\der^2\pt\der\eta_{\bar q}}\right|_{\mathrm{SE}1,\rm UV-fin.\times \rm LO, slow} + \left.\frac{\der\sigma^{\gamma_{\lambda}^*+A\to q\bar{q}+X}}{\der^2\kt\der\eta_q\der^2\pt\der\eta_{\bar q}}\right|_{\mathrm{SE}1',\rm UV-fin.\times \rm LO, slow} \nonumber \\
    & =\frac{\alpha_{\rm em}e_f^2N_c}{(2\pi)^6}\delta(1-z_q-z_{\bar q})\int\der\Pi_{\rm LO}\Rcal_{\mathrm{LO}}^{\lambda}(\rxyt,\rxytp)\nonumber\\
    &\times\frac{\alpha_s}{\pi^2}\ln\left(\frac{z_f}{z_0}\right)\int\der^2\zt\left[\frac{1}{\rzxt^2}\Xi_{\rm NLO,1}+\frac{1}{\rzyt^2}\Xi_{\rm NLO,2}-\frac{\rxyt^2}{\rzxt^2\rzyt^2}C_F\Xi_{\rm LO}\right] \,,
    \label{eq:dijet-NLO-slow-xsection2}
\end{align}
with the color structures,
\begin{align}
    \Xi_{\rm NLO,1}(\xt,\yt,\zt;\xt',\yt')&=\frac{1}{N_c}\left\langle\Tr\left[\Ccal_{\mathrm{SE}1}(\xt,\yt,\zt)\Ccal^\dagger_{\rm LO}(\xt',\yt')\right]\right\rangle_{Y}\nonumber\\
    &=  \frac{N_c}{2}\left\langle1-D_{y'x'}+Q_{zy,y'x'}D_{xz}-D_{xz}D_{zy}\right\rangle_{Y}-\frac{1}{2N_c}\Xi_{\rm LO}\,,\\
    \Xi_{\rm NLO,2}(\xt,\yt,\zt;\xt',\yt')&=\frac{1}{N_c}\left\langle\Tr\left[\Ccal_{\mathrm{SE}1'}(\xt,\yt,\zt)\Ccal^\dagger_{\rm LO}(\xt',\yt')\right]\right\rangle_{Y}\nonumber\\
    &= \frac{N_c}{2}\left\langle1-D_{y'x'}+Q_{xz,y'x'}D_{zy}-D_{xz}D_{zy}\right\rangle_{Y}-\frac{1}{2N_c}\Xi_{\rm LO}\,.
   \label{eq:dijet-NLO-XiSE2}
\end{align}
The complex conjugate ``$c.c$" to the term above, is obtained by replacing  $\xt\to\xt'$ and $\yt\to\yt'$, and then take the complex conjugate of these color structures. 

Eq.\,\eqref{eq:dijet-NLO-slow-xsection2} deserves further commentary. Firstly, as for  Eq.\,\eqref{eq:dijet-NLO-SE2-finite_term-slow}, the $\zt$ integral is free of UV divergences when $\zt\to\xt$ or $\zt\to \yt$ because in these limits, $\Xi_{\rm NLO,1}\to C_F\,\Xi_{\rm LO}$ (and likewise for $\Xi_{\rm NLO,2}$). On the other hand, the first two terms in the square brackets of Eq.\,\eqref{eq:dijet-NLO-slow-xsection2} are infrared divergent as $|\zt|\to\infty$. However we shall see that the IR divergence cancels in the sum of all the diagrams. As shown in \cite{Hatta:2005as}, this is a consequence of considering the small $x$ evolution of the gauge invariant operators in the leading order cross-section.  To avoid the IR divergence in the intermediate steps of the computation in the slow gluon limit, one can regulate it by multiplying each $\zt$-dependent color correlator $\Xi_{\rm NLO}$ (associated with a diagram in which the gluon scatters off the shock wave) by an exponentially decaying factor $e^{-\lambda^2\zt^2}$ and then taking the limit $\lambda\to0$ at the end of the calculation\footnote{A more rigorous way to proceed would be to compute all diagrams in $d=4-\varepsilon$ dimensions (even UV finite ones), take the slow gluon limit of this result, and finally the limit $\varepsilon\to0$ once all the diagrams are combined.}. Notice that the appearance of such IR divergences is specific to the slow gluon limit. Indeed for finite $z_g$, the $K_0$ function in Eq.\,\eqref{eq:dijet-NLO-SE2-finite_term_3} regulates the large $|\zt|$ behaviour. Therefore even if the finite term defined by \eqref{eq:dijet-NLO-slow-vs-finite} may have IR singularity term by term, the sum of all the contributions is also IR finite.

\subsubsection*{Free vertex correction after shock wave} The slow gluon limit of the amplitude $\mathrm{V}3$ is nontrivial to extract from the results in  Eq.\,\eqref{eq:dijet-NLO-virtual-V2-pert-final-spin} and Eq.\,\eqref{eq:dijet-NLO-virtual-V2-pert-final-trans-spin} where the transverse momentum integrations have been performed explicitly. If we undo these integrations, one can find formally the slow gluon limit expected from the JIMWLK factorization. The price to pay is that we lose analytic control over the impact factor -- the finite piece after subtraction of the slow divergence. For this reason, we will directly isolate from Eqs.\,\eqref{eq:dijet-NLO-virtual-V2-pert-final-spin}-\eqref{eq:dijet-NLO-virtual-V2-pert-final-trans-spin} the leading slow gluon divergence.  In Eq~\eqref{eq:dijet-NLO-virtual-V2-pert-final-spin}, the first two terms contribute to the logarithmic divergence when $z_g\to 0$.
Using the result Eq.\,\eqref{eq:dijet-NLO-app-Jdotslow} from Appendix \ref{app:jdotjcross}, one first shows that
\begin{align}
    \Jcal_{\odot}\left(\rxyt,\left(1-\frac{z_g}{z_q}\right)\Pt,\Delta_{\mathrm{V}3}\right)&=\ln\left(\frac{z_g}{2z_qz_{\bar q}}\right)+\frac{1}{2}\ln\left(\Pt^2\rxyt^2\right)-\frac{i\pi}{2}+\gamma_E\nonumber\\
    &+e^{i\Pt \cdot \rxyt}K_0(-iP_\perp r_{xy})+\mathcal{O}(z_g)\,.
\end{align}
Since in the first term of Eq.\,\eqref{eq:dijet-NLO-virtual-V2-pert-final-trans-spin}, $K_0(-i\Delta_{\mathrm{V}3}r_{xy})=K_0(-iP_\perp r_{xy})+\mathcal{O}(z_g)$, the slow gluon limit of $\mathrm{V}3$ reads
\begin{align}
    \left.\Mcal_{\mathrm{V}3}\right|_{\rm slow}&=\frac{ee_fq^-}{\pi}\int\der^{2}\xt\der^{2}\yt \,e^{-i\kt \cdot \xt -i\pt \cdot \yt}\left[t^aV(\xt)V^\dagger(\yt)t_a-C_F\mathbbm{1}\right]\Ncal_{\rm LO}(\rxyt)\nonumber\\
    &\times\frac{(-\alpha_s)}{\pi}\int_{z_0}^{z_f}\frac{\der z_g}{z_g}\left[2\ln\left(\frac{z_g}{2z_qz_{\bar q}}\right)+\ln\left(\Pt^2\rxyt^2\right)-i\pi+2\gamma_E\right]\,.
    \label{eq:dijet-NLO-slowlimit-V2}
\end{align}
Finally, the contribution of this amplitude to the cross-section level in the $z_0\to0$ limit can be expressed as 
\begin{align}
    &  \left.\frac{\der\sigma^{\gamma_{\lambda}^*+A\to q\bar{q}+X}}{\der^2\kt\der\eta_q\der^2\pt\der\eta_{\bar q}}\right|_{\mathrm{V}3\times \rm LO, slow}=\frac{\alpha_{\rm em}e_f^2N_c}{(2\pi)^6}\delta(1-z_q-z_{\bar q})\int\der\Pi_{\rm LO}\Rcal_{\mathrm{LO}}^{\lambda}(\rxyt,\rxytp)\nonumber\\
    &\times\frac{(-\alpha_s)}{\pi}\int_{z_0}^{z_f}\frac{\der z_g}{z_g}\left[2\ln\left(\frac{z_g}{2z_qz_{\bar q}}\right)+\ln\left(\Pt^2\rxyt^2\right)+2\gamma_E\right]\Xi_{\rm NLO,3}(\xt,\yt;\xt',\yt') \,,
    \label{eq:dijet-NLO-slow-xsection3}
\end{align}
with 
\begin{align}
    \Xi_{\rm NLO,3}(\xt,\yt;\xt',\yt')&=\frac{1}{N_c}\left\langle\Tr\left[\Ccal_{\mathrm{V}3}(\xt,\yt)\Ccal^\dagger_{\rm LO}(\xt',\yt')\right]\right\rangle_{Y}\nonumber\\
    &=\frac{N_c}{2}\left\langle 1-D_{xy}-D_{y'x'}+D_{xy}D_{y'x'}\right\rangle_{Y}-\frac{1}{2N_c}\Xi_{\rm LO}\,.
    \label{eq:dijet-NLO-color-V2xLO}
\end{align}
The remaining integral over $z_g$ in Eq.\,\eqref{eq:dijet-NLO-slow-xsection3} can be performed analytically but we will refrain from doing so in order to make more transparent the cancellation between various terms in this expression and the soft divergences in the real cross-section.  In fact, only the $\ln(\rxyt^2)$ term in the square bracket of the $z_g$ integral in Eq.\,\eqref{eq:dijet-NLO-slow-xsection3} remains after combining the slow gluon limit of $\rm V 3$ with the cross-term in the real cross-section where the emitted gluon does not scatter off the shock wave (the contribution $\rm R2\times \rm R2'$). The other terms are truly \textit{soft} divergences, in the sense that they originate from an integration domain where all the component of the virtual gluon are small.

\subsubsection*{Dressed vertex corrections} Finally, we will consider the slow gluon limit of the dressed vertex corrections $\mathrm{V1}$ and $\mathrm{V1'}$; we include here the diagram with gluon exchange from the quark to the antiquark. The cross product term in Eq.\,\eqref{eq:dijet-NLO-virtual-V3-final} is sub-leading in this limit and one easily sees that the transverse coordinate $\RtV$ simplifies to $\rxyt$ and $X_V\to\bar Q r_{xy}$ as $z_g\to0$. Eventually, one obtains the  simple result
\begin{align}
    &\left.\Mcal_{\mathrm{V}1}\right|_{\rm slow} + \left.\Mcal_{\mathrm{V}1'}\right|_{\rm slow} \nonumber \\
    &=\frac{e e_f q^-}{\pi} \int \der^2 \xt \der^2 \yt  e^{- i \kt \cdot \xt} e^{- i \pt \cdot \yt} \Ncal_{\mathrm{LO}}(\rxyt)   \nonumber \\
    &\times \frac{-\alpha_s}{\pi^2} \ln\left(\frac{z_f}{z_0}\right)  \int \der\zt \frac{\rzxt \cdot \rzyt}{\rzxt^2 \rzyt^2}\left[\Ccal_{\mathrm{V}1'}(\xt,\yt,\zt)+\Ccal_{\mathrm{V}1}(\xt,\yt,\zt)\right]\,,
\end{align}
which is valid for both longitudinal and transversely polarized photons like the other diagrams. The color factor $\Ccal_{\mathrm{V}1'}$ corresponds to the color structure of the dressed vertex correction with gluon exchange from the quark to the antiquark, and is related to $\Ccal_{\mathrm{V}1}$ by $\xt\leftrightarrow\yt$ interchange and taking the Hermitian conjugation of this expression. Note that by definition, $\Ccal_{\mathrm{V}1}=\Ccal_{\mathrm{SE}1}$, so that $\Ccal_{\mathrm{V}1'}=\Ccal_{\mathrm{SE}1'}$. At the cross-section level, the CGC correlators are then identical to the ones associated with the dressed self energy, and one obtains
\begin{align}
    & \left.\frac{\der\sigma^{\gamma_{\lambda}^*+A\to q\bar{q}+X}}{\der^2\kt\der\eta_q\der^2\pt\der\eta_{\bar q}}\right|_{\mathrm{V}1\times \rm LO, slow} + \left.\frac{\der\sigma^{\gamma_{\lambda}^*+A\to q\bar{q}+X}}{\der^2\kt\der\eta_q\der^2\pt\der\eta_{\bar q}}\right|_{\mathrm{V}1'\times \rm LO, slow} \nonumber \\ &=\frac{\alpha_{\rm em}e_f^2N_c}{(2\pi)^6}\delta(1-z_q-z_{\bar q})\int\der\Pi_{\rm LO}\Rcal_{\mathrm{LO}}^{\lambda}(\rxyt,\rxytp)\nonumber\\
    &\times\frac{(-\alpha_s)}{\pi^2}\ln\left(\frac{z_f}{z_0}\right)\int\der^2\zt\frac{\rzxt \cdot \rzyt}{\rzxt^2 \rzyt^2}\left[\Xi_{\rm NLO,1}+\Xi_{\rm NLO,2}\right] \,.
    \label{eq:dijet-NLO-slow-xsection4}
\end{align}
The same comment about the infrared divergence of the $\zt$ integral applies here as that for the UV finite piece of the dressed self energy.

In conclusion,  combining equations Eq.\,\eqref{eq:dijet-NLO-slow-xsection1}, Eq.\,\eqref{eq:dijet-NLO-slow-xsection2}, Eq.\,\eqref{eq:dijet-NLO-slow-xsection3}, Eq.\,\eqref{eq:dijet-NLO-slow-xsection4}, and their complex conjugated, one obtains the complete result for the virtual part of the NLO dijet cross-section in the slow gluon limit.

\subsubsection{Real unscattered contributions}

We turn now to the slow gluon limit of the real corrections. Amongst these, there are four terms that deserve special attention corresponding to the configurations where the real gluon (emitted by quark or antiquark) does not scatter off the shock wave \textit{both} in the amplitude and in the complex conjugate amplitude. For these four terms, which we shall now discuss, we will study the slow limit behaviour directly at cross-section level. The $\gamma^{*}\to q\bar qg+X$ differential cross-section reads
\begin{align}
    \frac{\der\sigma^{\gamma_{\lambda}^*+A\to q\bar{q}g+X}}{\der^2\kt\der\eta_q\der^2\pt\der\eta_{\bar q}\der^2\kgt\der\eta_g}&=\frac{1}{8(2\pi)^9}\frac{1}{2q^-}(2\pi)\delta(q^--k^--p^--k_g^-) \nonumber \\
    &\times \sum_{\bar{\lambda}\sigma\sigma',\mathrm{color}}
    \left \langle \mathcal{M}^{\lambda\bar{\lambda}\sigma\sigma' \dagger}_{\rm R}[\rho_A] \mathcal{M}^{\lambda\bar{\lambda}\sigma \sigma'}_{\mathrm{R}} [\rho_A] \right \rangle_{Y}\,.
\end{align}
For the inclusive dijet cross-section, the gluon is integrated over the dijet phase space, as will be discussed in the Section~\ref{sub:jetdef}. Here we focus on the gluon phase space with $z_g=k_g^-/q^-\ll 1$ and $\kgt$ finite. In this limit, one can set $z_g=0$ in the delta function. Formally, this approximation violates exact longitudinal momentum conservation, but this violation can be corrected for systematically.

\subsubsection*{Direct terms} We start with the unscattered direct terms corresponding to gluon emission from either quark or antiquark in both the amplitude and complex conjugate amplitude. To extract the slow gluon limit of their contribution to the cross-section, we shall use the expressions for $\mathrm{R}2$ and $\rm \mathrm{R}2'$ where no additional transverse momentum integral is introduced. We generalize the expressions obtained in Section~\ref{sec:real} to $d=4-\varepsilon$ dimensions in order to extract the infrared pole coming from slow and collinear gluon emissions.

In the limit $z_g\to0$, the amplitude for $\mathrm{R}2$ reduces to
\begin{align}
    \left.\Mcal_{\mathrm{R}2}\right|_{\rm slow}&=\frac{ee_fq^-}{\pi}\mu^{-2\varepsilon}\int\der^{2-\varepsilon}\xt\der^{2-\varepsilon}\yt e^{-i(\kt+\kgt) \cdot \xt-i\pt \cdot \yt}\Ncal_{\rm LO,\varepsilon}(\rxyt)\nonumber\\
    &\times\Ccal_{\mathrm{R}2}(\xt,\yt)(-2g)\frac{\et^{\bar\lambda*}\cdot(\kgt-z_g/z_q\kt)}{(\kgt-z_g/z_q\kt)^2}\,.
    \label{eq:dijet-NLO-real-R1-slow}
\end{align}
The slow gluon limit of $\rm \mathrm{R}2'$, with the gluon emitted from the quark, can be obtained from Eq.\,\eqref{eq:dijet-NLO-real-R1-slow} by exchanging $\xt\leftrightarrow\yt$ and $\kt\leftrightarrow\pt$, replacing the color factor $\Ccal_{\mathrm{R}2}$ by $\Ccal_{\rm \mathrm{R}2'}(\xt,\yt)=V(\xt)V^\dagger(\yt)t^a-t^a$ and multiplying the result by an overall minus sign.

As we shall see, it is important to keep the $z_g$ dependence of the transverse momentum  $\kgt-(z_g/z_q)\,\kt$ in the cross-term to properly account for the soft gluon phase space, where $\kgt\ll\kt$ is of the same order as $(z_g/z_q)\, \kt$.
Summing over gluon polarizations, integrating over the slow gluon phase space $z_0\le z_g\le z_f$, and shifting $\kgt\to\kgt-z_g/z_q\kt$ inside the $\kgt$ integral, one finds
\begin{align}
    & \left.\frac{\der\sigma^{\gamma_{\lambda}^*+A\to q\bar{q}+X}}{\der^2\kt\der\eta_q\der^2\pt\der\eta_{\bar q}}\right|_{\mathrm{R}2\times \mathrm{R}2, \rm slow}
    =\frac{\alpha_{\rm em}e_f^2N_c}{(2\pi)^6}\delta(1-z_q-z_{\bar q})\int\der\Pi_{\rm LO,\varepsilon}\Rcal_{\mathrm{LO},\varepsilon}^{\lambda}(\rxyt,\rxytp)\nonumber\\
    &\times(4\alpha_sC_F)\,\Xi_{\rm LO}(\xt,\yt;\xt',\yt')\int_{z_0}^{z_f}\frac{\der z_g}{z_g}e^{i\frac{z_g}{z_q}\kt \cdot \rxxtp} \mu^\varepsilon\int\frac{\der^{2-\varepsilon}\kgt}{(2\pi)^{2-\varepsilon}}\frac{e^{-i\kgt \cdot \rxxtp}}{\kgt^2}\,.
\end{align}
The exponential phase does not contribute to the logarithmic slow gluon divergence and can thus be neglected.
The last integral is infrared divergent in the $\kgt\to0$ limit. Since we have already assumed that $z_g$ is small, this divergence comes physically from the \textit{soft and collinear} phase space for the gluon emission. We have indeed $z_g\to0$ and $\kgt\to0$ so that all the components of the four momentum of the gluon are small, and moreover, the relative transverse momentum with respect to the quark, defined as $z_q\kgt-z_g\kt$, is also small.
Using the following identity (see Appendix~\ref{sub:gluonkernel}) for $\varepsilon<0$, 
\begin{equation}
    \mu^\varepsilon\int\frac{\der^{2-\varepsilon}\kgt}{(2\pi)^{2-\varepsilon}}\frac{e^{-i\kgt \cdot \boldsymbol{r}_{xx'}}}{\kgt^2}=-\frac{1}{4\pi}\left[\frac{2}{\varepsilon}+\ln\left(e^{\gamma_E}\pi\mu^2\rxxtp^2\right)+\mathcal{O}(\varepsilon)\right]\,,
\end{equation}
one can write the previous expression under a form which resembles  Eq.~\eqref{eq:dijet-NLO-slow-xsection1}:
\begin{align}
    & \left.\frac{\der\sigma^{\gamma_{\lambda}^*+A\to q\bar{q}+X}}{\der^2\kt\der\eta_q\der^2\pt\der\eta_{\bar q}}\right|_{\mathrm{R}2\times \mathrm{R}2, \rm slow}
    =\frac{\alpha_{\rm em}e_f^2N_c}{(2\pi)^6}\delta(1-z_q-z_{\bar q})\int\der\Pi_{\rm LO,\varepsilon}\Rcal_{\mathrm{LO},\varepsilon}^{\lambda}(\rxyt,\rxytp)\nonumber\\
    &\times\frac{(-\alpha_s)C_F}{\pi}\ln\left(\frac{z_f}{z_0}\right)\Xi_{\rm LO}(\xt,\yt;\xt',\yt')\left[\frac{2}{\varepsilon}+\ln\left(e^{\gamma_E}\pi\mu^2\rxxtp^2\right)+\mathcal{O}(\varepsilon)\right]\,.
    \label{eq:dijet-NLO-slow-xsection5}
\end{align}
In particular, if we compare this result with Eq.\,\eqref{eq:dijet-NLO-slow-xsection1}, we notice that the two $\varepsilon$ poles will cancel, demonstrating thus that the slow gluon limit is free of collinear divergences. The direct term associated with an unscattered gluon emission from the antiquark and absorbed by the antiquark, denoted as $\rm \mathrm{R}2'\times \rm \mathrm{R}2'$, can be obtained from Eq.\,\eqref{eq:dijet-NLO-slow-xsection5} with $\rxxtp\to\ryytp$ in the logarithm inside the square brackets.

\subsubsection*{Cross-terms} The cross-term, which we label as $\rm \mathrm{R}2\times \rm \mathrm{R}2'$, corresponds to an unscattered gluon emitted from the quark in the amplitude and absorbed by the antiquark in the complex conjugate amplitude. Using Eq.\,\eqref{eq:dijet-NLO-real-R1-slow}, and the corresponding expression for $\rm \mathrm{R}2'$, one finds
\begin{align}
    & \left.\frac{\der\sigma^{\gamma_{\lambda}^*+A\to q\bar{q}+X}}{\der^2\kt\der\eta_q\der^2\pt\der\eta_{\bar q}}\right|_{\mathrm{R}2\times \mathrm{R}2', \rm slow}
    =\frac{\alpha_{\rm em}e_f^2N_c}{(2\pi)^6}\delta(1-z_q-z_{\bar q})\int\der\Pi_{\rm LO}\Rcal_{\mathrm{LO}}^{\lambda}(\rxyt,\rxytp)\nonumber\\
    &\times(-4\,\alpha_s)\,\Xi_{\rm NLO,3}\int_{z_0}^{z_f}\frac{\der z_g}{z_g}\int\frac{\der^{2}\kgt}{(2\pi)^{2}}e^{-i\kgt \cdot \rxypt}\frac{(\kgt-(z_g/z_q)\,\kt)\cdot(\kgt-(z_g/z_{\bar q})\,\pt)}{(\kgt-(z_g/z_q)\,\kt)^2\,(\kgt-(z_g/z_{\bar q})\,\pt)^2}\,.
    \label{eq:dijet-NLO-real-R1R1'-slow}
\end{align}
Observe that the color structure $\Xi_{\rm NLO,3}$ here is the same as that in the $\mathrm{V}3\times \rm LO$ contribution to the dijet cross-section given by  Eq.~\eqref{eq:dijet-NLO-color-V2xLO}. Notice also that in this case, we have written the result directly in $d=4$ dimensions. Indeed the $\kgt$ integral is convergent in two dimensions as long as $z_g$ is not set exactly to 0. This explains why it is important to keep the $z_g$ dependence in $\kgt-(z_g/z_q)\,\kt$ and $\kgt-(z_g/z_{\bar q})\,\pt$. To extract the slow behaviour of this formula, we need then to understand how the $\kgt$ integral diverges as $z_g$ goes to 0. 

Towards that aim, we introduce the integral $\Jcal_{R}$ defined as 
\begin{align}
    \Jcal_{ R}(\rt,\Kt)=\int\frac{\der^2\lt}{(2\pi)^2}e^{-i\lt \cdot \rt}\frac{4\,\lt \cdot (\lt+\Kt)}{\lt^2(\lt+\Kt)^2}\,.
    \label{eq:dijet-NLO-intJcalR-def}
\end{align}
The $\kgt$ integral in Eq.~\eqref{eq:dijet-NLO-real-R1R1'-slow} can be expressed in terms of $\Jcal_R$ as 
\begin{align}
    &\int\frac{\der^{2}\kgt}{(2\pi)^{2}}e^{-i\kgt \cdot \rxypt}\frac{(\kgt-(z_g/z_q)\,\kt)\cdot(\kgt-(z_g/z_{\bar q})\,\pt)}{(\kgt-(z_g/z_q)\,\kt)^2(\kgt-(z_g/z_{\bar q})\,\pt)^2}\nonumber\\
    &=\frac{1}{4}e^{-i\frac{z_g}{z_q}\kt \cdot \rxypt}\Jcal_R\left(\rxypt,\frac{z_g}{z_qz_{\bar q}}\Pt\right)\,.
\end{align}
In Appendix~\ref{sub:JR-asymptot}, we show that 
\begin{align}
    \Jcal_R\left(\rt,\frac{z_g}{z_qz_{\bar q}}\Pt\right)=\frac{1}{2\pi}\left[-2\ln\left(\frac{z_g^2}{4z_q^2z_{\bar q}^2}\Pt^2\rt^2\right)-4\gamma_E-2i\pi\right]+\mathcal{O}(z_g)\,.
    \label{eq:dijet-NLO-JR-smallzg}
\end{align}
Hence with the help of these identities, the cross-section for  $\mathrm{R}2\times \mathrm{R}2'$ (plus its complex conjugate) in the slow gluon limit can be expressed as 
\begin{align}
& \left.\frac{\der\sigma^{\gamma_{\lambda}^*+A\to q\bar{q}+X}}{\der^2\kt\der\eta_q\der^2\pt\der\eta_{\bar q}}\right|_{\mathrm{R}2\times \mathrm{R}2', \rm slow}
    =\frac{\alpha_{\rm em}e_f^2N_c}{(2\pi)^6}\delta(1-z_q-z_{\bar q})\int\der\Pi_{\rm LO}\Rcal_{\mathrm{LO}}^{\lambda}(\rxyt,\rxytp)\nonumber\\
    &\times\frac{\alpha_s}{\pi}\int_{z_0}^{z_f}\frac{\der z_g}{z_g}\left[2\ln\left(\frac{z_g}{2z_qz_{\bar q}}\right)+\ln\left(\Pt^2\rxypt^2\right)+2\gamma_E\right]\Xi_{\rm NLO,3}(\xt,\yt;\xt',\yt') \,.
    \label{eq:dijet-NLO-slow-xsection6}
\end{align}
The attentive reader will notice the nice cancellations between  terms inside the square bracket of this expression with terms in the $\rm V 3\times\rm LO$ contribution in  Eq.\,\eqref{eq:dijet-NLO-slow-xsection3}. These cancellations will be discussed further in next subsection.

\subsubsection{Real scattered contributions} 

We end our discussion of  the slow gluon limit of the NLO corrections to the dijet cross-section from real gluon emissions by considering the case in which the gluon scatters off the shock wave in the amplitude or in the complex conjugate amplitude. Since there are overall 16 real emission diagrams, and 4 of these were just discussed, there are 12 such contributions. 

We will begin with the slow gluon limit of the amplitude with the gluon emitted from the quark before the shock wave, which we label as  $\mathrm{R}1$. In the $z_g\to0$ limit of the  amplitude given by Eqs.\,\eqref{eq:dijet-NLO-R2-Npert-long-final}-\eqref{eq:dijet-NLO-R2-Npert-trans-final}, one finds
\begin{align}
    \left.\Mcal^{\lambda}_{\mathrm{R}1}\right|_{\rm slow}&=\frac{ee_fq^-}{\pi}\int\der^2\xt\der^2\yt\der^2\zt e^{-i\kt \cdot \xt-i\pt \cdot \yt-i\kgt\cdot\zt}\nonumber \\
    &\times \Ncal^{\lambda}_{\rm LO}(\rxyt)\,
    \Ccal_{\mathrm{R}1}(\xt,\yt,\zt)\,\frac{ig}{\pi}\frac{\rzxt \cdot \et^{\bar\lambda*}}{\rzxt^2}\,.
\end{align}
The slow gluon limit of gluon emission before the shock wave from the antiquark, labeled as  $\rm \mathrm{R}1'$ is obtained from $q\leftrightarrow\bar q $ interchange, Hermitian conjugation of  $\Ccal_{\mathrm{R}1}$ and multiplication by an overall minus sign.

To compute the contribution to the cross-section of the product between real gluon emission before the shock wave and after, it is convenient to consider the slow gluon limit of the alternative expression we derived for the diagram $\mathrm{R}2$, i.e.\ Eqs.\,\eqref{eq:dijet-NLO-R1-Npert-long-final}-\eqref{eq:dijet-NLO-R1-Npert-trans-final}.
Again taking the brute force $z_g\to0$ limit of this result, one gets
\begin{align}
    \left.\Mcal^{\lambda}_{\mathrm{R}2}\right|_{\rm slow}&=\frac{ee_fq^-}{\pi}\int\der^2\xt\der^2\yt\der^2\zt e^{-i\kt \cdot \xt-i\pt \cdot \yt-i\kgt\cdot\zt}\nonumber \\
    &\times \Ncal^{\lambda}_{\rm LO}(\rxyt)
    \Ccal_{\mathrm{R}2}(\xt,\yt)\frac{(-ig)}{\pi}\frac{\rzxt \cdot \et^{\bar\lambda*}}{\rzxt^2}\,,
\end{align}
and similarly for $\rm \mathrm{R}2'$. 

With these results in hand, we can compute the product of the amplitudes and complex conjugate amplitudes (without double counting the $\mathrm{R}2\times\mathrm{R}2$, $\mathrm{R}2\times \mathrm{R}2'$, $\rm \mathrm{R}2'\times\rm \mathrm{R}2'$ and $\rm \mathrm{R}2'\times\rm \mathrm{R}2$ terms computed in the previous subsection). The sum over the gluon polarizations gives $\et^{\bar \lambda*,i}\et^{\bar\lambda,j}=\delta^{ij}$ and 
the integral over the gluon transverse momentum $\kgt$ gives a $\delta$ function which freezes its transverse coordinate in the amplitude and complex conjugate amplitude:
\begin{equation}
    \int\frac{\der^2\kgt}{(2\pi)^2}e^{-i\kgt \cdot (\zt-\zt')}=\delta^{(2)}(\zt-\zt')\,.
\end{equation}
Finally integrating over the slow gluon phase space $z_0\le z_g\le z_f$, one finds
\begin{align}
    &\left.\frac{\der\sigma^{\gamma_{\lambda}^*+A\to q\bar{q}+X}}{\der^2\kt\der\eta_q\der^2\pt\der\eta_{\bar q}}\right|_{\rm scatt., slow}=\frac{\alpha_{\rm em}e_f^2N_c}{(2\pi)^6}\delta(1-z_q-z_{\bar q})\int\der\Pi_{\rm LO}\Rcal_{\mathrm{LO}}^{\lambda}(\rxyt,\rxytp)\nonumber\\
    &\times
    \frac{\alpha_s}{\pi^2}\ln\left(\frac{z_f}{z_0}\right) \int \der^2 \zt \left \{  \left[\frac{\rzyt\cdot\rzxpt}{\rzyt^2\rzxpt^2}-\frac{\rzxt\cdot\rzxpt}{\rzxt^2\rzxpt^2}\right]\Xi_{\rm NLO,1}(\xt,\yt,\zt;\xt',\yt') \right.  \nonumber \\
    &+\left[\frac{\rzxt\cdot\rzypt}{\rzxt^2\rzypt^2}-\frac{\rzyt\cdot\rzypt}{\rzyt^2\rzypt^2}\right]\Xi_{\rm NLO,2}(\xt,\yt,\zt;\xt',\yt') \nonumber\\
    &\left. +\frac{1}{2}\left[\frac{\rzxt\cdot\rzxpt}{\rzxt^2\rzxpt^2}-\frac{\rzxt\cdot\rzypt}{\rzxt^2\rzypt^2}+\frac{\rzyt\cdot\rzypt}{\rzyt^2\rzypt^2}-\frac{\rzyt\cdot\rzxpt}{\rzyt^2\rzxpt^2}\right]\Xi_{\rm NLO,4}(\xt,\yt,\zt;\xt',\yt') \right \} +c.c \,.
    \label{eq:dijet-NLO-slow-xsection7}
\end{align}
The color correlators $\Xi_{\rm NLO,1}$ and $\Xi_{\rm NLO,2}$ were introduced in Eq.\,\eqref{eq:dijet-NLO-XiSE2}. Since the product of the amplitudes for gluon emission, from the quark before the shock wave and absorption by the quark after, has the same color topology as the dressed self energy, it is natural to find the same CGC correlator in both contributions. The factor $1/2$ in the term proportional to $\Xi_{\mathrm{NLO},4}$ is introduced to avoid over-counting when applying the complex conjugation.

The color correlator $\Xi_{\rm NLO,4}$, on the other hand, is new. It comes from the product of the color structure $\Ccal_{\mathrm{R}1}$ with itself, $\Xi_{\rm NLO,4}=\frac{1}{N_c}\Tr[\Ccal_{\mathrm{R}1}(\xt,\yt,\zt)\Ccal_{\mathrm{R}1}^\dagger(\xt',\yt',\zt)]$; in other words, an emission from the quark before and absorbed by the quark (or the antiquark) before the shock wave. Expressed in terms of dipoles and quadrupoles, it reads as 
\begin{align}
    \Xi_{\rm NLO,4}(\xt,\yt,\zt;\xt',\yt')  &=\frac{N_c}{2}\left\langle 1-D_{xz}D_{zy}-D_{y'z}D_{zx'}+D_{xx'}D_{y'y}\right\rangle_{Y} \nonumber \\
    & -\frac{1}{2N_c}\Xi_{\rm LO}(\xt,\yt;\xt',\yt')\,.
\end{align}

Eqs.\,\eqref{eq:dijet-NLO-slow-xsection5}, \eqref{eq:dijet-NLO-slow-xsection6} and \eqref{eq:dijet-NLO-slow-xsection7} constitute the slow gluon limit of the real corrections to the NLO dijet cross-section.

\subsection{Proof of JIMWLK factorization}
Now that we have extracted the leading slow gluon divergence of the NLO dijet cross-section, diagram by diagram, we can put our results together. After doing so,  we will demonstrate that the same result can be obtained  by applying the JIMWLK Hamiltonian to the leading order cross-section, thereby providing an explicit proof of rapidity factorization at leading logarithmic accuracy in $x$.

\subsubsection{Combined result for all diagrams in the slow gluon limit} 
\label{subsub:slow-combining}

Let us first sum the term coming from the divergent pieces of the virtual cross-section, (labeled previously as $\rm IR \times\rm LO$) with the real unscattered direct terms (labeled  $\rm \mathrm{R}2\times \rm \mathrm{R}2$ and $\rm \mathrm{R}2'\times \rm \mathrm{R}2'$). Adding Eq.\,\eqref{eq:dijet-NLO-slow-xsection1} (plus its complex conjugate) and Eq.\,\eqref{eq:dijet-NLO-slow-xsection5} (plus the $\rm \mathrm{R}2'\times \rm \mathrm{R}2'$ term), and taking the limit $\varepsilon\to0$, one finds
\begin{align}
    &\left.\frac{\der\sigma^{\gamma_{\lambda}^*+A\to q\bar{q}+X}}{\der^2\kt\der\eta_q\der^2\pt\der\eta_{\bar q}}\right|_{\rm IR \times LO, \rm slow} \!\!\!\!\!\!\!\! + \left.\frac{\der\sigma^{\gamma_{\lambda}^*+A\to q\bar{q}+X}}{\der^2\kt\der\eta_q\der^2\pt\der\eta_{\bar q}}\right|_{\rm LO \times IR , \rm slow} \!\!\!\!\!\!\!\!  +\left.\frac{\der\sigma^{\gamma_{\lambda}^*+A\to q\bar{q}+X}}{\der^2\kt\der\eta_q\der^2\pt\der\eta_{\bar q}}\right|_{\rm R,direct, \rm slow} \nonumber \\
    & =\frac{\alpha_{\rm em}e_f^2N_c}{(2\pi)^6}\delta(1-z_q-z_{\bar q}) \int\der\Pi_{\rm LO}\Rcal_{\mathrm{LO}}^{\lambda}(\rxyt,\rxytp) \nonumber \\
    & \times \frac{\alpha_sC_F}{\pi}\ln\left(\frac{z_f}{z_0}\right)\Xi_{\rm LO}\left[\ln\left(\frac{\rxyt^2}{\rxxtp^2}\right)+\ln\left(\frac{\rxytp^2}{\ryytp^2}\right)\right]\,.
\end{align}
where $\left. \der \sigma^{\gamma_{\lambda}^*+A\to q\bar{q}+X} \right|_{\rm R,direct, \rm slow} = \left. \der \sigma^{\gamma_{\lambda}^*+A\to q\bar{q}+X} \right|_{\mathrm{R}2\times \mathrm{R}2, \rm slow} + \left. \der \sigma^{\gamma_{\lambda}^*+A\to q\bar{q}+X} \right|_{\mathrm{R}2'\times \mathrm{R}2', \rm slow}$.

In order for the limit $\varepsilon\to 0$ from below to be unambiguous, it is important that both poles in Eqs.~\eqref{eq:dijet-NLO-slow-xsection1} and \eqref{eq:dijet-NLO-slow-xsection5} are infrared poles, so that $\varepsilon<0$ in both cases.
Using an identity derived in Appendix~\ref{sub:gluonkernel}, 
\begin{equation}
   \frac{1}{\pi} \int\der^2\zt\left[\frac{\rzxt\cdot\rzxpt}{\rzxt^2\rzxpt^2}-\frac{\rzxt\cdot\rzyt}{\rzxt^2\rzyt^2}\right]=\ln\left(\frac{\rxyt^2}{\rxxtp^2}\right)\,,
   \label{eq:dijet-NLO-jimwlk-kernel-identity}
\end{equation}
one can write the factor inside the square bracket instead as 
\begin{align}
    &\left.\frac{\der\sigma^{\gamma_{\lambda}^*+A\to q\bar{q}+X}}{\der^2\kt\der\eta_q\der^2\pt\der\eta_{\bar q}}\right|_{\rm IR\times LO, \rm slow} \!\!\!\!\!\!\!\! + \left.\frac{\der\sigma^{\gamma_{\lambda}^*+A\to q\bar{q}+X}}{\der^2\kt\der\eta_q\der^2\pt\der\eta_{\bar q}}\right|_{\rm LO \times IR, \rm slow} \!\!\!\!\!\!\!\!  +\left.\frac{\der\sigma^{\gamma_{\lambda}^*+A\to q\bar{q}+X}}{\der^2\kt\der\eta_q\der^2\pt\der\eta_{\bar q}}\right|_{\rm R,direct, \rm slow} \nonumber \\ 
    &=\frac{\alpha_{\rm em}e_f^2N_c}{(2\pi)^6}\delta(1-z_q-z_{\bar q}) \int\der\Pi_{\rm LO}\Rcal_{\mathrm{LO}}^{\lambda}(\rxyt,\rxytp)\frac{\alpha_sC_F}{\pi^2}\ln\left(\frac{z_f}{z_0}\right)\Xi_{\rm LO}(\xt,\yt;\xt',\yt')\nonumber\\
&\times\int\der^2\zt\left[\frac{\rzxt\cdot\rzxpt}{\rzxt^2\rzxpt^2}-\frac{\rzxt\cdot\rzyt}{\rzxt^2\rzyt^2}+\frac{\rzyt\cdot\rzypt}{\rzyt^2\rzypt^2}-\frac{\rzxpt\cdot\rzypt}{\rzxpt^2\rzypt}\right]\,.
\label{eq:dijet-NLO-slow-xsection8}
\end{align}

The same kind of cancellation occurs between the $\mathrm{V}3\times \rm LO$ terms and the unscattered real emission cross-terms. Combining Eqs.\,\eqref{eq:dijet-NLO-slow-xsection3} and \eqref{eq:dijet-NLO-slow-xsection6}, and using the identity Eq.\,\eqref{eq:dijet-NLO-jimwlk-kernel-identity}, we get
\begin{align}
    &\left.\frac{\der\sigma^{\gamma_{\lambda}^*+A\to q\bar{q}+X}}{\der^2\kt\der\eta_q\der^2\pt\der\eta_{\bar q}}\right|_{\mathrm{V}3 \times \mathrm{LO}, \rm slow}+\left.\frac{\der\sigma^{\gamma_{\lambda}^*+A\to q\bar{q}+X}}{\der^2\kt\der\eta_q\der^2\pt\der\eta_{\bar q}}\right|_{\mathrm{R}2\times\mathrm{R}2', \rm slow}=\frac{\alpha_{\rm em}e_f^2N_c}{(2\pi)^6}\delta(1-z_q-z_{\bar q})\nonumber\\
    &\times\int\der\Pi_{\rm LO}\Rcal_{\mathrm{LO}}^{\lambda}(\rxyt,\rxytp)\frac{\alpha_s}{\pi^2}\ln\left(\frac{z_f}{z_0}\right)\Xi_{\rm NLO,3}\int\der^2\zt\left[\frac{\rzxt\cdot\rzyt}{\rzxt^2\rzyt^2}-\frac{\rzxt\cdot\rzypt}{\rzxt^2\rzypt^2}\right] \,.
    \label{eq:dijet-NLO-slow-xsection9}
\end{align}
At this stage, we have managed to express all the terms in terms of a ``JIMWLK-like" kernel, even for the diagrams in which the gluon does not scatter of the shock wave. 

As we have just seen, such a result is highly nontrivial given the delicate cancellations which must occur between the different terms. 

We are now ready to combine Eqs.\,\eqref{eq:dijet-NLO-slow-xsection2}, \eqref{eq:dijet-NLO-slow-xsection4}, \eqref{eq:dijet-NLO-slow-xsection7}, \eqref{eq:dijet-NLO-slow-xsection8} and \eqref{eq:dijet-NLO-slow-xsection9}. There are five different color correlators at NLO represented by $C_F\Xi_{\rm LO}$ and $\Xi_{\rm NLO,i=1...4}$. They are summarized in Table~\ref{tab:NLO-color}.
\begin{table}[tbh]
    \centering
    \begin{tabular}{|c|c|}
    \hline
    $C_F\Xi_{\rm LO}(\xt,\yt;\xt',\yt')$ & $C_F\langle1-D_{xy}-D_{y'x'}+Q_{xy,y'x'}\rangle$\\
    \hline
    $\Xi_{\rm NLO,1}(\xt,\yt,\zt;\xt',\yt')$ & $\frac{N_c}{2}\langle1-D_{y'x'}+Q_{zy,y'x'}D_{xz}-D_{xz}D_{zy}\rangle-\frac{1}{2N_c}\Xi_{\rm LO}$\\
    \hline
    $\Xi_{\rm NLO,2}(\xt,\yt,\zt;\xt',\yt')$ & $\frac{N_c}{2}\langle1-D_{y'x'}+Q_{xz,y'x'}D_{zy}-D_{xz}D_{zy}\rangle-\frac{1}{2N_c}\Xi_{\rm LO}$\\
    \hline 
    $\Xi_{\rm NLO,3}(\xt,\yt;\xt',\yt')$ & $\frac{N_c}{2}\langle1-D_{xy}-D_{y'x'}+D_{xy}D_{y'x'}\rangle-\frac{1}{2N_c}\Xi_{\rm LO}$\\
    \hline  
    $\Xi_{\rm NLO,4}(\xt,\yt,\zt;\xt',\yt')$ & $\frac{N_c}{2}\langle1-D_{xz}D_{zy}-D_{y'z}D_{zx'}+D_{xx'}D_{y'y}\rangle-\frac{1}{2N_c}\Xi_{\rm LO}$\\
    \hline  
    $\Xi_{\rm NLO,4}(\xt,\yt,\zt;\xt',\yt',\zt')$ & $\frac{N_c}{2}\langle1-D_{xz}D_{zy}-D_{y'z}D_{zx'}+Q_{xz,z'x'}Q_{y'z',zy}\rangle-\frac{1}{2N_c}\Xi_{\rm LO}$\\
    \hline 
    \end{tabular}
    \caption{Color correlators contributing to the next-to-leading order cross-section. Only $\Xi_{\rm NLO,1}$ and $\Xi_{\rm NLO,2}$ are not invariant under complex conjugation and the $\xt\to\xt'$, $\yt\to\yt'$ transformations. The color correlator $\Xi_{\rm NLO,4}$ with six arguments does not appear in slow gluon limit, but it will be featured in the real emission contribution to the impact factor.  }\label{tab:NLO-color}
\end{table}
We will now sum up the $\zt$ kernels that appear in front of each correlator using the identity,
\begin{equation}
    \frac{\rzxt\cdot\rzyt}{\rzxt^2\rzyt^2}=\frac{1}{2}\Bigg[-\underbrace{\frac{\rxyt^2}{\rzxt^2\rzyt^2}}_{=\Kcal_{xy}}+\frac{1}{\rzxt^2}+\frac{1}{\rzyt^2}\Bigg]\label{eq:dijet-NLO-jimwk-ker-id}.
\end{equation}
The result of this calculation is summarized in Table~\ref{tab:slow-limit}.
\begin{table}[tbh]
    \centering
    \begin{tabular}{|c|c|c|}
    \hline
    Common factor & Color correlator & Kernel   \\
    \hline
    $\frac{\alpha_s N_c}{2\pi^2}\ln\left(\frac{z_f}{z_0}\right)\int\der^2\zt$ & $C_F\Xi_{\rm LO}(\xt,\yt;\xt',\yt')$ & $-\Kcal_{xx'}-\Kcal_{xy}-\Kcal_{yy'}-\Kcal_{x'y'}$\\
   \hline
    $\frac{\alpha_s N_c}{2\pi^2}\ln\left(\frac{z_f}{z_0}\right)\int\der^2\zt$ & $\Xi_{\rm NLO,1}(\xt,\yt,\zt;\xt',\yt')$ &  $\Kcal_{xy}-\Kcal_{x'y}+\Kcal_{xx'}$\\
    \hline
    $\frac{\alpha_s N_c}{2\pi^2}\ln\left(\frac{z_f}{z_0}\right)\int\der^2\zt$ & $\Xi^{*}_{\rm NLO,1}(\xt',\yt',\zt;\xt,\yt)$ &  $\Kcal_{x'y'}-\Kcal_{xy'}+\Kcal_{xx'}$\\ 
    \hline
    $\frac{\alpha_s N_c}{2\pi^2}\ln\left(\frac{z_f}{z_0}\right)\int\der^2\zt$ & $\Xi_{\rm NLO,2}(\xt,\yt,\zt;\xt',\yt')$ &  $\Kcal_{xy}-\Kcal_{xy'}+\Kcal_{yy'}$\\
    \hline
    $\frac{\alpha_s N_c}{2\pi^2}\ln\left(\frac{z_f}{z_0}\right)\int\der^2\zt$ & $\Xi^{*}_{\rm NLO,2}(\xt',\yt',\zt;\xt,\yt)$ &  $\Kcal_{x'y'}-\Kcal_{x'y}+\Kcal_{yy'}$\\
    \hline
    $\frac{\alpha_s N_c}{2\pi^2}\ln\left(\frac{z_f}{z_0}\right)\int\der^2\zt$ & $\Xi_{\rm NLO,3}(\xt,\yt;\xt',\yt')$ &  $-\Kcal_{xy}+\Kcal_{xy'}-\Kcal_{x'y'}+\Kcal_{x'y}$\\
    \hline
    $\frac{\alpha_s N_c}{2\pi^2}\ln\left(\frac{z_f}{z_0}\right)\int\der^2\zt$ & $\Xi_{\rm NLO,4}(\xt,\yt,\zt;\xt',\yt')$ &  $-\Kcal_{xx'}+\Kcal_{xy'}-\Kcal_{yy'}+\Kcal_{x'y}$\\
    \hline
    \end{tabular}
    \caption{Summary of the slow gluon limit color correlators and kernels.}\label{tab:slow-limit}
\end{table}
Each color structure contributing to the cross-section is a sum of a $1/N_c$ suppressed term and a term proportional to $N_c/2$. By summing all the rows in Table~\ref{tab:slow-limit}, we first notice that all the $1/N_c$ suppressed terms cancel. For the leading $N_c$ terms, it is convenient to organize the terms as follows:
\begin{align}
    & \left.\frac{\der\sigma^{\gamma_{\lambda}^*+A\to q\bar{q}+X}}{\der^2\kt\der\eta_q\der^2\pt\der\eta_{\bar q}}\right|_{\rm slow}=\frac{\alpha_{\rm em}e_f^2N_c}{(2\pi)^6}\delta(1-z_q-z_{\bar q})\ln\left(\frac{z_f}{z_0}\right)\frac{\alpha_sN_c}{4\pi^2}\int\der\Pi_{\rm LO}\Rcal_{\mathrm{LO}}^{\lambda}(\rxyt,\rxytp)\nonumber\\
    &\times \left\langle\int\der^2\zt\left\{\frac{\rxyt^2}{\rzxt^2\rzyt^2}(2D_{xy}-2D_{xz}D_{zy}+D_{zy}Q_{y'x',xz}+D_{xz}Q_{y'x',zy}-Q_{xy,y'x'}-D_{xy}D_{y'x'})\right.\right.\nonumber\\
    &\hspace{1.6cm}+\frac{\rxytp^2}{\rzxpt^2\rzypt^2}(2D_{y'x'}-2D_{y'z}D_{zx'}+D_{zx'}Q_{xy,y'z}+D_{y'z}Q_{xy,zx'}-Q_{xy,y'x'}-D_{xy}D_{y'x'})\nonumber\\
    &\hspace{1.8cm}+\frac{\rxxtp^2}{\rzxt^2\rzxpt^2}(D_{zx'}Q_{xy,y'z}+D_{xz}Q_{y'x',zy}-Q_{xy,y'x'}-D_{xx'}D_{y'y})\nonumber\\
    &\hspace{1.8cm}+\frac{\ryytp^2}{\rzyt^2\rzypt^2}(D_{y'z}Q_{xy,zx'}+D_{zy}Q_{y'x',xz}-Q_{xy,y'x'}-D_{xx'}D_{y'y})\nonumber\\
    &\hspace{1.8cm}+\frac{\rxypt^2}{\rzxt^2\rzypt^2}(D_{xx'}D_{y'y}+D_{xy}D_{y'x'}-D_{zx'}Q_{xy,y'z}-D_{zy}Q_{y'x',xz})\nonumber\\
    &\hspace{1.8cm}\left.\left.+\frac{\rxpyt^2}{\rzxpt^2\rzyt^2}(D_{xx'}D_{y'y}+D_{xy}D_{y'x'}-D_{y'z}Q_{xy,zx'}-D_{xz}Q_{y'x',zy})\right\}\right\rangle_{Y}\,.
\label{eq:dijet-NLO-full-slow}
\end{align}
Written in this form, it is clear that even though each kernel is UV divergent, the divergence is cured by the color structure which vanishes precisely at the location of the singularity. On the other hand, there is no infrared divergence in the equation above because the $\zt$-dependent kernel that appears in each line decays like $1/\zt^4$ at large $\zt$. A posteriori, it  explains why the infrared divergent $\zt$ integrals were not a concern in our diagram-per-diagram discussion of the slow gluon limit.

\subsubsection{The JIMWLK Hamiltonian}

One can now confirm that Eq.\,\eqref{eq:dijet-NLO-full-slow} can be obtained by applying the leading-log JIMWLK Hamiltonian,
\begin{align}
    \mathcal{H}_{\mathrm{JIMWLK}} = \frac{1}{2} \int \der^2 \ut \der^2 \vt \frac{\delta}{\delta A^{+,a}_{\rm cl}(\ut)} \eta^{ab}(\ut,\vt) \frac{\delta}{\delta A^{+,b}_{\rm cl}(\vt)}\,, \label{eq:JIMWLK-Hamiltonian}
\end{align}
acting on the leading order dijet cross-section. Note that in the above expression, 
\begin{align}
    \eta^{ab}(\ut,\vt)&=\frac{1}{\pi}\int\frac{\der^2\zt}{(2\pi)^2}\frac{(\ut-\zt) \cdot (\vt-\zt)}{(\ut-\zt)^2(\vt-\zt)^2}\nonumber\\
    &\times[1+U^\dagger(\ut)U(\vt)-U^\dagger(\ut)U(\zt)-U^\dagger(\zt)U(\vt)]^{ab}\,,
\end{align}
where the $U$ are lightlike Wilson lines in the adjoint representation. To apply $\mathcal{H}_{\rm JIMWLK}$ to the leading order cross-section in  Eq.\,\eqref{eq:dijet-LO-cross-section}, one needs the identity
\begin{equation}
    \frac{\delta V(\xt)}{\delta A^{+,a}_{\rm cl}(\ut)}=-ig\delta^{(2)}(\xt-\ut)V(\xt)t^a\,,\label{eq:JIMWLK-func-der}
\end{equation}
and the relations Eq.\,\eqref{sub:adjfund} between adjoint and fundamental Wilson lines in Appendix~\ref{app:convention}. After some algebraic manipulations, one recovers Eq.\,\eqref{eq:dijet-NLO-full-slow}, thereby establishing the non-trivial  result:
\begin{equation}
    \left.\frac{\der\sigma^{\gamma_{\lambda}^*+A\to q\bar{q}+X}}{\der^2\kt\der\eta_q\der^2\pt\der\eta_{\bar q}}\right|_{\rm slow}=\ln\left(\frac{z_f}{z_0}\right)  \mathcal{H}_{\mathrm{JIMWLK}}\otimes\left.\frac{\der\sigma^{\gamma_{\lambda}^*+A\to q\bar{q}+X}}{\der^2\kt\der\eta_q\der^2\pt\der\eta_{\bar q}}\right|_{\rm LO}.\label{eq:JIMWLKxLO}
\end{equation}
Even if not manifest in \eqref{eq:JIMWLKxLO}, this expression is of order $\mathcal{O}(\alpha_s)$ as the application of the JIMWLK Hamiltonian on the LO cross-section brings a $g^2$ power thanks to Eq.\,\eqref{eq:JIMWLK-func-der}.
The identity \eqref{eq:JIMWLKxLO} enables one to absorb the slow gluon logarithmic divergence into the rapidity evolution of the leading order cross-section. More precisely, the leading order cross-section depends on $z_0$ via the CGC average over color charge configurations (see Eq.\,\eqref{eq:CGC-expectation value}) inside the target at the scale $\Lambda^-_0=z_0\,q^-$. 

Evolving this cross-section with the help of the JIMWLK Hamiltonian up to the factorization scale $z_f$ enables one to cancel the $z_0$ dependence of the NLO cross-section, provided that the stochastic weight functional $W_{z_0}[\rho_A]$ that defines the CGC average satisfies the renormalization group (RG) equation,
\begin{equation}
    \frac{\partial W_{\Lambda^-}[\rho_A]}{\partial \ln(z_0)}=\mathcal{H}_{\rm JIMWLK}W_{\Lambda^-}[\rho_A]\,.
\end{equation}
This RG equation, combined with Eq.\,\eqref{eq:CGC-expectation value}, provides the essence of the CGC EFT. 

\section{Constructing the dijet cross-section in the small cone approximation}

\label{sec:small_cone}

In this section, we will show that our calculation leads to a cross-section for dijet production which is infrared finite. To achieve this, one has to define an infrared and collinear safe cross-section using jets instead of partons to define the final state. Another way to proceed would be to consider the dihadron cross-section. In this latter case, the remaining divergence left in the sum of the virtual $q\bar q+X$ and real $q\bar qg+X$ cross-section is absorbed into the evolution of the fragmentation function into hadrons of the quark and antiquark, as discussed for instance in \cite{Chirilli:2011km,Iancu:2020mos}. Our focus here will be on constructing infrared finite dijet cross-sections in our framework.  

\subsection{Structure of the parton-level NLO cross-section}
In the previous section, we introduced the factorization scale $z_f$, extracted the slow gluon $(z_0 \leq z_g \leq z_f)$ logarithmic divergence, and expressed it as the action of the JIMWLK Hamiltonian acting on the leading order result for the dijet cross-section.
In this section, we will explicitly show the cancellation of divergences when the gluon is fast $(z_f \leq z_g \leq z_q,z_{\bar{q}})$. This is required to demonstrate the finiteness of the impact factor. 

First, we observe that:
\begin{align}
    \left.\Mcal_{\rm IR}\right|_{\rm fast}&=\Mcal_{\rm IR} -  \left.\Mcal_{\rm IR}\right|_{\rm slow}
    \nonumber \\
    &=\frac{ee_fq^-}{\pi}\mu^{-2\varepsilon}\int\der^{2-\varepsilon}\xt\der^{2-\varepsilon}\yt e^{-i\kt \cdot \xt-i\pt \cdot \yt}\Ccal_{\rm LO}(\xt,\yt)\Ncal_{\rm LO,\varepsilon}(\rxyt)\nonumber\\
    &\times \frac{\alpha_sC_F}{2\pi}\left\{\left(\ln\left(\frac{z_q}{z_f}\right)+\ln\left(\frac{z_{\bar q}}{z_f}\right)-\frac{3}{2}\right)\left(\frac{2}{\varepsilon}+\ln(e^{\gamma_E}\pi\mu^2\rxyt^2)\right)+\frac{1}{2}\ln^2\left(\frac{z_{\bar q}}{z_q}\right)\right.\nonumber\\
    &\hspace{1.5cm}\left.-\frac{\pi^2}{6}+\frac{5}{2}-\frac{1}{2}\right\}\,.
\end{align}
Note the presence of the factorization scale $z_f$, which occurs since the slow gluon piece has been subtracted.

Thus there is still a $1/\varepsilon$  pole in the virtual cross-section. At this stage, 
we can summarize the full NLO calculation as: 
\begin{align}
    \alpha_s\left.\frac{\der \sigma^{\gamma_{\lambda}^*+A\to q\bar{q}+X}}{ \der^2 \kt \der^2 \pt \der \eta_q \der \eta_{\bar{q}}}\right|_{\rm NLO}\!\! &=\frac{\alpha_sC_F}{\pi}\left(\ln\left(\frac{z_q}{z_f}\right)+\ln\left(\frac{z_{\bar q}}{z_f}\right)-\frac{3}{2}\right)\times\frac{2}{\varepsilon}\times\left.\frac{\der \sigma^{\gamma_{\lambda}^*+A\to q\bar{q}+X}}{ \der^2 \kt \der^2 \pt \der \eta_q \der \eta_{\bar{q}}} \right|_{\rm LO}\nonumber\\
    &+\ln\left(\frac{z_f}{z_0}\right) \mathcal{H}_{\mathrm{JIMWLK}}\otimes\left.\frac{\der \sigma^{\gamma_{\lambda}^*+A\to q\bar{q}+X}}{ \der^2 \kt \der^2 \pt \der \eta_q \der \eta_{\bar{q}}} \right|_{\rm LO}\nonumber\\
    &+\left.\frac{\der \sigma^{\gamma_{\lambda}^*+A\to q\bar{q}+X}}{ \der^2 \kt \der^2 \pt \der \eta_q \der \eta_{\bar{q}}} \right|_{\rm real, fast}+\left.\frac{\der \sigma^{\gamma_{\lambda}^*+A\to q\bar{q}+X}}{ \der^2 \kt \der^2 \pt \der \eta_q \der \eta_{\bar{q}}} \right|_{\rm virtual, finite}\,.
     \label{eq:dijet-NLO-xsection-summary}
\end{align}
The first term corresponds to the $1/\varepsilon$ pole surviving in the virtual cross-section. The second term is the slow gluon divergence and its associated JIMWLK structure we discussed previously. The ``real, fast" term is the real contribution to the dijet cross-section, with the slow gluon phase space \textit{excluded} since it is already taken into account in $\mathcal{H}_{\rm JIMWLK}$. This contribution also contains a $1/\varepsilon$ pole as we will soon demonstrate. Finally, the last term in Eq.\,\eqref{eq:dijet-NLO-xsection-summary} is the finite piece of the virtual cross-section, namely, what is leftover after subtraction of the slow gluon divergence and the $1/\varepsilon$ pole. 

\subsection{Jet definition and small cone approximation}
\label{sub:jetdef}

We will discuss here how the dijet cross-section is obtained from the $q\bar q+X$ and $q\bar qg+X$ cross-sections at order $\alpha_s$. Without loss of generality, a jet algorithm is defined as a set of measurement functions on the $n$-body phase space $\der\Omega_n$. For instance, at order $\alpha_s$ (see e.g.\ \cite{Kunszt:1992tn,Catani:1996vz}),
\begin{align}
    \frac{\der\sigma^{\gamma_\lambda^{*}+A\to \dijet+X}}{\der\Omega_{ \rm dijet}}&=\int\der\Omega_2\frac{\der\sigma^{\gamma_\lambda^{*}+A\to q\bar q+X}}{\der\Omega_2}\mathcal{S}_{\rm jet;2}(k^\mu,p^\mu)\nonumber \\
    &+\int\der\Omega_3\frac{\der\sigma^{\gamma_\lambda^{*}+A\to q\bar qg+X}}{\der\Omega_3}\mathcal{S}_{\rm jet;3}(k^\mu,p^\mu,k_g^\mu)\,.
\end{align}
The jet definition is encoded in the functions $\mathcal{S}_{\rm jet;2,3}$ which relate the partonic phase space to the jet phase space. The functions $\Scal_{\textrm{jet};i}$ must  be  infrared and collinear safe. 

We shall now define our jet algorithm by specifying the form of $\mathcal{S}_{\rm jet;2}(k^\mu,p^\mu)$ and $\mathcal{S}_{\rm jet;3}(k^\mu,p^\mu,k_g^\mu)$. The dijet phase space $\der\Omega_{2,\rm jet} $ is given by 
\begin{equation}
    \der\Omega_{\rm dijet}=\der^2\ptj\,\der\eta_{J}\,\der^2\ptk\,\der\eta_K\,,
\end{equation}
where $\ptj$ ($\ptk$) and $\eta_J$ ($\eta_K$) are respectively the transverse momentum and rapidity of the two jets. The form of $\Scal_{\rm jet;2}$ is 
\begin{equation}
    \Scal_{\rm jet;2}(p^\mu,k^\mu)=\delta(\kt-\ptj)\delta(\eta_q-\eta_J)\delta(\pt-\ptk)\delta(\eta_{\bar q}-\eta_K)\label{eq:jetdef-S2}\,,
\end{equation}
which simply means that the two jets are identified with the two final state partons. The function $\Scal_{\rm jet;3}$ is more complicated, even though the physical interpretation is elementary: for each pair of partons, the pair is recombined into one jet if  the distance in the rapidity-azimuth plane between both partons to the jet $J$ is smaller than the parameter $R$, the jet radius. This corresponds to the conditions
\begin{align}
    \Delta R_{i,J}^2=\Delta\phi_{i,J}^2+\Delta\eta_{i,J}^2<R^2\,, \nonumber \\
    \Delta R_{k,J}^2=\Delta\phi_{k,J}^2+\Delta\eta_{k,J}^2<R^2\,,
    \label{eq:dijet-NLO-jetdef}
\end{align}
then partons $i$ and $k$ are recombined into the jet $J$ with
\begin{align}
  p_J^\mu=p_i^\mu+p_k^\mu\,,
  \label{eq:dijet-NLO-jetrecomb}
\end{align}
and the remaining third parton forms the jet $K$.

Our primary interest here is to demonstrate that the jet cross-section is finite. It is sufficient then to work in the ``small cone approximation" \cite{Ivanov:2012ms}, as previously also employed in \cite{Boussarie:2016ogo,Roy:2019cux,Roy:2019hwr}, neglecting powers of $R$ which are suppressed for small $R$. The final cross-section takes then the form $A\ln(R)+B$ where the $\ln(R)$ behaviour is the remnant of the singularity when the gluon becomes collinear to the quark or the antiquark. 

In the small $R$ limit, the condition in Eq.\,\eqref{eq:dijet-NLO-jetdef} for the quark and gluon to lie inside the same jet can be written in terms of the ``collinearity" variable 
\begin{equation}
    \boldsymbol{\Ccal_{qg,\perp}}=\frac{z_q}{z_J}\left(\kgt-\frac{z_g}{z_q}\kt\right)\,,
\end{equation}
which satisfies 
\begin{equation}
    \Ccal_{qg,\perp}^2\le R^2\ptj^2\,\textrm{min}\left(\frac{z_g^2}{z_J^2},\frac{(z_J-z_g)^2}{z_J^2}\right)+\mathcal{O}(R^4)\,.
    \label{eq:dijet-NLO-jetdef-smallR}
\end{equation}
A similar condition holds when the gluon is inside the same jet as the antiquark. 

The case where the $q\bar q$ pair forms a jet and the gluon another is sub-leading in the small cone approximation, because of the absence of the collinear singularity between the  quark and antiquark \cite{boussarie:tel-01468540,Roy:2019hwr}. Finally, for a three-jet event (where each out-going parton forms its own jet), one has to integrate over one of the jets, typically the softer one. (For instance, in a realistic dijet measurement, one might consider the leading and sub-leading jets only). 

Such a configuration leads to an infrared divergence associated with a soft singularity in real gluon emission. Strictly speaking, our jet definition does not cure this singularity. However we remind the reader that slow gluons with $z_g\le z_f$ have already been taken into account, via the rapidity evolution of the leading order cross-section. Since all soft gluons are also slow, the rapidity factorization scale $z_f$ acts as a natural infrared cut-off of the soft jet singularity, in such a way that no additional phase space constraint (such as a lower $p_T$ cut) is required to ensure the cross-section is infrared finite. This interplay between rapidity factorization and jet infrared safety, as previously noted in \cite{Roy:2019cux,Roy:2019hwr}, is at the heart of a powerful spacelike-timelike correspondence in high energy QCD~\cite{Hatta:2008st,Mueller:2018llt,Neill:2020bwv}. Indeed this correspondence was exploited in \cite{Caron-Huot:2016tzz} to compute conformal contributions to the NNLO BK kernel.

\subsection{Cancellation of the collinear divergence}

The infrared finiteness of the inclusive dijet cross-section at NLO relies on the cancellation of the collinear divergence between the real and virtual terms, established using the jet definition introduced in the previous subsection. Among all the real contributions, only the direct unscattered gluon emissions from the quark or the antiquark develop a collinear divergence. We will therefore focus on these terms here.
We reemphasize that when integrating over the phase space of the collinear gluon  the logarithmic phase space $z_0\le z_g \le z_f$ has already been taken into account in the 
real pieces contributing to the JIMWLK Hamiltonian.

We now apply the jet definitions in Eqs.\,\eqref{eq:dijet-NLO-jetdef-smallR} and \eqref{eq:dijet-NLO-jetrecomb} to the $\mathrm{R}2\times \mathrm{R}2$ contribution to the $q\bar qg$ cross-section. To isolate the collinear divergence, we work in $d=4-\varepsilon$ dimensions. Performing the the change of variables 
\begin{equation}
    (z_q,\kt,z_{\bar q},\pt,z_g,\kgt)\to(z_J=z_q+z_g,\ptj=\kt+\kgt,z_{K}=z_{\bar q},\ptk = \pt,z_g,\boldsymbol{\Ccal_{qg,\perp}})\,,
\end{equation}
the  $\mathrm{R}2\times \mathrm{R}2$ contribution with gluon and quark inside the same jet reads
\begin{align}
   & \left.\frac{\der\sigma^{\gamma_{\lambda}^*+A \rightarrow \, {\rm dijet}+X}}{\der^2\ptj\der\eta_j\der^2\ptk\der\eta_{K}}\right|_{\mathrm{R}2\times \mathrm{R}2, \mathrm{dijet}}
    =\frac{\alpha_{\rm em}e_f^2N_c}{(2\pi)^6}\delta(1-z_J-z_{K})\int\der\Pi_{\rm LO,\varepsilon}\Rcal_{\mathrm{LO},\varepsilon}^{\lambda}(\rxyt,\rxytp)\nonumber\\
    &\times \alpha_s C_F\,\Xi_{\rm LO}(\xt,\yt;\xt',\yt')\int^{z_J}\der z_g\left[4\left(\frac{1}{z_g}-\frac{1}{z_J}\right)+(2-\varepsilon)\frac{z_g}{z_J^2}\right]\nonumber\\
    &\times\mu^{\varepsilon}\int\frac{\der^{2-\varepsilon} \boldsymbol{\Ccal_{qg,\perp}}}{(2\pi)^{2-\varepsilon}}\frac{1}{\Ccal^2_{qg,\perp}}\Theta\left(\Ccal^2_{qg,\perp,\rm max}-\Ccal^2_{qg,\perp}\right)\,.
\end{align}
Note that we have restored the finite term in $\varepsilon$ coming from the Dirac algebra in the $\mathrm{R}2$ amplitude.
The upper limit of the $\Ccal_{qg,\perp}$ integration is set by the small cone condition. The lower limit of the $z_g$ integral was intentionally left unspecified. It should be $z_f$ for the logarithmically divergent term in $z_g$ (since the phase space $z_g\le z_f$ is part of the slow gluon limit) and $0$ for the finite piece; thus we can rewrite the above expression as 
\begin{align}
    & \left.\frac{\der\sigma^{\gamma_{\lambda}^*+A \rightarrow \, {\rm dijet}+X}}{\der^2\ptj\der\eta_J\der^2\ptk\der\eta_{K}}\right|_{\mathrm{R}2\times \mathrm{R}2, \mathrm{dijet}}=\frac{\alpha_{\rm em}e_f^2N_c}{(2\pi)^6}\delta(1-z_J-z_{K})\int\der\Pi_{\rm LO,\varepsilon}\Rcal_{\mathrm{LO},\varepsilon}^{\lambda}(\rxyt,\rxytp)\nonumber\\
    &\times \alpha_sC_F\Xi_{\rm LO}(\xt,\yt;\xt',\yt')\left\{4\int_{z_f}^{z_J}\frac{\der z_g}{z_g}\mu^{\varepsilon}\int\frac{\der^{2-\varepsilon} \boldsymbol{\Ccal_{qg,\perp}}}{(2\pi)^{2-\varepsilon}}\frac{1}{\Ccal^2_{qg,\perp}}\Theta\left(\Ccal^2_{qg,\perp,\rm max}-\Ccal^2_{qg,\perp}\right)\right.\nonumber\\
    &\left.+\int_0^{z_J}\der z_g\left[-\frac{4}{z_J}+(2-\varepsilon)\frac{z_g}{z_J^2}\right]\mu^{\varepsilon}\int\frac{\der^{2-\varepsilon} \boldsymbol{\Ccal_{qg,\perp}}}{(2\pi)^{2-\varepsilon}}\frac{1}{\Ccal^2_{qg,\perp}}\Theta\left(\Ccal^2_{qg,\perp,\rm max}-\Ccal^2_{qg,\perp}\right)   \right\}\,.
\end{align}
Using the result 
\begin{equation}
    \mu^{\varepsilon}\int^{\left.\Ccal_{qg,\perp}^2\right|_{\rm max}}\frac{\der^{2-\varepsilon}\boldsymbol{\Ccal_{qg,\perp}}}{(2\pi)^{2-\varepsilon}}\frac{1}{\Ccal^2_{qg,\perp}}=-\frac{1}{4\pi}\left[\frac{2}{\varepsilon}-\ln\left(\frac{\left.\Ccal_{qg,\perp}^2\right|_{\rm max}}{\tilde{\mu}^2}\right)\right]+\mathcal{O}(\varepsilon)\,,
\end{equation}
and integrating over $z_g$, then leads to the result
\begin{align}
    &\left.\frac{\der\sigma^{\gamma_{\lambda}^*+A \rightarrow \, {\rm dijet}+X}}{\der^2\ptj\der\eta_J\der^2\ptk\der\eta_{K}}\right|_{\mathrm{R}2\times \mathrm{R}2,\mathrm{dijet}}= \frac{\alpha_sC_F}{\pi}\left.\frac{\der\sigma^{\gamma_{\lambda}^*+A \rightarrow \, {\rm dijet}+X}}{\der^2\ptj\der\eta_J\der^2\ptk\der\eta_{K}}\right|_{\rm LO,\varepsilon}\times\left\{\left(\frac{3}{4}-\ln\left(\frac{z_J}{z_f}\right)\right)\frac{2}{\varepsilon}\right.\nonumber\\
    &\left.+\ln^2(z_J)-\ln^2(z_f)-\frac{\pi^2}{6}+\left(\ln\left(\frac{z_J}{z_f}\right)-\frac{3}{4}\right)\ln\left(\frac{R^2\ptj^2}{\tilde{\mu}^2z_J^2}\right)+\frac{1}{4}+\frac{3}{2}\left(1-\ln\left(\frac{z_J}{2}\right)\right)\right\}\,,   \label{eq:dijet-NLO-real-collinardiv}
\end{align}
where we used the condition in Eq.\,\eqref{eq:dijet-NLO-jetdef-smallR} which neglects terms that are power suppressed in the small cone approximation.

Combining this result with its $\mathrm{R}2'\times \mathrm{R}2'$ counterpart, obtained from $J\leftrightarrow K$ interchange, one sees that the $1/\varepsilon$ pole cancels with Eq.\,\eqref{eq:dijet-NLO-xsection-summary}. The $1/4$ term in the above expression comes from the linear term in $\varepsilon$ of the Dirac algebra multiplying the $1/\varepsilon$ pole. When combined with the same term in the $\rm R2'\times R2'$ contribution, it gives a factor $1/2$. This $1/2$ term cancels with the finite contribution from $\Mcal_{\rm IR}\Mcal^\dagger_{\rm LO}$ (plus its complex conjugate) given below by
\begin{align}
    &\left.\frac{\der \sigma^{\gamma_{\lambda}^*+A\to q\bar{q}+X}}{ \der^2 \kt \der^2 \pt \der \eta_q \der \eta_{\bar{q}}} \right|_{\rm IR\times LO,finite} + \left.\frac{\der \sigma^{\gamma_{\lambda}^*+A\to q\bar{q}+X}}{ \der^2 \kt \der^2 \pt \der \eta_q \der \eta_{\bar{q}}} \right|_{\rm LO\times IR,finite} \nonumber \\
    &=\frac{\alpha_{\rm em}e_f^2N_c}{(2\pi)^6}\delta(1-z_q-z_{\bar q})\int\der\Pi_{\rm LO}\Rcal_{\mathrm{LO}}^{\lambda}(\rxyt,\rxytp) \Xi_{\rm LO}(\xt,\yt;\xt',\yt')\nonumber\\
    &\times \frac{\alpha_sC_F}{2\pi}\left\{\left(\ln\left(\frac{z_q}{z_f}\right)+\ln\left(\frac{z_{\bar q}}{z_f}\right)-\frac{3}{2}\right)\ln(e^{\gamma_E}\pi\mu^2\rxyt^2)+\frac{1}{2}\ln^2\left(\frac{z_{\bar q}}{z_q}\right)-\frac{\pi^2}{6}+\frac{5}{2}-\frac{1}{2}\right\}+c.c.\,,
     \label{eq:dijet-NLO-finite-SE1-V1-SE2uv}
\end{align}
where one should set $z_q,\kt\to z_J,\ptj$ and $z_{\bar q},\pt\to z_{K}, \ptk$ to translate this partonic cross-section into a jet cross-section.
A similar cancellation occurs for the $\mu^2$ dependence which is what one would expect since our final results should not depend on this scale.
These two cancellations provide important cross-checks of our calculation.

In conclusion, we have demonstrated that our results provide an infrared finite dijet cross-section at NLO which can be expressed as 
\begin{align}
    &\alpha_s\left.\frac{\der\sigma^{\gamma_{\lambda}^*+A \rightarrow \, {\rm dijet}+X}}{\der^2\ptj\der\eta_J\der^2\ptk\der\eta_{K}}\right|_{\rm NLO}=\frac{\alpha_{\rm em}e_f^2N_c}{(2\pi)^6}\delta(1-z_J-z_{K})\int\der\Pi_{\rm LO}\Rcal_{\mathrm{LO}}^{\lambda}(\rxyt,\rxytp)\Xi_{\rm LO}\nonumber\\
    &\times\frac{\alpha_sC_F}{\pi}\left\{\left(\ln\left(\frac{z_J}{z_f}\right)-\frac{3}{4}\right)\ln\left(\frac{R^2\ptj^2r_{xy}r_{x'y'}}{4e^{-2\gamma_E}}\right)+\left(\ln\left(\frac{z_{K}}{z_f}\right)-\frac{3}{4}\right)\ln\left(\frac{R^2\ptk^2r_{xy}r_{x'y'}}{4e^{-2\gamma_E}}\right)\right.\nonumber\\
    &\hspace{2cm}\left.+2\ln(z_f)\ln\left(\frac{z_Jz_{K}}{z_f}\right)-\frac{1}{2}\ln^2(z_Jz_{K})+\ln(8)+\frac{11}{2}-\frac{\pi^2}{2}\right\}\nonumber\\
    &\hspace{2cm}+\ln\left(\frac{z_f}{z_0}\right)\mathcal{H}_{\rm JIMWLK}\otimes\left.\frac{\der\sigma^{\gamma_{\lambda}^*+A\to \dijet+X}}{\der^2\ptj\der\eta_J\der^2\ptk\der\eta_{K}}\right|_{\rm LO}+\textrm{other finite terms}\,,
    \label{eq:dijet-NLO-summary-final}
\end{align}
after combining Eqs.\,\eqref{eq:dijet-NLO-xsection-summary}, \eqref{eq:dijet-NLO-real-collinardiv} (plus the $\mathrm{R}2'\times \mathrm{R}2'$ contribution obtained by $q \leftrightarrow \bar{q}$ interchange) and \eqref{eq:dijet-NLO-finite-SE1-V1-SE2uv}. The other finite terms come from the diagrams which are not included in the $\rm IR\times LO + \mathrm{c.c.}$ term, and are displayed in the following section.

\section{Inclusive dijet impact factor at NLO}
\label{sub:final}

\begin{figure}[h]
    \centering
    \includegraphics[width=1.0\textwidth]{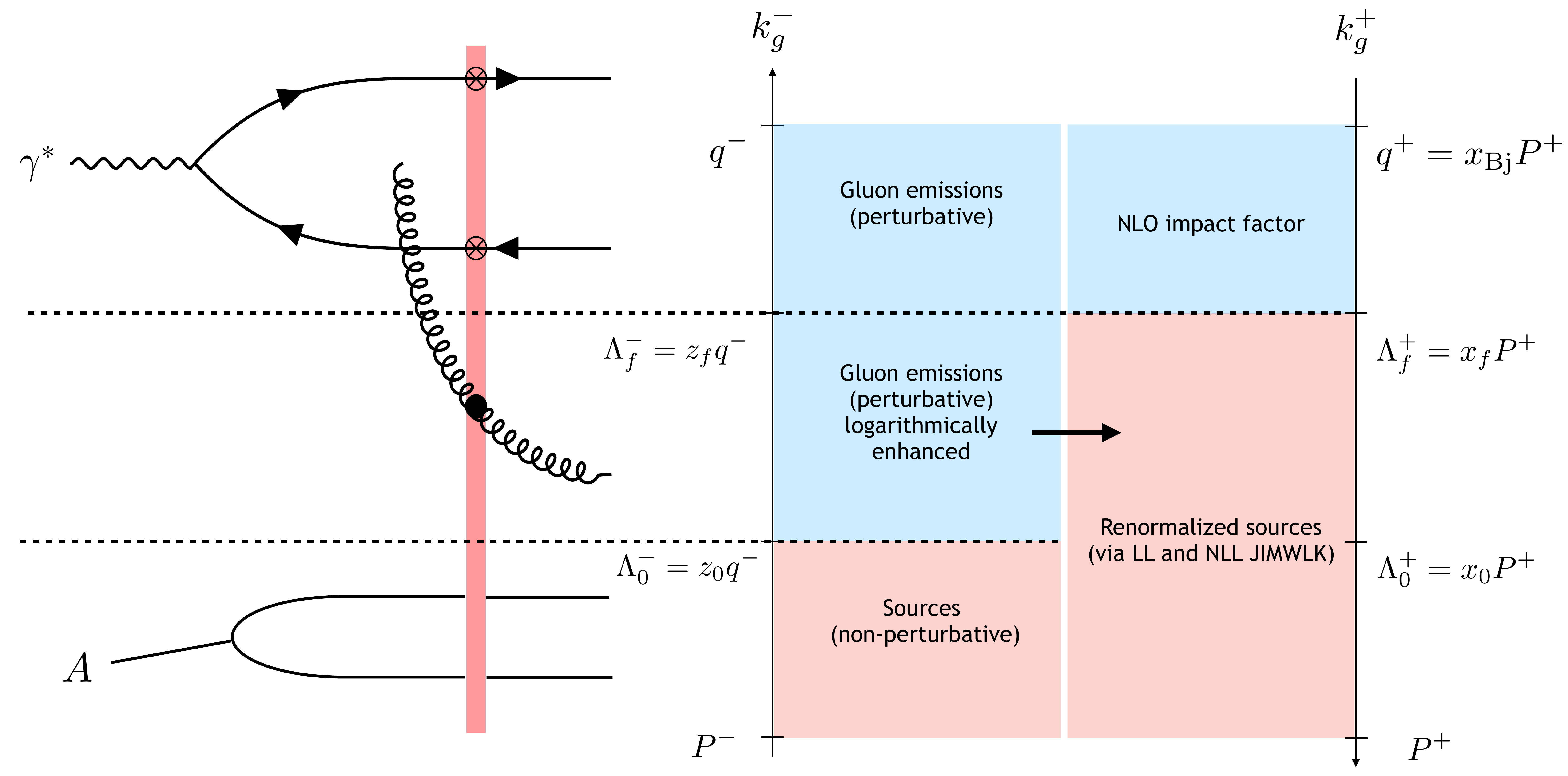}
    \caption{Gluon emission (real or virtual) phase space. Fluctuations carrying ``minus" (``plus") lightcone momentum $P^- <k_g^- < z_0 q^-$ ( $ P^+ > k_g^+ > x_0 P^+$) are accounted for in the sources. Emissions in the interval $z_0 q^-< k_g^- < z_f q^-$ (or equivalently $x_0 P^+ > k_g^+ > x_f P^+$) are logarithmically enhanced $\alpha_s \ln(z_f/z_0) \approx \alpha_s \ln(x_0/x_f)$ and are absorbed by renormalization of sources via JIMWLK evolution. The emissions in the interval $z_f q^- < k_g^- < q^-$ (or equivalently $x_f P^+ > k_g^+ > x_{\rm Bj} P^+$) are not logarithmically enhanced and are part of the NLO impact factor.  The value of $x_f$ can vary depending on the kinematics of the final state. With each successive order in perturbation theory, one expects the corresponding uncertainties to diminish allowing for increasingly quantitative comparisons with experiment.
}
\label{fig:impact-factor}
\end{figure}

In this section, we will summarize our calculation of the inclusive dijet cross-section at NLO in the CGC. We begin with a general discussion of the final result of the previous section given by Eq.\,\eqref{eq:dijet-NLO-summary-final}.  We represent pictorially in Fig.~\ref{fig:impact-factor} the different contributions to the DIS inclusive dijet contribution at next-to-leading order, identifying in particular the contributions of fast and slow gluons with respect to the rapidity phase space encompassed by the virtual photon and the nucleus. 

As noted, the slow gluon divergence can be absorbed into the JIMWLK rapidity evolution of the weight functional $W_{Y_0}^{\rm LL}[\rho_A]\to W_{Y_f}^{\rm LL}[\rho_A]$ with $Y_f=\ln(z_f/z_0)$. On the target side, this is equivalent to an evolution of the weight functional from the scale $x_0$ of the fast sources in the nucleus, up to a scale $x_f\sim \Lambda_\perp^2x_{\rm Bj}/(z_f Q^2)$ with $\Lambda_\perp$ some (perturbative) transverse scale, as obtained for instance in \cite{McLerran:1993ka,McLerran:1993ni} for a large nucleus. One can further assume rapidity factorization to 
hold to NLL accuracy, and promote $W_{Y_f}^{\rm LL}[\rho_A]\rightarrow W_{Y_f}^{\rm NLL}[\rho_A]$.  This enables one to correctly account for terms of order $\mathcal{O}(\alpha_s^{n+1}\ln^n(x_0/x_f))$, to all orders.

With this NLL small-$x$ resummation, the uncertainty in our result comes from two loop contributions not contained in NLL JIMWLK that correspond to the NNLO impact factor, of order $\mathcal{O}(\alpha_s^{2}\ln(x_f/x_{\rm Bj}))$. The factor $\ln(x_f/x_{\rm Bj})$ provides the upper bound for the magnitude of the NNLO impact factor, and it is understood that $\ln(x_f/x_{\rm Bj})$ should be a number of order one and parametrically smaller\footnote{If $x_f\rightarrow x_0$, then because 
$\alpha_s \ln(x_0/x_{\rm Bj})= {\cal O}(1)$, the introduction of a rapidity factorization scale to separate fast from slow modes is no longer meaningful.} than $ 1/\alpha_s$. This condition constrains the range of physical values for $x_f$ one should consider when evaluating the NLO impact factor.

Within this order of accuracy,  up to terms of order $\mathcal{O}(\alpha_s^{2}\ln(x_f/x_{\rm Bj}))$, our result can be expressed as 
\begin{align}
    &\der\sigma^{\gamma_{\lambda}^*+A\rightarrow \, {\rm dijet}+X}=\int\mathcal{D}\rho_A \ W_{Y_f}^{\rm NLL}[\rho_A]\left[\der\sigma^{\gamma_{\lambda}^*+A\to \rm dijet+X}_{\rm LO}+\alpha_s\der\sigma^{\gamma_{\lambda}^*+A\to \rm dijet+X}_{\rm NLO,i.f.}\right]\,,
    \label{eq:dijet-NLO-final}
\end{align}
where $Y_f=\ln(x_0/x_f)$.

The NLO impact factor is given by the sum of three contributions,
\begin{align}
    \alpha_s\left.\frac{\der\sigma^{\gamma_{\lambda}^*+A\rightarrow \, {\rm dijet}+X}}{\der^2\ptj\der\eta_J\der^2\ptk\der\eta_{K}}\right|_{\rm NLO,i.f.} & =\alpha_s\left. \frac{\der\sigma^{\gamma_{\lambda}^*+A\rightarrow \, {\rm dijet}+X}}{\der^2\ptj\der\eta_J\der^2\ptk\der\eta_{K}}\right|_{\rm IRC,i.f.}+\alpha_s\left. \frac{\der\sigma^{\gamma_{\lambda}^*+A\rightarrow \, {\rm dijet}+X}}{\der^2\ptj\der\eta_J\der^2\ptk\der\eta_{K}}\right|_{\rm V,i.f.}\nonumber \\
    &+\alpha_s\left. \frac{\der\sigma^{\gamma_{\lambda}^*+A\rightarrow \, {\rm dijet}+X}}{\der^2\ptj\der\eta_J\der^2\ptk\der\eta_{K}}\right|_{\rm R,i.f.}\,.
    \label{eq:dijet-NLO-full-impact-factor}
\end{align}

The first term in this expression is the finite piece that survives the cancellation of the collinear divergence of the $\rm{R2} \times \rm{R2}$ and $\rm{R2}' \times \rm{R2}'$ real contributions, and the IR divergent $\rm{IR}\times \rm{LO} + ``c.c."$ contribution. It depends on the polarization of the virtual photon via the LO wave function only. In the small $R$ limit, it is given by the first term in Eq.\,\eqref{eq:dijet-NLO-summary-final}:
\begin{align}
    & \alpha_s\left. \frac{\der\sigma^{\gamma_{\lambda}^*+A\rightarrow \, {\rm dijet}+X}}{\der^2\ptj\der\eta_J\der^2\ptk\der\eta_{K}}\right|_{\rm IRC,i.f.} = \frac{\alpha_{\rm em}e_f^2N_c}{(2\pi)^6}\delta(1-z_J-z_{K})\int\der\Pi_{\rm LO}\Rcal_{\mathrm{LO}}^{\lambda}(\rxyt,\rxytp)\Xi_{\rm LO}\nonumber\\
    &\times\frac{\alpha_sC_F}{\pi}\left\{\left(\ln\left(\frac{z_J}{z_f}\right)-\frac{3}{4}\right)\ln\left(\frac{R^2\ptj^2r_{xy}r_{x'y'}}{4e^{-2\gamma_E}}\right)+\left(\ln\left(\frac{z_{K}}{z_f}\right)-\frac{3}{4}\right)\ln\left(\frac{R^2\ptk^2r_{xy}r_{x'y'}}{4e^{-2\gamma_E}}\right)\right.\nonumber\\
    &\hspace{2cm}\left.+2\ln(z_f)\ln\left(\frac{z_Jz_{K}}{z_f}\right)-\frac{1}{2}\ln^2(z_Jz_{K})+\ln(8)+\frac{11}{2}-\frac{\pi^2}{2}\right\} \,.\label{eq:NLO-IRCsafe}
\end{align}

The second term is the contribution to the impact factor coming from the virtual diagrams which are not included in the $\rm IR\times LO+``c.c"$ contribution. These are the UV finite part of $\mathrm{SE}1$, $\mathrm{V}1$ and $\mathrm{V}3$. We will provide the  explicit expression for this term shortly.

The last term in Eq.\,\eqref{eq:dijet-NLO-full-impact-factor} is the contribution from real gluon emissions ({\it excluding finite pieces} from in-cone gluon emission) from the direct $\mathrm{R}2\times \mathrm{R}2$ and $\mathrm{R}2'\times \mathrm{R}2'$ diagrams. This is because, as noted,  their finite contributions are already included in the first term in in Eq.\,\eqref{eq:dijet-NLO-full-impact-factor}. More precisely, this term can be obtained by integrating the gluon over nonsingular regions of phase space which do not contain either slow or collinear divergences. For completeness, and future numerical implementation of our results, we write the full expression for the real $\gamma_\lambda^*+A\rightarrow \, {\rm dijet}+X$ cross-section:
\begin{align}
    \alpha_s\left.\frac{\der\sigma^{\gamma_\lambda^*+A\rightarrow \, {\rm dijet}+X}}{\der^2\ptj\der\eta_J\der^2\ptk\der\eta_K}\right|_{\rm R,i.f.}\!\!\!\!\!=
    &\int\der\Omega_3\left\{\alpha_s\left.\frac{\der\sigma^{\gamma_\lambda^*+A\to q\bar qg+X}}{\der^2\kt\der\eta_q\der^2\pt\der\eta_{\bar q}\der^2\kgt\der\eta_g}\right|_{\mathrm{R}2\times \mathrm{R}2} \!\!\!\!\!\!\!\! \mathcal{S}_{g\notin q-\rm jet}(k^\mu,p^\mu,k_g^\mu)\right.\nonumber\\
    &+\alpha_s\left.\frac{\der\sigma^{\gamma_\lambda^*+A\to q\bar qg+X}}{\der^2\kt\der\eta_q\der^2\pt\der\eta_{\bar q}\der^2\kgt\der\eta_g}\right|_{\mathrm{R}2'\times \mathrm{R}2'} \!\!\!\!\!\!\!\! \mathcal{S}_{g\notin \bar q-\rm jet}(k^\mu,p^\mu,k_g^\mu)\nonumber\\
    &\left.+\alpha_s\left.\frac{\der\sigma^{\gamma_\lambda^*+A\to q\bar qg+X}}{\der^2\kt\der\eta_q\der^2\pt\der\eta_{\bar q}\der^2\kgt\der\eta_g}\right|_{\rm other} \!\!\!\!\!\!\!\! \mathcal{S}_{\rm jet,3}(k^\mu,p^\mu,k_g^\mu)\right\}\,.
    \label{eq:NLO-real-infrared-finite}
\end{align} 
The expressions for the  $\gamma_\lambda^{*}\to q \bar q g+X$ cross-sections which enter in the integrands of  Eq.\,\eqref{eq:NLO-real-infrared-finite} are obtained by squaring the real amplitudes. The first two terms, labeled respectively $\rm R2\times R2$ and $\rm R2'\times R2'$ come from the modulus squares of the amplitude $\rm R2$ and $\rm R2'$.
The jet functions $\mathcal{S}_{g\notin q-\rm jet}$ ( $\mathcal{S}_{g\notin \bar q-\rm jet}$) select configurations in which the gluon does not belong to the quark (antiquark) jet, as such configurations are already accounted for in Eq.\,\eqref{eq:NLO-IRCsafe}, computed here in the small $R$ approximation.  The contribution labeled ``other" gathers all the remaining terms; those coming from the direct $\rm R1\times \rm R1$ and $\rm R1'\times \rm R1' $ diagrams, the cross-term diagrams $\mathrm{R}1\times\mathrm{R}1'$, $\mathrm{R}1\times\mathrm{R}2$, $\mathrm{R}1\times\mathrm{R}2'$, $\mathrm{R}1'\times\mathrm{R}2$, $\mathrm{R}1'\times\mathrm{R}2'$ (and their conjugates), and likewise the contribution from the diagrams $\mathrm{R}2\times\mathrm{R}2'$, $\mathrm{R}2'\times\mathrm{R}2$. The jet function $\mathcal{S}_{\rm jet;3}(k^\mu,p^\mu,k_g^\mu)$ was previously defined in Section \ref{sub:jetdef}. We remind the reader that the three functions $\mathcal{S}_{g\notin q-\rm jet}$, $\mathcal{S}_{g\notin \bar q-\rm jet}$ and $\mathcal{S}_{\rm jet;3}$ depend on the jet algorithm and the event selection. Therefore the 3-body phase space integral is usually performed numerically.

We will now gather the expressions for the various terms which enter in Eq.\,\eqref{eq:dijet-NLO-full-impact-factor} and Eq.\,\eqref{eq:NLO-real-infrared-finite} in the case of a longitudinally polarized photon. The corresponding results for a transversely polarized virtual photon, which are considerably lengthier, are given in appendix~\ref{app:final-res-trans}. To guide the reader, we also gather our intermediate notations in Table~\ref{tab:params_dijet}. 

\begin{table}[h]
    \centering
    \caption{Notations for the longitudinal DIS inclusive dijet cross-section}
    \label{tab:params_dijet}
    \begin{tabular}{ll}
    \hline \hline 
    $\bar Q^2=z_qz_{\bar q}Q^2$ &  effective virtuality squared in $\mathrm{LO}$\\
    $\bar Q_{\mathrm{R}2}^2=z_{\bar q}(1-z_{\bar q})Q^2$ & effective virtuality squared in $\mathrm{R2}$ \\
    $\bar Q_{\rm R2'}^2=z_{q}(1-z_{q})Q^2$ & effective virtuality squared in $\mathrm{R2'}$  \\
    $X_{\rm R}^2=z_qz_{\bar q}\rxyt^2+z_qz_g\rzxt^2+z_{\bar q}z_g\rzyt^2$ & effective $q{\bar q}g$ dipole size squared \\
    & in $\mathrm{R}1$, $\mathrm{R}1'$  \\ 
    $\wt=(z_q\xt+z_g\zt)/(z_q+z_g)$ & quark transverse coordinate \\
    & before gluon emission in $\mathrm{R}2$\\
    $\wtbar=(z_{\bar q}\yt+z_g\zt)/(z_{\bar q}+z_g)$ & antiquark transverse coordinate  \\
    & before gluon emission in $\mathrm{R}2'$\\
    $X_{\rm V}^2=z_{\bar q}(z_q-z_g)\rxyt^2+z_g(z_q-z_g)\rzxt^2+z_{\bar q}z_g\rzyt^2$ & effective $q\bar{q}g$ dipole size squared\\
    & in $\mathrm{SE}1$ and $V_1$ \\
    $X_{\rm V'}^2=z_{q}(z_{\bar q}-z_g)\rxyt^2+z_g(z_{\bar q}-z_g)\rzyt^2+z_{q}z_g\rzxt^2$ & effective $q\bar{q}g$ dipole size squared \\
    & in $\mathrm{SE}1'$ and $V_1'$ \\
    $\Pt=z_{\bar q}\kt-z_q\pt$ & quark-antiquark dijet relative \\ & transverse momentum\\
    $\vect{\Delta}=\kt+\pt$ & quark-antiquark dijet transverse \\
    &momentum imbalance \\
    $Q_{\mathrm{V}3}^2=(z_q - z_g)(z_{\bar{q}} + z_g)Q^2$ & \\
    $\Delta_{\mathrm{V}3}^2=\left(1-\frac{z_g}{z_q}\right)\left(1+\frac{z_g}{z_{\bar q}}\right)\Pt^2$ & \\
    \hline \hline
    \end{tabular}
\end{table}

The NLO impact factor contribution from the remaining virtual diagrams (in which the jet definition is trivial) is 
\begin{align}
    &\alpha_s\left. \frac{\der\sigma^{\gamma_{\rm L}^*+A\rightarrow \, {\rm dijet}+X}}{\der^2\ptj\der\eta_J\der^2\ptk\der\eta_{K}}\right|_{\rm V,i.f.}= \alpha_s\left. \frac{\der\sigma^{\gamma_{\rm L}^*+A\rightarrow \, {\rm dijet}+X}}{\der^2\ptj\der\eta_J\der^2\ptk\der\eta_{K}}\right|^{(a)}_{\rm V,i.f.} + \alpha_s\left. \frac{\der\sigma^{\gamma_{\rm L}^*+A\rightarrow \, {\rm dijet}+X}}{\der^2\ptj\der\eta_J\der^2\ptk\der\eta_{K}}\right|^{(b)}_{\rm V,i.f.} \,,
   \label{eq:dijet-NLO-long-virtual-final}
\end{align}
where
\begin{align}
    &\alpha_s\left.\frac{\der\sigma^{\gamma_{\rm L}^*+A\rightarrow \, {\rm dijet}+X}}{\der^2\ptj\der\eta_J\der^2\ptk\der\eta_{K}}\right|^{(a)}_{\rm V,i.f.} \!\!\!\!\! = \frac{\alpha_{\rm em}e_f^2N_c}{(2\pi)^6} \delta(1-z_J-z_K) \int\der\Pi_{\rm LO} 8z_J^3z_{K}^3Q^2K_0(\bar Qr_{x'y'}) \int_0^{z_J}\frac{\der z_g}{z_g} \nonumber\\
    & \times \frac{\alpha_s}{\pi}\int\frac{\der^2\zt}{\pi} \left\{\frac{1}{\rzxt^2}\left[\left(1-\frac{z_g}{z_J}+\frac{z_g^2}{2z_J^2}\right) e^{-i\frac{z_g}{z_J}\kt \cdot \rzxt} K_0(QX_V)-\Theta(z_f-z_g)K_0(\bar Qr_{xy})\right]\right. \Xi_{\rm NLO,1} \nonumber\\
    &-\frac{1}{\rzxt^2}\left[\left(1-\frac{z_g}{z_J}+\frac{z_g^2}{2z_J^2}\right) e^{-\frac{\rzxt^2}{\rxyt^2e^{\gamma_E}}}K_0(\bar Qr_{xy})-\Theta(z_f-z_g) e^{-\frac{\rzxt^2}{\rxyt^2e^{\gamma_E}}} K_0(\bar Qr_{xy})\right] C_F\Xi_{\rm LO}\nonumber\\
    &-\frac{\rzxt\cdot\rzyt}{\rzxt^2\rzyt^2}\Bigg[\left(1-\frac{z_g}{z_J}\right)\left(1+\frac{z_g}{z_K}\right)\left(1-\frac{z_g}{2z_J}-\frac{z_g}{2(z_K+z_g)}\right) e^{-i\frac{z_g}{z_J}\kt \cdot \rzxt} K_0(QX_V)  \nonumber \\
    & \quad \quad \quad \quad \quad \left. -\Theta(z_f-z_g)K_0(\bar Q r_{xy}) \Bigg] \Xi_{\rm NLO,1} +(J\leftrightarrow K) \right \}  + c.c. \,,
   \label{eq:Vif-a}
\end{align} 
contains the finite part of $\mathrm{SE}1$ and $\mathrm{V1}$, and
\begin{align}
    &\alpha_s\left.\frac{\der\sigma^{\gamma_{\rm L}^*+A\rightarrow \, {\rm dijet}+X}}{\der^2\ptj\der\eta_J\der^2\ptk\der\eta_{K}}\right|^{(b)}_{\rm V,i.f.} \!\!\!\!\! = \frac{\alpha_{\rm em}e_f^2N_c}{(2\pi)^6} \delta(1-z_J-z_K) \int\der\Pi_{\rm LO} 8z_J^3z_{K}^3Q^2K_0(\bar Qr_{x'y'}) \int_0^{z_J}\frac{\der z_g}{z_g}
    \nonumber \\
    & \times \frac{\alpha_s}{\pi} \left\{  K_0( \bar{Q}_{\mathrm{V3}} r_{xy})\left[\left(1-\frac{z_g}{z_J}\right)^2\left(1+\frac{z_g}{z_K}\right)(1+z_g) e^{i(\Pt+z_g\vect{\Delta})\cdot \rxyt} K_0(-i\Delta_{\mathrm{V}3}r_{xy})  \right. \right. \nonumber\\
    &\left.-\left(1-\frac{z_g}{2z_J}+\frac{z_g}{2z_K}-\frac{z_g^2}{2z_Jz_K}\right) e^{i\frac{z_g}{z_J}\kt \cdot \rxyt} \Jcal_{\odot}\left(\rxyt,\left(1-\frac{z_g}{z_J}\right)\Pt,\Delta_{\mathrm{V}3}\right)  \right] \nonumber\\
    & +\Theta(z_f-z_g) K_0(\bar{Q} r_{xy}) \ln\left(\frac{z_gP_\perp r_{xy}e^{\gamma_E}}{2z_Jz_K}\right)   +(J\leftrightarrow K) \Bigg\} \Xi_{\rm NLO,3}  + c.c. \,,
    \label{eq:Vif-b}
\end{align}
contains the finite part of $\mathrm{V}3$. Recall that the other virtual pieces are implicitly contained in Eq.\,\eqref{eq:NLO-IRCsafe}.

In these expression, we have set $z_q=z_J$, $z_{\bar{q}}= z_K$, $\kt=\ptj$ and $\pt=\ptk$ according to the jet definition in Eq.\,\eqref{eq:jetdef-S2} for the 2-body phase space.
We have omitted here the dependence on transverse coordinates in the color structures $\Xi_{\rm LO}$ and $\Xi_{\mathrm{NLO}, i}$. They are defined in Table~\ref{tab:NLO-color}. It should be understood that these color structures are defined without the CGC average $\langle...\rangle_Y$, since this average is performed within Eq.\,\eqref{eq:dijet-NLO-final}.

The $(J\leftrightarrow K)$ contribution is obtained after quark-antiquark interchange (which also applies to the boundaries of the $z_g$ integral) as explained in the previous sections for each diagram. In particular, the color structure $\Xi_{\rm NLO,1}$ becomes $\Xi_{\rm NLO,2}$ after this transformation, while $\Xi_{\rm NLO,3}$ is unchanged. We emphasize that these expressions are free of all divergences.

For the  $\gamma^{*}_{\rm L}\to q \bar q g+X$ cross-sections which enter in the integrand of Eq.\,\eqref{eq:NLO-real-infrared-finite}, the $\mathrm{R}2\times \mathrm{R}2$ term was computed previously:
\begin{align}
    &\alpha_s\left.\frac{\der\sigma^{\gamma_{\rm L}^*+A\to q\bar qg+X}}{\der^2\kt\der\eta_q\der^2\pt\der\eta_{\bar q}\der^2\kgt\der\eta_g}\right|_{\mathrm{R}2\times \mathrm{R}2}=\frac{\alpha_{\rm em}e_f^2N_c}{(2\pi)^8}\delta(1-z_q-z_{\bar q}-z_g)\alpha_s\int\der\Pi_{\rm LO}C_F\Xi_{\rm LO}\nonumber\\
    &\times\left\{ 32z_qz_{\bar q}^3(1-z_{\bar q})^2Q^2\left(1+\frac{z_g}{z_q}+\frac{z_g^2}{2z_q^2}\right)K_0(\bar Q_{\mathrm{R}2}r_{xy})K_0(\bar Q_{\mathrm{R}2}r_{x'y'})\frac{e^{-i\kgt \cdot (\xt-\xt')}}{(\kgt-\frac{z_g}{z_q}\kt)^2}\right\}\,.
    \label{eq:dijet-NLO-long-R1R1-final}
\end{align}
The $\rm R2'\times R2'$ contribution is obtained from the $\mathrm{R}2\times \mathrm{R}2$ contribution after  $q\leftrightarrow\bar q$ interchange. In contrast to the virtual pieces, we do not subtract the slow gluon behaviour in order to avoid lengthy expressions. One should therefore keep in mind that the slow behavior has to be subtracted from these expressions (using a ``+" prescription similar to what is done in Eq.\,\eqref{eq:dijet-NLO-long-virtual-final}), since slow gluons are already accounted for in the second term of Eq.\,\eqref{eq:dijet-NLO-summary-final}.

Finally, the ``other" contribution  in Eq.\,\eqref{eq:NLO-real-infrared-finite} is given by\footnote{For the $\rm R2\times \rm R2'$ term  included in this formula, we used the alternative ``symmetric" expression for real gluon emission after the shock wave derived in section \ref{sub:R1}. Furthermore, the prefactors of $1/2$ in the first, fourth and fifth terms in the brackets in Eq.\,\eqref{eq:dijet-NLO-long-real-other-final} are introduced to avoid over-counting due to quark-antiquark interchange and complex conjugation.}
\begin{align}
    &\alpha_s\left.\frac{\der\sigma^{\gamma_{\rm L}^*+A\to q\bar qg+X}}{\der^2\kt\der\eta_q\der^2\pt\der\eta_{\bar q}\der^2\kgt\der\eta_g}\right|_{\rm other}=\frac{\alpha_{\rm em}e_f^2N_c}{(2\pi)^8}\delta(1-z_q-z_{\bar q}-z_g)\,\alpha_s\,\int\der\Pi_{\rm LO}8z_q^3z_{\bar q}^3Q^2\nonumber\\
    &\times\int\frac{\der^2\zt}{\pi}\frac{\der^2\zt'}{\pi} e^{-i\kgt \cdot (\zt-\zt')}\left\{\Bigg[-\frac{1}{2}\frac{\rzxt\cdot\rzytp}{\rzxt^2\rzytp^2}K_0(\bar Q_{\mathrm{R}2}r_{wy})K_0(\bar Q_{\mathrm{R}2'}r_{\bar w' x'})\right.\nonumber\\
    &\times\left(1+\frac{z_g}{2z_q}+\frac{z_g}{2z_{\bar q}}\right)\Xi_{\rm NLO,3}(\wt,\yt;\xt',\wtbar')
    -\frac{\rzxt\cdot\rzxtp}{\rzxt^2\rzxtp^2}K_0(QX_{\rm R})K_0(\bar Q_{\mathrm{R}2}r_{w'y'})\nonumber\\
    &\times\left(1+\frac{z_g}{z_q}+\frac{z_g^2}{2z_{q}^2}\right)\Xi_{\rm NLO,1}(\xt,\yt,\zt;\wt',\yt')+\frac{\rzyt\cdot\rzxtp}{\rzyt^2\rzxtp^2}K_0(QX_{\rm R})K_0(\bar Q_{\mathrm{R}2'}r_{w'y'})\nonumber\\
    &\times\left(1+\frac{z_g}{2z_q}+\frac{z_g}{2z_{\bar q}}\right)\Xi_{\rm NLO,1}(\xt,\yt,\zt;\wt',\yt')+\frac{1}{2}\frac{\rzxt\cdot\rzxtp}{\rzxt^2\rzxtp^2}K_0(QX_{\rm R})K_0(QX'_R)\nonumber\\
    &\times\left(1+\frac{z_g}{z_q}+\frac{z_g^2}{2z_{ q}^2}\right)\Xi_{\rm NLO,4}(\xt,\yt,\zt;\xt',\yt',\zt')-\frac{1}{2}\frac{\rzyt\cdot\rzxtp}{\rzyt^2\rzxtp^2}K_0(QX_{\rm R})K_0(QX'_R)\nonumber\\
    &\times\left(1+\frac{z_g}{2z_q}+\frac{z_g}{2z_{\bar q}}\right)\Xi_{\rm NLO,4}(\xt,\yt,\zt;\xt',\yt',\zt')+(q\leftrightarrow\bar q) \Bigg] +c.c.\Bigg\}
    \label{eq:dijet-NLO-long-real-other-final}\,.
\end{align}

The expression in Eq.\,\eqref{eq:dijet-NLO-final} is the final result of our computation of the DIS inclusive dijet cross-section. To compute the cross-section specifically for longitudinally polarized photons, one needs the  LO cross-section, the expression for which is  given in Eq.\,\eqref{eq:dijet-LO-cross-section} and the NLO impact factor given by  Eq.\,\eqref{eq:dijet-NLO-full-impact-factor}.  
The individual terms of the latter are,
\begin{enumerate}[i]
    \item Eq.\,\eqref{eq:NLO-IRCsafe}\,,
    \item Eq.\,\eqref{eq:NLO-real-infrared-finite} (with explicit expressions for 
    its individual parts contained in Eq.\,\eqref{eq:dijet-NLO-long-R1R1-final} and Eq.\,\eqref{eq:dijet-NLO-long-real-other-final})\,,
    \item Eq. \eqref{eq:dijet-NLO-long-virtual-final} (with explicit expressions for its individual parts given in Eqs.\,\eqref{eq:Vif-a} and \eqref{eq:Vif-b}).
\end{enumerate}

\section{Summary and Outlook}

\label{sec:NLO_conclusion}

In this work, we performed the first complete next-to-leading order computation of inclusive dijet production in deeply inelastic electron-nucleus scattering at small $x_{\rm Bj}$. In these small $x_{\rm Bj}$ kinematics, the dominant contribution to inclusive dijet production at leading order comes from the splitting of longitudinally or transversely polarized virtual photons into a quark-antiquark pair, which scatters off gluon fields in the nucleus. At small momentum fractions $x$, the gluon fields have maximal occupancy characterized by a saturation scale $Q_s\gg \Lambda_{\rm QCD}$ characterized by  classical shock wave configurations $A_{\rm cl}[\rho]$. In the CGC EFT, these shock wave fields are coupled to static color sources $\rho$ at larger $x$ in the nuclear target, which are represented by a nonperturbative gauge invariant stochastic weight functional $W_Y[\rho]$. The LO computation to inclusive dijet production in this framework was first performed in \cite{Dominguez:2011wm} and was shown to be sensitive to both dipole and quadrupole Wilson line correlators which contain information on all-twist nonperturbative color correlations in the nuclear target. 

Following the covariant perturbation theory framework of the computation of the photon+dijet NLO impact factor in \cite{Roy:2019cux,Roy:2019hwr}, we computed here all real and virtual gluon emissions that constitute the NLO corrections to inclusive dijet production in the nuclear shock wave background. We showed that in the slow gluon limit one obtains rapidity divergences of order $\alpha_s \ln(x_0/x_f)$, which can in principle become $\mathcal{O}(1)$ at sufficiently high energies. Here $x_0 = \Lambda^+_0/P^+$ is a scale characterizing the target color sources at the longitudinal momentum scale $\Lambda^+_0$.  The rapidity factorization scale $x_f$ separates fast (relative to $q^-$) real and virtual gluon emissions  which accompany the dijet from slow ones that can be absorbed into the rapidity evolution of $W_{Y_f}[\rho]$ from its initial nonperturbative distribution specified at the momentum scale $\Lambda^+_0$. We showed explicitly that the JIMWLK Hamiltonian describes this rapidity evolution, which resums the stated leading rapidity logs (LLx) to all orders in perturbation theory. An immediate consequence is that one obtains the LLx evolution equations for the dipole and quadrupole Wilson line correlators in the Balitsky-JIMWLK hierarchy. 

We also showed, employing a small-cone approximation, that a collinear divergence that survives in the real emission diagrams and an infrared divergence that survives in the virtual emission diagrams\footnote{We demonstrated explicitly along the way that all other apparent ultraviolet, soft and collinear divergences cancel in the intermediates steps of our computation.} can be absorbed into infrared and collinear safe jet functions. The finite $\mathcal{O}(\alpha_s)$ terms that remain constitute the next-to-leading order inclusive dijet impact factor, for which we obtain explicit expressions that can be numerically evaluated. 
When combined with LLx JIMWLK evolution of slow gluons, we see that our computation of the dijet cross-section is of $\mathcal{O}(\alpha_s \ln(x_f/x_{\rm Bj}))$ accuracy. These results are necessary for quantitative comparisons of the CGC EFT to anticipated experimental results from the EIC. However if we assume that rapidity factorization holds at NNLO, our computation  is accurate, up to missing terms of  $\mathcal{O}(\alpha_s^2 \ln(x_f/x_{\rm Bj})$, if we combine our NLO impact factor results with  next-to-leading logs in $x$ (NLLx) JIMWLK/BK evolution. As we noted in the introduction, formal results for these evolution equations are available; their practical implementation has seen significant developments as well. Missing at this order of accuracy are $\mathcal{O}(\alpha_s^2)$ terms in the two-loop NNLO impact factor that combine with the LLx terms in the rapidity evolution. 

In particular, a measurement of great theoretical and  phenomenological interest is the limit in which the dijets are almost back-to-back in transverse space \cite{Dominguez:2011wm}. We anticipate a significant reduction in the complexity of the numerical evaluation of our NLO results in this kinematic limit. At small $x$, this regime\footnote{The connection between linear evolution at moderate $x$ and nonlinear evolution at small $x$ was explored in this context in \cite{Balitsky:2015qba,Balitsky:2016dgz}.} has been extensively explored at NLO within the transverse momentum dependent (TMD) parton distribution framework \cite{Xiao:2017yya,Hatta:2020bgy,Hatta:2021jcd,Hentschinski:2021lsh}.  Recent numerical studies at leading order \cite{Kotko:2017oxg,Fujii:2020bkl,vanHameren:2019ysa,Altinoluk:2021ygv,Boussarie:2021lkb} suggest the importance of kinematic power \cite{Kotko:2015ura,vanHameren:2016ftb,Petreska:2018cbf} and genuine saturation effects \cite{Altinoluk:2019fui,Altinoluk:2019wyu,Boussarie:2020vzf} at kinematics accessible at current and future colliders. 

These contributions are included in our framework (CGC EFT) but absent in the TMD formalism. With our results, it is therefore now feasible to promote these studies to next-to-leading order, enabling more accurate predictions of the effects of gluon saturation on azimuthal dijet correlations at the EIC \cite{Zheng:2014vka,Dumitru:2015gaa,Dumitru:2018kuw}. We can also systematically explore the appearance of Sudakov double logarithms (and their resummation~\cite{Mueller:2012uf,Mueller:2013wwa}). These arise from the imperfect cancellation (in back-to-back kinematics) of virtual and real contributions and one can study their interplay with the effects from gluon saturation. Further, in addition to such logarithmically enhanced terms, our expressions should contain  genuine $\alpha_s$ suppressed contributions from the impact factor. We will pursue these studies in a subsequent publication.

A further application of our results for the NLO impact factor is to the inclusive production of a dijet pair in ultraperipheral nuclear collisions at RHIC and the LHC; this limit of photon-nucleus collisions is obtained straightforwardly by taking the $Q^2 \rightarrow 0$ limit of our results. 

We conclude with an outlook on future theoretical studies that are suggested by our work. It is in principle straightforward  to extend our results for massless quarks to  massive quarks, whose collinear divergences are  regulated by their mass. This could pave the way to promote current LO studies of the inclusive production of open heavy flavors and quarkonia to NLO+NLL accuracy. Similar studies are being carried out for the computation of charm structure functions \cite{Beuf:2021qqa} and exclusive $J/\psi$ production \cite{Mantysaari:2021ryb}. Another interesting possibility is to integrate out one of the jets in our differential cross-section, and obtain the NLO impact factor for single inclusive jet production. This computation is 
very similar to the NLO studies of inclusive forward 
jet production in proton-nucleus collisions~\cite{Iancu:2020jch}.

A more ambitious program is to capitalize on the techniques employed in this paper to extend the computation to two-loop order. At this order in perturbation theory it will be possible to unambiguously test NLL JIMWLK factorization for semi-inclusive process, determine how the coupling runs as a function of $Q_s$, and extract the $\alpha_s^2$ suppressed NNLO impact factor. 

\section*{Acknowledgements}

We are grateful to Renaud Boussarie, Edmond Iancu, Yair Mulian, and Bowen Xiao for valuable discussions. We thank Björn Schenke for carefully reading our manuscript and providing feedback. F.S. also thanks  Kaushik Roy for helpful discussions at the early stages of this project. 

This material is based on work supported by the U.S. Department of Energy, Office of Science, Office of Nuclear Physics, under Contracts No. de-sc0012704 and (for R.V.) within the framework of the TMD Theory Topical Collaboration.
F.S is also supported by the joint Brookhaven National Laboratory-Stony Brook University Center for Frontiers in Nuclear Science (CFNS). 

\appendix    
    
\section{Conventions and useful identities}
\label{app:convention}
\subsection{Lightcone coordinates}
We work in lightcone coordinates,
\begin{align}
    x^+ = \frac{1}{\sqrt{2}}\left(x^0 + x^3 \right), \quad x^- = \frac{1}{\sqrt{2}}\left(x^0 - x^3 \right)\,,
\end{align}
with the transverse momenta components the same as Minkowski space.  Four-vectors are defined as $a^\mu = (a^+,a^-,\at)$, where $\at$ denote the two-dimensional transverse components. The magnitude of the two-dimensional vector $\at$ is denoted as $a_\perp$. Following these conventions, the scalar product of two vectors is $a_\mu b^\mu = a^+b^- + a^- b^+ - \at \cdot \bt$.

The same convention is used for the gamma matrices $\gamma^+$ and $\gamma^-$, with the anti-commutation relations satisfying
\begin{align}
    \left\{\gamma^\mu,\gamma^\nu \right\} = 2 g^{\mu\nu} \mathbbm{1}_{4}\,,
    \label{eq:ACR}
\end{align}
where the only non-zero entries in the metric are $g^{+-}=g^{-+}=1$ and $g^{ij}=-\delta^{ij}$.

As a consequence of Eq.\,\eqref{eq:ACR}, we have $(\gamma^-)^2 = (\gamma^+)^2 =0$, which will be repeatedly used in our computations. Another useful relation, resulting from the anti-commutation relations of gamma matrices, is $\gamma^-\slashed k\gamma^-=2k^-\gamma^-$.

\subsection{Feynman Rules}
We employ the standard Feynman rules of QCD+QED supplemented with the effective vertices for the propagation of quarks and gluons in the classical back-ground of the CGC shock wave. We choose to work in the lightcone gauge $A^-=0$ which drastically simplifies  computations (for nuclei with $P^+\rightarrow \infty$) in background fields \cite{Ayala:1995kg,Gelis:2005pt}.

We will label below spinor and vector indices respectively as $(\sigma,\sigma')$ and $(\mu,\nu)$, and color indices in SU(3) in the fundamental and adjoint representation as $(i,j)$ and $(a,b,c)$ respectively.

The free massless quark and gluon Feynman propagators are,
\begin{align}
    S^0_{\sigma\sigma',ij}(l) &= \frac{i\slashed{l}_{\sigma\sigma'}}{l^2 + i\epsilon} \delta_{ij} \,,\\
    G^0_{\mu\nu,ab}(l) &= \frac{i}{l^2+i\epsilon}\left( -g_{\mu \nu} + \frac{l_\mu n_\nu + n_\mu l_\nu }{n.l} \right) \delta_{ab}\,,
\end{align}
where the lightcone vector $n$ is defined as $n^\mu = (1,0,\vect{0})$ satisfying $n.A = A^-$. We also define the gluon polarization tensor $\Pi_{\mu\nu}$ which appears in the free gluon propagator as
\begin{equation}
    \Pi_{\mu\nu}(l)=-g_{\mu \nu} + \frac{l_\mu n_\nu + n_\mu l_\nu }{n.l}\,. \label{eq:gluon-pol-tensor-def}
\end{equation}

The polarization vector for a photon with zero transverse momentum $\qt=\vect{0}$ and virtuality $Q^2=-q^2$ is given by
\begin{align}
    \epsilon^{\mu}(q,\lambda=0) &= \left(\frac{Q}{q^-},0,\vect{0} \right) \,, \\
    \epsilon^{\mu}(q,\lambda=\pm 1) &= \left(0,0,\et^{\pm 1} \right) \,,
\end{align}
where $\lambda =0$ denotes the longitudinal polarization, $\lambda= \pm 1$ denote the two transverse (circular) polarizations, and the two-dimensional vector $\vect{\epsilon}^{\pm 1} = \frac{1}{\sqrt{2}} \left(1,\pm i \right) $. 

The polarization vector for an on-shell gluon with non-zero transverse momentum $\lt$ is 
\begin{align}
    \epsilon^{\mu}(l,\lambda=\pm 1) &= \left(\frac{\vect{\epsilon}^{\pm 1}\cdot \lt}{l^-},0,\vect{\epsilon}^{\pm 1} \right)\,,
\end{align}
where we only have the two physical transverse polarizations. The transverse polarization vector also satisfies the identity
\begin{equation}
    \epsilon^{ij}\et^{\lambda,j}=i\lambda\et^{\lambda,i}\,, 
    \label{eq:eigen-pol-epsilon}
\end{equation}
which turns out to be very useful in performing spinor contractions in our calculation.

The photon quark-antiquark and the gluon-quark-antiquark vertices read
\begin{align}
    V^{\gamma q\bar{q}}_{\mu,\sigma\sigma'} = -ieq_f (\gamma_\mu)_{\sigma\sigma'} \,\,;\,\,
    V^{g q\bar{q},a}_{\mu,\sigma \sigma',ij} = i g (\gamma_\mu)_{\sigma\sigma'} t^a_{ij} \,,
\end{align}
where $e$ is the electromagnetic coupling constant, $q_f$ is the fractional charge of the quark, $g$ is the strong coupling, and $t^a_{ij}$ is a SU(3) generator in the fundamental representation. At one loop order in our computation we do not need the cubic and quartic gluon vertices except in the cubic coupling of gluons to the background field, represented below by the gluon effective vertex.

The CGC effective vertices for the eikonal interaction of the quark (moving with large minus lightcone momentum component) with the background is given by
\begin{align}
    \mathcal{T}^q_{\sigma\sigma',ij}(l,l') &=  (2\pi) \delta(l^--l'^-) \gamma^-_{\sigma\sigma'} \,\mathrm{sgn}(l^-) \int \der^2\vect{z} e^{-i(\lt-\lt')\cdot \vect{z}} V_{ij}^{\mathrm{sgn}(l^-)}(\vect{z}) \,,
\end{align}
and similarly the eikonal interaction of the gluon (moving with large minus lightcone momentum component) with the background reads
\begin{align}
    \mathcal{T}^g_{\mu\nu,ab}(l,l') &=  -(2\pi) \delta(l^--l'^-) (2l^-)  g_{\mu\nu} \,\mathrm{sgn}(l^-) \int \der^2\vect{z} e^{-i(\lt-\lt')\cdot \vect{z}} U_{ab}^{\mathrm{sgn}(l^-)}(\vect{z})\,,
\end{align}
where $l$ and $l'$ are the outgoing and incoming momenta of the quark/gluon. The
superscript $\mathrm{sgn}(l^-)$ denotes the color matrix or its inverse  $V
^{+1}(\zt) = V(\zt)$ and $V^{-1}
(\zt) = V^\dagger
(\zt)$,
where the latter follows from the unitarity of $V(\zt)$, and similarly for $U(\zt)$.

The lightlike Wilson lines in the fundamental and adjoint representations appearing in the effective CGC vertices are given by the SU(3) matrices
\begin{align}
    V_{ij}(\vect{z}) &= \Pcal \exp{ \left( ig \int_{-\infty}^\infty dz^- A^{+,c}_{\mathrm{cl}}  (z^-,\vect{z}) t^c_{ij}  \right)}\,, \\
    U_{ab}(\vect{z}) &= \Pcal \exp{ \left( ig \int_{-\infty}^\infty dz^- A^{+,c}_{\mathrm{cl}}  (z^-,\vect{z}) T^c_{ab}  \right)} \,,
\end{align}
where $t^c_{ij}$ and $T^c_{ab}$ are the generators of SU(3) in the fundamental and adjoint representations respectively. $A^+_\mathrm{cl}$ is the back-ground gauge field of the classical small-$x$ gluon field in Lorenz gauge. Here $\Pcal$ stands for path ordering such that the operator at $z=-\infty$ is in the rightmost position, while that at $z=+\infty$ is in the leftmost position.

\subsection{Color identities}
The generators of SU($\rm{N_c}$) in the fundamental representation satisfy the  commutation relations
\begin{align}
    \left[t^a,t^b \right] = i f^{abc} t^c\,, \quad \left[T^a,T^b \right] = i f^{abc} T^c\,,
\end{align}
where $f^{abc}$ are the structure constants. We will not need the explicit expressions for these objects. We normalize the generators such that they satisfy
\begin{align}
    \Tr(t^at^b) = \frac{1}{2} \delta^{ab} \,\, ; \,\,
    \Tr(T^aT^b) = N_c \delta^{ab} \,.
\end{align}
Then we find
\begin{align}
    t^a t^a = C_{\mathrm{F}} \mathbbm{1}_{N_c}\,\,;\,\,
    T^aT^a = C_{\mathrm{A}} \mathbbm{1}_{N^2_c-1}\,,
\end{align}
where the Casimirs in the fundamental and adjoint representation are defined as
\begin{align}
     C_{\mathrm{F}} = \frac{N^2_c-1}{2N_c} \,\,;\,\, C_A =N_c\,.
\end{align}
A useful Fierz identity for the color structures follows from the completeness relation of Hermitian matrices ($ \{\mathbbm{1}_3,t^a\}$ form a complete set) that
\begin{align}
    \delta_{il} \delta_{jk} = 2t^a_{ij}t^a_{kl} + \frac{1}{N_c} \delta_{ij}\delta_{kl}\,.
\end{align}
The coefficients follow from the normalization. Then one can show that for any 3 by 3 matrices $C$ and $D$ satisfy:
\begin{align}
    \Tr(C)\Tr(D) &=   2\, \Tr(Ct^a Dt^a) + \frac{1}{N_c}\Tr(CD) \,,\\
    \Tr(CD) &= 2 \,\Tr(C t^a) \Tr(D t^a) + \frac{1}{N_c} \Tr(C)\Tr(D)\,.
\end{align}
We close this section with a useful identity that related Wilson lines in the fundamental and adjoint representations:
\begin{align}
    V^\dagger(\xt) t^a V(\xt) = t^b U^{ab}(\xt)\,. \label{sub:adjfund}
\end{align}

\section{Dijet cross-section for transversely polarized virtual photon}
\label{app:final-res-trans}

In this Appendix, we provide the explicit expressions that enter inside the NLO impact factor for transversely polarized photon, using the same notations as in the longitudinally polarized case in Sec.~\ref{sub:final}. The finite terms from the virtual diagrams of SE1 and V1 read:
\begin{align}
    &\alpha_s\left.\frac{\der\sigma^{\gamma^*_T+A\rightarrow \,{\rm dijet}+X}}{\der^2\ptj\der\eta_J\der^2\ptk\der\eta_K}\right|^{(a)}_{\rm V,i.f.}=\frac{\alpha_{\rm em}e_f^2N_c}{(2\pi)^6}\delta(1-z_J-z_K)\int\der\Pi_{\rm LO}2z_J^2z_K^2\frac{QK_1(\bar Qr_{x'y'})}{r_{x'y'}}\nonumber\\
    &\times\frac{\alpha_s}{\pi}\Bigg\{ \int_0^{z_J}\frac{\der z_g}{z_g}\int\frac{\der^2\zt}{\pi}\left\{e^{-i\frac{z_g}{z_J}\kt\cdot\rzxt}\frac{\bar Q K_1(QX_V)}{X_V}\Xi_{\rm NLO,1}\left[-\frac{z_g(z_g-z_J)^2z_K}{2 z_J^3}\frac{\rzxt\cdot\rxytp}{\rzxt^2}\right.\right.\nonumber\\
    &\left.+(z_J^2+z_K^2)\left(1-\frac{z_g}{z_J}+\frac{z_g^2}{2z_J^2}\right)\frac{\RtS\cdot\rxytp}{\rzxt^2}\right]-(z_J^2+z_K^2)e^{-\frac{\rzxt^2}{\rxyt^2e^{\gamma_E}}}\left(1-\frac{z_g}{z_J}+\frac{z_g^2}{2z_J^2}\right)\frac{\rxyt\cdot\rxytp}{\rzxt^2}\nonumber\\
    &\times QK_1(\bar Qr_{xy})C_F\Xi_{\rm LO}-e^{-i\frac{z_g}{z_J}\kt\cdot\rzxt}\frac{\bar QK_1( Q X_V)}{X_V}\Xi_{\rm NLO,1}\left[\frac{z_g(z_J-z_g)}{2(z_g+z_K)}\frac{\rzxt\cdot\rxytp}{\rzxt^2}\right.\nonumber\\
    &+[z_J(z_J-z_g)+z_K(z_K+z_g)]\left(1-\frac{z_g}{z_J}\right)\left(1+\frac{z_g}{z_K}\right)\left(1-\frac{z_g}{2z_J}-\frac{z_g}{2(z_K+z_g)}\right)\nonumber\\
    &\left.\left.\times\frac{(\RtV\cdot\rxytp)(\rzxt\cdot\rzyt)}{\rzxt^2\rzyt^2}+\frac{z_g(z_J-z_g)(z_g+z_K-z_J)^2}{2z_J^2z_K}\frac{(\RtV\times\rxytp)(\rzxt\times\rzyt)}{\rzxt^2\rzyt^2}\right]\right\}
    \nonumber\\
    &+(J\leftrightarrow K)\Bigg\}+c.c.\,,
\end{align}
where we introduce again the two transverse vectors:
\begin{align}
    \RtS&=\rxyt+\frac{z_g}{z_q}\rzxt \,,\\
    \RtV&=\rxyt-\frac{z_g}{z_{\bar q}+z_g}\rzyt \,,
\end{align}
and the finite part of V3 reads
\begin{align}
    &\alpha_s\left.\frac{\der\sigma^{\gamma^*_T+A\rightarrow \,{\rm dijet}+X}}{\der^2\ptj\der\eta_J\der^2\ptk\der\eta_K}\right|^{(b)}_{\rm V,i.f.}=\frac{\alpha_{\rm em}e_f^2N_c}{(2\pi)^6}\delta(1-z_J-z_K)\int\der\Pi_{\rm LO} \ 2z_Jz_K\frac{\bar QK_1(\bar Qr_{x'y'})}{r_{x'y'}}\nonumber\\
    &\times\frac{\alpha_s}{\pi}\Bigg\{\int_0^{z_J}\frac{\der z_g}{z_g}\frac{\bar{Q}_{\mathrm{V3}}K_1(\bar{Q}_{\mathrm{V3}} r_{xy})}{r_{xy}}\Xi_{\rm NLO,3}(\xt,\yt;\xt',\yt')\nonumber\\
    &\times \left[\left[z_J(z_J-z_g)+z_K(z_K+z_g)\right](1+z_g)\left(1-\frac{z_g}{z_J}\right)e^{i(\Pt+z_g\vect{\Delta})\cdot\rxyt} (\rxyt\cdot\rxytp)K_0(-i\Delta_{\rm V3}r_{xy})\right.\nonumber\\
    &-\left[z_J(z_J-z_g)+z_K(z_K+z_g)\right]\left(1-\frac{z_g}{2z_J}+\frac{z_g}{2z_K}-\frac{z_g^2}{2z_Jz_K}\right)e^{i\frac{z_g}{z_J}\kt\cdot\rxyt}(\rxyt\cdot\rxytp)\nonumber\\
    &\times\Jcal_{\odot}\left(\rxyt,\left(1-\frac{z_g}{z_J}\right)\Pt,\Delta_{\rm V3}\right)\nonumber\\
    &\left.-i\frac{z_g(z_g+z_K-z_J)^2}{z_Jz_K}e^{i\frac{z_g}{z_J}\kt\cdot\rxyt}(\rxyt\times\rxytp)\Jcal_{\otimes}\left(\rxyt,\left(1-\frac{z_g}{z_J}\right)\Pt,\Delta_{\rm V3}\right)\right]+(J\leftrightarrow K)\Bigg\}+c.c. \,.
\end{align}
Recall that, as in the longitudinally polarized case, the other virtual pieces are implicitly contained in Eq.\,\eqref{eq:NLO-IRCsafe}.
Contrary to \eqref{eq:dijet-NLO-long-virtual-final}, we have not included the subtraction terms of the slow gluon divergence in order to keep this formula relatively short. It is straightforward to put these terms back. 
The notations for the variables used in this expression are gathered in Table~\ref{tab:params_dijet} for the kinematic parameters and Table~\ref{tab:NLO-color} for the color structures. 

Following the notations of section~\ref{sub:final}, the real contributions to the $\gamma^*_{\rm T}\to q\bar q g+X$ cross-section read as,
\begin{align}
    &\left.\frac{\der\sigma^{\gamma_{\rm T}^*+A\to q\bar qg+X}}{\der^2\kt\der\eta_q\der^2\pt\der\eta_{\bar q}\der^2\kgt\der\eta_g}\right|_{\rm R2\times R2}=\frac{\alpha_{\rm em}e_f^2N_c}{(2\pi)^8}\delta(1-z_q-z_{\bar q}-z_g)\alpha_s\int\der\Pi_{\rm LO}\nonumber\\
    &\times C_F\Xi_{\rm LO}(\xt,\yt;\xt',\yt')\left\{ 8z_qz_{\bar q}\bar Q_{\rm R2}^2\left[z_{\bar q}^2+(1-z_{\bar q})^2\right]\left(1+\frac{z_g}{z_q}+\frac{z_g^2}{2z_q^2}\right)\frac{e^{-i\kgt \cdot (\xt-\xt')}}{(\kgt-\frac{z_g}{z_q}\kt)^2}\right.\nonumber\\
    &\left.\times\frac{\rxyt\cdot\rxytp}{r_{xy}r_{x'y'}}K_1(\bar Q_{\mathrm{R}2}r_{xy})K_1(\bar Q_{\mathrm{R}2}r_{x'y'})\right\}\,,
    \label{eq:dijet-NLO-trans-R2R2-final}
\end{align}
and
\begin{align}
    &\alpha_s\left.\frac{\der\sigma^{\gamma_{\rm T}^*+A\to q\bar qg+X}}{\der^2\kt\der\eta_q\der^2\pt\der\eta_{\bar q}\der^2\kgt\der\eta_g}\right|_{\rm other}= \Bigg[ \frac{1}{2}\alpha_s\left.\frac{\der\sigma^{\gamma_{\rm T}^*+A\to q\bar qg+X}}{\der^2\kt\der\eta_q\der^2\pt\der\eta_{\bar q}\der^2\kgt\der\eta_g}\right|_{\rm R2\times R2'}\nonumber\\
    &+\alpha_s\left.\frac{\der\sigma^{\gamma_{\rm T}^*+A\to q\bar qg+X}}{\der^2\kt\der\eta_q\der^2\pt\der\eta_{\bar q}\der^2\kgt\der\eta_g}\right|_{\rm R1\times R_2}+\alpha_s\left.\frac{\der\sigma^{\gamma_{\rm T}^*+A\to q\bar qg+X}}{\der^2\kt\der\eta_q\der^2\pt\der\eta_{\bar q}\der^2\kgt\der\eta_g}\right|_{\rm R1\times R_2'}\nonumber\\
     &+\frac{1}{2}\alpha_s\left.\frac{\der\sigma^{\gamma_{\rm T}^*+A\to q\bar qg+X}}{\der^2\kt\der\eta_q\der^2\pt\der\eta_{\bar q}\der^2\kgt\der\eta_g}\right|_{\rm R1\times R_1}+\frac{1}{2}\alpha_s\left.\frac{\der\sigma^{\gamma_{\rm T}^*+A\to q\bar qg+X}}{\der^2\kt\der\eta_q\der^2\pt\der\eta_{\bar q}\der^2\kgt\der\eta_g}\right|_{\rm R1\times R_1'}\nonumber\\
     &+(q\leftrightarrow\bar q) \Bigg] +c.c.\,,
    \label{eq:dijet-NLO-trans-other-final}   
\end{align}
with 
\begin{align}
    &\alpha_s\left.\frac{\der\sigma^{\gamma_{\rm T}^*+A\to q\bar qg+X}}{\der^2\kt\der\eta_q\der^2\pt\der\eta_{\bar q}\der^2\kgt\der\eta_g}\right|_{\rm R2\times R2'}=\frac{\alpha_{\rm em}e_f^2N_c}{(2\pi)^8}\delta(1-z_q-z_{\bar q}-z_g)\,\alpha_s\,\int\der\Pi_{\rm LO}\nonumber\\
    &\times\Xi_{\rm NLO,3}(\xt,\yt;\xt',\yt')\Bigg\{8z_qz_{\bar q}\bar Q_{\rm R2}K_1(\bar Q_{\mathrm{R}2}r_{xy})\bar Q_{\rm R2'}K_1(\bar Q_{\rm R2'}r_{x'y'})e^{-i\kgt\cdot(\xt-\yt')}\nonumber\\
    &\times \left[-\left(z_q+z_{\bar q}-2z_qz_{\bar q}\right)\left(1+\frac{z_g}{2z_q}+\frac{z_g}{2z_{\bar q}}\right)\frac{\left(\kgt-\frac{z_g}{z_q}\kt\right)\cdot\left(\kgt-\frac{z_g}{z_{\bar q}}\pt\right)}{\left(\kgt-\frac{z_g}{z_q}\kt\right)^2\left(\kgt-\frac{z_g}{z_{\bar q}}\pt\right)^2}\frac{\rxyt\cdot\rxytp}{r_{xy}r_{x'y'}}\right.\nonumber\\
    &\left.\left.+\frac{z_g}{2z_qz_{\bar q}}(z_q-z_{\bar q})^2\frac{\left(\kgt-\frac{z_g}{z_q}\kt\right)\times\left(\kgt-\frac{z_g}{z_{\bar q}}\pt\right)}{\left(\kgt-\frac{z_g}{z_q}\kt\right)^2\left(\kgt-\frac{z_g}{z_{\bar q}}\pt\right)^2}\frac{\rxyt\times\rxytp}{r_{xy}r_{x'y'}}\right]\right\}\,.
    \label{eq:dijet-NLO-trans-R2R2'-final}
\end{align}
For the other terms labeled $\rm R1\times R2$, $\rm R1\times R2'$, $\rm R1\times R1$ and $\rm R1 \times R1'$, we do not fully perform the spin-helicity sum, in order to avoid lengthy expressions. We refer the reader to \cite{Kolbe:2020tlq} for useful formulas related to the computation of such spin-helicity sums from Dirac traces. 
\begin{align}
    &\alpha_s\left.\frac{\der\sigma^{\gamma_{\rm T}^*+A\to q\bar qg+X}}{\der^2\kt\der\eta_q\der^2\pt\der\eta_{\bar q}\der^2\kgt\der\eta_g}\right|_{\rm R1\times R2}=\frac{\alpha_{\rm em}e_f^2N_c}{(2\pi)^8}\delta(1-z_q-z_{\bar q}-z_g)\,\int\der\Pi_{\rm LO}\nonumber\\
    &\times\int\der^2\zt\der^2\zt'e^{-i\kgt\cdot(\zt-\zt')}\Xi_{\rm NLO,1}(\xt,\yt,\zt;\wt',\yt')\nonumber\\
    &\times\frac{1}{8\pi}\sum_{\lambda,\bar\lambda,\sigma,\sigma'}\Ncal^{\lambda=\pm1,\bar\lambda\sigma\sigma'}_{\rm R1}(\rxyt,\rzxt)\Ncal^{\lambda=\pm1,\bar\lambda\sigma\sigma'\dagger}_{\rm R2}(\rwytp,\rzxtp)\,,
\end{align}
\begin{align}
    &\alpha_s\left.\frac{\der\sigma^{\gamma_{\rm T}^*+A\to q\bar qg+X}}{\der^2\kt\der\eta_q\der^2\pt\der\eta_{\bar q}\der^2\kgt\der\eta_g}\right|_{\rm R1\times R2'}=\frac{\alpha_{\rm em}e_f^2N_c}{(2\pi)^8}\delta(1-z_q-z_{\bar q}-z_g)\,\int\der\Pi_{\rm LO}\nonumber\\
    &\times\int\der^2\zt\der^2\zt'e^{-i\kgt\cdot(\zt-\zt')}\Xi_{\rm NLO,1}(\xt,\yt,\zt;\wt',\yt')\nonumber\\
    &\times\frac{1}{8\pi}\sum_{\lambda,\bar\lambda,\sigma,\sigma'}\Ncal^{\lambda=\pm1,\bar\lambda\sigma\sigma'}_{\rm R1}(\rxyt,\rzxt)\Ncal^{\lambda=\pm1,\bar\lambda\sigma\sigma'\dagger}_{\rm R2'}(\rwytp,\rzxtp)\,,
\end{align}
\begin{align}
    &\alpha_s\left.\frac{\der\sigma^{\gamma_{\rm T}^*+A\to q\bar qg+X}}{\der^2\kt\der\eta_q\der^2\pt\der\eta_{\bar q}\der^2\kgt\der\eta_g}\right|_{\rm R1\times R1}=\frac{\alpha_{\rm em}e_f^2N_c}{(2\pi)^8}\delta(1-z_q-z_{\bar q}-z_g)\,\int\der\Pi_{\rm LO}\nonumber\\
    &\times\int\der^2\zt\der^2\zt'e^{-i\kgt\cdot(\zt-\zt')}\Xi_{\rm NLO,3}(\xt,\yt,\zt;\xt',\yt',\zt')\nonumber\\
    &\times\frac{1}{8\pi}\sum_{\lambda,\bar\lambda,\sigma,\sigma'}\Ncal^{\lambda=\pm1,\bar\lambda\sigma\sigma'}_{\rm R1}(\rxyt,\rzxt)\Ncal^{\lambda=\pm1,\bar\lambda\sigma\sigma'\dagger}_{\rm R1}(\rxytp,\rzxtp)\,,
\end{align}
\begin{align}
    &\alpha_s\left.\frac{\der\sigma^{\gamma_{\rm T}^*+A\to q\bar qg+X}}{\der^2\kt\der\eta_q\der^2\pt\der\eta_{\bar q}\der^2\kgt\der\eta_g}\right|_{\rm R1\times R1'}=\frac{\alpha_{\rm em}e_f^2N_c}{(2\pi)^8}\delta(1-z_q-z_{\bar q}-z_g)\,\int\der\Pi_{\rm LO}\nonumber\\
    &\times\int\der^2\zt\der^2\zt'e^{-i\kgt\cdot(\zt-\zt')}\Xi_{\rm NLO,3}(\xt,\yt,\zt;\xt',\yt',\zt')\nonumber\\
    &\times\frac{1}{8\pi}\sum_{\lambda,\bar\lambda,\sigma,\sigma'}\Ncal^{\lambda=\pm1,\bar\lambda\sigma\sigma'}_{\rm R1}(\rxyt,\rzxt)\Ncal^{\lambda=\pm1,\bar\lambda\sigma\sigma'\dagger}_{\rm R1'}(\rxytp,\rzxtp)\,.
\end{align}
The expressions for the perturbative factors $\Ncal^{\bar\lambda=\pm1}_{\rm R1}$ and $\Ncal^{\bar\lambda=\pm1}_{\rm R2}$ are given respectively by Eq.\,\eqref{eq:dijet-NLO-R2-Npert-trans-final} and Eq.\,\eqref{eq:dijet-NLO-R1-Npert-trans-final}. Even though these formulas are very lengthy, they can be implemented on a computer program for their subsequent numerical evaluation as can Eqs.\,\eqref{eq:dijet-NLO-trans-R2R2-final} and \eqref{eq:dijet-NLO-trans-R2R2'-final}.

\section{Dirac Algebra}
\label{app:dirac}
In this section, we provide various  gamma matrix identities that are useful in the computations of the perturbative factors. 
\subsection{General identities}
\subsubsection{Product of transverse gamma matrices}
It is advantageous to decompose the product of two transverse gamma matrices into symmetric and anti-symmetric components. In 4 dimensions (2 transverse dimensions), for $i,j\in\{1,2\}$, one has
\begin{align}
    \gamma^i\gamma^j&=\frac{1}{2}\{\gamma^i,\gamma^j\}+\frac{1}{2}[\gamma^i,\gamma^j]
    = -\delta^{ij}-i\epsilon^{ij}\Omega\,,
    \label{eq:app-B-gigj}
\end{align}
with
\begin{equation}
    \Omega=\frac{i}{2}[\gamma^1,\gamma^2]\,.
\end{equation}
The matrix $\Omega$ satisfies $\Omega^2=\mathbbm{1}$ and $[\gamma^-,\Omega]=[\gamma^+,\Omega]=0$.

In $d=2+(2-\varepsilon)$ dimension, these identities are generalized to
 \begin{equation}
    \gamma^i\gamma^j=-\delta^{ij}+\omega^{ij}\,,
 \end{equation}
 with $\omega^{ij}=\frac{1}{2}[\gamma^i,\gamma^j]$. The matrices $\omega^{ij}$ satisfy
 \begin{align}
    \omega^{ij}\omega^{jk}&=(1-\varepsilon)\delta^{ik}+\varepsilon\omega^{ik}\label{eq:omega-product}\,,\\
    [\omega^{lm},\omega^{ij}]&=2(\omega^{jl}\delta^{im}+\omega^{mj}\delta^{il}+\omega^{li}\delta^{mj}+\omega^{im}\delta^{jl})\,,
\end{align}
and the commutation relation $[\gamma^{\pm},\omega^{ij}]=0$. The $\varepsilon$ terms in Eq.\,\eqref{eq:omega-product} come from $\delta^{ii}=d-2=2-\varepsilon$. In 4 dimensions, when $\omega^{ij}=-i\epsilon^{ij}\Omega$, one has $[\omega^{lm},\omega^{ij}]=0$. Therefore $[\omega^{lm},\omega^{ij}]=\mathcal{O}(\varepsilon)$. From these relations, one easily finds that
\begin{align}
    (A_1\delta^{ij}+&A_2\omega^{ij})\gamma^-\left(C_1\delta^{jk}+C_2\omega^{jk}\right)=\nonumber\\
    &\left[\left(A_1C_1+(1-\varepsilon)A_2C_2\right)\delta^{ik}+\left(A_1C_2+A_2C_1+\varepsilon A_2C_2\right)\omega^{ik}\right]\gamma^-\,,
    \label{eq:4-edimension-rel1}\\
    (A_1\delta^{ij}+&A_2\omega^{ij})\omega^{lm}\gamma^-\left(C_1\delta^{jk}+C_2\omega^{jk}\right)=\nonumber\\
    & \omega^{lm}\gamma^-\left[\left(A_1C_1+(1-\varepsilon)A_2C_2\right)\delta^{ik}+\left(A_1C_2+A_2C_1+\varepsilon A_2C_2\right)\omega^{ik}\right]+\mathcal{O}(\varepsilon)\,.
\label{eq:4-edimension-rel2}
\end{align}

\subsubsection{Spinor contractions}
\label{subsub:spin-contraction}

We begin by presenting explicit expressions for the gamma matrices in $4$ dimensions and explicit representation for the Dirac spinors.

We work in the Dirac basis for gamma matrices
\begin{align}
    \gamma^0 = \begin{bmatrix}
    \mathbbm{1} & 0 \\
    0 & -\mathbbm{1}
    \end{bmatrix} \ \ \  , \   \gamma^i= \begin{bmatrix}
    0 & \sigma^i \\
    -\sigma^i & 0
    \end{bmatrix} \ \ \ , \ \sigma^1 = \begin{bmatrix}
    0 & 1 \\
    1 & 0 
    \end{bmatrix} \ \ \ , \ \sigma^2 = \begin{bmatrix}
    0 & -i \\
    i & 0 
    \end{bmatrix} \ \ \ , \ \sigma^3 = \begin{bmatrix}
    1 & 0 \\
    0 & -1 
    \end{bmatrix} \,,
\end{align}
where $\mathbbm{1}$ is the two-by-two identity matrix.

The helicity operator $h$ is defined as 
\begin{align}
    h = \frac{2 \vec{k}\cdot\vec{S}}{|\vec{k}|} \,, \quad \vec{S} = \frac{1}{2}\begin{bmatrix} \vec{\sigma} & 0 \\
0 & \vec{\sigma} 
\end{bmatrix} \,,
\end{align}
where $\vec{k} = (\kt,k^3)$ is the three momentum. The (massless) Dirac equation reads
\begin{align}
    \slashed{k} u(k) = 0 \,.
\end{align}
It has the following solutions\footnote{In the massless case, the spinors corresponding to particle and anti-particle are the same, but correspond to opposite helicities.}
\begin{align}
    u_+(k) =v_{-}(k) =  \frac{1}{2^{1/4}}\begin{bmatrix}
    \sqrt{k^+} e^{-i\phi_k} \\
    \sqrt{k^-} \\
    \sqrt{k^+} e^{-i\phi_k} \\
    \sqrt{k^-}
    \end{bmatrix} \,, \ \ \ \, \ \ u_-(k) =v_{+}(k)  = \frac{1}{2^{1/4}}\begin{bmatrix}
    \sqrt{k^-}  \\
    -\sqrt{k^+} e^{i\phi_k} \\
    -\sqrt{k^-}  \\
    \sqrt{k^+} e^{i\phi_k}
    \end{bmatrix} \,,
\end{align}
where the subscripts $\pm$ denote the helicities\footnote{Note that in this manuscript, $p^-$ is the large component of the spinor momenta; thus $p^3 < 0$.}, $\phi_k$ is the azimuthal angle of $\kt$, and the normalization is chosen so that
\begin{align}
    \slashed{k} = \sum_{\sigma} u_\sigma(k) \bar{u}_\sigma(k) \,,
\end{align}
where the barred spinors are defined as usual by $\bar{u} = u^\dagger \gamma^0$.

We provide here the relevant formulas for performing spinor contractions based on our conventions. These relations are valid in 4 dimensions only.

In the LO computation, we need
\begin{align}
    \bar u(k,\sigma)\gamma^- v(p,\sigma')&=2\sqrt{k^-p^-}\delta^{\sigma,-\sigma'}\,,\\
    \bar u(k,\sigma)\gamma^-\Omega \,v(p,\sigma')&=-2\sqrt{k^-p^-}\,\sigma\,\delta^{\sigma,-\sigma'} \,,
\end{align}
leading to
\begin{align}
    \bar{u}(k,\sigma) \left[A_1 - \lambda A_2 \Omega \right] \gamma^- v(p,\sigma')
    & = 2 \sqrt{k^- p^-}  \Gamma_{\gamma^*_\mathrm{T}\to q\bar{q}}^{\sigma,\lambda} (A_2 - A_1, A_2 + A_1) \delta^{\sigma,-\sigma'}\,.
\end{align}
Recall that $\Gamma_{\gamma^*_\mathrm{T}\to q\bar{q}}^{\sigma,\lambda}$, defined in  Eq.\,\eqref{eq:dijet-NLO-spinhel-splitting}, is the spin-helicity-dependent splitting vertex.

For the NLO real emission computations, we have used
\begin{align}
    \bar{u}(k,\sigma) \left[B_1 + \bar{\lambda} B_2 \Omega \right] \gamma^- v(p,\sigma') 
    & = 2 \sqrt{k^- p^-}  \Gamma_{q \to qg }^{\sigma,\bar{\lambda}} (B_1 - B_2, B_1 + B_2) \delta^{\sigma,-\sigma'}\,,
\end{align}
and
\begin{align}
    \bar{u}(k,\sigma) &\left[B_1 + \bar{\lambda} B_2 \Omega \right] \left[A_1 - \lambda A_2 \Omega \right]\gamma^- v(p,\sigma') \nonumber \\
    & = 2 \sqrt{k^- p^-} \  \Gamma_{q \to qg }^{\sigma,\bar{\lambda}}(B_1 - B_2 ,B_1 + B_2)\  \Gamma_{\gamma^*_{\rm T} \to q\bar{q}}^{\sigma,\lambda} (A_2 - A_1, A_2 + A_1)\  \delta^{\sigma,-\sigma'}\,,
\end{align}
where we defined $\Gamma_{q \to qg }^{\sigma,\bar{\lambda}}$ in Eq.\,\eqref{eq:Gammaqtoqg}.

Finally, for the NLO virtual computations, the following relations are useful:
\begin{align}
    \bar{u}(k,\sigma)&\left[B_1 \delta^{ij} + i B_2 \epsilon^{ij}\Omega \right] \gamma^- \left[C_1 \delta^{kj} + i C_2 \epsilon^{kj}\Omega \right] v(p,\sigma') \nonumber \\
    & = 2 \sqrt{k^- p^-} \left[(B_1 C_1 - B_2 C_2)\delta^{ik} + i \sigma (B_1 C_2 - B_2 C_1) \epsilon^{ik} \right] \delta^{\sigma,-\sigma'}\,,
    \label{eq:dijet-NLO-useful_id_1}
\end{align}
and
\begin{align}
   & \bar{u}(k,\sigma)\left[B_1 \delta^{ij} + i B_2 \epsilon^{ij}\Omega \right]\left[A_1 - \lambda A_2 \Omega \right] \gamma^- \left[C_1 \delta^{kj} + i C_2 \epsilon^{kj}\Omega \right] v(p,\sigma') \nonumber \\
    & = 2 \sqrt{k^- p^-}  \left[(B_1 C_1 - B_2 C_2)\delta^{ik} + i \sigma (B_1 C_2 - B_2 C_1) \epsilon^{ik} \right] \Gamma_{\gamma^*_{\rm T} \to q\bar{q}}^{\sigma,\lambda} (A_2 - A_1, A_2 + A_1) \delta^{\sigma,-\sigma'} \,.
\end{align}

\subsubsection{Gluon tensor structure}

To simplify the Dirac algebra, it is very convenient to decompose the tensor structure for the free gluon propagator in terms of polarization vectors as 
\begin{align}
    \Pi_{\alpha \beta}(l) = \sum_{\bar{\lambda}=\pm 1} \epsilon_{\alpha}(l,\bar{\lambda}) \epsilon^{*}_{\beta}(l,\bar{\lambda}) + \frac{l^2}{(l^-)^2} n_{\alpha} n_{\beta}\,. \label{eq:gluon_tensor_decomp}
\end{align}
The product of two such structure is given by 
\begin{align}
    \Pi_{\alpha\beta}(l) \Pi^{\beta\delta} (l') & = -\sum_{\bar{\lambda} = \pm 1 } \epsilon_{\alpha}(l,\bar{\lambda}) \epsilon^{\delta*}(l',\bar{\lambda}) \,.
    \label{eq:gluon_tensor_squared_decomp}
\end{align}
Note the piece proportional to $n$ in Eq.\,\eqref{eq:gluon_tensor_decomp} drops out when inserted in Eq.\,\eqref{eq:gluon_tensor_decomp} since
\begin{align}
    n_{\beta} \Pi^{\beta\delta}(l')  = 0 \,\,\,\,\,{\rm and}\,\,\,\,\,
    \epsilon^*_{\beta}(l,\bar{\lambda}) \Pi^{\beta\delta}(l')   =  - \epsilon^{\delta*}(l',\bar{\lambda})\,.
\end{align}
Physically, this means that longitudinal/instantaneous piece of the propagator drops out in the product. 

\subsection{Useful Dirac algebra tricks for gluon emission and absorption numerators}

In this subsection, we will collect useful algebraic identities that contribute to a significant simplification of the Dirac numerators in the real and virtual amplitudes. They isolate the contributions of the instantaneous terms and depend on the transverse coordinates which also naturally appear in the contour integrations over ``plus" momentum components.
In the following, we will employ 
\begin{align}
    \epsilon^{\mu}(l_2,\bar{\lambda}) = \left( \frac{\et^{\bar{\lambda}}\cdot \lttwo}{l_2^-}, 0, \et^{\bar{\lambda}}\right)\,. 
\end{align}
We point out further that in all of the following relations, the first equality is expressed in a form (as per our previous discussions) that enables one to extract the $\mathcal{O}(\varepsilon)$ terms in the Dirac structures; they can then be used in $d=4-\varepsilon$ dimensions as well. However the expression in the second equality is valid in 4 dimension only since it makes use of the identity $\omega^{ij}=-i\epsilon^{ij}\Omega$. 

\subsubsection*{Gluon absorption from quark after the shock wave}
\begin{align}
    \bar{u}(k,\sigma) \slashed{\epsilon}(l_2,\bar{\lambda}) (\slashed{k} - \slashed{l}_2) \gamma^- &=  \frac{2}{x} \bar{u}(k,\sigma) \left(\delta^{ij} + \frac{x}{2}\gamma^i \gamma^j \right) \gamma^- \Lt^i \et^{\bar{\lambda},j}  \label{eq:gluon_abs_quark_afterSW-ddim}\\
    &= \frac{2 \Lt \cdot \et^{\bar{\lambda}}}{x}\  \bar{u}(k,\sigma)  \left[\left( 1-\frac{x}{2}\right) + \bar{\lambda} \frac{x}{2} \Omega  \right]\gamma^-\,,
    \label{eq:gluon_abs_quark_afterSW}
\end{align}
where $x = \frac{l_2^-}{k^-}$ and $\Lt = \lttwo - x \kt$.

\subsubsection*{Gluon absorption from quark before the shock wave}
\begin{align}
    &\gamma^- (\slashed{l}_1+\slashed{l}_2)  \slashed{\epsilon}(l_2,\bar{\lambda}) \slashed{l}_1   \nonumber \\
    &= \frac{2}{x}  \left(\delta^{ij} - \frac{x}{2} \gamma^j \gamma^i \right) \gamma^- \slashed{l}_1 \Lt^i \et^{\bar{\lambda},j}  - (1+x) l_1^2   \gamma^i \gamma^- \et^{\bar{\lambda},i}     \label{eq:gluon_abs_quark_beforeSW-ddim}\\
    & = \frac{2 \Lt \cdot \et^{\bar{\lambda}}}{x}\   \left[\left(1 + \frac{x}{2} \right) + \bar{\lambda} \frac{x}{2} \Omega \right] \gamma^- \slashed{l}_1 - (1+x)   \gamma^i \gamma^- \et^{\bar{\lambda},i} \ l_1^2\,,
    \label{eq:gluon_abs_quark_beforeSW}
\end{align}
where $x = \frac{l_2^-}{l_1^-}$ and $\Lt = \lttwo - x \ltone$.

\subsubsection*{Gluon emission from quark after the shock wave}
\begin{align}
    &\bar{u}(k,\sigma) \left[ \slashed{\epsilon}^*(k_g,\bar{\lambda}) (\slashed{k}+\slashed{k}_g) \right] \gamma^- \nonumber \\
    &= \frac{2}{x}  \bar{u}(k,\sigma) \left( \delta^{ij} - \frac{x}{2} \gamma^i \gamma^j  \right)  \gamma^- \Lt^i \et^{\bar{\lambda}*,j}  \\
    &= \frac{2 \Lt \cdot \et^{\bar{\lambda}*}}{x}  \bar{u}(k,\sigma) \left[\left(1 + \frac{x}{2} \right) +\bar{\lambda} \frac{x}{2} \Omega  \right]  \gamma^- \,,
    \label{eq:gluon_emi_quark_afterSW}
\end{align}
where $x = \frac{k_g^-}{k^-}$ and $\Lt = \kgt - x \kt$.
\subsubsection*{Gluon emission from quark before the shock wave}
\begin{align}
    &\gamma^- (\slashed{l}_1 -\slashed{l}_2) \slashed{\epsilon}^*(l_2,\bar{\lambda}) \slashed{l}_1 \nonumber \\
    &= \frac{2}{x}\left(\delta^{ij} + \frac{x}{2}\gamma^j\gamma^i \right) \gamma^- \slashed{l}_1 \Lt^i \et^{\bar{\lambda}*,j} - (1-x) l_1^2   \gamma^i \gamma^- \et^{\bar{\lambda}*,i}  \label{eq:gluon_emi_quark_beforeSW-ddim} \\
    & = \frac{2 \Lt \cdot \et^{\bar{\lambda}*}}{x} \ \left[\left( 1-\frac{x}{2}\right) + \bar{\lambda} \frac{x}{2} \Omega \right] \gamma^- \slashed{l}_1 - (1-x) \gamma^i \gamma^- \et^{\bar{\lambda}*,i} \ l_1^2\,,
    \label{eq:gluon_emi_quark_beforeSW}
\end{align}
where $x = \frac{l_2^-}{l_1^-}$ and $\Lt = \lttwo - x \ltone$.

\subsubsection*{Gluon emission from antiquark before the shock wave}
\begin{align}
    &(\slashed{l}_1-\slashed{q})  \slashed{\epsilon}^{*}(l_2,\bar{\lambda}) (\slashed{l}_1 -\slashed{q} +  \slashed{l}_2) \gamma^- \nonumber \\
    & = \frac{2}{x}  (\slashed{q} - \slashed{l}_1) \gamma^- \left(\delta^{ij} + \frac{x}{2}  \gamma^i \gamma^j \right) \Lt^i \et^{\bar{\lambda}*,j} -\left(1 - x \right)(q-l_1)^2 \gamma^- \gamma^i \et^{\bar{\lambda}*,i}  \label{eq:gluon_emi_antiquark_beforeSW-ddim} \\
    & = \frac{2 \Lt \cdot \et^{\bar{\lambda}*}}{x} \ (\slashed{q} - \slashed{l}_1) \gamma^- \left[\left(1- \frac{x}{2} \right) - \bar{\lambda} \frac{x}{2} \Omega \right] + \left(1 - x \right)  \gamma^i  \gamma^- \et^{\bar{\lambda}*,i} \ (q-l_1)^2\,,
    \label{eq:gluon_emi_antiquark_beforeSW}
\end{align}
where $x = \frac{l_2^-}{q^- - l_1^-}$ and $\Lt = \lttwo - x (\qt- \ltone)$.

\subsubsection*{Gluon emission from antiquark after the shock wave}
\begin{align}
    &\gamma^- (-\slashed{p}-\slashed{l}_2) \slashed{\epsilon}^*(l_2,\bar{\lambda}) v(p,\sigma') \nonumber \\
    &=-\frac{2}{x}  \gamma^- \left(\delta^{ij} - \frac{x}{2}\gamma^j \gamma^i \right) \Lt^i \et^{\bar{\lambda}*,j} v(p,\sigma')    \label{eq:gluon_emi_antiquark_afterSW-ddim} \\
    & = -\frac{2 \Lt \cdot \et^{\bar{\lambda}*}}{x} \ \gamma^- \left[ \left( 1 + \frac{x}{2}\right) - \bar{\lambda} \frac{x}{2} \Omega \right] v(p,\sigma')\,,
    \label{eq:gluon_emi_antiquark_afterSW}
\end{align}
where $x = \frac{l_2^-}{p^-}$ and $\Lt = \lttwo -  x \pt$.

\section{Contour integrals}
\label{app:contour}

In this appendix, we will provide details of the computation of several of the contour integrals appearing in the main text.

\subsection{Generic $l^+$ integrals}

\subsubsection{Two pole case}

We first consider the following integral with two poles:
\begin{align}
    \Ical = \int \frac{\der l^+}{(2\pi)} \frac{1}{\left[l^2 + i \epsilon \right] \left[(l-l')^2 + i \epsilon \right]}\,,
\end{align}
where $l'^->0$. The locations of the two poles follow from
\begin{align}
    l^2 + i\epsilon
    &= 2l^- (l^+ - l_a^+), \quad\mbox{ with } l_a^+ = \frac{\lt^2}{2l^-} - \frac{i\epsilon}{2l^-}\,,\\
    (l-l')^2 + i\epsilon 
    &= -2(l'^- - l^-) (l^+ - l_b^+), \quad\mbox{ with } l_b^+ = l'^+-\frac{(\lt-\lt')^2}{2(l'^- - l^-)} + \frac{i\epsilon}{2(l'^- - l^-)}\,.
\end{align}
We then have
\begin{align}
    \Ical &= -\frac{1}{4 l^- (l'^- -l^-)}\int \frac{\der l^+}{(2\pi)} \frac{1}{(l^+ - l_a^+)(l^+ - l_b^+)}\,.
\end{align}
It is not difficult to verify that the poles sit on opposite half-planes when $0<l^-<l'^-$. Closing the integral in the upper-half plane, enclosing $l_b^+$ , we find
\begin{align}
    \Ical &= -\frac{i}{4 l^- (l'^- -l^-)} \frac{\Theta(l^-)\Theta(l'^- - l^-)}{(l_b^+ - l_a^+)}\,.
\end{align}
It is advantageous to simplify the difference $l_b^+-l_a^+$ by introducing the transverse momentum vector $\Lt = \lt-x\lt'$, where $x= l^-/l'^-$, which gives
\begin{align}
    l_b^+ - l_a^+ 
    &=  \frac{1}{2 l^- (1-x)} \left[ x(1-x) (l'^2 + i \epsilon) - \Lt^2\right]\,.
\end{align}
where we redefined $\epsilon \to x(1-x)\epsilon$.

The final result reads then
\begin{align}
   \Ical &= -\frac{i}{2 l'^-} \frac{\Theta(l^-)\Theta(l'^- - l^-)}{\left[x(1-x) (l'^2 + i\epsilon)  -\Lt^2 \right]}\,.
   \label{eq:contour_ll'_generic}
\end{align}
In LCPT, these $\Theta$ functions are implicitly accounted for by considering different diagrams with lightcone time orderings of the propagating particles.

\subsubsection{Three pole case}

A slight generalization of the equation above is
\begin{align}
    \Ical_2 = \int \frac{\der l^+}{(2\pi)} \frac{1}{\left[l^2 + i \epsilon \right] \left[ \alpha_1 ((l-l')^2 + i \epsilon) + \beta_1 \right]\left[\alpha_2((l-l')^2 + i \epsilon) + \beta_2 \right]}\,,
\end{align}
where $\alpha_1,\alpha_2,\beta_2$ and $\beta_2$ do not depend on $l^+$, and as before $l'^->0$.
Using the definitions of $l_a^+$ and $l_b^+$ as above, we find
\begin{align}
    \Ical_2 &= \frac{1}{2 l^-}\int \frac{\der l^+}{(2\pi)} \frac{1}{(l^+ - l_a^+)\left[2 \alpha_1 (l^--l'^-)(l^+ - l_b^+)+\beta_1\right] \left[2 \alpha_2 (l^--l'^-)(l^+ - l_b^+)+\beta_2\right]}\,.
\end{align}
Closing the contour in the lower-half plane enclosing $l_a^+$, we find
\begin{align}
    \Ical_2 
    &= \frac{1}{2l^-} \frac{-i \Theta(l^-)\Theta(l'^- - l^-)}{\left[2 \alpha_1 (l'^--l^-)(l_b^+ - l_a^+)+\beta_1\right]\left[2 \alpha_2 (l'^--l^-)(l_b^+ - l_a^+)+\beta_2\right]} \nonumber \\
    &= \frac{1}{2l^-} \frac{-i \Theta(l^-)\Theta(l'^- - l^-)}{\left\{\frac{ \alpha_1}{x} \left[x(1-x)(l'^2 + i\epsilon) - \Lt^2 \right]+\beta_1\right\}\left\{\frac{ \alpha_2}{x} \left[x(1-x)(l'^2 + i\epsilon) - \Lt^2 \right]+\beta_2\right\}} \,,
\end{align}
where $x= l^-/l'^-$ and $\Lt = \lt - x\lt'$. Finally we end up with the expression for $\Ical_2$,
\begin{align}
   \Ical_2
    &= \frac{1}{2l^-} \frac{-i \Theta(l^-)\Theta(l'^- - l^-)}{\left\{\frac{ \alpha_1}{x} \left[ \Lt^2 - x(1-x)(l'^2 + i\epsilon) \right]-\beta_1\right\}\left\{\frac{ \alpha_2}{x} \left[ \Lt^2 -x(1-x)(l'^2 + i\epsilon) \right]-\beta_2\right\}} \,,
    \label{eq:contour_ll'_generic2} 
\end{align}

Another generalization is
\begin{align}
    \Ical_3 = \int \frac{\der l^+}{(2\pi)} \frac{1}{\left[ \alpha_1 (l^2 + i \epsilon) + \beta_1 \right]\left[ \alpha_2 (l^2 + i \epsilon) + \beta_2 \right] \left[(l-l')^2 + i \epsilon \right]}\,,
\end{align}
where $\alpha_1,\alpha_2,\beta_2$ and $\beta_2$ do not depend on $l^+$, and as before $l'^->0$. Following the same steps as for $\Ical_2$ but closing the contour in the upper-half plane to enclose $l_b^+$,
 we find
\begin{align}
   \Ical_3
    &= \frac{1}{2(l'^- - l^-)} \frac{-i \Theta(l^-)\Theta(l'^- - l^-)}{\left\{\frac{ \alpha_1}{1-x} \left[\Lt^2 - x(1-x)(l'^2 +i\epsilon) \right]-\beta_1\right\}\left\{\frac{ \alpha_2}{1-x} \left[\Lt^2 - x(1-x)(l'^2 +i\epsilon)  \right]-\beta_2\right\}}  \,,
    \label{eq:contour_ll'_generic3} 
\end{align}
where $x= l^-/l'^-$ and $\Lt = \lt - x\lt'$. 

\subsection{Application to self energy contribution $\mathrm{SE}1$ and vertex correction $\mathrm{V}1$}

\subsubsection{Contour integration for $\mathrm{SE}1$}
\label{subsub:contour-SE1}

We now compute the integrals in Eq.\,\eqref{eq:dijet-NLO-SE2-l+reg} and Eq.\,\eqref{eq:dijet-NLO-SE2-l+qins} for the regular and instantaneous perturbative factor of diagram $\mathrm{SE}1$ respectively. Using Eq.\,\eqref{eq:contour_ll'_generic}, one finds that
\begin{align}
    &\int\frac{\der l_2^+}{(2\pi)}\int\frac{\der l_3^+}{(2\pi)}\frac{(2q^-)^2}{[(l_2-l_1)^2+i\epsilon][l_2^2+i\epsilon][l_3^2+i\epsilon][(l_3-k)^2+i\epsilon]}\nonumber\\
    &=\frac{1}{z_q^2}\frac{\Theta(z_g)\Theta(z_q-z_g)}{\Ltthrex^2\left[x(1-x)(l_1^2+i\epsilon)-\Lttwox^2\right]}\,,
\end{align}
where $x=l_2^-/l_1^-=l_2^-/k^-=l_3^-/k^-$, $\Lttwox = \lttwo - x\ltone$ and $\Ltthrex = \ltthre - x\kt$.

Now we can use Eq.\,\eqref{eq:contour_ll'_generic3} with $l'=q$, $\alpha_1=1$, $\beta_1=0$, $\alpha_2 = x(1-x)$, and $\beta_2 = -\Lttwox^2$. One finds that
\begin{align}
    &\int \frac{\der l_1^+}{(2\pi)} \frac{(2q^-)}{\left[ l_1^2 + i \epsilon  \right]\left[ x(1-x)(l_1^2 + i \epsilon) - \Lttwox^2 \right] \left[(l_1-q)^2 + i \epsilon \right]} \nonumber \\
    & = \frac{-i \Theta(l_1^-)\Theta(q^- - l_1^-)}{\left[ \ltone^2 +  \bar{x}(1-\bar{x})Q^2 \right]\left\{\frac{ x(1-x)}{1-\bar{x}} \left[ \ltone^2 +\bar{x}(1-\bar{x})Q^2 \right]+\Lttwox^2\right\}} \,,
\end{align}
where $\bar{x} = l_1^-/q^-$. Since  $l_1^- = k^-$ and $q^- -l_1^- =p^-$, we have $\bar{x} = k^-/q^-$, $1-\bar{x}= p^-/q^-$, $x = l_3^-/k^-$, and $1-x = (k^- - l_3^-)/k^-$. Therefore, 
\begin{align}
    &\int \frac{\der l_1^+}{(2\pi)} \frac{(2q^-)}{\left[ l_1^2 + i \epsilon  \right]\left[ x(1-x)(l_1^2 + i \epsilon) - \Lttwox^2 \right] \left[(l_1-q)^2 + i \epsilon \right]} \nonumber \\
    &=  \frac{-i \Theta(z_q)\Theta(z_{\bar{q}})}{\left( \ltone^2 +  \bar{Q}^2 \right)\left[\omega_{\mathrm{SE}1}\left( \ltone^2 +\bar{Q}^2 \right)+\Lttwox^2\right]}\,,
    \label{eq:dijet-NLO-SE2-l1+reg}
\end{align}
where
\begin{align}
    \bar{Q}^2 = \frac{k^- p^-}{(q^-)^2} Q^2 = z_q z_{\bar{q}} Q^2\,\,\,\,\,{\rm and}\,\,\,\,\,
    \omega_{\mathrm{SE} 1} =\frac{l_3^- (k^- - l_3)q^-}{(k^-)^2p^-} = \frac{z_g (z_q - z_g)}{z_q^2 z_{\bar{q}}}\,,
\end{align}
which finally gives Eq.\,\eqref{eq:dijet-NLO-SE2-l+reg-2}.

For the instantaneous contribution, we use again Eq.\,\eqref{eq:contour_ll'_generic3} with $l'=q$, $\alpha_1=0$, $\beta_1=1$, $\alpha_2 = x(1-x)$, and $\beta_2 = -\Lttwox^2$, so that
\begin{align}
    &\int \frac{\der l_1^+}{(2\pi)}\frac{(2q^-)}{ \left[(q-l_1)^2 + i \epsilon \right]\left[x(1-x)(l_1^2 + i \epsilon) -\Lttwox^2 \right]}   \nonumber \\
    & =  \frac{2q^-}{2(q^- - l_1^-)} \frac{i \Theta(l_1^-)\Theta(q^- - l_1^-)}{\left\{\frac{ x(1-x)}{1-\bar{x}} \left[\ltone^2 + \bar{x}(1-\bar{x})Q^2  \right]+\Lttwox^2 \right\}} \nonumber \\
    &=\frac{1}{z_{\bar{q}}} \frac{i \Theta(z_q)\Theta(z_{\bar{q}})}{\left[ \omega_{\mathrm{SE}1} \left(\ltone^2 + \bar{Q}^2  \right)+\Lttwox^2 \right]}\,,
    \label{eq:dijet-NLO-SE2-l1+qins}
\end{align}
leading to Eq.\,\eqref{eq:dijet-NLO-SE2-l+qbarins-2}.

\subsubsection{Contour integration for $\mathrm{V}1$}

The equations Eq.\,\eqref{eq:dijet-NLO-V3-l+reg-2} and Eq.\,\eqref{eq:dijet-NLO-V3-l+qbarins-2} quoted in Sec.~\ref{sub:V3} can be obtained in a similar fashion. First, we find
\begin{align}
    &\int\frac{\der l_2^+}{(2\pi)}\int\frac{\der l_3^+}{(2\pi)}\frac{(2q^-)^2}{[(l_2-q+l_1)^2+i\epsilon][l_2^2+i\epsilon][l_3^2+i\epsilon][(l_3-k)^2+i\epsilon]}\nonumber\\
    &=\frac{1}{z_q (z_{\bar{q}}+z_g)}\frac{\Theta(z_g)\Theta(z_q-z_g)}{\Ltthrex^2\left[y(1-y)( (q-l_1)^2+i\epsilon)-\Lttwoy^2\right]}\,,
\end{align}
where $x = l_3^-/k^-$, $y = l_2^-/(q^- - l_1^-)$, $\Ltthrex = \ltthre - x \kt $ and $\Lttwoy = \lttwo + y \ltone$.

Then using eq.\,\eqref{eq:contour_ll'_generic2} with $l'=q$, $\alpha_1=1$, $\beta_1=0$, $\alpha_2 = y(1-y)$, and $\beta_2 = -\Lttwoy^2$, we have
\begin{align}
    &\int \frac{\der l_1^+}{(2\pi)} \frac{(2q^-)}{\left[l_1^2 + i \epsilon \right] \left[(q-l_1)^2 + i \epsilon \right]\left[y(1-y)((q - l_1)^2 + i \epsilon) -\Lttwoy^2 \right]}   \nonumber \\
    &=  \frac{2q^-}{2l_1^-} \frac{-i \Theta(l_1^-)\Theta(q^- - l_1^-)}{\left\{\frac{ 1}{\bar{x}} \left[ \ltone^2 + \bar{x}(1-\bar{x})Q^2 \right]\right\}\left\{\frac{y(1-y)}{\bar{x}} \left[ \ltone^2 +\bar{x}(1-\bar{x})Q^2 \right]+\Lttwoy^2\right\}} \nonumber \\
    &=   \frac{-i \Theta(l_1^-)\Theta(q^- - l_1^-)}{ \left[ \ltone^2 + \bar{x}(1-\bar{x})Q^2 \right]\left\{\frac{y(1-y)}{\bar{x}} \left[ \ltone^2 +\bar{x}(1-\bar{x})Q^2 \right]+\Lttwoy^2\right\}}\,,
\end{align}
where $\bar{x} = l_1^-/q^- $. For $\mathrm{V}1$, the delta functions constrain $l_1^-=k^- - l_3^-$ and $q^- - l_1^- = p^- + l_3^-$, giving
$\bar{x}=(k^- - l_3^-)/q^-$, $1-\bar{x}=(p^- + l_3^-)/q^-$, $y= l_3^-/(p^- + l_3^-)$ and $1-y= p^-/(p^- + l_3^-)$.
Finally, one gets
\begin{align}
    &\int \frac{\der l_1^+}{(2\pi)} \frac{(2q^-)}{\left[l_1^2 + i \epsilon \right] \left[(q-l_1)^2 + i \epsilon \right]}  \frac{1}{\left[y(1-y)((q - l_1)^2 + i \epsilon) -\Lttwoy^2 \right]} \nonumber \\
    &=  \frac{-i\Theta(z_q -z_g)\Theta(z_{\bar{q}} + z_g)}{\left(\ltone^2 + \Delta_{\mathrm{V}1}^2 \right) \left[ \omega_{\mathrm{V}1} \left(\ltone^2 + \Delta_{\mathrm{V}1}^2 \right) + \Lttwoy^2 \right]}\,,
    \label{eq:dijet-NLO-V3-l1+reg}
\end{align}
where
\begin{align}
    \Delta_{\mathrm{V}1}^2 &= \frac{(k^- - l_3^-)(p^- + l_3^-)}{(q^-)^2} Q^2 = (z_q - z_g)(z_{\bar{q}} + z_g) Q^2 \,, \nonumber \\
    \omega_{\mathrm{V}1} &= 
    \frac{l_3^- p^- q^-}{(k^- - l_3^-)(p^- + l_3^-)^2} = \frac{z_g z_{\bar{q}}}{(z_q - z_g)(z_{\bar{q}}+ z_g)^2}\,.
\end{align}
This gives the result in  Eq.~\eqref{eq:dijet-NLO-V3-l+reg-2}.
For the antiquark instantaneous contribution, we use  Eq.\,\eqref{eq:contour_ll'_generic2} with this time $l'=q$, $\alpha_1=0$, $\beta_1=1$, $\alpha_2 = y(1-y)$, and $\beta_2 = -\Lttwoy^2$, so that
\begin{align}
    &\int \frac{\der l_1^+}{(2\pi)} \frac{(2q^-)}{\left[l_1^2 + i \epsilon \right] }  \frac{1}{\left[y(1-y)((q - l_1)^2 + i \epsilon) -\Lttwoy^2 \right]} \nonumber \\
    &=  \frac{2q^-}{2l_1^-} \frac{i \Theta(l_1^-)\Theta(q^- - l_1^-)}{\left\{\frac{ y(1-y)}{\bar{x}} \left[ \ltone^2 + \bar{x}(1-\bar{x}) Q^2 \right] + \Lttwoy^2 \right\}} \nonumber \\
    &=  \frac{1}{z_q - z_g} \frac{i \Theta(z_q - z_g) \Theta(z_{\bar{q}} + z_g) }{\left[ \omega_{\mathrm{V}1} \left( \ltone^2 + \Delta_{\mathrm{V}1}^2 \right) + \Lttwoy^2 \right]}\,,
    \label{eq:dijet-NLO-V3-l1+qbarins}
\end{align}
where $\bar{x} = l_1^-/q^- $, which leads to Eq.\,\eqref{eq:dijet-NLO-V3-l+qbarins-2}.

\section{Useful transverse momentum integrals}
\label{app:transverse-int}

\subsection{Schwinger parametrization and multidimensional Gaussian integral}

All the integrals considered in this Appendix are computed using 
Schwinger's  parametrization of denominators as 
\begin{equation}
    \frac{1}{D^\beta}=\frac{1}{\Gamma(\beta)}\int_0^\infty \der s \ s^{\beta -1} \  e^{-sD}\,,
\end{equation}
for $D>0$ and $\beta>0$. We will also make use of the $2-\varepsilon$ dimensional Gaussian integral
\begin{equation}
    \int \der^{2-\varepsilon}\rt e^{-\rt^2}=\pi^{1-\varepsilon/2}.
\end{equation}
Finally, we will employ a very useful integral representation of modified Bessel functions of order $\nu$ that naturally emerge following the  Schwinger parametrization of denominators in the amplitudes:
\begin{equation}
    \int_0^\infty\der s \ s^{\nu-1}e^{-\rt^2/s}e^{-sQ^2}=2\left(\frac{Q}{r_\perp}\right)^{-\nu}K_{-\nu}(2Qr_\perp)\,,
\end{equation}
for $Q^2, \rt^2>0$. We will extensively use this result in the computations of the LO and NLO virtual photon ``wavefunctions".

\subsection{Fourier transforms}

\subsubsection{Gluon emission kernel}
\label{sub:gluonkernel}

The simplest transverse momentum integral encountered when evaluating self energies and vertex corrections is
\begin{align}
    \mu^\varepsilon\int\frac{\der^{2-\varepsilon}\lt}{(2\pi)^{2-\varepsilon}}\frac{e^{-i\lt \cdot \rt}}{\lt^2}&= \frac{1}{4\pi} (\pi \mu^2 \rt^2)^{\varepsilon/2} \Gamma\left(-\frac{\varepsilon}{2} \right) \nonumber \\
    &=-\frac{1}{4\pi}\left\{\frac{2}{\varepsilon}+\ln(e^{\gamma_E}\pi\mu^2\rt^2)+\mathcal{O}(\varepsilon)\right\}\,,
    \label{eq:Transverse_eps_1}
\end{align}
with $\varepsilon<0$.
Differentiating with respect to $\rt^i$, one easily gets the $2-\varepsilon$ integral representation of the gluon emission kernel
\begin{equation}
   \mu^\varepsilon\int\frac{\der^{2-\varepsilon}\lt}{(2\pi)^{2-\varepsilon}}\frac{\lt^ke^{-i\lt \cdot \rt}}{\lt^2}=\frac{1}{4\pi} (\pi \mu^2 \rt^2)^{\varepsilon/2} \varepsilon\Gamma\left(-\frac{\varepsilon}{2}\right)\frac{i\rt^k}{\rt^2}\,.
   \label{eq:Transverse_eps_2}
\end{equation}
In 2 dimensions, this reduces to 
\begin{align}
        \int \frac{\der^2 \lt}{(2\pi)} \frac{\lt^j e^{i \lt \cdot \rt}}{\lt^2} &= \frac{i \rt^j}{r_\perp^2}\,.
        \label{eq:Transverse_int_3} 
\end{align}
Eq.\,\eqref{eq:Transverse_eps_2} leads to a useful identity for the integral over $\zt$ of the JIMWLK kernel $\frac{\rzxt\cdot\rzyt}{\rzxt^2\rzyt^2}$ in coordinate space. Using twice  Eq.\,\eqref{eq:Transverse_eps_2}, integrating over $\zt$, and using Eq.\,\eqref{eq:Transverse_eps_1}, one gets
\begin{align}
    \frac{\mu^{\varepsilon}}{\pi}\int\der^{2-\varepsilon}\zt\frac{\rzxt\cdot\rzyt}{[\rzxt^2]^{1-\frac{\varepsilon}{2}}[\rzyt^2]^{1-\frac{\varepsilon}{2}}}&=\frac{4}{\varepsilon^2\Gamma\left(-\frac{\varepsilon}{2}\right)}\left(\frac{\pi}{\mu^2\rxyt^2}\right)^{-\varepsilon/2} \nonumber \\
    &=-\frac{2}{\varepsilon}-\ln\left(\frac{\rxyt^2\mu^2}{e^{\gamma_E}\pi}\right)+\mathcal{O}(\varepsilon)\,.
    \label{eq:dijet-NLO-dimreg-JIMWLK-kernel}
\end{align}
One can then show that the integral of the difference between two JIMWLK kernels reads as 
\begin{equation}
    \frac{1}{\pi}\int\der^2\zt\left[\frac{\rzxt\cdot\rzxpt}{\rzxt^2\rzxpt^2}-\frac{\rzxt\cdot\rzyt}{\rzxt^{2}\rzyt^{2}}\right]=\ln\left(\frac{\rxyt^2}{\rxxtp^2}\right)\,.
\end{equation}
This relation follows from Eq.\,\eqref{eq:dijet-NLO-dimreg-JIMWLK-kernel} after taking the limit $\varepsilon\to 0^-$ on both sides of the equation.

We conclude this subsection with a proof of the identity in  Eq.\,\eqref{eq:dijet-NLO-slow-jimwlk-identity}. Using the Schwinger parametrization, the integral of the UV singular JIMWLK kernel is 
\begin{align}
    \frac{\mu^{-\varepsilon}}{\pi}\int\der^{2-\varepsilon}\zt \ \frac{\rxyt^2}{\rzxt^2\rzyt^2}&=\mu^{-\varepsilon}\pi^{-\varepsilon/2}\rxyt^2\int_0^\infty \der s \ e^{-s\rxyt^2}\int_s^\infty\der u \ \frac{e^{s^2(\rxyt)^2/u}}{u^{1-\varepsilon/2}}\nonumber\\
    &=\mu^{-\varepsilon}\pi^{-\varepsilon/2}[\rxyt^{2}]^{1+\frac{\varepsilon}{2}}i^\varepsilon\int_0^\infty\der s\ e^{-s\rxyt^2}s^\varepsilon\left[\Gamma\left(-\frac{\varepsilon}{2}\right)-\Gamma\left(-\frac{\varepsilon}{2},-s\rxyt^2\right)\right]\nonumber\\
    &=\mu^{-\varepsilon}\pi^{-\varepsilon/2}[\rxyt^2]^{-\varepsilon/2}\Gamma\left(-\frac{\varepsilon}{2}\right)\Gamma(1+\varepsilon)(i^\varepsilon+i^{-\varepsilon}) \nonumber \\
    &=2\left\{-\frac{2}{\varepsilon}+\ln(e^{\gamma_E}\pi\mu^2\rxyt^2)+\mathcal{O}(\varepsilon)\right\}\,,
\end{align}
where we have used $\int_{0}^{\infty} \der s \ e^{-s}   s^{\varepsilon}   \Gamma\left(-\frac{\varepsilon}{2} , -s \right) = -(i)^{-2\varepsilon} \Gamma\left(-\varepsilon/2 \right) \Gamma\left(1+\varepsilon \right)$.

On the other hand, the integral of the $\xi$ dependent regulator reads, for $\varepsilon<0$,
\begin{align}
    \frac{\mu^{-\varepsilon}}{\pi}\int\der^{2-\varepsilon}\zt \ \frac{1}{\rzxt^2}e^{-\frac{\rzxt^2}{2\xi}}&=-\mu^{-\varepsilon}(2\xi\pi)^{-\varepsilon/2}\frac{2}{\varepsilon} \nonumber \\
    &=-\frac{2}{\varepsilon}+\ln(2\pi\mu^2\xi)+\mathcal{O}(\varepsilon)\,.
    \label{eq:int_xi_reg}
\end{align}
Combining these equations, and choosing $\xi=\rxyt^2e^{\gamma_E}/2$, one gets
\begin{equation}
    \frac{1}{\pi}\int\der^2\zt\left[\frac{\rxyt^2}{\rzxt^2\rzyt^2}-\frac{1}{\rzxt^2}e^{-\frac{\rzxt^2}{\rxyt^2e^{\gamma_E}}}-\frac{1}{\rzyt^2}e^{-\frac{\rzyt^2}{\rxyt^2e^{\gamma_E}}}\right]=0\,,
\end{equation}
in the limit $\varepsilon\to0^-$.

\subsubsection{LO wavefunctions}

The LO amplitude in $4-\varepsilon$ dimensions involves the Fourier transforms:
\begin{align}
    \mu^\varepsilon\int\frac{\der^{2-\varepsilon}\lt}{(2\pi)^{2-\varepsilon}}\frac{e^{i\lt \cdot \rt}}{\lt^2+\Delta^2}&=\frac{1}{(2\pi)^{1-\varepsilon/2}}\left(\frac{\Delta}{\mu^2r_\perp}\right)^{-\varepsilon/2}K_{-\varepsilon/2}(\Delta r_\perp)\,,
    \label{eq:dijet-LO-wf-dimreg-long}\\
    \mu^\varepsilon\int\frac{\der^{2-\varepsilon}\lt}{(2\pi)^{2-\varepsilon}}\frac{\lt^j e^{i\lt \cdot \rt}}{\lt^2+\Delta^2}&=\frac{1}{(2\pi)^{1-\varepsilon/2}}\left(\frac{\Delta}{\mu^2r_\perp}\right)^{-\varepsilon/2}\frac{i\Delta \rt^j}{r_\perp} K_{1-\varepsilon/2}(\Delta r_\perp)\,.
    \label{eq:dijet-LO-wf-dimreg-trans}
\end{align}
In two dimensions, one recovers the familiar expressions
\begin{align}
     \int \frac{\der^2 \lt}{(2\pi)} \frac{e^{i \lt \cdot \rt}}{\lt^2 + \Delta^2} &= K_0(\Delta r_\perp)\,,
     \label{eq:Transverse_int_1} \\
    \int \frac{\der^2 \lt}{(2\pi)} \frac{\lt^j e^{i \lt \cdot \rt}}{\lt^2 + \Delta^2} &= \frac{i \Delta \rt^j}{r_\perp}K_1(\Delta r_\perp)\,. \label{eq:Transverse_int_2}
\end{align}

\subsubsection{NLO wavefunctions}
\label{sub:NLOwf}
\subsubsection*{Virtual diagrams}
For the dressed self energies and vertex corrections, one needs results which are straightforward to obtain using Schwinger's parametrization. (A more detailed derivation is given in \cite{Hanninen:2017ddy,Beuf:2016wdz}). These are summarized here:
\begin{align}
    &\mu^{2\varepsilon}\int\frac{\der^{2-\varepsilon}\ltone}{(2\pi)^{2-\varepsilon}}\int\frac{\der^{2-\varepsilon}\lttwo}{(2\pi)^{2-\varepsilon}}
    \frac{\lttwo^k e^{i\ltone\cdot\rtone}e^{i\lttwo\rttwo}}{\left(\ltone^2+\Delta^2\right)\left(\lttwo^2+\omega(\ltone^2+\Delta^2)\right)}\nonumber\\
    &=\frac{\mu^\varepsilon}{2}\frac{(\mu^2\rttwo^2)^{\varepsilon/2}}{(2\pi)^{2-\varepsilon}}\frac{i\rttwo^k}{\rttwo^2}\int_0^\infty \frac{\der s}{s^{1-\varepsilon/2}}e^{-s \Delta^2}e^{-\frac{\rtone^2}{4s}}\Gamma\left(1-\frac{\varepsilon}{2},\frac{\omega \rttwo^2}{4s}\right)\,,
    \label{eq:Transverse_eps_3}
\end{align}
and 
\begin{align}
    &\mu^{2\varepsilon}\int\frac{\der^{2-\varepsilon}\ltone}{(2\pi)^{2-\varepsilon}}\int\frac{\der^{2-\varepsilon}\lttwo}{(2\pi)^{2-\varepsilon}} \frac{\ltone^j\lttwo^k e^{i\ltone\cdot\rtone}e^{i\lttwo\rttwo}}{\left(\ltone^2+\Delta^2\right)\left(\lttwo^2+\omega(\ltone^2+\Delta^2)\right)}\nonumber\\
    &=-\frac{\mu^\varepsilon}{4}\frac{(\mu^2\rttwo^2)^{\varepsilon/2}}{(2\pi)^{2-\varepsilon}}\frac{\rtone^j\rttwo^k}{\rttwo^2}\int_0^\infty \frac{\der s}{s^{2-\varepsilon/2}}e^{-s \Delta^2}e^{-\frac{\rtone^2}{4s}}\Gamma\left(1-\frac{\varepsilon}{2},\frac{\omega \rttwo^2}{4s}\right)\,.
    \label{eq:Transverse_eps_4}
\end{align}
where $\Gamma(a,x)$ the incomplete gamma function defined by $\Gamma(a,x)=\int_x^\infty\der t \ t^{a-1}e^{-t}$.

In two dimensions ($\varepsilon=0$), the remaining integral over $s$ can be expressed in terms of modified Bessel functions, as in the LO case:
\begin{align}
    \int \frac{\der^{2} \ltone}{(2\pi)^2} \int \frac{\der^{2} \lttwo }{(2\pi)^2}  \frac{\lttwo^k e^{i \ltone \cdot \rtone} e^{i \lttwo \cdot \rttwo} }{\left(\ltone^2 + \Delta^2 \right)  \left[ \lttwo^2 + \omega \left(\ltone^2 + \Delta^2 \right)  \right]} &= \frac{1}{(2\pi)^2}\frac{i \rttwo^k}{\rttwo^2} K_0 \left(\Delta \sqrt{\rtone^2+ \omega \rttwo^2} \right)\,, \label{eq:Transverse_int_4}
\end{align}
\begin{align}
    &\int \frac{\der^{2} \ltone}{(2\pi)^2} \int \frac{\der^{2} \lttwo }{(2\pi)^2}  \frac{\ltone^j\lttwo^k e^{i \ltone \cdot \rtone} e^{i \lttwo \cdot \rttwo} }{\left(\ltone^2 + \Delta^2 \right)  \left[ \lttwo^2 + \omega \left(\ltone^2 + \Delta^2 \right)  \right]} \nonumber\\
    &=-\frac{1}{(2\pi)^2}\frac{\rtone^j \rttwo^k}{\rttwo^2}\frac{\Delta}{\sqrt{\rtone^2+\omega\rttwo^2}}K_1 \left(\Delta \sqrt{\rtone^2+ \omega \rttwo^2} \right)\,. \label{eq:Transverse_int_5}
\end{align}
For instantaneous terms, we need
\begin{equation}
    \int\frac{\der^2\ltone}{(2\pi)^2}\frac{\der^2\lttwo}{(2\pi)^2}\frac{e^{i\ltone\rtone+i\lttwo\rttwo}}{\lttwo^2+\omega(\ltone^2+\Delta^2)}=\frac{1}{(2\pi)^2}\frac{\Delta}{\sqrt{\rtone^2+\omega\rttwo^2}}K_1 \left(\Delta \sqrt{\rtone^2+ \omega \rttwo^2} \right)\,.\label{eq:dijet-NLO-transint-inst}
\end{equation}
For the free self energies, the following results are useful (with $\varepsilon>0$):
\begin{align}
    &\mu^{2\varepsilon}\int\frac{\der^{2-\varepsilon}\ltone}{(2\pi)^{2-\varepsilon}}\int\frac{\der^{2-\varepsilon}\lttwo}{(2\pi)^{2-\varepsilon}}\frac{ e^{i\ltone\cdot\rt}}{\left(\ltone^2+Q^2\right)\left(\lttwo^2+\omega(\ltone^2+Q^2)\right)}\nonumber\\
    &=\frac{1}{(2\pi)^{2-\varepsilon}}\left(\frac{\omega Q^2}{\mu^4\rt^2}\right)^{-\varepsilon/2}\frac{1}{\varepsilon}K_{-\varepsilon}(Qr_\perp) \nonumber \\
    &=\frac{1}{(2\pi)}\left(\frac{Q}{(2\pi)\mu^2r_\perp}\right)^{-\varepsilon/2}K_{-\varepsilon/2}(Q r_\perp)\times\frac{1}{4\pi}\left\{\frac{2}{\varepsilon}+\frac{1}{2}\ln\left(\frac{4\pi^2\mu^4\rt^2}{Q^2\omega^2}\right)+\mathcal{O}(\varepsilon)\right\}\,,
    \label{eq:Transverse_int_UVpole}
\end{align}
and
\begin{align}
    &\mu^{2\varepsilon}\int\frac{\der^{2-\varepsilon}\ltone}{(2\pi)^{2-\varepsilon}}\int\frac{\der^{2-\varepsilon}\lttwo}{(2\pi)^{2-\varepsilon}}\frac{\ltone^i e^{i\ltone\cdot\rt}}{\left(\ltone^2+Q^2\right)\left(\lttwo^2+\omega(\ltone^2+Q^2)\right)}\nonumber\\
    &=\frac{1}{(2\pi)^{2-\varepsilon}}\frac{iQ\rt^i}{r_\perp}\left(\frac{\omega Q^2}{\mu^4\rt^2}\right)^{-\varepsilon/2}\frac{1}{\varepsilon}K_{1-\varepsilon}(Qr_\perp) \nonumber \\
    &=\frac{1}{(2\pi)}\frac{iQ\rt^i}{r_\perp}\left(\frac{Q}{(2\pi)\mu^2r_\perp}\right)^{-\varepsilon/2}K_{1-\varepsilon/2}(Q r_\perp)\nonumber\\
    &\times\frac{1}{4\pi}\left\{\frac{2}{\varepsilon}+\frac{1}{2}\ln\left(\frac{4\pi^2\mu^4\rt^2}{Q^2\omega^2}\right)-\frac{1}{Qr_\perp}\frac{K_0(Qr_\perp)}{K_1(Qr_\perp)}+\mathcal{O}(\varepsilon)\right\}\label{eq:Transverse_int_UVpole-trans}\,.
\end{align}
To obtain the last line in each of the two expressions above, we used the identities
\begin{align}
    K_{-\varepsilon}(Qr_\perp)&=K_{-\varepsilon/2}(Qr_\perp)+\mathcal{O}(\varepsilon^2)\,, \nonumber \\
    K_{1-\varepsilon}(Qr_\perp)&=K_{1-\varepsilon/2}(Qr_\perp)-\frac{\varepsilon}{2}\left.\partial_{\nu}K_\nu(Qr_\perp)\right|_{\nu=1}+\mathcal{O}(\varepsilon^2) \nonumber\\
    &=K_{1-\varepsilon/2}(Qr_\perp)-\frac{\varepsilon}{2}\frac{1}{Qr_\perp}K_0(Qr_\perp)+\mathcal{O}(\varepsilon^2)\,.
\end{align}
\subsubsection*{Real diagrams} We turn now to the NLO wavefunctions appearing in the real diagrams. In particular, we will derive the striking result in Eq.\,\eqref{eq:dijet-NLO-R2-transverse-integral}, which appears in \cite{Beuf:2011xd} without detailed proof. It can be derived from the integral
\begin{align}
    I_1^m(\rtone,\rttwo) &= \int \frac{\der^2 \ltone}{2\pi} \int \frac{\der^2 \lttwo}{2\pi} \frac{\left( \frac{\lttwo^m}{z_2} + \frac{\ltone^m}{1-z_1} \right)e^{i \ltone \cdot \rtone } e^{i \lttwo \cdot \rttwo }}{\left(z_1(1-z_1) Q^2 +\ltone^2 \right)\left(Q^2 + \frac{(\ltone + \lttwo)^2}{z_3} + \frac{\ltone^2}{z_1} + \frac{\lttwo^2}{z_2} \right)} \,,
\end{align}
where $z_1 +z_2 + z_3 = 1$.
One can obtain $I_1^m(\rtone,\rttwo)$ by taking derivatives of the scalar integral:
\begin{align}
    I_0(\rtone,\rttwo)  = \int \frac{\der^2 \ltone}{2\pi} \int \frac{\der^2 \lttwo}{2\pi} \frac{e^{i \ltone \cdot \rtone } e^{i \lttwo \cdot \rttwo }}{\left(z_1(1-z_1) Q^2 +\ltone^2 \right)\left(Q^2 + \frac{(\ltone + \lttwo)^2}{z_3} + \frac{\ltone^2}{z_1} + \frac{\lttwo^2}{z_2} \right)} \,.
\end{align}
Then
\begin{align}
    I_1^m(\rtone,\rttwo)  = -i \left( \frac{1}{z_2} \frac{\partial}{\partial \rttwo^m} + \frac{1}{1-z_1} \frac{\partial}{\partial \rtone^m} \right) I_0(\rtone,\rttwo) \,. \label{eq:I0toI1}
\end{align}
 The scalar integral $I_0(\rtone,\rttwo)$ does not have a fully analytical solution but  $I_1^m(\rtone,\rttwo)$ will have one. After Schwinger parametrizing the two denominators and performing the two Gaussian integrals over $\ltone$ and $\lttwo$, one gets 
 \begin{align}
    I_0(\rtone,\rttwo) &= \frac{z_1 z_2 z_3}{4z_1(1-z_1)}\int_0^\infty \der s \ e^{-s  Q^2} \int_0^\infty \frac{\der t}{t} \ \frac{e^{-t  Q^2} e^{-\frac{z_2 z_3 \rttwo^2}{4 t (1-z_1)}}}{(s + t)} e^{-\frac{z_1(1-z_1)\left( \rtone - \frac{z_2}{1-z_1} \rttwo\right)^2}{4(s + t)}} \,.
\end{align}
This expression suggests the obvious change of variables $u = s + t$ and $v=t$, then we have 
\begin{align}
    I_0(\rtone,\rttwo)&= \frac{z_1 z_2 z_3 }{4z_1(1-z_1)}\int_0^\infty \frac{\der u}{u} \ e^{-u Q^2} e^{-\frac{z_1(1-z_1)}{4u}\left( \rtone - \frac{z_2}{1-z_1} \rttwo\right)^2} \int_0^u \frac{\der v}{v} \  e^{-\frac{z_2 z_3 \rttwo^2}{4  (1-z_1) v}} \,.
\end{align}
The integral over $v$ leads to an incomplete gamma function with $u$ as an argument and this renders the analytic computation of the integral over $u$ intractable. Therefore we switch to $I_1^m(\rtone,\rttwo)$ by taking the corresponding derivatives in Eq.\,\eqref{eq:I0toI1}. We need:
\begin{align}
    &\left( \frac{1}{z_2} \frac{\partial}{\partial \rttwo^m} + \frac{1}{1-z_1} \frac{\partial}{\partial \rtone^m} \right) \left[\frac{z_1(1-z_1)}{4u}\left( \rtone - \frac{z_2}{1-z_1} \rttwo\right)^2+\frac{z_2 z_3 \rttwo^2}{4  (1-z_1) v}\right] \nonumber \\
    &= \left( \frac{1}{z_2} \frac{\partial}{\partial \rttwo^m} + \frac{1}{1-z_1} \frac{\partial}{\partial \rtone^m} \right) \left[\frac{z_2 z_3 \rttwo^2}{4  (1-z_1) v}\right] = \frac{z_3 \rttwo^m}{2(1-z_1)v} \,.
\end{align}
Amazingly, the derivatives acting on the first term cancel each other due to the $1/z_2$ and $1/(1-z_1)$ weights, and only the action on the second term survives. This brings a factor of $\frac{1}{v}$, which now helps us 
obtain an analytic solution!
We find
\begin{align}
    I_{1}^{m}(\rtone,\rttwo) &= i\frac{z_2 z_3^2 \rttwo^m}{8(1-z_1)^2}\int_0^\infty \frac{\der u}{u} \ e^{-u Q^2} e^{-\frac{z_1(1-z_1)}{4u}\left( \rtone - \frac{z_2}{1-z_1} \rttwo\right)^2} \int_0^u \frac{\der v}{v^2}  \  e^{-\frac{z_2 z_3 \rttwo^2}{4  (1-z_1) v}} \,.
\end{align}
The $v$ and $u$ integrals can be performed in terms of usual functions:
\begin{align}
    I_{1}^m(\rtone,\rttwo) = i\frac{z_3 \rttwo^m}{(1-z_1) \rttwo^2} K_0(QX) \,,
\end{align}
where
\begin{align}
    X^2 &=  z_1(1-z_1) \left( \rtone - \frac{z_2}{1-z_1} \rttwo\right)^2 + \frac{z_2 z_3}{(1-z_1)} \rttwo^2 \nonumber \\    
    & = z_1 z_3 \rtone^2 + z_1 z_2(\rtone - \rttwo)^2 + z_2 z_3 \rttwo^2  \,.
\end{align}
In the last equality, we  used $1-z_1 =z_2 +z_3$.

\section{The integrals $\Jcal_{\odot}$, $\Jcal_{\otimes}$ and $\Jcal_R$}
\label{app:jdotjcross}

\subsection{Integral representation of $\Jcal_{\otimes}$ and $\Jcal_{\odot}$}

The two integrals $\Jcal_{\odot}$ and $\Jcal_{\otimes}$
can be derived from the scalar integral
\begin{align}
    \Jcal(\rt,\Kt,\Delta)&=\mu^{\varepsilon}\int\frac{d^{2-\varepsilon}\lt}{(2\pi)^{2-\varepsilon}}\frac{ e^{i\lt \cdot \rt}}{\lt^2\left[ (\lt-\Kt)^2-\Delta^2 - i \epsilon \right]}\,,
\end{align}
which has to be computed in dimensional regularization since it is infrared divergent in two dimensions. For the numerical evaluation of this integral in the physical domain $\Delta^2>\Kt^2$, Feynman parametrization, supplemented by the analytic continuation to physical values of $\Delta^2$, turns out to be the most convenient method. After Feynman parametrization of the denominators, and completing the squares, one gets
\begin{align}
    \Jcal(\rt,\Kt,\Delta)&=\int_0^1\der u \ e^{iu\Kt \cdot \rt}\mu^\varepsilon\int\frac{\der^{2-\varepsilon}\Lt}{(2\pi)^{2-\varepsilon}}\frac{e^{i\Lt \cdot \rt}}{\left[\Lt^2+u(1-u)\Kt^2-u\Delta^2 - ui\epsilon \right]^2}.
\end{align}
For $\Delta^2 > \Kt^2$ and $u\in [0,1]$ one has $u(1-u)\Kt^2-u\Delta^2<0$. Let us define
\begin{align}
    \delta_{\rm{V}3} = \sqrt{|u(1-u)\Kt^2-u\Delta^2|} \,.
\end{align}
Then using we have
\begin{align}
    \Jcal(\rt,\Kt,\Delta) &=\frac{2\mu^\varepsilon}{(4\pi)^{1-\varepsilon/2}}\int_0^1\der u \ i (-1)^{-\varepsilon/4} \left(\frac{\rt^2}{4 \delta_{\mathrm{V}3}^2}\right)^{\frac{1}{2}+\frac{\varepsilon}{4}}e^{iu\Kt \cdot \rt} K_{-1-\varepsilon/2}\left(-i\delta_{\rm{V}3}r_\perp\right)\,,
\end{align}
where we have analytically continued the modified Bessel function to imaginary values of its argument (see the discussion below Eq.\,\eqref{eq:BesselK_imagarg} for a careful demonstration of the analytic continuation).
Differentiating with respect to $\rt^i$, and setting $\varepsilon=0$ since the integral is now convergent in two dimensions, one obtains the result,
\begin{align}
    \int\frac{d^{2}\lt}{(2\pi)^{2}}\frac{\lt^i e^{i\lt \cdot \rt}}{\lt^2\left[ (\lt-\Kt)^2-\Delta^2 \right]}&=\frac{1}{4\pi}\int_0^1\der u  \ e^{iu\Kt \cdot \rt}K_0\left(- i \delta_{\rm{V}3} r_\perp\right)i\rt^i\nonumber\\
    &+\frac{1}{4\pi}\int_0^1\der u \ \frac{ur_\perp e^{iu\Kt \cdot \rt}}{(-i\delta_{\mathrm{V}3})}K_1\left(- i \delta_{\mathrm{V}3}r_\perp\right)\Kt^i \,.
    \label{eq:dijet-NLO-Appendix-Jsi}
\end{align}
In order to simplify the term proportional to $\rt^i$, we used the  identity
\begin{equation}
    K_0(x)+\frac{1}{x}K_1(x)=\frac{1}{2}\left(K_0(x)+K_2(x)\right).
\end{equation}
The integrals $\Jcal_{\otimes}$ and $\Jcal_{\odot}$ can then be derived by taking the cross product or the dot product of Eq.\,\eqref{eq:dijet-NLO-Appendix-Jsi} with $-2\pi i\Kt^i$ and $(4\pi)\Kt$ respectively, giving
\begin{align}
    \Jcal_{\otimes}(\rt,\Kt,\Delta)=\frac{\rt\times\Kt}{2}\int_0^1\der u  \ e^{iu\Kt \cdot \rt}K_0\left(- i \delta_{\mathrm{V}3} r_\perp\right)\,,
    \label{eq:dijet-NLO-app-Jcrossfinal}
\end{align}
and
\begin{align}
    \Jcal_{\odot}(\rt,\Kt,\Delta)&=i\rt\cdot\Kt\int_0^1\der u \ e^{iu\Kt \cdot \rt}K_0\left(-i \delta_{\mathrm{V}3} r_\perp\right)\nonumber\\
    &+\Kt^2\int_0^1\der u \ e^{iu\Kt \cdot \rt}\frac{u r_\perp K_1\left( -i \delta_{\mathrm{V}3} r_\perp\right)}{(-i\delta_{\mathrm{V}3})}\,.
    \label{eq:dijet-NLO-app-Jdotfinal2}
\end{align}
These two formulas are well suited for numerical evaluation.

\subsection{Slow gluon limit of $\Jcal_{\odot}$}

Eq.~\eqref{eq:dijet-NLO-app-Jdotfinal2} is not very practical if we want to extract the behaviour of $\Jcal_{\odot}$ near $\Delta^2=\Kt^2$.
In this limit, the second term of Eq.\,\eqref{eq:dijet-NLO-app-Jdotfinal2} diverges because of the $u=0$ logarithmic divergence of the modified Bessel function $K_1$, but finding the asymptotic expansion around $\Delta^2=\Kt^2$ is difficult. Therefore we would like to find another analytic expression which enables the extraction of the behaviour of $\Jcal_{\odot}$ as $\Delta^2\to \Kt^2$. 
 
Our starting point is again the integral $\Jcal$ but now computed using the Schwinger parametrization of the denominator. Thus rigorously speaking, our calculation is valid for $\Delta^2<\Kt^2$ only. Nevertheless, the asymptotic behaviour near $\Delta^2=\Kt^2$ of Eq.\,\eqref{eq:dijet-NLO-app-Jdotfinal2} that we obtain after analytic continuation is the correct one. From the Schwinger parametrization of the two denominators (and computing the corresponding Gaussian integrals), one gets 
\begin{align}
    \Jcal(\rt,\Kt,\Delta)&=\frac{\mu^\varepsilon}{(4\pi)^{1-\varepsilon/2}}\int_0^{\infty}\der t \ e^{-t(\Kt^2 -\Delta^2)}\int_0^\infty \der s \ \frac{\exp\left(\frac{(i\rt+2t\Kt)^2}{4(s+t)}\right)}{(s+t)^{1-\varepsilon/2}} \nonumber \\
    &=\frac{1}{4\pi} \Gamma\left(-\frac{\varepsilon}{2}\right)\int_0^\infty \der t \ \frac{e^{-t(\Kt^2 -\Delta^2)}}{\left[\pi \mu^2(\rt-2it\Kt)^2\right]^{-\varepsilon/2}}\nonumber\\
    &-\frac{1}{4\pi}\int_0^\infty \der t \ e^{-t(\Kt^2-\Delta^2)}\Gamma\left(0,\frac{(\rt-2it\Kt)^2}{4t}\right)+\mathcal{O}(\varepsilon)\,.
    \end{align}
Expanding the first term in powers of $\varepsilon$,
\begin{equation}
    \frac{1}{4\pi}\Gamma\left(-\frac{\varepsilon}{2}\right)\left(\frac{1}{\pi \mu^2(\rt-2it\Kt)^2}\right)^{-\varepsilon/2}=-\left[\frac{2}{\varepsilon}+\ln\left(e^{\gamma_E}\pi \mu^2(\rt-2it\Kt)^2\right)+\mathcal{O}(\varepsilon)\right]\,,
\end{equation}
one finds
\begin{align}
   \Jcal(\rt,\Kt,\Delta)&=-\frac{1}{4\pi}\frac{1}{\Kt^2 - \Delta^2}\left[\frac{2}{\varepsilon}+\ln\left(e^{\gamma_E}\pi\mu^2\rt^2\right)\right]\nonumber\\
   &-\frac{1}{4\pi}\frac{1}{\Kt^2-\Delta^2}\int_0^\infty \der t\,e^{-t}\ln\left[\frac{\left(\rt-\frac{2i\Kt}{\Kt^2-\Delta^2}t\right)^2}{\rt^2}\right]\nonumber\\
   &-\frac{1}{4\pi}\int_0^\infty \der t \ e^{-t(\Kt^2-\Delta^2)}\Gamma\left(0,\frac{(\rt-2it\Kt)^2}{4t}\right)+\mathcal{O}(\varepsilon)\,.
   \label{eq:I_21}
\end{align}
One can compute analytically the integral in the second term,
\begin{align}
    \int_0^\infty \der t \ e^{-t}\ln\left[\frac{\left(\rt-\frac{2i\Kt}{\Kt^2-\Delta^2}t\right)^2}{\rt^2}\right]=e^{\chi_+}\Gamma\left(0,\chi_+\right)+e^{\chi_-}\Gamma\left(0,\chi_-\right)\,,
    \label{eq:dijet-NLO-Jdot-usefulint}
\end{align}
with
\begin{align}
    \chi_+(\rt,\Kt,\Delta)=\frac{\Kt^2-\Delta^2}{2\Kt^2}\left(i\Kt \cdot \rt+\sqrt{\Kt^2\rt^2-(\rt\cdot\Kt)^2}\right)\,,\\
    \chi_-(\rt,\Kt,\Delta)=\frac{\Kt^2-\Delta^2}{2\Kt^2}\left(i\Kt \cdot \rt-\sqrt{\Kt^2\rt^2-(\rt\cdot\Kt)^2}\right)\,. \label{eq:dijet-NLO-Jdto-chidef}
\end{align}
Note that the Cauchy-Schwarz inequality, $\Kt^2\rt^2\ge (\Kt \cdot \rt)^2$ ensures that the square-root is real.
Gathering all the pieces together, one finds
\begin{align}
    \Jcal(\rt,\Kt,\Delta)&=-\frac{1}{4\pi}\frac{1}{\Kt^2-\Delta^2}\left[\frac{2}{\varepsilon}+\ln\left(e^{\gamma_E}\pi\mu^2\rt^2\right)+e^{\chi_+}\Gamma\left(0,\chi_+\right)+e^{\chi_-}\Gamma\left(0,\chi_-\right)\right]\nonumber\\
    &-\frac{1}{4\pi}\int_0^\infty \der t \ e^{-t(\Kt^2-\Delta^2)}\Gamma\left(0,\frac{(\rt-2it\Kt)^2}{4t}\right)+\mathcal{O}(\varepsilon) \,.
\end{align}
With this expression, it is now straightforward to compute the integral $\Jcal_{\odot}$. Representing the scalar product between $\lt$ and $\Kt$ as $2\lt \cdot \Kt=-[(\lt-\Kt)^2-\Delta^2]+[\Kt^2 -\Delta^2]+\lt^2$, one gets
\begin{align}
   \Jcal_{\odot}(\rt,\Kt,\Delta)
   &=-(2\pi)\mu^\varepsilon\int\frac{d^{2-\varepsilon}\lt}{(2\pi)^{2-\varepsilon}}\frac{ e^{i\lt \cdot \rt}}{\lt^2}+(2\pi)(\Kt^2 -\Delta^2)\Jcal(\rt,\Kt,\Delta)\nonumber\\
   &+(2\pi)\mu^\varepsilon\int\frac{d^{2-\varepsilon}\lt}{(2\pi)^{2-\varepsilon}}\frac{ e^{i\lt \cdot \rt}}{\left[(\lt-\Kt)^2-\Delta^2\right]}\,.
\end{align}
Further, using  
\begin{equation}
   \mu^\varepsilon\int\frac{d^{2-\varepsilon}\lt}{(2\pi)^{2-\varepsilon}}\frac{ e^{i\lt \cdot \rt}}{\lt^2}= -\frac{1}{4\pi}\left[\frac{2}{\varepsilon}+\ln\left(e^{\gamma_E}\pi\mu^2\rt^2\right)\right]+\mathcal{O}(\varepsilon)\,,
\end{equation}
one sees that the infrared pole in $1/\varepsilon$ cancels, as it should, and one is left with a finite expression in two dimensions
\begin{align}
    \Jcal_{\odot}(\rt,\Kt,\Delta)&=-\frac{1}{2}\left[e^{\chi_+}\Gamma\left(0,\chi_+\right)+e^{\chi_-}\Gamma\left(0,\chi_-\right)\right]+e^{i\Kt \cdot \rt}K_0(-i\Delta r_\perp)\nonumber\\
    &-\frac{\Kt^2-\Delta^2}{2}\int_0^\infty\der t \ e^{-t(\Kt^2-\Delta^2)}\Gamma\left(0,\frac{\rt^2}{4t}-i\Kt \cdot \rt-t\Kt^2\right) \,.
    \label{eq:dijet-NLO-app-Jdotfinal}
\end{align}
We can now determine the asymptotic behaviour of this expression, as a function of $\Delta$, in the limit $\Delta^2\to\Kt^2$. The last term vanishes in this this limit and $\chi_{\pm}\to 0$. Using
\begin{equation}
    \Gamma(0,x)=-\gamma_E-\ln(x)+\mathcal{O}(x)\,,
    \label{eq:dijet-NLO-gammaexpansion}
\end{equation}
one finds
\begin{align}
    \Jcal_{\odot}(\rt,\Kt,\Delta)&=\ln\left(\frac{\Kt^2 - \Delta^2}{2\Kt^2}\right)+\frac{1}{2}\ln(\Kt^2\rt^2)+\frac{i\pi}{2}+\gamma_E+e^{i\Kt \cdot \rt}K_0(-i K_\perp r_\perp) \nonumber \\
    &+\mathcal{O}(\Kt^2-\Delta^2)\,,
    \label{eq:dijet-NLO-app-Jdotslow}
\end{align}
with the imaginary part determined modulo $2\pi$.
We have checked numerically that this is the expected asymptotic behaviour of Eq.~\eqref{eq:dijet-NLO-app-Jdotfinal2}.

\subsection{Slow gluon limit of $\mathcal{J}_R$}
\label{sub:JR-asymptot}

In this section, we will prove the identity stated in Eq.\,\eqref{eq:dijet-NLO-JR-smallzg}. We recall that the transverse momentum  integral $\Jcal_R$ is defined by
\begin{equation}
    \Jcal_R(\rt,\Kt)=\int\frac{\der^2\lt}{(2\pi)^2}e^{-i\lt \cdot \rt}\frac{4\lt \cdot (\lt+\Kt)}{\lt^2(\lt+\Kt)^2}\,,
\end{equation}
for  $K_\perp>0$. We are looking for the asymptotic behaviour of this function as $\Kt$ goes to 0. Let us first write the numerator as a sum of square using $4\lt \cdot (\lt+\Kt)=2(\lt^2+(\lt+\Kt)^2-\Kt^2)$. Plugging this identity inside the definition of $\Jcal_R$ gives
\begin{equation}
    \Jcal_R(\rt,\Kt)=2\left[\left(1+e^{i\Kt \cdot \rt}\right)\mu^\varepsilon\int\frac{\der^{2-\varepsilon}\lt}{(2\pi)^{2-\varepsilon}}\frac{e^{-i\lt \cdot \rt}}{\lt^2}-\Kt^2\mu^\varepsilon\int\frac{\der^{2-\varepsilon}\lt}{(2\pi)^{2-\varepsilon}}\frac{e^{-i\lt \cdot \rt}}{\lt^2(\lt+\Kt)^2}\right]\,, \label{eq:dijet-NLO-JR-dec}
\end{equation}
where each integral has been analytically continued to $2-\varepsilon$ dimensions (with $\varepsilon<0$). Indeed, even though the integral $\Jcal_R$ is convergent in  2 dimensions as long as $\Kt$ is non-zero, each term in the expression above is IR divergent in 2 dimensions. The second integral in Eq.~\eqref{eq:dijet-NLO-JR-dec} can be evaluated using Schwinger parametrization:
\begin{align}
    \mu^\varepsilon\int\frac{d^{2-\varepsilon}\lt}{(2\pi)^{2-\varepsilon}}\frac{e^{-i\lt \cdot \rt}}{\lt^2(\lt+\Kt)^2}&=\frac{2^{-1-\varepsilon/2}}{(2\pi)^{1-\varepsilon/2}}\Gamma\left(-\frac{\varepsilon}{2}\right)\int_0^\infty \der t\frac{e^{-t\Kt^2}}{(\mu^2(\rt-2it\Kt)^2)^{-\varepsilon/2}}\nonumber\\
    &-\frac{2^{-1-\varepsilon/2}}{(2\pi)^{1-\varepsilon/2}}\int_0^\infty \der t \frac{e^{-t\Kt^2}}{(\mu^2(\rt-2it\Kt)^2)^{-\varepsilon/2}}\Gamma\left(-\frac{\varepsilon}{2},\frac{(\rt-2it\Kt)^2}{4t}\right)\,.
    \label{eq:dijet-NLO-JR-step1}
\end{align}
The first term was computed in the previous section-see Eq.\,\eqref{eq:dijet-NLO-Jdot-usefulint}. Setting $\Delta=0$ in the latter equation, one gets
\begin{align}
    &\frac{2^{-1-\varepsilon/2}}{(2\pi)^{1-\varepsilon/2}}\Gamma\left(-\frac{\varepsilon}{2}\right)\int_0^\infty \der t \frac{e^{-t\Kt^2}}{(\mu^2(\rt-2it\Kt)^2)^{-\varepsilon/2}}\nonumber\\
    &=-\frac{1}{4\pi}\frac{1}{\Kt^2}\left(\frac{2}{\varepsilon}+\ln\left(e^{\gamma_E}\pi\mu^2\rt^2\right)\right)\nonumber\\ 
    &-\frac{1}{4\pi}\frac{1}{\Kt^2}\left[e^{\chi_+(\rt,\Kt,0)}\Gamma(0,\chi_+(\rt,\Kt,0))+(\chi_+\leftrightarrow\chi_-)\right]
    +\mathcal{O}(\varepsilon)\,,
\end{align}
with the functions $\chi_{\pm}$ defined in Eq.\,\eqref{eq:dijet-NLO-Jdto-chidef}.
The second term in Eq.\,\eqref{eq:dijet-NLO-JR-step1} requires special care because it also contains a $1/\varepsilon$ pole. To extract this pole, we write it as
\begin{align}
    &\frac{2^{-1-\varepsilon/2}}{(2\pi)^{1-\varepsilon/2}}\int_0^\infty \der t \ e^{-t\Kt^2} \frac{}{}\left\{\frac{\Gamma\left(-\frac{\varepsilon}{2},\frac{(\rt-2it\Kt)^2}{4t}\right)}{(\mu^2(\rt-2it\Kt)^2)^{-\varepsilon/2}}-e^{i\Kt \cdot \rt}\frac{\Gamma\left(-\frac{\varepsilon}{2},-t\Kt^2\right)}{(2it\mu K_\perp)^{-\varepsilon}}\right\}\nonumber\\
    &+\frac{2^{-1-\varepsilon/2}}{(2\pi)^{1-\varepsilon/2}}e^{i\Kt \cdot \rt}\int_0^\infty \der t \frac{e^{-t\Kt^2}}{(2it\mu K_\perp)^{-\varepsilon}}\Gamma\left(-\frac{\varepsilon}{2},-t\Kt^2\right)\nonumber\\
    &=\frac{1}{4\pi}\int_0^\infty \der t \  e^{-t\Kt^2}\left\{\Gamma\left(0,\frac{(\rt-2it\Kt)^2}{4t}\right)-e^{i\Kt \cdot \rt}\Gamma\left(0,-t\Kt^2\right)\right\}\nonumber\\
    &+\frac{2^{-1-\varepsilon/2}}{(2\pi)^{1-\varepsilon/2}}e^{i\Kt \cdot \rt}\int_0^\infty \der t\frac{e^{-t\Kt^2}}{(2it\mu K_\perp)^{-\varepsilon}}\Gamma\left(-\frac{\varepsilon}{2},-t\Kt^2\right)+\mathcal{O}(\varepsilon)\,.
\end{align}
To obtain the last line, we used the fact that the first integral is convergent when $\varepsilon=0$. As a consequence, one can set $\varepsilon=0$ up to terms of order $\mathcal{O}(\varepsilon)$.
Combining all of these results inside Eq.\,\eqref{eq:dijet-NLO-JR-dec}, performing a  change of variable $t\Kt^2\to t$ inside the $t$ integrals, one gets
\begin{align}
    \Jcal_R(\rt,\Kt)&=\lim\limits_{\varepsilon\to0} \ \frac{1}{2\pi}\left\{-e^{i\Kt \cdot \rt}\left(\frac{2}{\varepsilon}+\ln(e^{\gamma_E}\pi\mu^2\rt^2)\right)+e^{\chi_+}\Gamma(0,\chi_+)+e^{\chi_-}\Gamma(0,\chi_-)\right.\nonumber\\
    & +\int_0^\infty \der t \  e^{-t}\left[\Gamma\left(0,\frac{\rt^2\Kt^2}{4t}-i\rt\Kt-t\right)-e^{i\Kt \cdot \rt}\Gamma\left(0,-t\right)\right]\nonumber\\
    &\left.+\left(\frac{K_\perp}{2i\mu\pi^2}\right)^{-\varepsilon}e^{i\Kt \cdot \rt}\int_0^\infty \der t \ e^{-t} \ t^\varepsilon \ \Gamma\left(-\frac{\varepsilon}{2},-t\right)\right\}\,.
    \label{eq:dijet-NLO-JR-step2}
\end{align}
The integral in the third term of the r.h.s of this expression contains a pole of the form $2e^{i\Kt \cdot \rt}/\varepsilon$ which exactly cancels the pole in the first term. This is to be expected since $\Jcal_R(\rt,\Kt)$ has no IR nor UV divergences. We have indeed 
\begin{align}
    \left(\frac{K_\perp}{2i\mu\pi^2}\right)^{-\varepsilon}\int_0^\infty\der t \ e^{-t} \ t^\varepsilon \ \Gamma\left(-\frac{\varepsilon}{2},-t\right)=\frac{2}{\varepsilon}-2\gamma_E-\ln\left(\frac{\Kt^2}{4\pi e^{\gamma_E}\mu^2}\right)-i\pi+\mathcal{O}(\varepsilon)\,,
\end{align}
so our final expression for $\Jcal_R(\rt,\Kt)$ reads
\begin{align}
    \Jcal_R(\rt,\Kt)&=\frac{1}{2\pi}\left\{-e^{i\Kt \cdot \rt}\left[\ln\left(\frac{\Kt^2\rt^2}{4}\right)+2\gamma_E+i\pi\right]+e^{\chi_+}\Gamma(0,\chi_+)+e^{\chi_-}\Gamma(0,\chi_-)\right.\nonumber\\
  &\left.  +\int_0^\infty \der t \  e^{-t}\left[\Gamma\left(0,\frac{\rt^2\Kt^2}{4t}-i\rt\Kt-t\right)-e^{i\Kt \cdot \rt}\Gamma\left(0,-t\right)\right]\right\}\,,
\end{align}
with $\chi_{\pm}=\chi_{\pm}(\rt,\Kt,0)$. It is now straightforward to obtain the asymptotic behaviour of $\Jcal_R(\rt,\Kt)$ in the limit $\Kt\to0$. The integral in the second line goes to $0$ at small $K_\perp$. Using Eq.\,\eqref{eq:dijet-NLO-gammaexpansion}, one gets finally
\begin{align}
    \Jcal_R(\rt,\Kt)&=\frac{1}{2\pi}\left\{-2\ln\left(\frac{\Kt^2\rt^2}{4}\right)-4\gamma_E-2i\pi+\mathcal{O}(K_\perp)\right\}\,,
\end{align}
which completes our proof of Eq.\,\eqref{eq:dijet-NLO-JR-smallzg}.

\section{Details of the computation of diagram $\mathrm{R}2$}
\label{app:R1}
 The scattering amplitude for $q\bar{q}+g$  emission from a quark after scattering from the shock wave is given by
\begin{align}
    \Scal^{\lambda\bar{\lambda}\sigma\sigma'}_{\mathrm{R}2} &= \int \frac{\der^4 l}{(2\pi)^4} \bar{u}(k,\sigma) \left( ig t^a \slashed{\epsilon}^{*}(k_g,\bar{\lambda}) \right) S^0(k+k_g) \Tcal^q(k+k_g,l)  S^0(l) \left(-iee_f \slashed{\epsilon}(q,\lambda) \right) \nonumber\\
    &\times S^0(l-q) \Tcal^q(l-q,-p) v(p,\sigma')\,.
\end{align}
As usual, we have subtracted the noninteracting term and factored out the overall $2\pi \delta(q^--k^--p^--k_g^-)$ function, to obtain the physical amplitude
\begin{align}
    \Mcal^{\lambda\bar{\lambda}\sigma\sigma'}_{\mathrm{R}2} = \frac{ee_fq^-}{\pi} \int \der^2 \wt \der^2 \yt e^{-i (\kt + \kgt )\cdot \wt } e^{-i \pt \cdot \yt } \Ccal_{\mathrm{R}2}(\wt,\yt)  \mathcal{N}_{\mathrm{R}2}^{\lambda\bar{\lambda}\sigma\sigma'}(\rwyt)\,,
\end{align}
with the color factor
\begin{align}
    \Ccal_{\mathrm{R}2}(\wt,\yt) = \left[ t^a V(\wt)V^\dagger(\yt) -t^a \right]\,,
\end{align}
and the perturbative factor
\begin{align}
    \mathcal{N}_{\mathrm{R}2}^{\lambda\bar{\lambda}\sigma\sigma'} (\rwyt) =  ig\int \frac{\der^4 l}{(2\pi)^2} e^{i \lt \cdot \rwyt} \frac{ (2q^-)N_{\mathrm{R}2}^{\lambda\bar{\lambda}\sigma\sigma'}(l)   \delta(l^- - q^- + p^-) }{ (l^2 + i \epsilon) ((l-q)^2 + i \epsilon)}\,,
    \label{eq:dijet-NLO_real1-pert}
\end{align}
with the Dirac numerator
\begin{align}
    N_{\mathrm{R}2}^{\lambda\bar{\lambda}\sigma\sigma'}(l) = \frac{1}{(2q^-)^2} \frac{1}{(2 k.k_g)}  \left[\bar{u}(k,\sigma) \slashed{\epsilon}^*(k_g,\bar{\lambda}) (\slashed{k}+\slashed{k}_g) \gamma^- \slashed{l} \slashed{\epsilon}(q,\lambda) (\slashed{l}-\slashed{q}) \gamma^- v(p,\sigma') \right] \,.
    \label{eq:dijet-NLO-real1-dirac}
\end{align}

In the perturbative factor, the integration over $l^-$ is trivial due to the presence of the delta function $\delta(l^- - q^- + p^-)$, while the integration over $l^+$ is performed using contour integration employing  Cauchy's theorem. We note that due the location of the $\gamma^-$, the Dirac structure $N_{\mathrm{R}1}$ is independent of $l^+$. Therefore the contour integral is the same as in the LO calculation. We find then that Eq.\,\eqref{eq:dijet-NLO_real1-pert} becomes
\begin{align}
    \mathcal{N}_{\mathrm{R}2}^{\lambda\bar{\lambda}\sigma\sigma'} (\rwyt) =  -g\int \frac{\der^2 \lt}{(2\pi)}  \frac{ N_{\mathrm{R}2}^{\lambda\bar{\lambda}\sigma\sigma'}(l) e^{i \lt \cdot \rwyt} }{\lt^2 + \bar Q_{\mathrm{R}2}^2}\,,
    \label{eq:dijet-NLO-R1-perturbative1}
\end{align}
where $\bar Q_{\mathrm{R}2}^2=z_{\bar q}(1-z_{\bar q})Q^2$.
To proceed, we reexpress the Dirac structure in Eq.\,\eqref{eq:dijet-NLO-real1-dirac} employing the identity in Eq.\,\eqref{eq:gluon_emi_quark_afterSW} and find
\begin{align}
    N_{\mathrm{R}2}^{\lambda\bar{\lambda}\sigma\sigma'}(l) &=
    \frac{\left(z_q \kgt - z_g \kt \right)\cdot\et^{\bar{\lambda}*}}{\left( z_q \kgt - z_g \kt\right)^2}   \left \{\bar{u}(k,\sigma)\left[(2z_q+z_g)+z_g\bar\lambda\Omega\right] \mathcal{D}^{\lambda}_{\rm LO}(l)  v(p,\sigma') \right \}\,.
    \label{eq:dijet-NLO-R1-diracnum}
\end{align}
Combining Eq.\,\eqref{eq:dijet-NLO-R1-diracnum} and Eq.\,\eqref{eq:dijet-NLO-R1-perturbative1} gives  Eq.~\eqref{eq:dijet-NLO-R1-perturbative} in the main text.

\section{Details of the computation of diagram $\mathrm{SE}2$}
\label{app:SE1}

The quark free self energy before the shock wave is UV divergent in 4 dimensions; we will therefore  compute it using dimensional regularization.
The amplitude in $d=4-\varepsilon$ dimensions reads
\begin{align}
    \Scal^{\lambda\sigma \sigma'}_{\mathrm{SE}2} &= \mu^{2\varepsilon}\int \frac{\der^{4-\varepsilon} l_1}{(2\pi)^{4-\varepsilon}} \frac{\der^{4-\varepsilon} l_2}{(2\pi)^{4-\varepsilon}}  \left[\bar{u}(k,\sigma) \Tcal^q(k,l_1) S^0(l_1) (ig \gamma^\mu t^a) S^0(l_1-l_2)(ig \gamma^\nu t^a) S^0(l_1) \right. \nonumber \\ &\times (-iee_f\slashed{\epsilon}(q,\lambda)) 
   \left. S^0(l_1 - q) \Tcal^q(l_1-q,-p) v(p,\sigma') \right] G^0_{\mu\nu}(l_2)\,.
\end{align}
Subtracting the noninteracting piece, and factoring out the overall delta function $2\pi \delta(q^- - k^- - p^-)$, we find
\begin{align}
    \Mcal^{\lambda\sigma \sigma'}_{\mathrm{SE}2} = \frac{ee_fq^-}{\pi} \mu^{-2\varepsilon}\int \der^{2-\varepsilon} \xt \der^{2-\varepsilon} \yt e^{-i \kt \cdot \xt } e^{-i \pt \cdot \yt } \Ccal_{\mathrm{SE}2}(\xt,\yt)  \mathcal{N}_{\mathrm{SE}2}^{\lambda\sigma \sigma'}(\rxyt) \,,
\end{align}
with the color structure
\begin{align}
    \Ccal_{\mathrm{SE}2}(\xt,\yt) = C_F\left[V(\xt) V^\dagger(\yt) - \mathbbm{1} \right]\,,
\end{align}
and the perturbative factor
\begin{align}
    \Ncal_{\mathrm{SE}2}^{\lambda\sigma \sigma'}(\rxyt) &=  g^2 \mu^{2\varepsilon} \int \frac{\der^{4-\varepsilon}  l_1}{(2\pi)^{3-\varepsilon} } \frac{\der^{4-\varepsilon}  l_2}{(2\pi)^{3-\varepsilon} } e^{i \ltone \cdot \rxyt } \nonumber \\
    &\times \frac{(2q^-)N_{\mathrm{SE}2}^{\lambda\sigma \sigma'}(l_1,l_2) \delta(k^- - l_1^-) }{(l_1^2 + i \epsilon)((l_1 -l_2)^2 + i \epsilon)(l_1^2 + i \epsilon) ((l_1 -q)^2 + i \epsilon)(l_2^2 + i \epsilon)} \,.
    \label{eq:dijet-NLO-SE1-pert-1}
\end{align}
The Dirac numerator for this diagram reads
\begin{align}
    N_{\mathrm{SE}2}^{\lambda\sigma \sigma'} = \frac{1}{(2q^-)^2} \left[ \bar{u}(k,\sigma) \gamma^- \slashed{l}_1 \gamma^\mu (\slashed{l}_1 - \slashed{l}_2) \gamma^\nu \slashed{l}_1 \slashed{\epsilon}(q,\lambda) (\slashed{l}_1 -\slashed{q})\gamma^- v(p,\sigma')\right] \Pi_{\mu\nu}(l_2) \,.
    \label{eq:dijet-NLO-SE1-dirac-1}
\end{align}

\subsubsection*{Dirac numerator $ N_{\mathrm{SE}2}(l_1,l_2)$ in $d=4-\varepsilon$ dimensions}

Using the definition in Eq.\,\eqref{eq:gluon-pol-tensor-def} of the gluon polarization tensor $\Pi^{\mu\nu}$, one finds that
\begin{align}
    &\gamma_\mu(\slashed l_1-\slashed l_2)\gamma_\nu \Pi^{\mu\nu}(l_2)=\gamma_\mu\left[2(l_{1,\nu}-l_{2,\nu})-\gamma_\nu(\slashed l_1-\slashed l_2)\right]\Pi^{\mu\nu}(l_2)\nonumber\\
    &=2\gamma_\mu(l_{1,\nu}-l_{2,\nu})\left(-g^{\mu\nu}+\frac{1}{l_2^-}(l_2^\mu n^\nu+l_2^\nu n^\mu)\right)
    -\gamma_{\mu}\gamma_{\nu}\left(-g^{\mu\nu}+\frac{1}{l_2^-}(l_2^\mu n^\nu+l_2^\nu n^\mu)\right)(\slashed l_1-\slashed l_2)\nonumber\\
    &=-2(\slashed l_1-\slashed l_2)+\frac{2}{l_2^-}\left(\slashed l_2(l_1^- -l_2^-)+\gamma^-(l_1-l_2)l_2\right)
    +\left(4-\varepsilon-\frac{1}{l_2^-}(\slashed l_2 \gamma^-+\gamma^-\slashed l_2)\right)(\slashed l_1-\slashed l_2)\nonumber\\
    &=\frac{2}{l_2^-}\left(\slashed l_2(l_1^- -l_2^-)+\gamma^-(l_1-l_2)l_2\right)-\varepsilon (\slashed l_1- \slashed l_2)\,.
\end{align}
In this expression, the $\mathcal{O}(\varepsilon)$ term comes from $\gamma^\mu\gamma_\mu=d=4-\varepsilon$.
Then using $\slashed l_1 \slashed l_2\slashed l_1=2 l_1 l_2\slashed l_1-l_1^2\slashed l_2$ and $2l_1l_2=l_1^2+l_2^2-(l_1-l_2)^2$, the Dirac structure reads
\begin{align}
    N^{\lambda\sigma \sigma'}_{\mathrm{SE}2}&=\left\{-\left(\frac{4l_1^-}{l_2^-}-2+\varepsilon\right)(l_1-l_2)^2+\left(\frac{4l_1^-}{l_2^-}-2\right)l_1^2-(2-\varepsilon)l_2^2\right\}\bar{u}(k,\sigma)\mathcal{D}^{\lambda}_{\rm LO}(l_1) v(p,\sigma')\nonumber\\
   & -l_1^2\left(\frac{2l_1^-}{l_2^-}-2+\varepsilon\right)\frac{\left[\bar{u}(k,\sigma)\gamma^-\slashed l_2\slashed \epsilon(q,\lambda)(\slashed l_1-\slashed q)v(p,\sigma')\right]}{(2q^-)^2}\,.
\end{align}
Finally, decomposing the $l_2$ four-vector in the second term according to $l_2=\left(l_2-\frac{l_2^-}{l_1^-}l_1\right)+\frac{l_2^-}{l_1^-}l_1$, one gets
\begin{align}
   N^{\lambda\sigma \sigma'}_{\mathrm{SE}2}&=\left\{\left(2-\varepsilon-\frac{4l_1^-}{l_2^-}\right)(l_1-l_2)^2+\left(\frac{4l_1^-}{l_2^-}-4+(2-\varepsilon)\frac{l_2^-}{l_1^-}\right)l_1^2-(2-\varepsilon)l_2^2\right\} \nonumber \\
   &\times \bar{u}(k,\sigma)\mathcal{D}^{\lambda}_{\rm LO}(l_1) v(p,\sigma') \nonumber\\
   & +l_1^2\Lttwox^i\left(\frac{2l_1^-}{l_2^-}-2+\varepsilon\right)\frac{\left[\bar{u}(k,\sigma)\gamma^-\gamma^i\slashed \epsilon(q,\lambda)(\slashed l_1-\slashed q)v(p,\sigma')\right]}{(2q^-)^2}\,,
\end{align}
with $\Lttwox=\lttwo-\frac{l_2^-}{l_1^-}\ltone$. To sum up, the Dirac numerator of the diagram $\mathrm{SE}2$ can be expressed as 
\begin{equation}
    N_{\mathrm{SE}2}=l_1^2N_{\mathrm{SE}2,\rm reg}+l_1^2N_{\mathrm{SE}2,q\mathrm{inst}1}+(l_1-l_2)^2N_{\mathrm{SE}2,q\mathrm{inst}2}+l_2^2N_{\mathrm{SE}2,g\mathrm{inst}}\,.
    \label{eq:dijet-NLO-SE1-dirac-dec}
\end{equation}
The last two terms cancel the corresponding propagators in the denominator of the perturbative factor. They do not contribute to the amplitude since they vanish after integration over $l_2^+$ and $l_1^+$. The regular term $N_{\mathrm{SE}2,\rm reg}$ can be written as
\begin{align}
     N^{\lambda\sigma \sigma'}_{\mathrm{SE}2,\rm reg}
     &=\frac{4z_q}{z_g}\left[1-\frac{z_g}{z_q}+\left(1-\frac{\varepsilon}{2}\right)\frac{z_g^2}{2z_q^2}\right]\bar{u}(k,\sigma)\mathcal{D}^{\lambda}_{\rm LO}(l_1) v(p,\sigma')\,,
     \label{eq:dijet-NLO-SE1-dirac-0}
\end{align}
with $z_q=k^-/q^-$ and $z_g=l_2^-/q^-$. The quark instantaneous term $ N^{\lambda\sigma \sigma'}_{\mathrm{SE}2,q \rm inst,1}$ is proportional to $\Lttwox^i$ and vanishes due to rotational invariance of the transverse integral. We do not discuss its structure further. This term corresponds to the instantaneous quark self energy diagram in LCPT, and is indeed identically zero by rotational invariance, as shown in \cite{Beuf:2016wdz}. The decomposition given by Eq.\,\eqref{eq:dijet-NLO-SE1-dirac-dec} is slightly different from the general decomposition of the Dirac structure outlined in Sec.\,\ref{sec:gen-strat} since both the regular and the quark instantaneous contributions are multiplied by a factor $l_1^2$. The reason is that strictly speaking, our discussion in Sec.\,\ref{sec:gen-strat} applies for diagrams without double propagators, which is not the case of diagram $\rm SE2$. In the regular term, the $l_1^2$ factor cancels the quark propagator after the virtual gluon emission, while in the instantaneous quark term $N^{\lambda\sigma \sigma'}_{\mathrm{SE}2,q \rm inst,1}$, the $l_1^2$ factor cancels the quark propagator before the virtual gluon emission.

\subsubsection*{Pole structure}

As mentioned, the quark instantaneous term vanishes for both longitudinal and transverse virtual photon polarization, because of rotational invariance of the $\Lttwox$ integral.
On the other hand, the regular perturbative factor does not depend on $l_1^+$ nor $l_2^+$ in the numerator. One can thus easily perform the $l_1^+$ and $l_2^+$ integrals by contour integration, using the results of Appendix~\ref{app:contour}. The regular perturbative factor is then 
\begin{align}
 \Ncal_{\mathrm{SE}2,\rm reg}(\rxyt)&=\frac{g^2}{2}\int \der z_g \ \mu^{2\varepsilon}\int\frac{\der^{2-\varepsilon}\ltone}{(2\pi)^{2-\varepsilon}}\frac{\der^{2-\varepsilon}\lttwo}{(2\pi)^{2-\varepsilon}}e^{i\ltone\cdot\rxyt}\Ical_{\mathrm{SE}2,\rm reg}N_{\mathrm{SE}2,\rm reg}\,,
\end{align}
with
\begin{align}
     \Ical_{\mathrm{SE}2,\rm reg}&=\int\frac{\der l_1^+}{(2\pi)}\int\frac{\der l_2^+}{(2\pi)}\frac{(2q^-)^2}{[l_1^2+i\epsilon][(l_1-q)^2+i\epsilon][(l_1-l_2)^2+i\epsilon][l_2^2+i\epsilon]}\\
     &=-\frac{1}{z_q}\frac{1}{\left(\ltone^2+\bar{Q}^2\right)}\frac{\Theta(z_g)\Theta(z_q-z_g)}{\left[\Lttwox^2+\omega_{\mathrm{SE}2}(\ltone^2+\bar{Q}^2)\right]}\,,
\end{align}
where $\bar Q^2=z_qz_{\bar q}Q^2$ and $\omega_{\mathrm{SE}2}=z_g(z_q-z_g)/(z_q^2z_{\bar q})$.

\subsubsection*{Transverse momentum integration} The next step consists in calculating the remaining transverse momeentum integrals over $\ltone$ and $\lttwo$ in $2-\varepsilon$ dimensions. For a longitudinal virtual photon ($\lambda=0)$, the expression  reads as 
\begin{align}
    &\Ncal^{\lambda=0,\sigma \sigma'}_{\mathrm{SE}2}(\rxyt)=2g^2z_qz_{\bar q}Q\int_0^{z_q}\frac{\der z_g}{z_g}\left[1-\frac{z_g}{z_q}+\left(1-\frac{\varepsilon}{2}\right)\frac{z_g^2}{2z_q^2}\right]\frac{[\bar{u}(k,\sigma)\gamma^-v(p,\sigma')]}{q^-}\nonumber\\
    &\hspace{2cm}\times \int\frac{\der^{2-\varepsilon}\ltone}{(2\pi)^{2-\varepsilon}}\frac{\der^{2-\varepsilon}\lttwo}{(2\pi)^{2-\varepsilon}}\frac{e^{i\ltone\cdot\rxyt}}{\left(\ltone^2+\bar{Q}^2\right)\left[\Lttwox^2+\omega_{\mathrm{SE}2}(\ltone^2+\bar{Q}^2)\right]}\label{eq:dijet-NLO-SE1-Ncal-step1}\\
    &=-\frac{\alpha_s}{\pi}\mathcal{N}^{\lambda=0,\sigma \sigma'}_{\rm LO,\varepsilon}(\rxyt)\int_0^{z_q}\frac{\der z_g}{z_g}\left[1-\frac{z_g}{z_q}+\left(1-\frac{\varepsilon}{2}\right)\frac{z_g^2}{2z_q^2}\right]\left\{\frac{2}{\varepsilon}+\frac{1}{2}\ln\left(\frac{4\pi^2\mu^4\rxyt^2}{\bar Q^2\omega^2_{\mathrm{SE}2}}\right)+\mathcal{O}(\varepsilon)\right\}\,.
    \label{eq:dijet-NLO-SE1-Ncal-step2}
\end{align}
To go from Eq.~\eqref{eq:dijet-NLO-SE1-Ncal-step1} to Eq.\,\eqref{eq:dijet-NLO-SE1-Ncal-step2}, we used the formula Eq.\,\eqref{eq:Transverse_int_UVpole}. Similarly, for a transversely polarized photon ($\lambda=\pm1$), one finds 
\begin{align}
     \Ncal^{\lambda=\pm1,\sigma \sigma'}_{\mathrm{SE}2}(\rxyt)&=g^2\int_0^{z_q}\frac{\der z_g}{z_g}\left[1-\frac{z_g}{z_q}+\left(1-\frac{\varepsilon}{2}\right)\frac{z_g^2}{2z_q^2}\right]\frac{[\bar{u}(k,\sigma)((z_{\bar q}-z_q)\delta^{ij}+\omega^{ij})\gamma^-v(p,\sigma')]}{q^-}\et^{\lambda,i}\nonumber\\
    &\times \int\frac{\der^{2-\varepsilon}\ltone}{(2\pi)^{2-\varepsilon}}\frac{\der^{2-\varepsilon}\lttwo}{(2\pi)^{2-\varepsilon}}\frac{\ltone^je^{i\ltone\cdot\rxyt}}{\left(\ltone^2+\bar{Q}^2\right)\left[\Lttwox^2+\omega_{\mathrm{SE}2}(\ltone^2+\bar{Q}^2)\right]}\label{eq:dijet-NLO-SE1-Ncal-step1-trans}\\
    &=-\frac{\alpha_s}{\pi}\mathcal{N}^{\lambda=\pm1,\sigma \sigma'}_{\rm LO,\varepsilon}(\rxyt)\int_0^{z_q}\frac{\der z_g}{z_g}\left[1-\frac{z_g}{z_q}+\left(1-\frac{\varepsilon}{2}\right)\frac{z_g^2}{2z_q^2}\right]\nonumber\\
    &\times\left\{\frac{2}{\varepsilon}+\frac{1}{2}\ln\left(\frac{4\pi^2\mu^4\rxyt^2}{\bar Q^2\omega^2_{\mathrm{SE}2}}\right)-\frac{K_0(\bar Qr_{xy})}{\bar Q r_{xy} K_1(\bar Qr_{xy})}+\mathcal{O}(\varepsilon)\right\}\,,
    \label{eq:dijet-NLO-SE1-Ncal-step2-trans}
\end{align}
where we used Eq.\,\eqref{eq:Transverse_int_UVpole-trans} to obtain the second line. Computing  the remaining $z_g$ integrals results in  Eq.~\eqref{eq:dijet-NLO-SE2-final} and Eq.\,\eqref{eq:dijet-NLO-SE2-final-transverse} in the main text.

\section{Details of the computation of diagram $\mathrm{V}2$}
\label{app:V1}
This diagram is UV divergent, therefore we compute it in $4-\varepsilon$ dimensions in order to extract the $1/\varepsilon$ pole (when a given diagram is UV convergent, we compute it directly in 4 dimensions). The amplitude of the free vertex correction before the shock wave is given by 
\begin{align}
    \Scal_{\mathrm{V}2}^{\lambda\sigma\sigma'} &= \mu^{2\varepsilon}\int \frac{\der^{4-\varepsilon} l_1}{(2\pi)^{4-\varepsilon}} \frac{\der^{4-\varepsilon} l_2}{(2\pi)^{4-\varepsilon}}  \left[ \bar{u}(k,\sigma) \Tcal^q(k,l_1) S^0(l_1) (ig \gamma^\mu t^a ) S^0(l_1-l_2) (-ie e_f \slashed{\epsilon}(q,\lambda))  \nonumber \right. \nonumber \\
    &\times \left.S^0(l_1 -l_2- q ) (ig t^b \gamma^\nu)  S^0(l_1-q) \Tcal^q(l_1-q,-p) v(p,\sigma') \right] G^{0,ab}_{\mu\nu}(l_2)\,.
\end{align}
Subtracting the noninteracting piece, and factoring out the overall delta function $2\pi \delta(q^- - k^- - p^-)$, we find
\begin{align}
    \Mcal^{\lambda\sigma\sigma'}_{\mathrm{V}2} = \frac{e e_f q^-}{\pi} \mu^{-2\varepsilon}\int \der^{2-\varepsilon} \xt \der^{2-\varepsilon} \yt e^{-i \kt \cdot \xt-i \pt \cdot \yt } \Ccal_{\mathrm{V}2}(\xt,\yt) \Ncal_{\mathrm{V}2}^{\lambda\sigma\sigma'}(\rxyt)\,,
\end{align}
with the color structure
\begin{align}
    \Ccal_{\mathrm{V}2}(\xt,\yt,\zt) = C_F \left[V(\xt) V^\dagger(\yt) -\mathbbm{1} \right]\,,
\end{align}
and the perturbative factor
\begin{align}
    \Ncal_{\mathrm{V}2}^{\lambda\sigma\sigma'}(\rxyt) &= g^2 \mu^{2\varepsilon}\int \frac{\der^{4-\varepsilon} l_1}{(2\pi)^{3-\varepsilon}} \frac{\der^{4-\varepsilon} l_2}{(2\pi)^{3-\varepsilon}}e^{i\ltone\cdot \rxyt}\nonumber\\
    &\times\frac{ (2q^-) \delta(k^- - l_1^-) N_{\mathrm{V}2}^{\lambda\sigma\sigma'}(l_1,l_2) }{\left[l_1^2 + i\epsilon \right] \left[ (l_1-l_2)^2 + i \epsilon \right] \left[ (l_1-l_2 -q)^2 + i \epsilon \right] \left[(l_1-q)^2 + i \epsilon \right] \left[ l_2^2 + i \epsilon \right] }\,,
    \label{eq:dijet-NLO-V1-pert}
\end{align}
where the Dirac numerator is given by
\begin{align}
    N_{\mathrm{V}2}^{\lambda\sigma\sigma'}= \frac{1}{(2q^-)^2} \left[ \bar{u}(k,\sigma) \gamma^-\slashed{l}_1  \gamma^\mu (\slashed{l}_1-\slashed{l}_2)  \slashed{\epsilon}(q,\lambda) (\slashed{l}_1-\slashed{l}_2 - \slashed{q}) \gamma^\nu (\slashed{l}_1 - \slashed{q}) \gamma^- v(p,\sigma') \right] \Pi_{\mu\nu}(l_2) \label{eq:dijet-NLO-V1-dirac-1}.
\end{align}
As usual, the integration over $l_1^-$ is trivial with the delta function constraint resulting in $l_1^- = k^-$.

\subsubsection*{Dirac numerator $N_{\mathrm{V}2}(l_1,l_2)$ in $4-\varepsilon$ dimensions} The Dirac algebra of diagram $\mathrm{V}2$ is by far the most difficult of those encountered thus far. We  used the formula provided in Appendix~\ref{app:dirac} to write the gluon emission and absorption pieces in Eq.\,\eqref{eq:dijet-NLO-V1-dirac-1}. To use the identities in Appendix~\ref{app:dirac}, one first expresses the gluon polarization tensor $\Pi^{\mu\nu}(l_2)$ using Eq.~\eqref{eq:gluon_tensor_decomp}. Then the gluon absorption piece of $N^\lambda_{\mathrm{V}2}(l_1,l_2)$, given by $\gamma^-\slashed l_1\slashed \epsilon(l_2,\bar\lambda)(\slashed l_1-\slashed l_2)$, is simplified using Eq.\,\eqref{eq:gluon_abs_quark_beforeSW} (after the change of variables $l_1\to l_1-l_2$). Similarly, the gluon emission piece $(\slashed l_1-\slashed l_2-\slashed q)\slashed \epsilon^*(q,\bar\lambda)(\slashed l_1-\slashed q)\gamma^-$ is simplified using Eq.\,\eqref{eq:gluon_emi_antiquark_beforeSW}. The resulting expression for the Dirac numerator in Eq.\,\eqref{eq:dijet-NLO-V1-dirac-1} can be decomposed into 
\begin{align}
    N_{\mathrm{V}2} &= N_{\mathrm{V}2,\mathrm{reg}} + l_2^2 N_{\mathrm{V}2,g \mathrm{inst}} + (l_1-l_2)^2 N_{\mathrm{V}2,q \mathrm{inst}}+(l_1-l_2-q)^2 N_{\mathrm{V}2,\bar{q} \mathrm{inst}}\,. \label{eq:dijet-NLO-V1-dirac-2}
\end{align}

This expression, in mathematical terms, is the statement that we  made in Section~\ref{sec:gen-strat} about the correspondence between our calculation and the LCPT approach. In the latter, different instantaneous diagrams appear, either with an instantaneous gluon, quark or antiquark. They correspond to the last three terms in the expression Eq.\,\eqref{eq:dijet-NLO-V1-dirac-2}. The Dirac algebra computation associated with each contribution yields 
\begin{align}
    N_{\mathrm{V}2,\mathrm{reg}}^{\lambda\sigma\sigma'} &= -\frac{4(1+x)(1-y)}{xy} \Lttwox^i  \Lttwoy^k \nonumber\\
    &\times \left\{ \bar{u}(k,\sigma)   \left[\left(1+\frac{x}{2}\right)\delta^{ij} +\frac{x}{2}\omega^{ij}\right] \mathcal{D}^\lambda_{\rm LO}(l_1-l_2)\left[\left(1-\frac{y}{2}\right)\delta^{jk}-\frac{y}{2}\omega^{jk}\right] v(p,\sigma') \right\} \,,\\
    N_{\mathrm{V}2,g \mathrm{inst}}^\lambda &= -\frac{4z_qz_{\bar q}}{z_g^2}\left[ \bar{u}(k,\sigma)\mathcal{D}^\lambda_{\rm LO}(l_1-l_2)v(p,\sigma') \right]\,,\\
    N_{\mathrm{V}2,q \mathrm{inst}}^{\lambda\sigma\sigma'} &= \frac{2(1+x)(1-y)}{y(2q^-)^2} \left[ \bar{u}(k,\sigma)   \gamma^j \gamma^-  \slashed{\epsilon}(q,\lambda)   (\slashed{l}_1-\slashed{l}_2-\slashed{q}) \gamma^- \left(\delta^{ij} + \frac{y}{2}  \gamma^i \gamma^j \right)  v(p,\sigma') \right]  \Lttwoy^i
    \,,\label{eq:dijet-NLO-V1-Nqins}\\
    N_{\mathrm{V}2,\bar{q} \mathrm{inst}}^{\lambda\sigma\sigma'} &= - \frac{2(1+x)(1-y)}{x(2q^-)^2} \left[ \bar{u}(k,\sigma) \left(\delta^{ij} - \frac{x}{2} \gamma^j \gamma^i \right) \gamma^- (\slashed{l}_1-\slashed{l}_2) \slashed{\epsilon}(q,\lambda) \gamma^- \gamma^j v(p,\sigma') \right]  \Lttwox^i \,.
\end{align}
Here $x=l_2^-/(l_1^- - l_2^-)=z_g/(z_q -z_g)$, $\Lttwox = \lttwo -\frac{z_g}{z_q}\ltone$, $y = l_2^-/(q^- -l_1^-+l_2^-)=z_g/(z_{\bar{q}}+z_g)$ and $\Lttwoy = \lttwo +\frac{z_g}{z_{\bar q}}\ltone$.

For a longitudinal photon ($\lambda=0$), it is clear that the quark and antiquark instantaneous Dirac numerators vanish, because of the identity $\gamma^-\slashed\epsilon(q,\lambda=0)=0$. In the transverse polarization case, even though $N^{\lambda=\pm1}_{\mathrm{V}2,q\rm inst}$ is non-zero, the corresponding perturbative factor will vanish after the transverse momentum  integration; this is a consequence of rotational symmetry since the numerator\footnote{When $\lambda=\pm1$, one has $\gamma^-\slashed \epsilon(q,\lambda)(\slashed l_1-\slashed l_2-\slashed q)\gamma^-=2(l_1^- - l_2^- -q^-)\et^{\lambda,k}\gamma^k\gamma^-$ and therefore, the Dirac structure inside the square bracket of Eq.\,\eqref{eq:dijet-NLO-V1-Nqins} does not depend on the transverse momentum vector $\Lttwoy$.} is proportional to $\Lttwoy^i$. Therefore we do not discuss further the quark and antiquark instantaneous Dirac numerators.
Using
\begin{align}
    \mathcal{D}_{\rm LO}^{\lambda=0}(l_1-l_2)&=-(z_q-z_g)(z_{\bar q}+z_g)Q\frac{\gamma^-}{q^-}\,,\\
    \mathcal{D}_{\rm LO}^{\lambda=\pm1}(l_1-l_2)&= \frac{1}{2}\et^{\lambda,l}(\ltone^m-\lttwo^m)((z_{\bar q}-z_q+2z_g)\delta^{lm}+\omega^{lm})\frac{\gamma^-}{q^-}\,,
\end{align}
and the identities Eq.\,\eqref{eq:4-edimension-rel1} and Eq.\,\eqref{eq:4-edimension-rel2}, one finds that the instantaneous gluon and regular Dirac numerators for a longitudinal and transverse photon in $d=2+2-\varepsilon$ dimensions are given by 
\begin{align}
    &N_{\mathrm{V}2,g \mathrm{inst}}^{\lambda=0,\sigma\sigma'}=\frac{4z_qz_{\bar q}Q(z_q-z_g)(z_{\bar q}+z_g)}{z_g^2}\frac{[\bar{u}(k,\sigma)\gamma^-v(p,\sigma')]}{q^-}\,,\\
    &N_{\mathrm{V}2,g \mathrm{inst}}^{\lambda=\pm1,\sigma\sigma'}=\frac{-2z_qz_{\bar q}}{z_g^2}\et^{\lambda,l}(\ltone^m-\lttwo^m)\bar{u}(k,\sigma)\left\{(z_{\bar q}-z_q+2z_g)\delta^{lm}\frac{\gamma^-}{q^-}+\omega^{lm}\frac{\gamma^-}{q^-}\right\}v(p,\sigma')\,,
\end{align}
\begin{align}
    &N_{\mathrm{V}2, \mathrm{reg}}^{\lambda=0,\sigma\sigma'}=\frac{4z_qz_{\bar q}Q(z_q-z_g)(z_{\bar q}+z_g)}{z_g^2}\Lttwox^i\Lttwoy^k \nonumber \\
    &\times\bar{u}(k,\sigma)\left\{\left[1+\frac{z_g}{2(z_q-z_g)}-\frac{z_g}{2(z_{\bar q}+z_g)}-\left(1-\frac{\varepsilon}{2}\right)\frac{z_g^2}{2(z_q-z_g)(z_{\bar q}+z_g)}\right]\delta^{ik}\frac{\gamma^-}{q^-}\right.\nonumber\\
    &\left.+\left[\frac{z_g}{2(z_q-z_g)}-\frac{z_g}{2(z_{\bar q}+z_g)}-\left(1+\frac{\varepsilon}{2}\right)\frac{z_g^2}{2(z_q-z_g)(z_{\bar q}+z_g)}\right]\omega^{ik}\frac{\gamma^-}{q^-}\right\}v(p,\sigma')\,,\\
    &N_{\mathrm{V}2, \mathrm{reg}}^{\lambda=\pm1,\sigma\sigma'}=\frac{-2z_qz_{\bar q}}{z_g^2}\et^{\lambda,l}(\ltone^m-\lttwo^m)\Lttwox^i\Lttwoy^k\bar{u}(k,\sigma)\left\{\left[(z_{\bar q}-z_q+2z_g)\delta^{lm}\frac{\gamma^-}{q^-}+\omega^{lm}\frac{\gamma^-}{q^-}\right]\right.\nonumber\\ 
    &\times\left[\left(1+\frac{z_g}{2(z_q-z_g)}-\frac{z_g}{2(z_{\bar q}+z_g)}-\left(1-\frac{\varepsilon}{2}\right)\frac{z_g^2}{2(z_q-z_g)(z_{\bar q}+z_g)}\right)\delta^{ik}\right.\nonumber\\
    &\left.\left.+\left(\frac{z_g}{2(z_q-z_g)}-\frac{z_g}{2(z_{\bar q}+z_g)}-\left(1+\frac{\varepsilon}{2}\right)\frac{z_g^2}{2(z_q-z_g)(z_{\bar q}+z_g)}\right)\omega^{ik}\right]\right\}v(p,\sigma')+\mathcal{O}(\varepsilon)\label{eq:dijet-NLO-diracnum-V2-trans}\,.
\end{align}
We do not perform the spinor contraction explicitly since we want our formulas to be valid in $d=4-\varepsilon$ dimensions. It is nevertheless straightforward to do so using $\omega^{ij}=-i\epsilon^{ij}\Omega$, the identity Eq.\,\eqref{eq:eigen-pol-epsilon} and the formulas provided in Appendix~\ref{subsub:spin-contraction}. In Eq.\,\eqref{eq:dijet-NLO-diracnum-V2-trans}, the term of order $\mathcal{O}(\varepsilon)$ arises from the commutator of $[\omega^{lm},\omega^{ij}]$ when using \eqref{eq:4-edimension-rel2}. The complete set of $\mathcal{O}(\varepsilon)$ terms for the Dirac structure of $\rm V2$ in the transversely polarized case can be found in \cite{Beuf:2016wdz}.

\subsubsection*{Pole structures of the perturbative factor} Given Eq.\,\eqref{eq:dijet-NLO-V1-dirac-2}, the perturbative factor can be decomposed as 
\begin{equation}
 \Ncal_{\mathrm{V}2}(\rxyt)=\Ncal_{\mathrm{V}2,\rm reg}(\rxyt)+\Ncal_{\mathrm{V}2,g\rm inst }(\rxyt)+\Ncal_{\mathrm{V}2,q\rm inst }(\rxyt)+\Ncal_{\mathrm{V}2,\bar q\rm inst }(\rxyt)\,,
\end{equation}
with 
\begin{align}
 \Ncal_{\mathrm{V}2,\rm reg}(\rxyt)&=\frac{g^2}{2}\int \der z_g \ \mu^{2\varepsilon}\int\frac{\der^{2-\varepsilon}\ltone}{(2\pi)^{2-\varepsilon}}\frac{\der^{2-\varepsilon}\lttwo}{(2\pi)^{2-\varepsilon}}e^{i\ltone\cdot\rxyt}\Ical_{\mathrm{V}2,\rm reg}N_{\mathrm{V}2,\rm reg}\,,\\
 \Ncal_{\mathrm{V}2,\rm g inst}(\rxyt)&=\frac{g^2}{2}\int \der z_g \ \mu^{2\varepsilon}\int\frac{\der^{2-\varepsilon}\ltone}{(2\pi)^{2-\varepsilon}}\frac{\der^{2-\varepsilon}\lttwo}{(2\pi)^{2-\varepsilon}}e^{i\ltone\cdot\rxyt}\Ical_{\mathrm{V}2,g\rm inst}N_{\mathrm{V}2,g\rm inst}\,, \\
  \Ncal_{\mathrm{V}2,\rm qins}(\rxyt)&=\frac{g^2}{2}\int \der z_g \ \mu^{2\varepsilon}\int\frac{\der^{2-\varepsilon}\ltone}{(2\pi)^{2-\varepsilon}}\frac{\der^{2-\varepsilon}\lttwo}{(2\pi)^{2-\varepsilon}}e^{i\ltone\cdot\rxyt}\Ical_{\mathrm{V}2,q \rm inst}N_{\mathrm{V}2,q\rm inst}\,, \\
  \Ncal_{\mathrm{V}2,\rm \bar{q}\mathrm{inst}}(\rxyt)&=\frac{g^2}{2}\int \der z_g \ \mu^{2\varepsilon}\int\frac{\der^{2-\varepsilon}\ltone}{(2\pi)^{2-\varepsilon}}\frac{\der^{2-\varepsilon}\lttwo}{(2\pi)^{2-\varepsilon}}e^{i\ltone\cdot\rxyt}\Ical_{\mathrm{V}2,\rm \bar{q}\mathrm{inst}}N_{\mathrm{V}2,\rm \bar{q}\mathrm{inst}}\,.
\end{align}
The contour integrals $\Ical_{\mathrm{V}2}$ are performed using Cauchy's theorem:
\begin{align}
    \Ical_{\mathrm{V}2,\rm reg}&=\int\frac{\der l_1^+}{(2\pi)}\frac{\der l_2^+}{(2\pi)}\frac{(2q^-)^2}{\left[l_1^2 + i\epsilon \right] \left[ (l_1-l_2)^2 + i \epsilon \right] \left[ (l_1-l_2 -q)^2 + i \epsilon \right] \left[(l_1-q)^2 + i \epsilon \right] \left[ l_2^2 + i \epsilon \right]}\nonumber\\
    &=\frac{(z_q-z_g)}{z_q}\frac{\Theta(z_g)\Theta(z_q-z_g)}{\left[\ltone^2+\bar Q^2\right]\left[\left(\lttwo-\ltone\right)^2+\Delta_{\mathrm{V}2,g\rm inst}^2\right]\left[\Lttwox^2+\Delta_{\mathrm{V}2,\bar{q}\rm{inst}}^2\right]}\nonumber\\
    &+\frac{(z_g+z_{ \bar q})}{z_{\bar q}}\frac{\Theta(-z_g)\Theta(z_g+z_{\bar q})}{\left[\ltone^2+\bar Q^2\right]\left[\left(\lttwo-\ltone\right)^2+\Delta_{\mathrm{V}2,g\rm inst}^2\right]\left[\Lttwoy^2+\Delta_{\mathrm{V}2,q \rm inst}^2\right]}\,,\\
    \Ical_{\mathrm{V}2,g\rm inst}&=\int\frac{\der l_1^+}{(2\pi)}\frac{\der l_2^+}{(2\pi)}\frac{(2q^-)^2}{\left[l_1^2 + i\epsilon \right] \left[ (l_1-l_2)^2 + i \epsilon \right] \left[ (l_1-l_2 -q)^2 + i \epsilon \right] \left[(l_1-q)^2 + i \epsilon \right]} \nonumber\\
    &=-\frac{\Theta(z_q-z_g)\Theta(z_{\bar q}+z_g)}{\left[\ltone^2+\bar Q^2\right]\left[(\lttwo-\ltone)^2+\Delta_{\mathrm{V}2,g\rm inst}^2
    \right]}\,,\\
    \Ical_{\mathrm{V}2,q \rm inst}&=\int\frac{\der l_1^+}{(2\pi)}\frac{\der l_2^+}{(2\pi)}\frac{(2q^-)^2}{\left[l_1^2 + i\epsilon \right] \left[ (l_1-l_2 -q)^2 + i \epsilon \right] \left[(l_1-q)^2 + i \epsilon \right] \left[ l_2^2 + i \epsilon \right]}\nonumber\\
    &=\frac{z_g+z_{\bar q}}{z_{\bar q}(z_g-z_q)}\frac{\Theta(-z_g)\Theta(z_{\bar q}+z_g)}{\left[\ltone^2+\bar Q^2\right]\left[\Lttwoy^2+\Delta_{\mathrm{V}2,q \rm inst}\right]}\,,\\
    \Ical_{\mathrm{V}2,\rm \bar{q}\mathrm{inst}}&=\int\frac{\der l_1^+}{(2\pi)}\frac{\der l_2^+}{(2\pi)}\frac{(2q^-)^2}{\left[l_1^2 + i\epsilon \right] \left[ (l_1-l_2)^2 + i \epsilon \right] \left[(l_1-q)^2 + i \epsilon \right] \left[ l_2^2 + i \epsilon \right]}\nonumber\\
    &=\frac{z_g-z_{q}}{z_{q}(z_g+z_{\bar q})}\frac{\Theta(z_g)\Theta(z_{q}-z_g)}{\left[\ltone^2+\bar Q^2\right]\left[\Lttwox^2+\Delta_{\mathrm{V}2,\bar{q}\mathrm{inst}}\right]}\,.
\end{align}
Here 
\begin{align}
    \Delta^2_{\mathrm{V}2,\rm g inst}&=(z_q-z_g)(z_{\bar q}+z_g)Q^2 \,,\\
   \Delta^2_{\mathrm{V}2,q \rm inst}&=-\frac{z_g(z_{\bar q}+z_g)}{z_{\bar q}^2z_q}\left[\ltone^2+\bar Q^2\right] \,, \\
   \Delta^2_{\mathrm{V}2,\bar q \rm inst}&=\frac{z_g(z_q-z_g)}{z_{\bar q}z_q^2}\left[\ltone^2+\bar Q^2\right]\,.
\end{align}
\subsubsection*{Transverse momentum integration} Combining the results for $\Ical_{\mathrm{V}2,q\rm inst}$ and $\Ical_{\mathrm{V}2,\bar{q}\mathrm{inst}}$ with the corresponding Dirac numerators $N_{\mathrm{V}2, q\rm inst}$ and $N_{\mathrm{V}2,\bar{q}\mathrm{inst}}$, one verifies that the quark and antiquark instantaneous diagrams vanish due to rotational invariance. For the same reason, the instantaneous gluon diagram vanishes when the virtual photon has transverse polarization $\lambda=\pm 1$.
To perform the transverse integrals in the gluon instantaneous contribution for longitudinal photon and in the regular contribution, we have followed the same method as in \cite{Beuf:2016wdz} and employed the Passarino-Veltman tensor reduction \cite{Passarino:1978jh} of the $\lttwo$ integrals. After the remaining integration over $z_g$, one gets \eqref{eq:dijet-NLO-V1-final-long} and \eqref{eq:dijet-NLO-V1-final-trans}.

\section{Details of the computation of diagram $\mathrm{V}3$}
\label{appp:V3}
From the Feynman diagram in Fig.\,\ref{fig:NLO-dijet-V23}-right, one readily writes the amplitude
\begin{align}
    S_{\mathrm{V}3}^{\lambda\sigma\sigma'}  &=\int \frac{\der^{4} l_1}{(2\pi)^{4}} \frac{\der^{4} l_2}{(2\pi)^{4}} \left[\bar{u}(k,\sigma) (ig \gamma_\mu t^a) S^0(k-l_2) \Tcal^q(k-l_2,l_1) S^0(l_1) \right. \nonumber \\
    &\left. (-i e e_f\slashed{\epsilon}(q,\lambda)) S^0(l_1-q)  \Tcal^q(l_1-q,-p-l_2) S^0(-p-l_2) (ig \gamma_\nu t^b) v(p,\sigma') \right] G^{0,\mu\nu}_{ab}(l_2) \,.
\end{align}
Subtracting the noninteracting piece, and factoring out the overall delta function $2\pi \delta(q^- - k^- - p^-)$, we find
\begin{align}
    \Mcal^{\lambda\sigma\sigma'}_{\mathrm{V}3}=\frac{ee_fq^-}{\pi}\int\der^{2}\xt\der^{2}\yt \,e^{-i\kt \cdot \xt -i\pt \cdot \yt}\Ccal_{\mathrm{V}3}(\xt,\yt)\Ncal^{\lambda\sigma\sigma'}_{\mathrm{V}3}(\rxyt) \,,
\end{align}
with color structure
\begin{equation}
    \Ccal_{\mathrm{V}3}(\xt,\yt)=t^aV(\xt)V^\dagger(\yt)t_a-C_F\mathbbm{1} \,,
\end{equation}
and perturbative factor
\begin{align}
    \Ncal^{\lambda\sigma\sigma'}_{\mathrm{V}3}(\rxyt)=g^2&\int\frac{\der^{4}l_1}{(2\pi)^{3}}\frac{\der^{4}l_2}{(2\pi)^{3}}e^{i(\ltone+\lttwo) \cdot \rxyt}\delta(k^- - l_2^--l_1^-)\nonumber\\
    &\times\frac{(2q^-)N^{\lambda\sigma\sigma'}_{\mathrm{V}3}(l_1,l_2)}{(l_1^2+i\epsilon)((l_1-q)^2+i\epsilon)((k-l_2)^2+i\epsilon)((l_2+p)^2+i\epsilon)(l_2^2+i\epsilon)} \,.
\end{align}
The Dirac structure of this diagrams reads
\begin{align}
 N_{\mathrm{V}3}^{\lambda\sigma\sigma'}&=\frac{1}{(2q^-)^2}\left[ \bar{u}(k,\sigma) \gamma_\mu (\slashed{k} - \slashed{l}_2) \gamma^- \slashed{l}_1 \slashed{\epsilon}(q,\lambda) (\slashed{l}_1 - \slashed{q}) \gamma^- (-\slashed{p}-\slashed{l}_2) \gamma_\nu v(p,\sigma') \right] \Pi^{\mu\nu}(l_2) \,.
\end{align}

\subsubsection*{Dirac structure}
Using the identities \eqref{eq:gluon_tensor_decomp}, \eqref{eq:gluon_abs_quark_afterSW}, and \eqref{eq:gluon_emi_antiquark_afterSW}, one decomposes the Dirac numerator as
\begin{align}
    N_{\mathrm{V}3}&=N_{\mathrm{V}3,\rm reg}+l_2^2N_{\mathrm{V}3,g \rm inst} \,,
    \label{eq:dijet-NLO-V2-dirac_dec}
\end{align}
with
\begin{align}
    N^{\lambda\sigma\sigma'}_{\mathrm{V}3,\rm reg} &= \frac{-4\Lttwox^i \Lttwoy^k}{xy} \nonumber \\
    &\times \left\{ \bar{u}(k,\sigma) \left[\left(1-\frac{x}{2}\right)\delta^{ij}-i\frac{x}{2}\epsilon^{ij}\Omega\right]\mathcal{D}_{\rm LO}(l_1)\left[\left(1+\frac{y}{2}\right)\delta^{jk}+i\frac{y}{2}\epsilon^{jk}\Omega\right]  v(p,\sigma') \right\}\,,\\
    N^{\lambda\sigma\sigma'}_{\mathrm{V}3,g\rm inst} &=\frac{-4(1-x)(1+y)}{xy}\left[\bar{u}(k,\sigma)\mathcal{D}_{\rm LO}(l_1)v(p,\sigma')\right] \,,
\end{align}
with $x=l_2^-/k^-$, $y=l_2^-/p^-$, $\Lttwox=\lttwo-x\kt$, $\Lttwoy=\lttwo-y\pt$. For a longitudinally polarized photon, with $\slashed{\epsilon}(q,\lambda=0) =\frac{Q}{q^-} \gamma^-$, we find for the gluon instantaneous term
\begin{align}
    N_{\mathrm{V}3,g\mathrm{inst}}^{\lambda=0,\sigma\sigma'} &=\frac{8Q(z_g-z_q)^2(z_g+z_{\bar q})^2}{z_g^2} (z_qz_{\bar q})^{1/2}\delta^{\sigma,-\sigma'} \,,
\end{align}
and for the regular term, using Eq~ \eqref{eq:dijet-NLO-useful_id_1},
\begin{align}
    N^{\lambda=0,\sigma\sigma'}_{\mathrm{V}3,\textrm{reg}}&=\frac{-8(z_qz_{\bar q})^{3/2}Q(z_g-z_q)(z_g+z_{\bar q})}{z_g^2}\Lttwox^i\Lttwoy^k\nonumber\\
    &\times\left\{\left[1-\frac{z_g}{2z_q}+\frac{z_g}{2z_{\bar q}}-\frac{z_g^2}{2z_qz_{\bar q}}\right]\delta^{ik}+i\sigma\left[\frac{z_g}{2z_q}-\frac{z_g}{2z_{\bar q}}+\frac{z_g^2}{2z_qz_{\bar q}}\right]\epsilon^{ik}\right\} \,.
\end{align}
The transverse polarized case can be worked out in a similar fashion. The gluon instantaneous Dirac structure reads 
\begin{align}
    N^{\lambda = \pm 1,\sigma\sigma'}_{\mathrm{V}3,g\mathrm{inst}}&=\frac{8(z_g-z_q)(z_g+z_{\bar q})}{z_g^2}(\ltone\cdot\et^\lambda) (z_q z_{\bar{q}})^{1/2}\Gamma^{\sigma,\lambda}_{\gamma_{\rm T}^*\to q \bar q}(z_q-z_g,z_{\bar q}+z_g)\delta^{\sigma,-\sigma'} \,,
\end{align}
while the regular Dirac structure can be written as
\begin{align}
    N^{\lambda = \pm 1,\sigma\sigma'}_{\mathrm{V}3,\mathrm{reg}}&  =-\frac{ 8 (z_qz_{\bar q})^{3/2}}{z_g^2} \Lttwox^i\Lttwoy^k(\ltone \cdot \et^{\lambda})\Gamma^{\sigma,\lambda}_{\gamma_{\rm T}^*\to q \bar q}(z_q-z_g,z_{\bar q}+z_g)\delta^{\sigma,-\sigma'}\nonumber\\
    &\times\left\{\left[1-\frac{z_g}{2z_q}+\frac{z_g}{2z_{\bar q}}-\frac{z_g^2}{2z_qz_{\bar q}}\right]\delta^{ik}+i\sigma\left[\frac{z_g}{2z_q}-\frac{z_g}{2z_{\bar q}}+\frac{z_g^2}{2z_qz_{\bar q}}\right]\epsilon^{ik}\right\} \,.
\end{align}

\subsubsection*{Pole structure of the instantaneous and regular pieces}

Using the decomposition in Eq.~\eqref{eq:dijet-NLO-V2-dirac_dec}, we can express the perturbative factor as 
\begin{align}
\Ncal_{\mathrm{V}3}(\rxyt)&=\Ncal_{\mathrm{V}3,\rm reg}(\rxyt)+\Ncal_{\mathrm{V}3,g\rm inst}(\rxyt) \,,
\end{align}
with
\begin{align}
    \Ncal_{\mathrm{V}3,\rm reg}(\rxyt)&=\frac{g^2}{2}\int\der z_g \ \int\frac{\der^{2}\ltone}{(2\pi)^{2}}\frac{\der^{2}\lttwo}{(2\pi)^{2}}e^{i(\ltone+\lttwo)\cdot\rxyt}\Ical_{\mathrm{V}3,\rm reg}N_{\mathrm{V}3,\rm reg} \,,\\
    \Ncal_{\mathrm{V}3,g\rm inst}(\rxyt)&=\frac{g^2}{2}\int\der z_g \ \int\frac{\der^{2}\ltone}{(2\pi)^{2}}\frac{\der^{2}\lttwo}{(2\pi)^{2}}e^{i(\ltone+\lttwo)\cdot\rxyt}\Ical_{\mathrm{V}3,g\rm inst}N_{\mathrm{V}3,g\rm inst} \,,
\end{align}
with the pole structure:
\begin{align}
    \Ical_{\mathrm{V}3,\rm reg}&=\int\frac{\der l_1^+}{(2\pi)}\frac{\der l_2^+}{(2\pi)}\frac{(2q^-)^2}{(l_1-q)^2l_1^2(k-l_2)^2(p+l_2)^2l_2^2} \,, \\
    \Ical_{\mathrm{V}3,g\rm inst}&=\int\frac{\der l_1^+}{(2\pi)}\frac{\der l_2^+}{(2\pi)}\frac{(2q^-)^2}{(l_1-q)^2l_1^2(k-l_2)^2(p+l_2)^2} \,,
\end{align}
where we have again omitted the $+i\epsilon$ prescription for the propagators.
The computation of these integrals is straightforward using Cauchy's theorem. After a little bit of algebra, we write them in the form:
\begin{align}
    \Ical_{\mathrm{V}3,\rm reg}&=\Ical^{>}_{\mathrm{V}3,\rm reg}\Theta(z_g)+\Ical^{<}_{\mathrm{V}3,\rm reg}\Theta(-z_g)\,, \\
    \Ical^{>}_{\mathrm{V}3,\rm reg}&=\frac{-(z_g-z_q)}{z_q}\frac{1}{\ltone^2+\bar{Q}_{\mathrm{V3}}^2}\frac{\Theta(z_q-z_g)}{\Lttwox^2\left[\left(\Lttwox-\left(1-\frac{z_g}{z_q}\right)\Pt\right)^2-\Delta_{\mathrm{V}3}^2 - i \epsilon\right]}\,, \label{eq:dijet-NLO-V2-reg-polestructure}
    \\
    \Ical^{<}_{\mathrm{V}3,\rm reg}&=\frac{(z_g+z_{\bar q})}{z_{\bar q}}\frac{1}{\ltone^2+\bar{Q}_{\mathrm{V3}}^2}\frac{\Theta(z_{\bar q}+z_g)}{\Lttwoy^2\left[\left(\Lttwoy-\left(1+\frac{z_g}{z_{\bar q}}\right)\Pt\right)^2-\Delta_{\mathrm{V}3}^2 - i \epsilon \right]}\,, 
\end{align}
with
\begin{align}
    \bar{Q}_{\mathrm{V3}}^2 & = (z_q-z_g)(z_{\bar{q}}+z_g) Q^2 \,, \nonumber \\
    \Pt&=z_{\bar q}\kt-z_q\pt \,, \\
    \Delta_{\mathrm{V}3}^2&=\left(1-\frac{z_g}{z_q}\right)\left(1+\frac{z_g}{z_{\bar q}}\right)\Pt^2\,.
\end{align}
The two contributions with either $z_q\ge z_g \ge0$ or $-z_{\bar q}\le z_g\le0$ are in one-to-one correspondence with the two possible time ordering of the exchanged gluon in lightcone perturbation theory.
Similarly, the contour integration of the instantaneous term gives
\begin{align}
    \Ical_{\mathrm{V}3,g\rm inst}
    &=-\frac{1}{\ltone^2+\bar{Q}_{\mathrm{V3}}^2}\frac{\Theta(z_q-z_g)\Theta(z_{\bar q}+z_g)}{\left[\left(\Lttwox-\left(1-\frac{z_g}{z_q}\right)\Pt\right)^2-\Delta_{\mathrm{V}3}^2 - i \epsilon \right]} \,.
\end{align}
It is convenient to decompose the instantaneous term into two pieces depending on the sign of $z_g$, as for the regular contribution,
\begin{align}
    \Ical_{\mathrm{V}3,g \rm inst}&=\Ical^{>}_{\mathrm{V}3,\rm gins} \Theta(z_g)+\Ical^{<}_{\mathrm{V}3,g\rm inst} \Theta(-z_g) \,, \\
    \Ical^{>}_{\mathrm{V}3,g\rm inst}&=-\frac{1}{\ltone^2+\bar{Q}_{\mathrm{V3}}^2}\frac{\Theta(z_q-z_g)}{\left[ \left(\Lttwox-\left(1-\frac{z_g}{z_q}\right)\Pt\right)^2-\Delta_{\mathrm{V}3}^2 - i \epsilon \right]}\,, \\
    \Ical^{<}_{\mathrm{V}3,g\rm inst}&=-\frac{1}{\ltone^2+\bar{Q}_{\mathrm{V3}}^2}\frac{\Theta(z_{\bar q}+z_g)}{\left[\left(\Lttwoy-\left(1+\frac{z_g}{z_{\bar q}}\right)\Pt\right)^2-\Delta_{\mathrm{V}3}^2 - i \epsilon \right]} \,.
\end{align}

\subsubsection*{Transverse momentum integration}

We consider only the $>$ component of the perturbative factor since one can relate the $<$ piece by $q \leftrightarrow \bar q$ interchange. One should first notice that given the topology of the free vertex correction after shock wave, the integrals over $\ltone$ and $\lttwo$ decouple. In the $>$ component, it is convenient to do the change of variable $\lttwo\to\Lttwox$ so that the exponential phase becomes
\begin{equation}
    e^{i(\ltone+\lttwo)\cdot \rxyt}=e^{i\frac{z_g}{z_q}\kt \cdot \rxyt}e^{i\ltone\cdot\rxyt}e^{i\Lttwox\cdot\rxyt} \,.
\end{equation}
This change of variable enables to simplify the Dirac numerator $N^{\lambda}_{\mathrm{V}2,\rm reg}$ as well, since $\Lttwoy=\Lttwox+\frac{z_g}{z_qz_{\bar q}}\Pt$. Indeed, in the cross-product term $\Lttwox^i\Lttwoy^k\epsilon^{ik}$, only $\Lttwox\times\Pt$ contributes. On the other hand, the dot product term $\Lttwox^i\Lttwoy^k\delta^{ik}$ gives two contributions, one proportional to $\Lttwox^2$ and the other proportional to $\Lttwox\cdot \Pt$. In the former, the square cancels against the same square in the denominator of Eq.\,\eqref{eq:dijet-NLO-V2-reg-polestructure} so that the transverse momentum integral takes the form
\begin{align}
    \mathcal{K}(\rxyt, \Kt, \Delta_{\rm{V}3}) = \int \frac{\der^2 \Lttwox}{(2\pi)^2} \frac{e^{i \Lttwox \cdot \rxyt}}{\left[\left(\Lttwox-\Kt \right)^2-\Delta_{\mathrm{V}3}^2 - i \epsilon \right]} \,,
    \label{eq:BesselK_imagarg}
\end{align}
where $\Kt = \left(1-\frac{z_g}{z_q}\right)\Pt$.

Some care must be exercised to compute the integral in Eq.\,\eqref{eq:BesselK_imagarg}. First we shift the integration variable $\Lttwox \to \Lttwox + \Kt $, and perform the angular integral
\begin{align}
    \mathcal{K}(\rxyt, \Kt, \Delta_{\rm{V}3}) =\frac{e^{i \Kt \cdot \rxyt}}{2\pi} \int_0^\infty  \der L_{2x,\perp}  \frac{L_{2x,\perp}}{\left[L_{2x,\perp}^2 -\Delta_{\mathrm{V}3}^2 - i \epsilon \right]} J_0(L_{2x,\perp} r_{xy}) \,.
\end{align}
The resulting integral over $L_{2x,\perp}$ can be evaluated with the help of the identity
\begin{align}
    \frac{1}{x \pm i \epsilon} &= \mathrm{PV} \frac{1}{x} \mp i \pi \delta(x) \,,
\end{align}
where $\mathrm{PV}$ denotes Cauchy's principal value\footnote{For a real integral Cauchy's principal value is defined as $\mathrm{PV} \int_0^\infty \der x \ \frac{f(x)}{x-a} = \lim_{\epsilon \to 0} \left[ \int_0^{a-\epsilon} \der x \ \frac{f(x)}{x-a} + \int_{a+\epsilon}^\infty \der x \ \frac{f(x)}{x-a} \right]$, where $a>0$.}. The ensuing integrals are then
\begin{align}
    \mathrm{PV} \int_0^\infty  \der L_{2x,\perp}  \frac{L_{2x,\perp}}{\left[L_{2x,\perp}^2 -\Delta_{\mathrm{V}3}^2\right]} J_0(L_{2x,\perp} r_{xy}) & =  - \frac{\pi}{2} Y_0(\Delta_{\rm{V}3} r_{xy}) \,, \\
    \int_0^\infty  \der L_{2x,\perp} \delta(L_{2x,\perp}^2 - \Delta_{\mathrm{V}3}^2) L_{2x,\perp} J_0(L_{2x,\perp} r_{xy}) & =  \frac{1}{2} J_0(\Delta_{\rm{V}3} r_{xy}) \,,
\end{align}
with $\Delta_{\rm{V}3}\ge 0$.
Thus, we find
\begin{align}
    \mathcal{K}(\rxyt, \Kt, \Delta_{\rm{V}3}) &= \frac{e^{i \Kt \cdot \rxyt}}{2\pi} \left[ - \frac{\pi}{2} Y_0(\Delta_{\rm{V}3} r_{xy}) + \frac{i\pi}{2} J_0(\Delta_{\rm{V}3} r_{xy}) \right] \nonumber \\
    &= \frac{e^{i \Kt \cdot \rxyt}}{2\pi} K_0 (-i \Delta_{\mathrm{V}3} r_{xy}) \,,
\end{align}
where the second equality defines the modified Bessel function $K_0$ for imaginary values.

For the contributions arising from $\Lttwox\cdot\Pt$ and $\Lttwox\times\Pt$ terms we have not found similar expressions in terms of standard functions.
In terms of the functions $\Jcal_{\odot}$ and $\Jcal_{\otimes}$ defined respectively in Eq.\,\eqref{eq:Jdot-def} and Eq.\,\eqref{eq:Jtimes-def}, and computed using Feynman parametrization in Appendix~\ref{app:jdotjcross}, the regular perturbative factor for a longitudinal photon reads
\begin{align}
    \Ncal^{>,\lambda=0,\sigma\sigma'}_{\mathrm{V}3,\rm reg}&=\frac{\alpha_s}{\pi}\int_0^{z_q}\frac{\der z_g}{z_g} \ (-2)(z_qz_{\bar q})^{3/2}\delta^{\sigma,-\sigma'}QK_0(\bar{Q}_{\mathrm{V3}} r_{xy})\left(1-\frac{z_g}{z_q}\right)\left(1+\frac{z_g}{z_{\bar q}}\right)\nonumber\\
    \times &\left\{\frac{2z_{\bar q}(z_g-z_q)}{z_g}\left[1-\frac{z_g}{2z_q}+\frac{z_g}{2z_{\bar q}}-\frac{z_g^2}{2z_qz_{\bar q}}\right]e^{i(\Pt+z_g\vect{\Delta})\cdot\rxyt}K_0(-i\Delta_{\mathrm{V}3}r_{xy})\right.\nonumber\\
    &-\left[1-\frac{z_g}{2z_q}+\frac{z_g}{2z_{\bar q}}-\frac{z_g^2}{2z_qz_{\bar q}}\right]e^{i\frac{z_g}{z_q}\kt \cdot \rxyt}\Jcal_{\odot}\left(\rxyt,\left(1-\frac{z_g}{z_q}\right)\Pt,\Delta_{\mathrm{V}3}\right)\nonumber\\
    &\left.+\sigma \left[\frac{z_g}{z_q}-\frac{z_g}{z_{\bar q}}+\frac{z_g^2}{z_qz_{\bar q}}\right]e^{i\frac{z_g}{z_q}\kt \cdot \rxyt}\Jcal_{\otimes}\left(\rxyt,\left(1-\frac{z_g}{z_q}\right)\Pt,\Delta_{\mathrm{V}3}\right)\right\} \,, \label{eq:dijet-NLO-virtual-V2-reg-pert}
\end{align}
with $\vect{\Delta}=\kt+\pt$.
In the first term of this expression, one notices a power divergence in the $z_g$ integral, due to an overall $1/z_g^2$ factor. In fact, this power divergence cancels against a similar divergence in the instantaneous piece $\Ncal^{>}_{\mathrm{V}3,g\rm inst}$, which reads
\begin{align}
    \Ncal^{>,\lambda=0,\sigma\sigma'}_{\mathrm{V}3,g\rm inst} &=\frac{\alpha_s}{\pi}\int_0^{z_q}\frac{\der z_g}{z_g} \  (-2)(z_qz_{\bar q})^{3/2}\delta^{\sigma,-\sigma'}QK_0(\bar{Q}_{\mathrm{V3}} r_{xy})\left(1-\frac{z_g}{z_q}\right)\left(1+\frac{z_g}{z_{\bar q}}\right)\nonumber\\
    &\times\left[\frac{2(z_q-z_g)(z_g+z_{\bar q})}{z_g}\right] e^{i(\Pt+z_g\vect{\Delta})\cdot\rxyt}K_0(-i\Delta_{\mathrm{V}3}r_{xy}) \,.
\end{align}
Combining these two expressions together, and the $<$ component related by $q\leftrightarrow\bar q$ interchange, one finds Eq.\,\eqref{eq:dijet-NLO-virtual-V2-pert-final-spin}. The transversely polarized  photon case is worked out in a similar fashion, leading to Eq.~\eqref{eq:dijet-NLO-virtual-V2-pert-final-trans-spin}.

\bibliographystyle{utcaps}
\bibliography{nlo-dijet-ref}

\end{document}